\documentclass[11pt,twoside]{book}

\pdfoutput=1 

\usepackage[ngerman,english]{babel}
\usepackage{graphicx}
\usepackage{amsmath}
\usepackage{amsthm}
\usepackage{amsfonts}
\usepackage{amssymb}
\usepackage{multicol}
\usepackage{tikz}
\usepackage{calc}
\usepackage{hyperref}
\usepackage{fancyhdr}
\usepackage[a4paper, bottom=3cm]{geometry}
\newcommand{\p}{\cdot}

\newtheorem{thm}{Theorem}[section]

\newtheorem{lem}{Lemma}[section]
\newtheorem{cor}{Corollary}[section]
\theoremstyle{definition}

\theoremstyle{remark}
\newtheorem{rmk}{Remark}[section]

\begin{document}
\frontmatter
\thispagestyle{empty}
\begin{center}
{\huge Microscopic derivation of Vlasov equations with singular potentials}
\end{center}
\vspace*{1cm}
 \begin{center}
{\large Dissertation\\ an der Fakult\"at f\"ur Mathematik, Informatik und Statistik\\ der Ludwig-Maximilians-Universit\"at M\"unchen}
\end{center}
\vspace*{\stretch{1}}
\begin{center}
{\large Eingereicht von\\ [0,2cm]
Phillip Gra{\ss}\\ [0,2cm]
am 07.11.2018}
\end{center}
\newpage
\thispagestyle{empty}
\vspace*{\stretch{1}}
\begin{flushleft}
\large {\bf 1. Gutachter:} Prof. Dr. Peter Pickl \\[1mm]
\large {\bf 2. Gutachter:} Prof. Dr. Dr. h.c. Herbert Spohn \\[1mm]
\large {\bf 3. Gutachter:} Prof. Dr. Pierre-Emmanuel Jabin  \\[3mm]
\large {\bf Tag der m\"undlichen Pr\"ufung:} 04.01.2019
\end{flushleft}

\newpage
\noindent
{\Large \bf Abstract}\\\\
\noindent The Vlasov-Poisson equation is a classical example of an effective equation which shall describe the coarse-grained time evolution of a system consisting of a large number of particles which interact by Coulomb or Newton's gravitational force. Although major progress concerning a rigorous justification of such an approach was made recently, there are still substantial steps necessary to obtain a completely convincing result. The main goal of this work is to yield further progress in this regard. \\ To this end, we consider on the one hand $N$-dependent forces $f^N$ (where $N$ shall denote the particle number) which converge pointwise to Coulomb or alternatively Newton`s gravitational force. More precisely, the interaction fulfills $f^N(q)=\pm\frac{q}{|q|^3}$ for $|q|>N^{-\frac{7}{18}+\epsilon}$ and has a cut-off at $|q|= N^{-\frac{7}{18}+\epsilon}$ where $\epsilon>0$ can be chosen arbitrarily small. We prove that under certain assumptions on the initial density $k_0$ the characteristics of Vlasov equation provide typically a very good approximation of the $N$-particle trajectories if their initial positions are i.i.d. with respect to density $k_0$. Interestingly, the cut-off diameter is of smaller order than the average distance of a particle to its nearest neighbor. Nevertheless, the cut-off is essential for the success of the applied approach and thus we consider additionally less singular forces scaling like $|f(q)|=\frac{1}{|q|^\alpha}$ where $\alpha\in (1,\frac{4}{3}]$. In this case we are able to show a corresponding result even without any regularization. Although such forces are distinctly less interesting than for instance Coulomb interaction from a physical perspective, the introduced ideas for dealing with forces where even the related potential is singular might still be helpful for attaining comparable results for the arguably most interesting case $\alpha=2$.

\newpage
\noindent
{\Large \bf Zusammenfassung}\\\\
\begin{otherlanguage}{ngerman}
\noindent Die Vlasov-Poisson Gleichung ist wohl eine der bekanntesten effektiven Gleichungen, deren L\"osungen die zeitliche Entwicklung von Vielteilchensystemen von einer makroskopischen Perspektive beschreiben sollen, wobei die betrachteten Teilchen mittels Coulombkraft oder Newton`s Gravitationskraft miteinander wechselwirken. Obwohl in der letzten Zeit gro{\ss}e Fortschritte erzielt wurden, die Anwendbarkeit einer solchen Beschreibung rigoros zu begr\"unden, bleiben immer noch bedeutende L\"ucken zu einem vollst\"andig zufriedenstellenden Ergebnis bestehen. Das Kernziel dieser Arbeit ist es, bereits vorhandene Resultate in verschiedener Hinsicht auszubauen. \\ Zu diesem Zweck betrachten wir einerseits Zweiteilchenwechselwirkungen $f^N$, die von der Teilchenzahl $N$ abh\"angen und punktweise gegen die Coulombkraft oder gegen Newton's Gravitationskraft konvergieren. Genauer gesagt, besitzen die Kr\"afte die Form $f^N(q)=\pm\frac{q}{|q|^3}$, falls $|q|>N^{-\frac{7}{18}+\epsilon},\  \epsilon>0$ und sind in einem Bereich der Gr\"o{\ss}enordnung $|q|= N^{-\frac{7}{18}+\epsilon}$ um die Singularit\"at in geeigneter Weise regularisiert. Wir werden zeigen, dass unter gewissen Annahmen an die Anfangsdichte $k_0$ die Charakteristiken der Vlasov-Gleichung die Trajektorien der Teilchen in sehr guter N\"aherung beschreiben, falls ihre Startpositionen unabh\"angig und identisch verteilt sind bez\"uglich der Dichte $k_0$. Interessanterweise ist die Gr\"o{\ss}enordnung des Bereichs, in dem die Wechselwirkung regularisiert wird, bedeutend kleiner als der durchschnittliche Abstand eines Teilchens zu seinem n\"achsten Nachbarn in der betrachteten Situation. Leider ist die Regularisierung in dem beschriebenen Fall trotzdem essentiell f\"ur den Erfolg der verwendeten Methode. Deshalb betrachten wir zus\"atzlich weniger singul\"are Kr\"afte der Form $|f(q)|=\frac{1}{|q|^\alpha}$ f\"ur $|q|>0$ sowie $\alpha\in (1,\frac{4}{3}]$ und zeigen, dass in diesem Fall entsprechende Resultate auch ohne Regularisierung bewiesen werden k\"onnen. Obwohl solche Kr\"afte von einem physikalischen Standpunkt aus gesehen weitaus weniger interessant erscheinen als z.B. Coulomb-Wechselwirkung, k\"onnte das pr\"asentierte Vorgehen trotzdem hilfreich dabei sein, letztendlich vergleichbare Resultate f\"ur den wohl relevantesten Fall $\alpha=2$ zu erzielen.
\end{otherlanguage}
\newpage
\noindent
{\Large \bf Acknowledgements}\\\\
First of all, I would like to express my gratitude to my advisor Prof. Dr. Peter Pickl for the great support in recent years. I would like to thank him for confiding me this fascinating and significant topic for my Ph.D. thesis. In particular, I am grateful for his exceptionally patient guidance which allows independent working assisted by brilliant creative input whenever necessary and additionally never evokes any feeling of pressure.\\
Moreover, I would like to thank Prof. Dr. Detlef D\"urr for the remarkable commitment to all of his students which also constitutes the foundation of the pleasant 
research conditions in our working group.\\
I am grateful to Dr. Dirk-Andr\'{e} Deckert for helpful conversations and advice.\\
Furthermore, I am very honored that Prof. Dr. Herbert Spohn and Prof. Dr. Pierre--Emmanuel Jabin agreed to co-supervise
this thesis.\\
I would like to thank Dr. Nikolai Leopold for his company and for always keeping the motivation high. It was a pleasure to share the office with great colleagues like Dr. Max Jeblick and Dr. David Mitrouskas and I am grateful to all remaining nice people of our working group for the good times in previous years. \\
Finally, a very special thanks is due to my family for their invaluable support.

\newpage

\tableofcontents

\newpage
\newpage
\mainmatter
\setcounter{page}{1}
\chapter{Introduction}
\section{Introduction to the basic objective}
The essential aim of this work is to provide a mathematical rigorous justification for the application of Vlasov equation to describe the time evolution of certain microscopic systems from a coarse-grained perspective. In the following sections we will explain what exactly is meant by the last statement and give additionally a first impression to which extend we can meet these expectations.\\
Consider a system of $N$ identical particles (e.g.) in $3$-dimensional space evolving by Newtonian dynamics. Moreover, let $f$ be some pair interaction, then the related particle trajectories are determined by the following system of coupled differential equations
\begin{align}
i\in \{1,...,N\},\ \begin{cases}
& {\dot{Q}}_i=\frac{P_i}{m}\\
& \dot{P}_i=\sum_{j\neq i}f(Q_i-Q_j) 
\end{cases} \label{gen.micr.sys.}
\end{align}
where $m>0$ shall denote the particle mass. The particle numbers of real systems are in many cases huge so that solving these equations is an extremely complicated or even unfeasible problem. However, by heuristic arguments it is often possible to find an effective equation (resp. a PDE) so that solutions to this equation determine a continuous approximation of the particle distribution for a certain time span (where usually the position and/or time variables must be rescaled in a suitable way first). And while previously we argued that the huge number of particles causes problems, the converse is true for this purpose since the approximation typically improves as the particle number increases. In contrast to heuristic derivations, our aim is to show with mathematical rigor that such an approach is justified for one of the most classical examples which is the Vlasov-Poisson system. In this case the pair interaction is given by the Coulomb force or Newton's gravitational force $f(q)= a\frac{q}{|q|^3},\ a\in \{-1,1\}$ for $|q|>0$ (where in our units the coupling constant which includes the remaining physical quantities is set equal to $1$) and the related effective equation is the Vlasov equation
\begin{align}
\frac{d}{\partial t}k_t+ \frac{p}{m}\cdot \nabla_q k_t+f*(\int_{\mathbb{R}^3}k_t(\cdot,p)d^3p)\cdot \nabla_p k_t=0 \label{def.unreg.Vlas.}
\end{align}
where `$*$' denotes the convolution and $m>0$ is a parameter (which, however, in our considerations will always be equal to $1$). In case of these special interactions the equation is usually referred to as Vlasov-Poisson equation. The precise sense in which the solutions to this equation shall be related to the particle distribution will be introduced in section \ref{Prop.Chaos}.\\ The topic of effective equations in general as well as the justification of their respective application is a widespread research field and in particular issues concerning Vlasov equation (being one of the classical examples) are well documented in literature. Hence, we limit the introduction to aspects which are crucial or helpful for understanding the approach which is applied in the thesis and refer the reader to some very enlightening sources in which basically all relevant aspects concerning the current topic are discussed. In this regard, the lecture notes of Jabin \cite{Jabin lecture notes} are very noteworthy. In addition to an introduction to the conceptual framework, also insight into (classical and current) techniques is provided which can be applied for the derivation of Vlasov equation and other effective equations. Furthermore, the current work is at least partly build on results of the thesis of Lazarovici \cite{thesisDustin} and essentially all basic issues which will be of importance are discussed there to a wide extent. A slightly broader insight into the topic is determined by the advisable course of Golse \cite{Golse lecture notes}. Finally, for a general overview about issues concerning the effective description of large particle systems the famous book of Spohn \cite{SpohnBook} is a very instructive source.\\
We continue by introducing the essential systems and equations which will be relevant in this work. \\ 
\section{The microscopic system}
As mentioned at the beginning, we always consider systems of $N$ identical particles in $3$-dimensional space which evolve according to Newton's laws. The trajectories of the particles shall be given by the following system of differential equations 
\begin{align}
 i\in \{1,...,N\},\ \begin{cases}
& {\dot{Q}}_i=V_i\\
& \dot{V}_i=\frac{1}{N}\sum_{j\neq i}f(Q_i-Q_j) 
\end{cases} \label{Def.micro.sys.}
\end{align} 
where we consider (homogeneous) force kernels of the form stated in \eqref{def.lim.force} and a related regularized version (see \eqref{def.force}). In our units the mass of each particle shall be equal to $1$ so that velocity and momentum coincide (which explains the equality in the first line of \eqref{Def.micro.sys.}). Moreover, as indicated in the previous section, usually a certain rescaling of time, positions and/or momenta is necessary so that the regarded system shows interesting behavior which approximately can be described by a related effective equation. The prefactor $\frac{1}{N}$ appearing in equation \eqref{Def.micro.sys.} constitutes such a scaling factor and the applied scaling is generally called `mean-field scaling' in literature. Taking a closer look at equations \eqref{Def.micro.sys.} the factor $\frac{1}{N}$ seems to be the most obvious choice, where an interesting behavior of the system may be expected, which can be captured by an effective equation. This is, for instance, also noted in \cite{Jabin lecture notes}. If the particle number $N$ increases, then this prefactor basically compensates the rising number of addends appearing in the force term so that in total the force on the average particle is expected to keep of order 1. If, on the other hand, the prefactor takes the form $\frac{1}{N^{\beta}}$ where $\beta>1$, then for large $N$ the force term is expected to become negligibly small for most particles which in turn leads to an almost free time evolution and thus to a rather uninteresting behavior. The situation is more complicated for a prefactor of the form $\frac{1}{N^{\beta}}$ where $\beta<1$. However, if the distribution of the particles is such that order $N$ particles `contribute' in a relevant way to the total force acting on an average particle, then this force will likely become arbitrarily large as $N$ increases which results in a correspondingly big acceleration of the particles, so that in this case no meaningful behavior should be expected in the limit. \\
 Let for $X_i=(^1X_i,{^2X_i})\in \mathbb{R}^6$ the first component $^1X_i\in \mathbb{R}^3$ denote the position of the $i$-th particle in space and $^2X_i\in \mathbb{R}^3$ its velocity. Then $X:=(X_1,...,X_N)\in \mathbb{R}^{6N}$ shall denote the whole $N$-particle configuration where for convenience we omit to make the $N$-dependence explicit in the notation since it will always be clear from the context anyway. Moreover, we define for $\alpha,c>0,\ a\in \{-1,1\}$ and $N\in \mathbb{N}$ the following interaction force which has a cut-off that becomes arbitrarily small as $N$ grows to $\infty$:
\begin{align}
f_c^N:\mathbb{R}^3\to \mathbb{R}^3, \ q\mapsto \begin{cases}  a
N^{(\alpha+1) c}q &\text{ , if }|q|\le N^{-c} \\ 
a\frac{q}{|q|^{\alpha+1}}  &\text{ , if }|q|> N^{-c} \end{cases} \label{def.force}
\end{align}
For the whole thesis the parameter $\alpha$ will be element of $(1,2]$. Most estimates can be done for general $\alpha$ in the respectively considered range and for situations where a case analysis is necessary we will explicitly point out which values are being considered. Thus, we will omit to make the dependence of the force on this parameter explicit in the notation. Furthermore, we also omit to indicate the dependence on $a\in \{-1,1\}$ because it will not matter for the estimates if the force is attractive or repulsive. We already point out that the `singularity parameter' $\alpha$ and the `cut-off parameter' $c$ are crucial quantities which will occur throughout the whole thesis and which are particularly important for the interpretation of results. Furthermore, we remark that the stated form of the cut-off is only one of arbitrarily many possibilities. It is only important that the Lipschitz constant related to the `inner area' of the force $|q|\le N^{-c}$ is not of distinctly larger order than the Lipschitz constant related to the `outer part'.  \\  
Moreover, in correspondence to the notation for the regularized force it makes sense to denote    
\begin{align}
f^{\infty}:\mathbb{R}^3\to \mathbb{R}^3, \ q\mapsto \begin{cases}  0 &\text{ , if }|q|= 0\\ 
a\frac{q}{|q|^{\alpha+1}}  &\text{ , if }|q|> 0 \end{cases} \label{def.lim.force}
\end{align}
where for convenience we often identify $f:=f^{\infty}$. Also for the non-regularized force the respectively considered values of $\alpha\in (1,2]$ will always be clear from the context.\\ 
Actually, (at least in our view) the most interesting choice for the singularity parameter is $\alpha=2$ where $f^\infty$ is given by the physically relevant Coulomb or Newton's gravitational force. However, the results which are attainable by our approach still rely on a cut-off for this choice of $\alpha$ so that we consider additionally structural similar but less singular force kernels where stronger results can be proven.\\
The Newtonian flow provided by the solutions to \eqref{Def.micro.sys.} for the regularized force $f^N_c$ will be denoted by $(\Psi_{s,t}^{N,c})_{s,t\in \mathbb{R}}$ which means that the map $\Psi_{\cdot,t}^{N,c}(X)=\big(^1{\Psi_{\cdot,t}^{N,c}(X)},{^2\Psi_{\cdot,t}^{N,c}(X)}\big)$ shall solve equations \eqref{Def.micro.sys.} if $f:=f^N_c$ and $\Psi_{t,t}^{N,c}(X)=X$ for $X\in \mathbb{R}^{6N}$, $t\in \mathbb{R}$. Hence, if $\Psi_{s,t}^{N,c}(X)$ indicates the positions of the particles in phase space, then their positions in physical space will be denoted by $^1\Psi_{s,t}^{N,c}(X)=([^1\Psi_{s,t}^{N,c}(X)]_1,...,[^1\Psi_{s,t}^{N,c}(X)]_{N})$ and their velocities/momenta by $^2\Psi_{s,t}^{N,c}(X)=([^2\Psi_{s,t}^{N,c}(X)]_1,..., [^2\Psi_{s,t}^{N,c}(X)]_N)$. We remark that (just like introduced here) the left superscript will generally be applied to distinguish between coordinates describing velocities on the one hand and positions in physical space on the other hand.   
\section{The Vlasov equation \label{sec.Vlas.eq}}
Furthermore, we consider the differential equation
\begin{align}
\frac{d}{\partial t}k_t+ v\cdot \nabla_x k_t+f*(\int_{\mathbb{R}^3}k_t(\cdot,v)d^3v)\cdot \nabla_v k_t=0 \label{Def.Vlas.eq}
\end{align}
which we already introduced as Vlasov equation (see \eqref{def.unreg.Vlas.}). For the regularized interaction $f^N_c$ the solution theory to this equation is standard because in this case the force is Lipschitz continuous. For the non-regularized, singular force $f^{\infty}$ the situation is in principle much less obvious. However, fortunately there already exist many results in literature that we can rely on. Since Vlasov-Poisson equation is the related effective equation to the microscopic system where the singularity parameter $\alpha$ equals $2$ (which we already designated as the physically most relevant option), the solution theory concerning this special case is very well studied. It is well known that under suitable conditions on the initial density $k_0$ there exist global, classical solutions to this equation (see e.g. the papers of Lions and Perthame \cite{lions1991} or Pfaffelmoser \cite{pfaffelmoser1992}). For our purposes a result of Horst seems to be best suited since it provides global existence of (unique) classical solutions under conditions which are very similar to the assumptions we need anyway for the proof of our Theorem \ref{thm1} (see \cite{Horst}). More precisely, it is shown that for arbitrary $T>0$ and any $k_0\in \mathcal{L}^1(\mathbb{R}^6)$ which is non-negative, continuously differentiable and fulfills for a suitable constant $C>0$, some $\delta>0$ and all $(q,v)\in \mathbb{R}^6$ the conditions
\begin{align} 
(i)& \ \ k_0(q,v)\le  \frac{C}{(1+|v|)^{3+\delta}} \notag \\
(ii)& \ \ |\nabla k_0(q,v)| \notag \le  \frac{C}{(1+|v|)^{3+\delta}}\\
(iii)& \ \ \int_{\mathbb{R}^6}|v|^2 k_0(q,v) d^6(q,v)<\infty\label{ass.sol.Horst}
\end{align}
there exists a continuously differentiable map $k:[0,T]\times \mathbb{R}^6\to [0,\infty)$ which satisfies Vlasov-Poisson equation and $k(0,\cdot)=k_0$. What Horst basically shows is that for $k_t(q,v):=k(t,(q,v))$ the spatial density
\begin{align}
\widetilde{k}_t(q):=\int_{\mathbb{R}^3}k_t(q,v)d^3v \label{Def.dens.unreg.}
\end{align}
keeps bounded for arbitrary times where the solution exists. More specifically, if for each existence interval $[0,T)$ there exists $C(T,k_0)>0$ such that
\begin{align}
\sup_{0\le s < T}\|\widetilde{k}_s\|_{\infty}<C(T,k_0) \label{bound.dens.},
\end{align} then for any interval $[0,T']\subseteq [0,\infty)$ a unique solution to Vlasov-Poisson equation with initial data $k_0$ exists (see for example \cite{Rein}). We will show in section \ref{sol.unbound.kin.} that the boundedness of the kinetic energy (resp. condition $(iii)$ in \eqref{ass.sol.Horst}) may be dropped and the existence of global solutions is still guaranteed. \\
In correspondence to the notation introduced for the non-regularized system we will denote solutions to the regularized Vlasov equation by $k^{N,c}_t(x)$ or $k^{N,c}_t(q,v)$ for $t\in \mathbb{R}$ and $x=(q,v)\in \mathbb{R}^6$ as well as the spatial density by
\begin{align}
\widetilde{k}_t^{N,c}(q):=\int_{\mathbb{R}^3}k_t^{N,c}(q,v)d^3v. \label{Def.dens.}
\end{align} 
Furthermore, the characteristics of Vlasov equation are given by the following system of differential equations
\begin{align}
 \begin{cases}
&\dot{q}=v  \\
&\dot{v}=f*\widetilde{k}_t(q). \label{def.flow}
\end{cases}
\end{align}
where $\widetilde{k}_t$ denotes the previously introduced `spatial density'. If $k_0$ fulfills assumptions \eqref{ass.sol.Horst}, then according to the results of Horst system \eqref{def.flow} is uniquely solvable on any interval $[0,T]$ and provides us the flow $(\varphi^{\infty}_{s,t})_{s,t\in \mathbb{R}}$ which for convenience will be often denoted by $(\varphi_{s,t})_{s,t\in \mathbb{R}}$. More specifically, the map $\varphi_{\cdot,s}(x)=(^1\varphi_{\cdot,s}(x),{^2\varphi_{\cdot,s}(x)})$ shall solve equations \eqref{def.flow} where $\varphi_{s,s}(x)=x$ for any $x\in \mathbb{R}^{6}$ and $s\in \mathbb{R}$ which in addition yields that $\varphi_{t,s}(x)=\varphi_{t,r}(\varphi_{r,s}(x))$ for any $r,s,t\in [0,T]$. Furthermore, it holds that $k_t(x)=k_0(\varphi_{0,t}(x))$ for arbitrary $s,t\in [0,T]$, $x\in \mathbb{R}^6$ and the map
\begin{align}
\varphi_{t,s}:\mathbb{R}^6\to \mathbb{R}^6,\ x\mapsto \varphi_{t,s}(x) \label{Liouville}
\end{align} 
is a Lebesgue-measure preserving diffeomorphism which will be applied on different occasions for the computation of integrals over $\mathbb{R}^6$. The flow for the regularized system (where the effective force field $f*\widetilde{k}_t$ in \eqref{def.flow} is replaced by $f^N_c*\widetilde{k}^{N,c}_t$) will be denoted by $(\varphi^{N,c}_{s,t})_{s,t\in \mathbb{R}}$ and fulfills corresponding properties.\\ 
Finally, we `lift' the effective flow to the $N$-particle phase space by the following identification 
\begin{align*}
\Phi^{N,c}_{s,t}(X):=(\varphi^{N,c}_{s,t}(X_1),...,\varphi^{N,c}_{s,t}(X_N))
\end{align*} for $X\in \mathbb{R}^{6N}$ and $s,t\in [0,T]$ while the corresponding notation for the non-regularized system shall be given by
\begin{align}
\Phi^{\infty}_{s,t}(X):=(\varphi^\infty_{s,t}(X_1),...,\varphi^\infty_{s,t}(X_N)). \label{lifted flow}
\end{align} 
The relevance of this definition will become clear shortly.\\
As a concluding remark for this section we point out that the result of Horst is actually only formulated for this most interesting case $\alpha=2$. However, the estimates applied in the proof of Horst also work for weaker singularities and in particular for $\alpha\in (1,2)$. In fact, the transition from $\alpha=2$ to $\alpha<2$ simplifies the estimates in many ways since in this case showing the Lipschitz continuity of the mean-field force is distinctly easier to achieve. Hence, we will simply assume that the mean-field flow exists and that the related spatial density fulfills property \eqref{bound.dens.} (which is all we need for the proofs). In fact, it should be a straightforward application of the results of Lemma \ref{lem1} together with relation \eqref{dist.effect.flow.limit} and the estimates applied in section \ref{sol.unbound.kin.} to prove this explicitly. However, since the value of stating all details seems to be rather limited we omit to make this explicit and continue by introducing in which sense the microscopic system shall be described by the effective equation. \\
\section{Propagation of Chaos \label{Prop.Chaos}} As mentioned before the (vary vague formulated) desired result of this thesis is to justify the effective description of the previously introduced microscopic system by solutions to Vlasov equation. This justification usually takes place by proving Propagation of Chaos for the considered system and is often referred to as `derivation' of the respective effective equation. In a formal sense the concept of Propagation of Chaos was introduced by Kac (see \cite{Kac}) and we refer the reader to the famous book of Sznitman \cite{Sznitman} for a deeper insight into this topic. By this time there are a number of different, partially equivalent definitions for this expression. One classical version (presented for the currently considered setting) is the following: Assume that the particles are initially i.i.d. with respect to the $N$-fold product of the density $k_0$. Thus, the density of the $N$-particle system at time $t\in [0,T]$ is given by 
\begin{align}
F^{N,c}_t(X):=\prod_{i=1}^N k_0([\Psi_{0,t}^{N,c}(X)]_i) \text{ for } X\in \mathbb{R}^{6N}. \label{Def.N.part.dens.}
\end{align}  
Moreover, let $\rho_1,\rho_2\in \mathcal{L}^1(\mathbb{R}^n)$ be two probability densities, then the bounded Lipschitz distance between them is given by
$$d_{L}(\rho_1,\rho_2):=\sup_{h\in \mathcal{L}}\big| \int_{\mathbb{R}^n}h(x)(\rho_1(x)-\rho_2(x))d^nx\big|. $$
where $\mathcal{L}$ shall denote the space of functions $h:\mathbb{R}^n\to \mathbb{R}$ fulfilling 
$$h\in \mathcal{L} \Leftrightarrow \|h\|_{\infty}=\sup_{x\neq y}\frac{|h(x)-h(y)|}{\|x-y\|_2}=1.$$ 
Propagation of Chaos holds if with respect to this distance the $n$-marginal of $F^{N,c}_t$ converges to the $n$-fold solution of the considered effective equation denoted by $ (k_t)^{\otimes n} $ for arbitrary $n\in \mathbb{N}$ and $t\in [0,T]$ respectively
$$d_L({F}^{(N,n),c}_{t},(k_t)^{\otimes n})\stackrel{N\to \infty}{\to}0$$
where $$F^{(N,n),c}_{t}(X_1,...,X_n):=\int_{\mathbb{R}^{6(N-n)}}{F^{N,c}_t}(X_{1},...,X_N)d^{6(N-n)}(X_{n+1},...,X_N).$$
In the current work we show a different statement which, however, implies this version of Propagation of Chaos. More precisely, we also assume that the initial data of the particles are i.i.d. with respect to the density $k_0 $ and show that for certain $\beta_1,\beta_2>0$ there exists a constant $C>0$ such that for all $N\in \mathbb{N}$
\begin{align*}
\mathbb{P}\big(X\in \mathbb{R}^{6N}:\sup_{0\le s \le T}\|\Psi_{s,0}^{N,c}(X)-\Phi_{s,0}^\infty(X)\|_{\infty}> N^{-\beta_1} \big)\le CN^{-\beta_2} 
\end{align*}
where by this notation we mean the probability with respect to the i.i.d. initial data. It is shown for example in \cite{PeterDustin} that the previously introduced version of Propagation of Chaos follows if this relation is fulfilled. However, independent of this circumstance it is straightforward to see why such a statement provides a very good justification for the effective description of the considered systems: Basically, the trajectories of all particles are predicted up to vanishing deviations by the effective flow for typical initial data. Hence, while usually one is satisfied if the effective equation provides information about the macroscopic time evolution of the particles (which for certain settings also is the strongest result that can be hoped for), the currently considered version even yields information about their respective trajectories for typical initial data. We point out that the notion `typical initial data' is supposed to mean that the probability related to the complement of these configurations gets arbitrarily small as the particle number $N$ increases. 
\section{Discussion of previous results}
In the following we want to discuss under which constraints on the interaction a (so called) `derivation' of Vlasov equation in the mean-field scaling was possible so far by means of some selected publications. After introducing certain classical results, the focus of this overview shall be on findings which target the Coulomb case (in correspondence to our aspired aim). We decided to keep the framework of the presented results a little closer in order to discuss those results which seem particularly relevant for our objectives slightly more detailed.    \\
For the first systems where Vlasov equation could be derived with mathematical rigor Lipschitz continuous forces were considered. To our knowledge publication \cite{NeunzertWick} of Neunzert and Wick in 1974 was the first result for such settings. Perhaps better known are the publications of Braun and Hepp \cite{BraunHepp} as well as Dobrushin \cite{Dobrushin} and the proof presented in the book of Spohn (see \cite{SpohnBook}). These results, however, are in a certain aspect stronger than what can be expected for the systems which we consider since  Lipschitz continuous interaction fits perfectly well to the basic idea of a mean-field approach. Broadly speaking, what they show is that if the initial particle distribution is (in an appropriate weak sense) close to the density $k_0$, then the particle distribution will also be close to $k_t$ at later times (where $k_t$ is the solution to Vlasov equation with initial data $k_0$). The crucial reason why this works lies in the circumstance that the particle structure does not come `into play' in a relevant way for Lipschitz continuous forces. It does barley matter for the related force field if a certain mass is concentrated at one point or if it is smeared out to a little `cloud' around the same position. Hence, if the closeness assumption between the initial particle distribution and $k_0$ is fulfilled, then the initial effective force field should be close to the microscopic force (resp. $f*\widetilde{k}_0(q)\approx \frac{1}{N}\sum_{j=1}^N f(q-q_j)$). This yields in turn that the dynamics of both systems are such that the closeness between them is maintained.\\  
The situation, however, is different if the considered interaction is singular and it took quite some time until Propagation of Chaos was shown  by Hauray and Jabin (see \cite{Hauray2007} and \cite{Hauray2013}) for force kernels $f$ which satisfy $|f(q)|\le \frac{C}{|q|^{\alpha}}$, $|\nabla f(q)|\le \frac{C}{|q|^{\alpha+1}}$ and $\alpha<1$. In this case, the force between individual particles can become arbitrarily large. The considered particle distribution might be an excellent discrete approximation of the initial density $k_0$ and still the previous relation $f*\widetilde{k}_0(q)\approx \frac{1}{N}\sum_{j=1}^N f(q-q_j)$ is not true anymore in general since the singularity of the pair interaction leads to large deviations of the force field around the positions of the particles. Hence, if two particles are extremely close to each other, the force between them might completely dominated the force exerted by all remaining particles. Consequently, in contrast to the previous case, here the particle structure does matter (at least in principle). Hauray and Jabin applied the second order nature of the dynamics to solve this problem: Although the force is singular, particles which have a sufficiently big relative velocity keep only close for short time periods so that the impact they have on each other still keeps small. Nevertheless, one has to abandon such strong Propagation of Chaos statements like applied in previous works. For the current systems the statements must rather take a form like: For typical initial particle distributions which provide a good discrete approximation of $k_0$ also the time-evolved distributions will keep `close' to the solution of Vlasov equation.\\ 
Moreover, Hauray and Jabin were able to handle forces converging pointwise to even more singular interactions but having an $N$-dependent cut-off for all $N\in \mathbb{N}$ (like introduced previously). Although their result is far more general, perhaps most notably also forces having a cut-off radius $c_N=N^{-\frac{1}{6}}$ and scaling like $\frac{1}{|q|^{\alpha}}$ (in $3$-dimensional space) where $\alpha $ can be arbitrarily close but still smaller than $2$ are included (see \cite{Hauray2013}).\\ Introducing a further method Boers and Pickl were able to shrink the cut-off for such forces to a size of order $N^{-\frac{1}{3}}$ which interestingly is the average distance of a particle to its nearest neighbor in $3$-dimensional space for the considered setting (see \cite{Peter}). Eventually, Lazarovici and Pickl managed to include the Coulomb case as well, but to do so they had to slightly increase the cut-off size to order $N^{-c}$ where $c <\frac{1}{3}$ (\cite{PeterDustin}). Another considerable result was obtained recently by Jabin and Wang where Propagation of chaos was shown for arbitrary $\mathcal{L}^{\infty}$-forces by application of a new approach which aims to control the relative entropy between the $N$-particle density and the product of solutions to Vlasov equation (see \cite{Wang}). It is also noteworthy that their method is particularly well-suited to deal with models where an additional stochastic term (resp. a Brownian motion) appears in the equations defining the dynamics (see \cite{Wang1}).\\  Furthermore, very recently Serfaty and Duerinckx presented an approach which can handle the case $\alpha=2$ for repulsive interaction (respectively the Coulomb case) and monokinetic solutions (see \cite{Serfaty}). In the setting they consider the initial situation shall be such that there exists a regular velocity field $u:\mathbb{R}^3\to \mathbb{R}^3$ which provides a continuous approximation of the initial velocity distribution of the particles in dependence on their positions, respectively $u(^1X_i)\approx {^2X_i}$ for all $i\in \{1,...,N\}$ (where we note that their result is not restricted to $3$-dimensional space). They introduce a functional denoted as {\em total modulated energy} which constitutes a measure of closeness between the empirical density associated to the $N$-particle trajectory and the monokinetic solution. If at the initial time the functional is sufficiently small, then it stays `small' on the time span where the solution exists. Due to the very recent date of appearance, the time did not permit to study why the approach in its current form is restricted to monokinetic initial data. However, heuristically it appears quite hard to define a functional which contains all the information such that initial `smallness' implies later `smallness' for the considered system and general initial data. We already discussed that deviations are expected if the particle structure comes into play in a significant way. While it is possible to find functionals which `reveal' if the initial state has this property, it is on the other hand not obvious how to define a reasonable condition such that this property propagates in time (deterministically). If there are not quite strict constraints on the initial velocities, then it should be possible to construct initial states where a considerable number of particles approaches each other with large relative velocities which in the Coulumb case leads to `strong' collisions and thereby to correspondingly large deviations to the effective dynamics. The monokinetic setting appears heuristically to be a very reasonable choice where such constellations can be excluded under certain constraints. As long as the monokinetic solution provides a good approximation of the particle distribution, small inter-particle distances should be connected with small relative velocity values which for repulsive interaction suppresses the appearance of `hard collisions'. On the other hand, the gap between heuristics and a rigorous proof is often huge, however, the result of Serfaty and Duerinckx closes this gap for the considered problem.
\section{Main objectives of the work}
Achieving the final aim which is showing Propagation of Chaos for the non-regularized Coulomb force (respectively Newton's gravitational force) for general initial densities still seems a long way off. Thus, the purpose of this work is proving further intermediate results which hopefully contribute to accomplish this aim. \\ 
The first obvious option where some of the previous results can be extended is a further shrinking of the cut-off size (for fixed $N$) resp. an increase of the cut-off parameter $c$. This objective is pursued in chapter \ref{sec.main1}. More precisely, we want to show Propagation of Chaos in the sense introduced in section \ref{Prop.Chaos} for a cut-off size of order $N^{-\frac{7}{18}+\epsilon},\ \epsilon>0$. The mere numbers perhaps do not create the impression that this yields a relevant improvement to the cut-off size $N^{-\frac{1}{3}+\epsilon}$ considered in \cite{PeterDustin}. However, a slightly more detailed analysis shows that this little increase of the cut-off parameter $c$ indeed has a certain effect on the interpretation of the result because it leads to a cut-off size which is distinctly below the order of the typical inter-particle distance $N^{-\frac{1}{3}}$. Let us assume that at a given point in time most particles are more or less homogeneously distributed over a volume of order $1$. If the cut-off radius is $N^{-\frac{7}{18}+\epsilon}$, then its (spatial) volume is of order $N^{-\frac{7}{6}+3\epsilon}$. Hence, the expected relative share of particles which at the given moment have a further particle inside their `cut-off area' should be roughly of order $N N^{-\frac{7}{6}+3\epsilon}=N^{-\frac{1}{6}+3\epsilon}$ which for large $N\in \mathbb{N}$ and sufficiently small $\epsilon>0$ is a vanishingly small amount. This yields that a typical particle will feel the `full' non-regularized force from all remaining particles (not for all but still) for practically all the time. Although proving Propagation of Chaos for such a system for sure is no rigorous reason that a corresponding statement holds if the cut-off is removed completely, it might nevertheless be seen as a further step in this direction.\\
Chapter \ref{sec.main2} is concerned with the second obvious problem which is proving Propagation of Chaos for interactions where not only the force but also the potential is singular (however still distinctly less singular than the Coulomb potential).
It is straightforward to see the new problem which arises by this: Even by application of the second order nature of the dynamics it is not directly possible to conclude that the impact of single particles on each other becomes negligible for large $N$. If the spatial distance between two particles is sufficiently small at a certain point in time, then even under the assumption of a large relative velocity the effect of this event on the dynamics of the particles can not be ignored. Handling this new issue is the main challenge. By the presented approach forces scaling like $\frac{1}{|q|^{\alpha}}$ where $1<\alpha \le \frac{4}{3}$ will be considered which unfortunately is still far away from the Coulomb case. We remark that basically nothing crucial goes wrong as the singularity parameter $\alpha$ attains $\frac{4}{3}$. However, without substantial modifications of the approach, the case $\alpha=2$ remains clearly out of reach and thus we decided to limit ourselves to this range of values for a slightly more convenient presentation. In addition, we will require stronger restrictions on the initial densities than in the second chapter, in order that we are able to show Propagation of Chaos. Very roughly speaking, the area where the density changes distinctly faster than its current value (resp. where $\nabla k_0$ has a value of far larger order than $k_0$) needs to have a small probability (with respect to the measure related to $k_0$). \\
Finally, chapter \ref{sol.unbound.kin.} is concerned with a slightly different topic and yields a rather secondary result. As mentioned in section \ref{sec.Vlas.eq}, we will show that global classical solutions to Vlasov-Poisson equation still exist if the limitation to initial densities $k_0$ with bounded kinetic energy is dropped from the set of assumptions \eqref{ass.sol.Horst}. Since the existence of global solutions to Vlasov equation and of the related effective flow as well as their properties are, of course, crucial for our main results presented in chapters \ref{sec.main1} and \ref{sec.main2}, this topic still fits well in the conceptual framework of the thesis.
\section{Notation} \label{Notation}
Before we start with the main part of the thesis, we introduce some important remarks concerning the applied notation.
{\allowdisplaybreaks
\begin{enumerate} 
\item[(i)] All probabilities throughout the thesis are meant with respect to the $n$-fold product of probability densities $k_0\in \mathcal{L}^1(\mathbb{R}^6)$ and we denote for any Borel-measurable set $A\subseteq \mathbb{R}^{6n}$
$$\mathbb{P}(X=(X_1,...,X_n)\in A)=\mathbb{P}(A):=\int_{\mathbb{R}^{6n}}\mathbf{1}_{A}(X)\prod_{i=1}^n k_0(X_i)d^{6n}X$$
\item[(ii)] To avoid clumsy expressions all constants which we apply are simply denoted by $C$ (or on rare occasions by $K$) where we call a positive real number a constant if it depends only on objects or values which are basically fixed during proofs like the initial density $k_0$ or the length of the considered time span $T$. However, they may never depend on variables like the particle number $N$ or the considered configuration. Furthermore, constants may differ from step to step during estimates without making this explicit. 
\item[(iii)] For ease of notation sometimes the indices which are clear from the context and not necessary to comprehend the estimates are dropped.
\item[(iv)] $|\cdot|:\mathbb{R}^n\to \mathbb{R}_{\geq 0}$ shall denote the euclidean norm for $n\in \mathbb{N}$. Moreover, we apply slightly modified versions of the usual $1$- and $\infty$-norm which shall be defined as follows:  
$$|X|_{\infty}:=\max_{i\in \{1,...,N\}}|X_i|\ \land \ |X|_{1}:=\sum_{i=1}^N|X_i|$$ where $X:=(X_1,...,X_N)\in \mathbb{R}^{6N}$. In the whole work the notations $|\cdot|_{\infty}$ and $|\cdot|_1$ will always refer to these definitions.
\end{enumerate}}
\chapter{A derivation of Vlasov(-Poisson) equation as the mean-field limit of particle systems with regularized interaction \label{sec.main1}}
 Our first main result is the following:
\begin{thm} \label{thm1}
Let $T>0$ and $k_0\in \mathcal{L}^1(\mathbb{R}^{6})$ be a continuously differentiable probability density fulfilling
\begin{align}
&k_0(x)\le C_0\frac{1}{(1+|x|)^{4+\delta}} \ \forall\  x \in \mathbb{R}^{6} \label{ass.dens.1}\\
& |\nabla k_0(x)|\le C_0 \frac{1}{(1+|x|)^{3+\delta}}\  \forall\  x \in \mathbb{R}^{6} \label{ass.dens.2} \\
& \int_{\mathbb{R}^6}|v|^2k_0(q,v)d^6(q,v)\le C_0 \label{ass.dens.3}
\end{align}
for some $C_0,\delta>0$. Moreover, let $(\Phi^{\infty}_{t,s})_{t, s\in \mathbb{R}}$ be the related lifted effective flow defined in \eqref{lifted flow} as well as $({\Psi}^{N,c}_{t,s})_{t,s\in \mathbb{R}}$ the $N$-particle flow defined in \eqref{Def.micro.sys.} for $\alpha\in (1,2]$ and $c>0$. 
\begin{itemize}
\item[(i)] If $\alpha=2,$ $\sigma>0$ and $c=\frac{7}{18}-\sigma$, then for any $\gamma>0$ there exists $C_1>0$ such that for all $N\in \mathbb{N}$ it holds that
\begin{align}
\mathbb{P}\big(X\in \mathbb{R}^{6N}:\sup_{0\le s \le T}|\Psi_{s,0}^{N,c}(X)-{\Phi_{s,0}^{\infty}}(X)|_{\infty}>  N^{-\frac{2}{9}} \big)\le C_1N^{-\gamma}.\label{result1}
\end{align}
\item[(ii)] If $\alpha \in (1,\frac{4}{3}]$, $c=\frac{2}{3}$ and $\sigma, \epsilon>0$, then there exists $C_2>0$ such that for all $N\in \mathbb{N}$ it holds that
\begin{align}
\mathbb{P}\big(X\in \mathbb{R}^{6N}:\sup_{0\le s \le T}|\Psi_{s,0}^{N,c}(X)-{\Phi_{s,0}^{\infty}}(X)|_{\infty}>  N^{-\frac{1}{2}+\sigma} \big)\le C_2N^{-\frac{1}{9}+\epsilon}.\label{result2}
\end{align}
\end{itemize} 
\end{thm} 
\begin{rmk} \label{rmk.1}  \mbox{}\vspace{-\topskip}
\vspace{0,4cm}
\begin{itemize}
\item[(i)] As mentioned in the previous section \ref{Notation}, which is concerned with the notation, the stated probability is meant with respect to the law given by the $N$-fold product of densities $k_0$.
\item[(ii)] As discussed in section \ref{Prop.Chaos}, statements $(i)$ and $(ii)$ of the Theorem imply certain classical notions of Propagation of Chaos like the version presented there.
\item[(iii)] In section \ref{sec.Vlas.eq} the results are stated on which we rely such that the existence and uniqueness of the effective flow is assured on $[0,T]$. Moreover, we recall the important feature that under the stated constraints on $k_0$ the related `spatial density' fulfills
\begin{align}
\sup_{N\in \mathbb{N}}\sup_{0\le s \le T}\|\widetilde{k}^{N,c}_s\|_{\infty}<\infty . \label{bound.dens.0}
\end{align}
\item[(iv)] Statement $(ii)$ of the Theorem, which considers smaller values of the `singularity parameter' $\alpha$, is basically an interim result and its further development is the essential aim of chapter \ref{sec.main2}.
\end{itemize}
\end{rmk}

\section{Proof of the first main result}
\subsection{Heuristic proceeding}
Roughly speaking, the Theorem states that for typical initial data the interacting particles evolve up to a small deviation as if they were `driven' by the effective force field. As mentioned, the applied notion of closeness was originally introduced in \cite{Peter} and thus also the current approach strongly relies on the basic ideas introduced there although the exact form of the implementation might look quite different. We introduce an auxiliary trajectory $\Phi^{N,c}_{\cdot,0}(X)=(\varphi^{N,c}_{\cdot,0}(X_1),...,\varphi^{N,c}_{\cdot,0}(X_N))$ which starts at the same initial data as the interacting particles, however, evolves according to the mean-field flow. For convenience we will introduce the notion of `mean-field particles' which shall be pictured as (fictive) particles whose positions are determined by this auxiliary trajectory. Since the `mean-field particles' are initially i.i.d. with respect to the $N$-fold product of $k_0$ and are subject to the effective flow (which obviously yields that they do not interact with each other), it follows that they are also i.i.d. at later times but with respect to the $N$-fold product of $k_t=k_0(\varphi^{N,c}_{0,t}(\cdot))$. Now as long as the deviation $\sup_{0\le s \le t}|\Psi^{N,c}_{s,0}(X)-\Phi^{N,c}_{s,0}(X)|_{\infty}$ is small enough, we are able to `transfer' a lot of the information which we have about the distribution and the dynamics of the `mean-field particles' on the system of interacting particles which in turn will help us to show that the smallness of the deviation is preserved. Everything written so far might as well be a heuristic introduction to the approach applied in \cite{Peter}. One crucial difference is that in the current case the possibility to `transfer' information from the `mean-field particles' to their related `partners' of the interacting system is applied to a distinctly larger extend. A second difference is that we will make full use of the second order nature of the dynamics. More precisely, the deviation between the `microscopic force' and the effective force field evaluated at the position of a given particle is heuristically supposed to pass through significant fluctuations. Every time a particle comes exceptional close to the `observed' one a correspondingly large deviation should be expected. However, such peaks are usually only of a very limited duration since the particles will just fly apart shortly after. Hence, it is reasonable to compare the dynamics on longer time periods so that the deviations between them do not become overestimated. This already clarifies how the second order nature will be relevant to us. But the statement that particles will keep close only for short times is typically only valid for the vast majority, but not for all collisions that take place (where a `collision' is supposed to describe an event in which two particles come close to each other and fly apart again). As long as the microscopic and their related auxiliary particles are close in phase space, the `types' of collisions corresponding particles of these two systems experience are expected to be very similar. Consequently, it seems to be reasonable to divide the particles into a `good' and a `bad' class (or a finer subdivision) in dependence on their `mean-field particle partners': If a hard collision is to be expected for a certain particle pair according to their auxiliary trajectories, then they will be labeled `bad'. Since for such `bad' particles larger deviations are supposed to occur, we will also `allow' larger distances to their related `mean field particles' than for the `good' ones. In this way a condition is imposed on the proximity between corresponding trajectories which can be complied with, and at the same time the information provided by the `mean-field particles' about the `real' ones is kept as big as possible. More details about the importance of the `mean field particles' will be introduced in the preliminary studies and the proof of the main result.\\
For implementing the proposed strategy, it is obviously necessary to first derive certain results for the auxiliary system which in turn shall be `transferred' on the microscopic system. The purpose of the following subsection is primarily this topic. 
\subsection{Preliminary studies}
First, we introduce a versatile applicable variant of Gronwall's Lemma which will be used on several occasions. By a slight abuse of notation we indicate for $n\in \mathbb{N}$ the multiple integral
\begin{align}
\int_{0}^{t_1}...\int_0^{t_n}f(t)dtdt_{n}...dt_2 \notag 
\end{align} 
which, however, for the special case $n=1$ shall obviously describe
\begin{align}
\int_{0}^{t_1}f(t)dt\ \ \text{ resp.  }\int_{0}^{t_1}\int_0^{t_2}f(t)dtdt_2 \text{ for }n=2. \notag
\end{align}
\begin{lem} \label{Gron.lem.}
Let $u:[0,\infty)\to [0,\infty)$ be a continuous and monotonously increasing map as well as $l,f_1:\mathbb{R}\to [0,\infty)$ and $f_2:\mathbb{R}\times \mathbb{R}\to [0,\infty)$ continuous maps such that for some $n\in \mathbb{N}$ and for all $t_1>0,\ x_1,x_2\geq 0$
\begin{itemize}
\item[(i)]
$\begin{aligned}&
x_1< x_2 \Rightarrow f_2(t_1,x_1)\le f_2(t_1,x_2)
\end{aligned}$
\item[(ii)]
$ \exists K_1,\delta>0: \sup\limits_{\substack{x,y  \in [f_1(0),f_1(0)+\delta]\\s \in [0,\delta]}}|f_2(s,x)-f_2(s,y)|\le K_1|x-y|$.
\item[(iii)]
$ \begin{aligned}
& f_1(t_1)+ \int_0^{t_1}...\int_0^{t_n} f_2(s,u(s))dsdt_n...dt_2<  u(t_1) \ \land \\ 
& f_1(t_1)+\int_0^{t_1}...\int_0^{t_n}  f_2(s,l(s))dsdt_n...dt_2 \geq l(t_1),
\end{aligned}$
\end{itemize} 
then it holds for all $t\geq 0$ that $l(t)\le u(t)$.
\end{lem}
\vspace{0,4cm}
\noindent The proof to this lemma is not important for the comprehension of the remaining part but can be found in the appendix (resp. in chapter \ref{sec.app.}). \\\\
\noindent Before we start with the relevant lemmas, we first have to introduce a function which will be very important on many occasions throughout the paper and start with some preliminary considerations.\\
It holds for $\alpha\in (1,2]$, $\delta,q\in \mathbb{R}^3,\ |q|\geq 3N^{-c}$ and $|\delta|\le \frac{2}{3}|q|$ that 
$$| f^N_c(q)-f^N_c(q+\delta)|\le \alpha 3^{\alpha+1} \frac{1}{|q|^{\alpha+1}}|\delta|$$
since 
\begin{align*}
 |\frac{q}{|q|^{\alpha+1}}-\frac{q+\delta}{|q+\delta|^{\alpha+1}}|\le & \sup_{t\in [0,1]}\frac{\alpha}{|q+t\delta|^{\alpha+1}}|\delta| \le \frac{\alpha|\delta|}{(|q|-|\delta|)^{\alpha+1}}\\
\le & \alpha 3^{\alpha+1}\frac{|\delta|}{|q|^{\alpha+1}}.
\end{align*}
Moreover, it is easy to see that $\alpha N^{c(\alpha+1)}$ constitutes a Lipschitz-constant for $f^N_c$ (see \eqref{def.force}) and thus
$$|f^N_c(q)-f^N_c(q+\delta)|\le \alpha N^{c(\alpha+1)}|\delta|.$$
We define 
\begin{align}
g_c^{N}:\mathbb{R}^3\to \mathbb{R}, \ q\mapsto \begin{cases}  \alpha N^{c(\alpha+1)} &\text{ , if }|q|\le 3N^{-c} \\ 
\alpha 3^{\alpha+1}\frac{1}{|q|^{\alpha+1}}  &\text{ , if }|q|> 3N^{-c} \end{cases}, \label{Def.g^N}
\end{align}
Due to the previous consideration it holds that
\begin{align}
|f_c^N(q)-f_c^N(q+\delta)|\le g_c^N(q)|\delta| \label{ineq.g^N}
\end{align}
if $|\delta|\le \frac{2}{3}|q|$ or alternatively $|q|\le 3N^{-c}$. This, however, yields that inequality \eqref{ineq.g^N} is in particular fulfilled for arbitrary $q\in \mathbb{R}^6$ if $|\delta|\le 2N^{-c}$. Moreover, according to the previous reasoning it is quite obvious that for $q_1,q_2,q_3\in \mathbb{R}^3$ where $|q_1|\le \min(|q_2|,|q_3|)$ the following relation holds 
\begin{align}
|f_c^N(q_2)-f_c^N(q_3)|\le g_c^N(q_1)|q_2-q_3| \label{ineq.g^N+}
\end{align}
because for $|q_1|\le 3N^{-c}$ the factor $g_c^N(q_1)= \alpha N^{c(\alpha+1)}$ constitutes a Lipschitz-constant for $f^N_c$ and for larger values again some mean value argument applies.\\
This concludes the considerations concerning the map $g_c^N$.\\
For ease of notation we allow in the following the range $ \mathbb{N}\cup \{\infty \}$ for the index $N$ in $(\varphi^{N,c}_{t,s})_{t, s\in \mathbb{R}}$ where $(\varphi^{\infty, c}_{t,s})_{t, s\in \mathbb{R}}$ shall simply denote the non-regularized flow  $(\varphi^{\infty}_{t,s})_{t, s\in \mathbb{R}}$ and correspondingly $k^{\infty,c}_{t,s}:=k^{\infty}_{t,s}$ as well as $f^{\infty}_c:=f^{\infty}$. Moreover, we remark that while the statement of the subsequent lemma will be crucial on many occasions, its (slightly elongated) proof can be skipped without missing something relevant for the main part. The same applies for basically all preliminary lemmas.
\begin{lem} \label{lem1}
Let $T>0$ and $k_0$ be a probability density fulfilling the assumptions of Theorem \ref{thm1} where $(\varphi^{N,c}_{t,s})_{t, s\in \mathbb{R}}$ shall be the related effective flow defined in \eqref{def.flow} for $\alpha\in (1,2]$ and $c>0$. Then there exist  $C_1,C_2>0$ such that for all configurations $X,Y\in \mathbb{R}^6,\ N\in \mathbb{N}\cup \{\infty\}$ and $t,t_0\in [0,T]$ it holds that
\begin{align*}
|\varphi_{t,t_0}^{N,c}(X)-\varphi_{t,t_0}^{N,c}(Y)| \le |X-Y|e^{C_1|t-t_0|}
\end{align*}
and 
\begin{align*}
|f_c^N*\widetilde{k}^{N,c}_t(^1X)-f^N_c*\widetilde{k}^{N,c}_t(^1Y)| \le C_2|^1X-{^1Y}|.
\end{align*}
\end{lem}
\vspace{0,4cm}
\begin{proof}
For ease of notation we omit to make the index $c$ explicit during the subsequent estimates. Moreover, the proof is a straightforward application of Gronwall's Lemma if $\alpha\in (1,2)$ and thus we will limit ourselves to state the estimates for the less obvious case $\alpha=2$.\\
We define for $N\in \mathbb{N}\cup \{\infty\}, X,Y\in \mathbb{R}^6$ and $R>0$ a set which basically constitutes a (generally) thick-walled spherical shell in space around $^1X$ $$\mu^{R,N}_{X,Y}:=\{Z\in\mathbb{R}^6:3\max(N^{-c},|^1X-{^1Y}|)\le |^1Z-{^1X}|\le R\}$$ where for the case $N=\infty$ the factor $N^{-c}$ shall simply be replaced by $0$. It holds that
{\allowdisplaybreaks \begin{align}
& |f^{N}*\widetilde{k}^N_{t}(^1X)-f^{N}*\widetilde{k}^N_{t}(^1Y)| \notag\\
\le & \int_{|Z-{^1X}|\le 3\max\big(N^{-c},|^1X-{^1Y}|\big)}\big(|f^{N}({^1X}-Z)-f^{N}({^1Y}-Z)|\big)\widetilde{k}^N_t(Z)d^3Z\big| \notag \\
& +\big|\int_{\mu^{R,N}_{X,Y}}\big(f^{N}({^1X}-{^1Z})-f^{N}({^1Y}-{^1Z})\big)k^N_t(Z)d^6Z\big| \notag  \\
 &+  |^1X-{^1Y}|\int_{|Z-{^1X}|\geq \max(3|^1X-{^1Y}|,R)}g^N({^1X}-Z)\widetilde{k}^N_t(Z)d^3Z \notag  
\end{align}}
\noindent where the estimate for the third term follows by application of the properties of $g^N$ (see \eqref{Def.g^N} resp. \eqref{ineq.g^N}). Since $g^N(q)\le C\min\big(N^{3c},\frac{1}{|q|^3} \big)$ for all $q\in \mathbb{R}^3$, we obtain that the third term is bounded by
$$C\|\widetilde{k}^N_t\|_{\infty}\ln^+(\frac{1}{R})|{^1X}-{^1Y}|$$
where $\ln^+(x):=\max\big(\ln(x),1\big),\ \forall x>0$. On the other hand for the first term the subsequent estimates hold
\begin{align*}
&\int_{|Z-{^1X}|\le 3\max\big(N^{-c},|^1X-{^1Y}|\big)}\big(|f^{N}({^1X}-Z)-f^{N}({^1Y}-Z)|\big)\widetilde{k}^N_t(Z)d^3Z \\
\le &|^1X-{^1Y}|\int_{|Z-{^1X}|\le 3N^{-c}}\underbrace{g^{N}(Z-{^1X})}_{\le CN^{3c}}\widetilde{k}^N_t(Z)d^3Z\\
+& \int_{|Z-{^1X}|\le 3|^1Y-{^1X}|}\big(  \underbrace{|f^{N}({^1X}-Z)|}_{ \displaystyle \le |{^1X}-Z|^{-2}}+|f^{N}({^1Y}-Z)|\big)\widetilde{k}^N_t(Z)d^3Z \\ 
\le & C\|\widetilde{k}^N_t\|_{\infty}|{^1Y}-{^1X}| 
\end{align*}
where the first of the two addends appearing after the first step shall be applied as upper bound if $N^{-c}\geq |^1X-{^1Y}|$ and the second in the alternative case.\\
It remains to determine a suitable upper bound for the second term where the mass related to the spherical shell `between' these two sets is taken into account. Of course, this set might also be empty if $R \le 3\max(N^{-c},|{^1X}-{^1Y}|)$ but we care for the more interesting situation where this is not the case. In the following we want to utilize that the contribution of mass (or charge) related to the set $\mu^{R,N}_{X,Y}$ to the force field at positions ${^1X}$ or ${^1Y}$ cancels out to a significant amount. For a more convenient comprehension we first recall the definition
 $$\mu^{R,N}_{X,Y}:=\{Z\in\mathbb{R}^6:3\max(N^{-c},|^1X-{^1Y}|)\le |^1Z-{^1X}|\le R\}$$
 and conclude subsequently by triangle inequality that
\begin{align*}
& \big|\int_{\mu^{R,N}_{X,Y}}\big(f^{N}({^1X}-{^1Z})-f^{N}({^1Y}-{^1Z})\big){k}^N_t(Z)d^6Z\big|\\
\le &  \big|\int_{ \mu^{R,N}_{X,Y}}\big(f^{N}({^1X}-{^1Z})-f^{N}({^1Y}-{^1Z})\big)\big(k^N_t(Z)-k^N_t(({^1X},{^2Z}))\big)d^6Z\big|\\
&+ \big|\int_{3\max(N^{-c},|^1X-{^1Y}|)\le |^1Z-{^1X}|\le R}f^{N}({^1Y}-{^1Z})\widetilde{k}^N_t({^1X})d^3(^1Z)\big|\\
&+ \big|\int_{3\max(N^{-c},|^1X-{^1Y}|)\le |^1Z-{^1X}|\le R}f^{N}({^1X}-{^1Z})\widetilde{k}^N_t({^1X})d^3(^1Z)\big|.
\end{align*}
By Newton`s shell Theorem, the spherically symmetry of the integration set and the circumstance that the density $\widetilde{k}^N_t({^1X})$ does not depend on the integration variable the last two terms vanish completely since all the `mass` or `charge' lies around ${^1}X$ and ${^1Y}$. For estimating the first addend we define 
$$\Delta^N(t):=\sup_{X,Y\in \mathbb{R}^6: X\neq Y}\sup_{r,s\in [0,t] }\frac{|\varphi^N_{r,s}(X)-\varphi^N_{r,s}(Y)|}{|X-Y|}.$$
While the existence of this variable is obvious for the regularized system, it is at least straightforward to see that for the non-regularized system it exists for sufficiently small times. We will first apply this quantity for the estimates and show afterwards that it is bounded by some ($T$-dependent) constant on $[0,T]$. Due to the mean value theorem (applied for the densities) and the properties of the map $g^N$ (see \eqref{Def.g^N}) it holds that
\begin{align}
&  \big|\int_{ \mu^{R,N}_{X,Y}}\big(f^{N}({^1X}-{^1Z})-f^{N}({^1Y}-{^1Z})\big)\big(k^N_t(Z)-k^N_t({^1X},{^2Z})\big)d^6Z\big| \notag \\
\le &  \int_{ \mu^{R,N}_{X,Y}}g^{N}({^1X}-{^1Z})|^1X-{^1Y}| \notag \\
&\cdot\big( \sup_{\widetilde{Z}\in \overline{\varphi^N_{0,t}(^1X,{^2Z})\varphi^N_{0,t}(Z)}}|\nabla k_0(\widetilde{Z})||\varphi^N_{0,t}(Z)-\varphi^N_{0,t}(^1X,{^2}Z)|\big)d^6Z \notag \\
\le &\Delta^N(t)|^1X-{^1Y}| \int_{ 
 |^1Z-{^1X}|\le R}\underbrace{g^{N}({^1X}-{^1Z}) |{^1X}-{^1Z}|}_{\le C|{^1X}-{^1Z}|^{-2}}d^3({^1Z})\notag \\
& \cdot \int_{\mathbb{R}^3}\sup_{Z'\in \mathbb{R}^3}\sup_{\widetilde{Z}\in \overline{\varphi^N_{0,t}(^1X,{^2Z})\varphi^N_{0,t}(Z',{^2Z})}}|\nabla k_0(\widetilde{Z})|d^3({^2Z}) \notag\\
\le & C\Delta^N(t)R|^1X-{^1Y}| \label{est.grad.dens.} .
\end{align}
where $\overline{X_1X_2}:=\{(1-\lambda)X_1+\lambda X_2\in \mathbb{R}^6: \lambda\in [0,1]\}$ for $X_1,X_2\in \mathbb{R}^6$.\\
In the last step we applied that due to the upper bound on $\|\widetilde{k}^N_t\|_{\infty}$ (see \eqref{bound.dens.0}) obviously also
$$f_{max}:=\sup_{N\in \mathbb{N}}\sup_{0\le s \le T }\|f^N*\widetilde{k}^N_s\|_{\infty}<\infty$$
which yields for any $Z'\in \mathbb{R}^6$ where $|^2Z'|\geq 2f_{max}T$ and $t\in [0,T]$ that
\begin{align*}
|^2\varphi^{N}_{0,t}(Z')|\geq |^2Z'|-f_{max}t\geq \frac{|^2Z'|}{2} .
\end{align*}
Consequently, it follows according to the assumption on the decay of $|\nabla k_0|$ (see \eqref{ass.dens.2}) that \begin{align*}
&  \int_{\mathbb{R}^3}\sup_{Z'\in \mathbb{R}^3}\sup_{\widetilde{Z}\in \overline{\varphi^N_{0,t}(^1X,{^2Z})\varphi^N_{0,t}(Z',{^2Z})}}|\nabla k_0(\widetilde{Z})|d^3({^2Z})\\
\le & \int_{\mathbb{R}^3}\sup_{Z'\in \mathbb{R}^3}\sup_{\widetilde{Z}\in \overline{\varphi^N_{0,t}(^1X,{^2Z})\varphi^N_{0,t}(Z',{^2Z})}}\frac{C}{(1+|{\widetilde{Z}}|)^{3+\delta}}d^3({^2Z})\\
\le & C\int_{|^2Z|\le 2f_{max}T}d^3({^2Z})+\int_{|^2Z|> 2f_{max}T}\frac{C}{(1+\frac{ |^2Z|}{2 })^{3+\delta}}d^3({^2Z})\\
\le &  C.
\end{align*}
In total we obtain
\begin{align}
& |f^{N}*\widetilde{k}^N_{t}(^1X)-f^{N}*\widetilde{k}^N_{t}(^1Y)| \le C\big(\ln^+(\frac{1}{R})+\Delta^N(t)R\big)|^1X-{^1Y}| \label{est.force}
\end{align}
where we again regarded the upper bound on the `spatial density' (see \eqref{bound.dens.0}). Hence, it remains to control the growth of $\Delta^N(t)$. Let to this end be $s,t\in [0,T]$ as well as $X,Y\in \mathbb{R}^6$ be given where $X\neq Y$. If we choose $R:=\Delta^N(t)^{-1}$ and omit to make the $N$-dependence of $\Delta^N(t)$ explicit for the estimates, then application of \eqref{est.force} in the first step shows that
\begin{align*}
& \sup_{s\le r \le t}|^2\varphi^N_{r,s}(X)-{^2\varphi^N_{r,s}}(Y)-(^2X-{^2Y})|\\
\le &  
 \int_{s}^t\underbrace{| f^{N}*\widetilde{k}^N_{u}(^1\varphi^N_{u,s}(X))-f^{N}*\widetilde{k}^N_{u}(^1\varphi^N_{u,s}(Y))|}_{\le C\ln^+(\Delta(u))|{^1\varphi^N_{u,s}}(X)-{^1\varphi^N_{u,s}}(Y)|}du\\
 \le &  C\ln^+(\Delta(t))\int_{s}^t |^1X+{^1Y}|+ \Big(\int_s^u\sup_{s\le r' \le r}|{^2\varphi^N_{r',s}}(X)-{^2\varphi^N_{r',s}}(Y)|dr\Big)du\\
\le &  C\ln^+(\Delta(t))\big(|^1X+{^1Y}|+|^2X+{^2Y}|(t-s)\big)(t-s)\\
&+  C\ln^+(\Delta(t))\int_{s}^t\int_s^u\sup_{s\le r' \le r}|{^2\varphi^N_{r',s}}(X)-{^2\varphi^N_{r',s}}(Y)-(^2X-{^2Y})|drdu.
\end{align*}
Now one easily verifies by application of Gronwall lemma \ref{Gron.lem.} that 
\begin{align}
 &\sup_{s\le r \le t}|^2\varphi^N_{r,s}(X)-{^2\varphi^N_{r,s}}(Y)-(^2X-{^2Y})|\notag \\
 \le&  \underbrace{ C\ln^+(\Delta(t))\big(|^1X+{^1Y}|+|^2X+{^2Y}|(t-s)\big)(t-s)}_{=:b(t)}e^{\sqrt{ C\ln^+(\Delta(t))}(t-s)} \label{est.lem1,0}
 \end{align} 
 because for $0\le s<t'\le t$ it holds that
\begin{align*}
& b(t')+C\ln^+(\Delta(t))\int_{s}^{t'}\int_s^u b(r)e^{\sqrt{ C\ln^+(\Delta(t))}(r-s)}drdu\\
< &b(t')\big(1+\sqrt{ C\ln^+(\Delta(t))}\int_{s}^{t'} e^{\sqrt{ C\ln^+(\Delta(t))}(u-s)}du \big) \\
= & b(t')e^{\sqrt{ C\ln^+(\Delta(t))}(t'-s)}.
\end{align*}
This yields additionally that 
\begin{align}
 &\sup_{s\le r \le t}|^1\varphi^N_{r,s}(X)-{^1\varphi^N_{r,s}}(Y)-(^1X-{^1Y})|\notag \\
 \le & \int_s^t|^2\varphi^N_{r,s}(X)-{^2\varphi^N_{r,s}}(Y)|dr \notag \\
 \le & \big(|^2X-{^2Y}|+b(t)e^{\sqrt{ C\ln^+(\Delta(t))}(t-s)}\big)(t-s) \label{est.lem1,1}
 \end{align} 
Now the upper bounds \eqref{est.lem1,0} and \eqref{est.lem1,1} imply that for $X\neq Y$ and $s,t\in [0,T]$
\begin{align*}
& \frac{1}{|X-Y|}\sup_{s\le r \le t}|\varphi^N_{r,s}(X)-\varphi^N_{r,s}(Y)|\\
\le & 1+ \sup_{s\le r \le t}\frac{|\varphi^N_{r,s}(X)-{\varphi^N_{r,s}}(Y)-(X-{Y})|}{|X-Y|}\\
\le & 1+ 2\max_{k\in \{1,2\}}\sup_{s\le r \le t}\frac{|^k\varphi^N_{r,s}(X)-{^k\varphi^N_{r,s}}(Y)-(^kX-{^kY})|}{|X-Y|}\\
\le & 1+ C\ln^+(\Delta(t))te^{\sqrt{ C\ln^+(\Delta(t))}(t-s)}.
\end{align*}
A corresponding relation for $\frac{1}{|X-Y|}\sup_{t\le r \le s}|\varphi^N_{r,s}(X)-\varphi^N_{r,s}(Y)|$ can be obtained by analogous estimates for the time reversed trajectories $(^1\varphi^N_{s-t,s}(Z),-{^2\varphi^N_{s-t,s}}(Z))$. Taking the supremum with respect to $s,t\in [0,t']\subseteq [0,T]$ and $X\neq Y$ (while regarding additionally the time reversal symmetry) finally yields that an inequality of the subsequent form holds for $t'\in  [0,T]$ provided that $\Delta(t')\geq e$:
\begin{align}
\Delta(t')&\le 1+ C\ln(\Delta(t'))t'e^{\sqrt{ C\ln(\Delta(t'))}t'}=1+  C\ln(\Delta(t'))t'\Delta(t')^{\frac{\sqrt{ C}}{\sqrt{\ln(\Delta(t'))}}t'}
\end{align}
This inequality provides us an ($N$-independent) upper bound for the growth of $\Delta^N(t)$ and implies in particular $\Delta^N(T)<C$. By regarding additionally relation \eqref{est.force}, this also yields the Lipschitz continuity of the mean-field force and thereby one directly obtains the existence of a constant $C_1>0$ such that for arbitrary $X,Y\in \mathbb{R}^6$, $t_0\in [0,T]$ and $t\in [0,T-t_0]$
\begin{align*}
& |\varphi^N_{t_0+t,t_0}(X)-\varphi^N_{t_0+t,t_0}(Y)|\le C_1\int_{0}^t|\varphi^N_{s+t_0,t_0}(X)-\varphi^N_{s+t_0,t_0}(Y)|ds+|X-Y| 
\end{align*}
as well as 
\begin{align*}
& |\varphi^N_{t_0-t,t_0}(X)-\varphi^N_{t_0-t,t_0}(Y)|\le C_1\int_{0}^t|\varphi^N_{t_0-s,t_0}(X)-\varphi^N_{t_0-s,t_0}(Y)|ds+|X-Y| 
\end{align*}
for $t\in [0,t_0]$ which by application of Gronwall's Lemma completes the proof.
\end{proof}
\noindent 
This lemma basically tells us that the distance in phase space between `mean-field particles' stays of the same order. Thus, if we know that at some point in time two `mean-field particles' are particularly close or far apart in phase space, then their distance is of the same order on the whole interval $[0,T]$.  \\\\
In the proof of our main result we will only consider configurations where the `mean-field particles' are always very close to their corresponding particle of the microscopic system (and it will turn out that these are in fact the typical configurations). As mentioned before, the expression `corresponding' in this context is supposed to mean that the `particles' start at the same initial data but evolve with respect to different dynamics. Consequently, all properties we show for the auxiliary system can to some extend be transferred on the system of interacting particles as long as the closeness between corresponding particles is maintained.\\
The most crucial point is to control the number and the `impact' of certain collisions. For this purpose we prove two further lemmas.\\
The first of them implies that on possibly short (in relation to $T$) but $N$-independent time intervals the trajectories of the `mean-field particles' are close to trajectories of freely evolving particles which will be crucial for the collision estimates later and yields additionally that the number of collisions two `mean-field particles' can in principle have with each other on $[0,T]$ is bounded by some constant. Since `particles' of the auxiliary system do not interact, the expression `collision' might sound confusing but should be understood as an event where two such `particles' come close in space and move apart afterwards. 
\begin{lem} \label{lem2}
Let $T>0$ and $k_0$ be a probability density fulfilling the assumptions of Theorem \ref{thm1} as well as $(\varphi^{N,c}_{t,s})_{t, s\in \mathbb{R}}$ the related effective flow defined in \eqref{def.flow} for $c>0$ and $\alpha\in (1,2]$. Then there exists $C_1>0$ such that for all $N\in \mathbb{N}$, $t,t_0\in [0,T]$ where $|t-t_0|\le 1$ and $X,Y\in \mathbb{R}^6$ it holds that
\begin{align*}
(i)\ \ \ & |{^1\varphi^{N,c}_{t,t_0}}(X)-{^1\varphi^{N,c}_{t,t_0}}(Y)-(^1X-{^1Y})-(^2X-{^2Y})(t-t_0)|\\
 \le & C_1(t-t_0)^2\big(|^1X-{^1Y}|+|^2X-{^2Y}||t-t_0|\big) \\
(ii)\ \ \ & \big|{^2\varphi^{N,c}_{t,t_0}}(X)-{^2\varphi^{N,c}_{t,t_0}}(Y)-({^2X}-{^2Y})\big|\\
\le & C_1|t-t_0|\big(|^1X-{^1Y}|+|^2X-{^2Y}|  |t-t_0|\big).
\end{align*}  
\end{lem}
\vspace{0,4cm}
\begin{proof} Again we omit to make the index $c$ explicit during the subsequent estimates and for ease of notation we only consider the case $t_0=0,\ t>0$ (since the remaining cases can be handled analogously). Applying the ($N$-independent) Lipschitz continuity of the mean-field force derived in Lemma \ref{lem1} it follows that
\begin{align}
& \Delta^N(X,Y,t):= \sup_{0\le s \le t}|{^1\varphi^{N}_{s,0}}(X)-{^1\varphi^{N}_{s,0}}(Y)-(^1X-{^1Y})-(^2X-{^2Y})s| \notag \\
\le & \int_{0}^t\int_{0}^s|f^{N}*\widetilde{k}^{N}_r({^1\varphi^{N}_{r,0}}(X))-f^{N}*\widetilde{k}^{N}_r({^1\varphi^{N}_{r,0}}(Y))|drds \notag \\
\le & C\int_{0}^t\int_{0}^s|{^1\varphi^{N}_{r,0}}(X)-{^1\varphi^{N}_{r,0}}(Y)|drds  \notag\\
\le & C \int_0^t\int_0^s|{^1\varphi^{N}_{r,0}}(X)-{^1\varphi^{N}_{r,0}}(Y)-(^1X-{^1Y})-(^2X-{^2Y})r|drds  \notag\\
&  +C\frac{t^2}{2}\big(|^1X-{^1Y}|+|^2X-{^2Y}|\frac{t}{3}\big.)\label{est.lem.2}
\end{align}
By application of Lemma \ref{Gron.lem.} this easily yields that
\begin{align*}
& \Delta^N(X,Y,t)\le C\frac{t^2}{2}\big(|^1X-{^1Y}|+|^2X-{^2Y}|\frac{t}{3}\big)e^{\sqrt{C}t},\ \forall t\in [0,1]
\end{align*} 
since for $t\in (0,1]$ it holds that 
{\allowdisplaybreaks
\begin{align*}
& C \int_0^t\int_0^s\big(  C\frac{r^2}{2}\big(|^1X-{^1Y}|+|^2X-{^2Y}|\frac{r}{3}\big)e^{\sqrt{C}r} \big)drds\\
&  +C\frac{t^2}{2}\big(|^1X-{^1Y}|+|^2X-{^2Y}|\frac{t}{3}\big)\\
 < & C\frac{t^2}{2}\big(|^1X-{^1Y}|+|^2X-{^2Y}|\frac{t}{3}\big) \Big(C\int_0^t\int_0^s e^{\sqrt{C}r} drds
  +1\Big)\\
< & C\frac{t^2}{2}\big(|^1X-{^1Y}|+|^2X-{^2Y}|\frac{t}{3}\big) e^{\sqrt{C}t} 
\end{align*}}
where for this single estimate we obviously had to keep the constant $C$ exceptionally fixed to check the assumptions of Lemma \eqref{Gron.lem.}.\\
This implies in turn for $t\in [0,1]$ that
\begin{align}
& \big|{^2\varphi^{N}_{t,0}}(X)-{^2\varphi^{N}_{t,0}}(Y)-({^2X}-{^2Y})\big| \notag \\
\le &\int_{0}^t |f^{N}*\widetilde{k}^{N}_s(^1\varphi^{N}_{s,0}(X))-f^{N}*\widetilde{k}^{N}_s(^1\varphi^{N}_{s,0}(Y))|ds \notag  \\
\le &C\int_{0}^t |^1\varphi^{N}_{s,0}(X)-{^1\varphi^{N}_{s,0}}(Y)|ds \notag  \\
\le &  C\int_{0}^t\big(|^1X-{^1Y}|+|^2X-{^2Y}|  s\big)\big(1+C\frac{s^2}{2}e^{\sqrt{C}s}\big)ds  \notag  \\
\le & Ct\big(|^1X-{^1Y}|+|^2X-{^2Y}|  t\big) \label{low.bound.v}
\end{align} 
\end{proof}
\noindent The last two lemmas will provide us the basis to introduce the following very important Corollary which yields us a tool to derive an upper bound for the impact two particles can in principle have on each other on $[0,T]$ in dependence on simple values like their minimal distance in space and their relative velocity at the moment when this minimum is attained. At first sight statement (i) of this Corollary seems to provide only information about the hypothetical impact of `collisions' between `mean-field particles'. But as long as related trajectories of micro- and macrosystem are close the derived upper bounds can (usually) be transferred from these auxiliary particles to the `real' ones. \\
\begin{cor}\label{cor1}
Let $k_0$ be a probability density fulfilling the assumptions of Theorem \ref{thm1} and $(\varphi^{N,c}_{t,s})_{t, s\in \mathbb{R}}$ be the related effective flow defined in \eqref{def.flow} as well as $(\Psi^{N,c}_{t,s})_{t,s\in \mathbb{R}}$ the $N$-particle flow defined in \eqref{Def.micro.sys.} for $\alpha\in (1,2]$ and $c>0$. Let additionally for $N,n\in \mathbb{N}$, $1<\widetilde{\alpha}\le 3$, $C_0>0$ and $c_N>0$ $h_N:\mathbb{R}^{3}\to \mathbb{R}^n$ be a continuous map fulfilling $$ |h_N(q)|\le\begin{cases} C_0 c_N^{-\widetilde{\alpha}},&\ |q|\le c_N\\\frac{C_0}{|q|^{\widetilde{\alpha}}},&\   |q|> c_N \end{cases}.$$
\begin{itemize}
\item[(i)] Let for $Y,Z\in \mathbb{R}^6$ $t_{min}\in [0,T]$ be a point in time where 
\begin{align*}
 \min_{0\le s \le T}|^1\varphi^{N,c}_{s,0}(Z)-{^1\varphi^{N,c}_{s,0}}(Y)|=&|^1\varphi^{N,c}_{t_{min},0}(Z)-{^1\varphi^{N,c}_{t_{min},0}}(Y)|=:\Delta r>0 \ \land \\
 &|^2\varphi^{N,c}_{t_{min},0}(Z)-{^2\varphi^{N,c}_{t_{min},0}}(Y)|=:\Delta v>0,
\end{align*}
then there exists $C_1>0$ (independent of $Y,Z\in \mathbb{R}^{6}$ and $N\in \mathbb{N}$) such that
\begin{align*}
& \int^{T}_{0}|h_N(^1\varphi^{N,c}_{s,0}(Z)-{^1\varphi^{N,c}_{s,0}}(Y))|ds
\le  C_1 \min\big(\frac{1}{\Delta r^{\widetilde{\alpha}}},\frac{1}{c_N^{\widetilde{\alpha}-1}\Delta v},\frac{1}{\Delta r^{\widetilde{\alpha}-1}\Delta v}\big).
\end{align*}
\item[(ii)] Let $T>0$, $i,j\in \{1,...,N\}$, $i\neq j$, $X\in \mathbb{R}^{6N}$ and $Y,Z\in \mathbb{R}^6$ be given such that for some $\delta>0$
$$N^{\delta}|^1\varphi^{N,c}_{t_{min},0}(Y)-{^1\varphi^{N,c}_{t_{min},0}}(Z)|\le |^2\varphi^{N,c}_{t_{min},0}(Y)-{^2\varphi^{N,c}_{t_{min},0}}(Z)|=:\Delta v$$
and
\[\sup_{0\le s \le T}|\varphi^{N,c}_{s,0}(Y)-[{\Psi^{N,c}_{s,0}}(X)]_i|\le N^{-\delta}\Delta v\land \sup_{0\le s \le T}|\varphi^{N,c}_{s,0}(Z)-[{\Psi^{N,c}_{s,0}}(X)]_j|\le N^{-\delta}\Delta v\]
where $t_{min}$ shall fulfill the same conditions as in item (i). Then there exist $N_0\in \mathbb{N}$ and $C_2>0$ (independent of $X\in \mathbb{R}^{6N}$, $Y,Z\in \mathbb{R}^6$) such that for all $N\geq N_0$ 
\begin{align*}
&\int^{T}_{0}|h_N([^1\Psi^{N,c}_{s,0}(X)]_i-[^1\Psi^{N,c}_{s,0}(X)]_j)|ds\\
\le & C_2\min\big(\frac{1}{c_N^{\widetilde{\alpha}-1}\Delta v}, \frac{1}{\min\limits_{0\le s\le T}|[^1\Psi^{N,c}_{s,0}(X)]_i-[^1\Psi^{N,c}_{s,0}(X)]_j|^{\widetilde{\alpha}-1}\Delta v}\big).
\end{align*}
\end{itemize}
\end{cor}
\vspace{0,4cm}
\begin{proof}
We put the proof for part (ii) of the Corollary in the appendix since the basic ideas are already included in the proof of case (i). Like in previous proofs the index $c$ will not be made explicit.\\
In a first step we want to derive an appropriate upper bound for the relative velocity between (mean-field) particles at times when they are `close' to each other. It will turn out by application of Lemma \ref{lem1} that the variables $\Delta r$ and $\Delta v$ which we introduced in the assumptions of the Corollary are sufficient to determine such a bound. To this end, we remark that according to Lemma \ref{lem1} there exists a constant $C_0\geq 1 $ such that for all $t\in [0,T]$ and $N\in \mathbb{N}$
\begin{align}
|\varphi^{N}_{t,0}(Z)-\varphi^{N}_{t,0}(Y)|\le C_0\min_{0\le s \le T}|\varphi^{N}_{s,0}(Z)-\varphi^{N}_{s,0}(Y)| . \label{cons.0}
\end{align}
Thus, it holds for arbitrary $t_1,t_2\in [0,T]$ that the condition 
\begin{align*}
&|^1\varphi^{N}_{t_1,0}(Z)-{^1\varphi^{N}_{t_1,0}}(Y)|\le  |^2\varphi^{N}_{t_1,0}(Z)-{^2\varphi^{N}_{t_1,0}}(Y)| 
\end{align*}
implies
\begin{align*}
|\varphi^{N}_{t_2,0}(Z)-{\varphi^{N}_{t_2,0}}(Y)|\le C_0|\varphi^{N}_{t_1,0}(Z)-{\varphi^{N}_{t_1,0}}(Y)|\le 2C_0 |^2\varphi^{N}_{t_1,0}(Z)-{^2\varphi^{N}_{t_1,0}}(Y)| . 
\end{align*}
Hence, in any case it holds that 
\begin{align*}
&\max\big(|^1\varphi^{N}_{t_1,0}(Z)-{^1\varphi^{N}_{t_1,0}}(Y)|,|^2\varphi^{N}_{t_1,0}(Z)-{^2\varphi^{N}_{t_1,0}}(Y)|\big)\\
\geq  &\max\big(|^1\varphi^{N}_{t_1,0}(Z)-{^1\varphi^{N}_{t_1,0}(Y)|,\frac{1}{2C_0}|^2\varphi^{N}_{t_2,0}}(Z)-{^2\varphi^{N}_{t_2,0}}(Y)|\big) \\
\geq &\ \frac{1}{ 2C_0 }  \max\big(\min_{0\le s \le T}|^1\varphi^{N}_{s,0}(Z)-{^1\varphi^{N}_{s,0}}(Y)|,|^2\varphi^{N}_{t_2,0}(Z)-{^2\varphi^{N}_{t_2,0}}(Y)|\big). 
\end{align*}
Let for $Y,Z\in \mathbb{R}^6$ $t_{min}\in [0,T]$ be a point in time where 
$$\min_{0\le s \le T}|^1\varphi^N_{s,0}(Z)-{^1\varphi^N_{s,0}}(Y)|=|^1\varphi^N_{t_{min},0}(Z)-{^1\varphi^N_{t_{min},0}}(Y)|=:\Delta r $$
as well as
 $$|^2\varphi^{N,c}_{t_{min},0}(Z)-{^2\varphi^{N,c}_{t_{min},0}}(Y)|=:\Delta v,$$
then the previous considerations (applied for $t_2=t_{min}$) yield that for any $t_1\in [0,T]$ the relation
\begin{align}
& \max\big(|^1\varphi^{N}_{t_1,0}(Z)-{^1\varphi^{N}_{t_1,0}}(Y)|,|^2\varphi^{N}_{t_1,0}(Z)-{^2\varphi^{N}_{t_1,0}}(Y)|\big)  \geq  \frac{1}{ 2C_0 }  \max\big(\Delta r, \Delta v\big) \label{cons.1}
\end{align}
is fulfilled which will be important shortly.\\  According to Lemma \ref{lem2} there exists $C_1>0$ (independent of $X,X'\in \mathbb{R}^6$ and $N$) such that for arbitrary $0\le t_0,t \le T$ where $|t-t_0|\le 1$ 
 \begin{align}
 & \big|{^2\varphi^N_{t,t_0}}(X)-{^2\varphi^N_{t,t_0}}(X')-({^2X}-{^2X'})\big| \notag\\
\le & C_1|t-t_0|\big(|{^1X }-{^1X'}|+|{^2X }-{^2X'}|  |t-t_0|\big) \label{lem3.free.eq.}.
\end{align} 
Let for $t\in [0,T]$ and $C_2\geq 0$ $t'_{min}\in [t,\min\big(t+\frac{1}{C_2},T\big)]=:I_{t}$ denote (one of) the point(s) in time where 
$$\min_{s\in I_{t}}|^1\varphi^N_{s,0}(Z)-{^1\varphi^N_{s,0}}(Y)|=|^1\varphi^N_{t'_{min},0}(Z)-{^1\varphi^N_{t'_{min},0}}(Y)| $$ and for a compact notation we abbreviate additionally
$$\widetilde{Z}:=\varphi^{N}_{t'_{min},0}(Z) \text{ and }\widetilde{Y}:=\varphi^{N}_{t'_{min},0}(Y).$$
If we choose $C_2=\lceil 2\sqrt{C_1} \rceil$ and regard the choice of $t'_{min}$ (in the third step), then relation \eqref{lem3.free.eq.} (applied for $t_0=t'_{min}$) yields that for $s\in I_t$
\begin{align*}
& |^1\varphi^{N}_{s,t'_{min}}(\widetilde{Z})-{^1\varphi^{N}_{s,t'_{min}}}(\widetilde{Y})| \\
\geq & \big| ({^1\widetilde{Z}}-{^1\widetilde{Y}})+({^2\widetilde{Z}}-{^2\widetilde{Y}})(s-t'_{min})\big|\\
& -\big|\int_{t'_{min}}^s\big( ^2\varphi^{N}_{r,t'_{min}}(\widetilde{Z})-{^2\varphi^{N}_{r,t'_{min}}}(\widetilde{Y})\big)-({^2\widetilde{Z}}-{^2\widetilde{Y}})dr\big| \\
\geq &\big|({^1\widetilde{Z}}-{^1\widetilde{Y}})+({^2\widetilde{Z}}-{^2\widetilde{Y}})(s-t'_{min})\big|\\
& - C_1\underbrace{|s-t'_{min}|^2}_{\le( \frac{1}{C_2})^2\le \frac{1}{4C_1}}\big(|{^1\widetilde{Z}}-{^1\widetilde{Y}}|+|{^2\widetilde{Z}}-{^2\widetilde{Y}}||s-t'_{min}|\big)\\
\geq & \max\big(|{^1\widetilde{Z}}-{^1\widetilde{Y}}|,|{^2\widetilde{Z}}-{^2\widetilde{Y}}||s-t'_{min}| \big)-\frac{1}{4}(|{^1\widetilde{Z}}-{^1\widetilde{Y}}|+|{^2\widetilde{Z}}-{^2\widetilde{Y}}||s-t'_{min}|)\\
\geq & \frac{1}{2} \max\big(|{^1\widetilde{Z}}-{^1\widetilde{Y}}|,|{^2\widetilde{Z}}-{^2\widetilde{Y}}||s-t'_{min}| \big)
\end{align*}
which implies that
\begin{align}
&\int^{\min(t+\frac{1}{C_2},T)}_{t}|h_N(^1\varphi^{N}_{s,0}(Z)-{^1\varphi^{N}_{s,0}}(Y))|ds \notag \\
\le &C\int^{\frac{1}{C_2}}_{0}\min\Big(\frac{1}{\max\big(|^1\widetilde{Z}-{^1\widetilde{Y}}|,|^2\widetilde{Z}-{^2\widetilde{Y}}|s\big)^{\widetilde{\alpha}}},c_N^{-\widetilde{\alpha}}\Big)ds \notag \\
\le &C\min\Big(\frac{1}{c_N^{\widetilde{\alpha}-1}|^2\widetilde{Z}-{^2\widetilde{Y}}|},\frac{1}{|^1\widetilde{Z}-{^1\widetilde{Y}}|^{\widetilde{\alpha}-1}|^2\widetilde{Z}-{^2\widetilde{Y}}|}\Big) \label{coll.est.1}
\end{align}
where we used the properties of $h_N$ stated in the assumptions of the Corollary. \\
Moreover, the constraints on $h_N$ directly imply that
$$\int^{t+\frac{1}{C_2}}_{t}|h_N(^1\varphi^{N}_{s,0}(Z)-{^1\varphi^{N}_{s,0}}(Y))|ds\le \frac{1}{C_2}\min\big(\frac{1}{|^1\widetilde{Z}-{^1\widetilde{Y}}|^{\widetilde{\alpha}}},\frac{1}{c_N^{\widetilde{\alpha}}}\big).$$
After merging the upper bounds it follows that 
\begin{align}
& \int^{\min(t+\frac{1}{C_2},T)}_{t}|h_N(^1\varphi^{N}_{s,0}(Z)-{^1\varphi^{N}_{s,0}}(Y))|ds \notag \\
\le & C\min\big(\frac{1}{|^1\widetilde{Z}-{^1\widetilde{Y}}|^{\widetilde{\alpha}}},\frac{1}{c_N^{\widetilde{\alpha}}},\frac{1}{c_N^{\widetilde{\alpha}-1}|^2\widetilde{Z}-{^2\widetilde{Y}}|},\frac{1}{|^1\widetilde{Z}-{^1\widetilde{Y}}|^{\widetilde{\alpha}-1}|^2\widetilde{Z}-{^2\widetilde{Y}}|}\big). \label{est.Cor}
\end{align}
What we have done so far is deriving an upper bound for the integral over the desired function but only for a (possibly) short interval $[t,\min(t+\frac{1}{C_2},T)]$ belonging to $[0,T]$ where the starting point $t$ can be selected arbitrarily. However, we recall that its length $\frac{1}{C_2}$ can be chosen independent of $N$ and the considered configurations.  Moreover, it will turn out that for any such interval the respective upper limit can be bounded itself by application of $\Delta r$ and $\Delta v$ which will finally enable us to show the desired result. More precisely, we apply that according to relation \eqref{cons.1}
$$\max\big(|^1\widetilde{Z}-{^1\widetilde{Y}}|,|^2\widetilde{Z}-{^2\widetilde{Y}}|\big)  \geq  \frac{1}{ 2C_0 }  \max\big(\Delta r, \Delta v\big)$$ as well as $|^1\widetilde{Z}-{^1\widetilde{Y}}|\geq\Delta r$ to obtain:
{\allowdisplaybreaks
\begin{align*}
& \min\big(\frac{1}{|^1\widetilde{Z}-{^1\widetilde{Y}}|^{\widetilde{\alpha}}},\frac{1}{c_N^{\widetilde{\alpha}}},\frac{1}{c_N^{\widetilde{\alpha}-1}|^2\widetilde{Z}-{^2\widetilde{Y}}|},\frac{1}{|^1\widetilde{Z}-{^1\widetilde{Y}}|^{\widetilde{\alpha}-1}|^2\widetilde{Z}-{^2\widetilde{Y}}|}\big)\\ 
=  & \min\big(\frac{1}{|^1\widetilde{Z}-{^1\widetilde{Y}}|^{\widetilde{\alpha}-1}},\frac{1}{c_N^{\widetilde{\alpha}-1}}\big)\min\big(\frac{1}{c_N},\frac{1}{|^1\widetilde{Z}-{^1\widetilde{Y}}|},\frac{1}{|^2\widetilde{Z}-{^2\widetilde{Y}}|} \big)\\
\le &   \min\big(\frac{1}{\Delta r^{\widetilde{\alpha}-1}},\frac{1}{c_N^{\widetilde{\alpha}-1}}\big)\min\big(\frac{1}{c_N},\frac{2C_0}{\Delta r},\frac{2C_0}{\Delta v} \big)\\  
\le & C\min\big(\frac{1}{\Delta r^{\widetilde{\alpha}}},\frac{1}{c_N^{\widetilde{\alpha}-1}\Delta v},\frac{1}{\Delta r^{\widetilde{\alpha}-1}\Delta v}\big)
\end{align*} }
Thus, estimates \eqref{est.Cor} and the previous comment concerning the constant $C_2$ imply that 
\begin{align*}
& \int^{T}_{0}|h_N(^1\varphi^{N}_{s,0}(Z)-{^1\varphi^{N}_{s,0}}(Y))|ds\\
\le &\big\lceil\frac{T}{\frac{1}{C_2}}\big\rceil \sup_{0\le t\le T}\int^{\min(t+\frac{1}{C_2},T)}_{t}|h_N(^1\varphi^{N}_{s,0}(Z)-{^1\varphi^{N}_{s,0}}(Y))|ds\\
\le &  C\min\big(\frac{1}{\Delta r^{\widetilde{\alpha}}},\frac{1}{c_N^{\widetilde{\alpha}-1}\Delta v},\frac{1}{\Delta r^{\widetilde{\alpha}-1}\Delta v}\big).
\end{align*}
which completes the proof of statement (i). 
\end{proof}
\noindent As mentioned before, this Corollary will be essential for the collision estimates later. If we choose for example $h_N:=|f^N|$, then it basically tells us that it suffices to know the minimal distance between two (mean-field) particles and their relative velocity at a point in time when this minimum is attained, to determine a suitable upper bound for the impact they have on each other on the whole time interval $[0,T]$ (or in case of the `mean-field particles' rather for the value of the integral which is considered in the Corollary since they do not interact).\\\\
\noindent Thus, it remains to determine an appropriate upper bound for the probability of the different kinds of collisions before we can start with the proofs of the main results.
\begin{samepage}
\begin{lem} \label{lem3}
Let $k_0$ be a probability density fulfilling the assumptions of Theorem \ref{thm1} and $(\varphi^{N,c}_{t,s})_{t, s\in \mathbb{R}}$ be the related effective flow defined by \eqref{def.flow} for $1<\alpha\le 2$ and $c>0$. Then there exists $C_1>0$ such that for all $\Delta x,\Delta v >0$, $N\in \mathbb{N}$,  $Y\in \mathbb{R}^6$ and $[t_1,t_2] \subseteq  [0,T]$ it holds that 
\begin{itemize}
\item[(i)]
$\begin{aligned}&\mathbb{P}\Big(X\in \mathbb{R}^6:(\exists t\in [t_1,t_2]:|^1\varphi^{N,c}_{t,0}(X)-{^1\varphi^{N,c}_{t,0}}(Y)|\le \Delta x\  \land \\
&\hspace{4,36cm} |^2\varphi^{N,c}_{t,0}(X)-{^2\varphi^{N,c}_{t,0}}(Y)|\le \Delta v)\Big)\\
 &\le   C_1\big(\Delta x^2\Delta v^4(t_2-t_1)+\Delta x^3\max\big( \Delta x,\Delta v\big)^3\big)
\end{aligned}$
\item[(ii)]
$ \begin{aligned}
& \mathbb{P}\Big(X\in \mathbb{R}^6:\min_{0\le s\le T}|^1\varphi^{N,c}_{s,0}(X)-{^1\varphi^{N,c}_{s,0}}(Y)|\le \Delta x\Big)\le C_1 \Delta x^2
\end{aligned}$
\item[(iii)]
$ \begin{aligned} & \mathbb{P}\Big((Z,X)\in \mathbb{R}^{12}:\big(\exists t\in [t_1,t_2]:|^1\varphi^{N,c}_{t,0}(X)-{^1\varphi^{N,c}_{t,0}}(Z)|\le \Delta x \big) \Big)\\ 
& \le C_1 \big(\Delta x^3+\Delta x^2(t_2-t_1)\big).
\end{aligned}$
\end{itemize}
\end{lem}
\end{samepage}
\vspace{0,4cm}
\begin{proof}
We only make the proof related to item $(i)$ explicit here since the reasoning for the remaining statements is essentially very similar but still quite elongated. However, we will give at least a short heuristic reasoning which shall suggest that the statements of items $(ii)$ and $(iii)$ are plausible. The detailed proofs, on the other hand, can be found in the appendix (resp. in chapter \ref{sec.app.}).\\
The constants which are applied in the proof do not depend on $t_1,t_2,\Delta x$ or $\Delta v$ and (as always) not on $N$ or the considered configuration. Since the value of the cut-off parameter $c$ has no relevance for the proof, we omit to indicate it in the notation of the different objects.\\ If, in accordance with the assumption belonging to $(i)$, there exists a point in time $t\in [t_1,t_2]$ such that
\[|^1\varphi^{N}_{t,0}(X)-{^1\varphi^{N}_{t,0}}(Y)|\le \Delta x \ \land \ |^2\varphi^{N}_{t,0}(X)-{^2\varphi^{N}_{t,0}}(Y)|\le \Delta v,\]
then it follows in turn by Lemma \ref{lem1} that
\begin{align}
\sup_{0\le s \le T}|\varphi^{N}_{s,0}(X)-{\varphi^{N}_{s,0}}(Y)|\le C\max(\Delta v,\Delta x) \label{upp.b.mean.flow-lem3}
\end{align}
for some appropriate constant $C>0$.\\ We consider different cases: The first possibility is that already at the starting time of the interval $[t_1,t_2]$ the positions of the `particles' fulfill $|^1\varphi^{N}_{t_1,0}(X)-{^1\varphi^{N}_{t_1,0}}(Y)|\le \Delta x $. The estimates for this case are straightforward and respecting \eqref{upp.b.mean.flow-lem3} it follows that the probability of configurations fulfilling this assumption is bounded by
\begin{align*}
& \int_{\mathbb{R}^6}\mathbf{1}_{\{Z\in \mathbb{R}^{6}:|^1Z-{^1\varphi^{N}_{t_1,0}}(Y)|\le \Delta x   \land  |^2Z-{^2\varphi^{N}_{t_1,0}}(Y)|\le C\max(\Delta v,\Delta x) \}}(X)k^{N}_{t_1}(X)d^6X\\
\le & C\|k_{0}\|_{\infty}\big(\Delta x ^3 \max (\Delta x, \Delta v)^3\big).
\end{align*} 
This bound is obviously small enough so that it complies with statement $(i)$.\\
Otherwise the assumption belonging to $(i)$ implies together with \eqref{upp.b.mean.flow-lem3} that for $M \in \mathbb{N}$ there exists $n \in \{0,...,M-1\}$ such that
\begin{align*}
 \Delta x \le & |{^1\varphi^{N}_{t'_n,0}}(X)-{^1\varphi^{N}_{t'_n,0}}(Y)| \le \Delta x +C\max(\Delta x,\Delta v) \frac{t_2-t_1}{M} \\
& |{^2\varphi^{N}_{t'_n,0}}(X)-{^2\varphi^{N}_{t'_n,0}}(Y)|  \le C \max(\Delta x,\Delta v) 
\end{align*}
where we abbreviated $t'_n:=t_1+n\frac{t_2-t_1}{M}$.\\
If we abbreviate additionally $\Delta_{x,v}:=C \max(\Delta x,\Delta v)$, then we get the following upper limit for the probability of configurations fulfilling this constraint:
\begin{align*}
&\sum_{n=0}^{M-1}\int\limits_{\mathbb{R}^{6}}\mathbf{1}_{\{Z\in \mathbb{R}^3:\Delta x \le |Z-{^1\varphi^{N}_{t'_n,0}}(Y)|\le \Delta x+\Delta_{x,v}   \frac{t_2-t_1}{M}    \}}({^1X})\\
& \cdot \mathbf{1}_{\{Z\in \mathbb{R}^3:  |Z-{^2\varphi^{N}_{t'_n,0}}(Y)|\le \Delta_{x,v}  \}}({^2X})k^{N}_{t'_n }(X)d^6X\\
\le & C M\|k_0\|_{\infty} \big((\Delta x+\Delta_{x,v}  \frac{t_2-t_1}{M} )^3-\Delta x^3 \big)\Delta_{x,v}^3 
\end{align*}
By choosing $M$ sufficiently large it follows that this term is bounded by 
\begin{align*}
& C\Delta x^2 \Delta_{x,v}^4(t_2-t_1)=C\Delta x^2 \max(\Delta x,\Delta v)^4(t_2-t_1) \\
\le &   C\big(\Delta x^6+\Delta x^2\Delta v^4\big)(t_2-t_1)
\end{align*} 
and the first part of the lemma follows.\\
As mentioned before, the proofs to items $(ii)$ and $(iii)$ can be found in the appendix. At this point we give at least a short heuristic idea to make these statements plausible. The statement related to item $(ii)$ might seem wrong at first sight since the upper bound for the probability is independent of the velocity of the considered configuration $Y$ though a fast traveling `particle' should  potentially be able to come close to a certain amount of `mass' in a shorter time span than a slower one. However, on the other hand the mean-field force is bounded and the considered initial density $k_0$ decays quite fast. Thus, a particle with a very large relative velocity can be pictured like a bullet flying through a slowly evolving concentrated `dust cloud'. It will move inevitably to areas of lower density so that after a certain time the amount of `mass' which it approaches becomes negligible.\\
As mentioned before, for short time spans a rapidly moving particle should in principle be able to come close to a larger amount of `mass' than a slower one. We argued that above a certain value a further increase of the velocity has basically no effect in this regards if the considered time span is long enough because the density decreases too fast. But for short time spans the previous reasoning does not work anymore. Hence, the statement of item $(iii)$ also seems to be wrong at first sight. Here, however, in contrast to item $(ii)$ both initial configurations are chosen randomly. Due to the bounded kinetic energy related to $k_0$ very fast `particles' are much less likely than slower ones (or a bit more formally the total mass related to very fast characteristics is distinctly smaller) so that these two effects basically cancel and the upper bound for the stated probability can indeed be indicated in the current form.
\end{proof} 
\newpage
\subsection{Implementation of the proof}
\begin{figure}[!h]
   \centering
   \def\svgscale{1.2}
 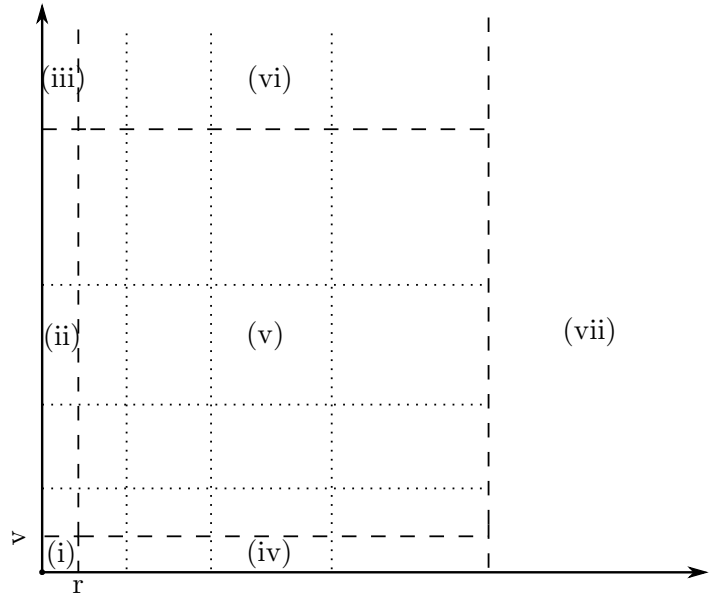
   \caption{Schematic diagram of the cover}
   \label{partition}
\end{figure}

\noindent After having completed the preliminary considerations we are finally able to start with the proof of the first main result.\\\\
\noindent \textit{{\bf Proof of Theorem \ref{thm1}:}}\\
Since large parts of the estimates are very similar for statement $(i)$ and $(ii)$, we want to present a unified proof. For the parts where the estimates differ we simply treat both cases separately. During the whole proof the cut-off parameter will take the value $c:=\frac{2}{3}$ if $\alpha \in (1,\frac{4}{3}]$ (see item (ii) of Theorem \ref{thm1}) and $c:=\frac{7}{18}-\sigma$ if $\alpha =2$ (see item (i) of Theorem \ref{thm1}). Thus, we drop related indices in the notation but it should become obvious from the context which of the two possibilities is considered in the respective situation (at least if this is relevant at all). It is obvious that the statements of the Theorem are stronger (or in case of $(i)$ at least harder to prove) the smaller the appearing parameter $\sigma>0$ is. Hence, it is not surprising that many objects and sets appearing in the proof have to be be defined $\sigma$-dependent. What we actually will show is that there exists $\sigma^*>0$ such that for any given $\sigma\in (0, \sigma^*]$ the statements related to items $(i)$ and $(ii)$ are valid. Consequently, we will assume on certain occasions that $\sigma>0$ is `small' if this is beneficial for the estimates. Although this will not be made explicit, it is straightforward to see that the remaining cases where the given $\sigma$ shall be larger than $\sigma^*$ can be handled by essentially the same proof, with the slight difference that $\sigma$ needs to be replaced by (the fixed value) $\sigma^*$ in the definitions of all introduced $\sigma$-dependent objects (respectively sets). The largest part of the proof is  concerned with showing that $\sup_{0\le s \le T}|\Psi_{s,0}^{N}(X)-\Phi^{N}_{s,0}(X)|_{\infty}$ keeps typically sufficiently small for large enough $N$. In a concluding step we will show that a corresponding statement holds for $\sup_{x\in \mathbb{R}^6}\sup_{0\le s \le T}|\varphi^{\infty}_{s,0}(x)-\varphi^N_{s,0}(x)|$ which together yields the desired result.\\ 
First, we define certain sets which can be understood as `collision classes' and which are very important throughout the proofs. Let for $r,R,v,V\in \mathbb{R}_{\geq 0}\cup \{\infty\}$, $t_1,t_2\in [0,T]$ and $Y\in \mathbb{R}^6$ the sets $M^{N,(t_1,t_2)}_{(r,R),(v,V)}(Y)\subseteq \mathbb{R}^{6}$ be defined as follows
\begin{align}
\begin{split}
& Z\in M^{N,(t_1,t_2)}_{(r,R),(v,V)}(Y)\subseteq  \mathbb{R}^6 \\
\Leftrightarrow &Z\neq Y\ \land\ \exists t\in [t_1,t_2]: \\ &r\le \min_{t_1\le s \le t_2}|^1\varphi^N_{s,0}(Z)-{^1\varphi^N_{s,0}}(Y)|=|^1\varphi^{N}_{t,0}(Z)-{^1\varphi^{N}_{t,0}}(Y)|\le R \ \land  \\
&  v\le |^2\varphi^{N}_{t,0}(Z)-{^2\varphi^{N}_{t,0}}(Y)|\le  V. \label{def.coll.cl.1}
\end{split}
\end{align} 
As mentioned, this set will also appear in many of the subsequent proofs and $(\varphi^N_{s,r})_{r,s\in \mathbb{R}}$ shall always be understood as the flow related to the respectively considered initial density (which is always referred to as $k_0$) where the cut-off parameter takes value $c=\frac{2}{3}$ if $\alpha\in (1,\frac{4}{3}]$ and $c=\frac{7}{18}-\sigma$ if $\alpha=2$ (unless explicitly stated otherwise). For ease of notation we abbreviate certain special cases:
\begin{align*}
M^{N,(t_1,t_2)}_{R,V}(Y)&:=M^{N,(t_1,t_2)}_{(0,R),(0,V)}(Y)\\
M^{N}_{(r,R),(v,V)}(Y)&:=M^{N,(0,T)}_{(r,R),(v,V)}(Y)\\
M^{N}_{R,V}(Y)&:=M^{N,(0,T)}_{(0,R),(0,V)}(Y)
\end{align*}
Moreover, we will define a set $G^N(Y)\subseteq \mathbb{R}^6$ by application of such `collision classes' to distinguish between problematic and unproblematic collisions. Unfortunately, we need different definitions for the set $G^N(Y)$ depending whether the case $\alpha=2$ or $\alpha \in (1,\frac{4}{3}]$ is considered. Thus, we delay stating the respective definitions to the moment where they are needed for the first time.\\
Next, we split the particles in two groups: A `bad' group where (for our purposes) hard collisions are expected to happen and a group of the remaining `good' particles:  \\
\begin{align}
\begin{split}
& \mathcal{M}^N_g(X):=\{i\in \{1,...,N\}:\big(\forall j\in \{1,...,N\}\setminus\{i\}: X_j\in  G^N(X_i)\big)\} \\
& \mathcal{M}^N_b(X):=\{1,...,N\}\setminus \mathcal{M}^N_g(X)
\end{split} 
\end{align}
Since $G^N(Y)$ will be defined solely by application of the `collision classes', it turns out that it depends only on their corresponding `mean-field particle', whether a particle is considered `good' or `bad'.\\
Now we define the following stopping times
\begin{align}
&\tau^N_g(X):=\sup\{t\in [0,T]:\max_{i\in \mathcal{M}^N_g(X)}\sup_{0\le s \le t}|[\Psi^{N}_{s,0}(X)]_i-{\varphi}^{N}_{s,0}(X_i)|\le \delta^N_g\}  \notag \\
&\tau^N_b(X):=\sup\{t\in [0,T]:\max_{i\in \mathcal{M}^N_b(X)}\sup_{0\le s \le t}|[\Psi^{N}_{s,0}(X)]_i-{\varphi}^{N}_{s,0}(X_i)|\le \delta^N_b\} \label{Def.stopping}
\end{align}
as well as $\tau^N(X):=\min(\tau^N_g(X),\tau^N_b(X))$ where
$$\delta^N_g=\delta^N_b= N^{-\frac{1}{2}+\sigma} \text{ if } \alpha\in (1,\frac{4}{3}]$$ and
$$\delta^N_g= N^{-c}=N^{-\frac{7}{18}+\sigma} \text{ and }\delta^N_b:=N^{-\frac{2}{9} -\sigma} \text{ if } \alpha=2.$$
If we are able to show that the probability of configurations fulfilling $\tau^N (X)< T$ becomes sufficiently small for large values of $N$, then Theorem \ref{thm1} follows.\\ 
 Moreover, we remark that in principle the distinction between `good' and `bad' particles is only necessary for the proof of statement $(i)$ where the value of the singularity parameter fulfills $\alpha=2$. As a consequence the confusing choice $\delta^N_g=\delta^N_b$ arises in the case $\alpha\in (1,\frac{4}{3}]$ which yields that the allowed deviation between `real' and related `mean-field particle' according to the stopping times is the same. Thus, no advantage for the proof is obtained by the distinction but this is simply a consequence of the decision to present the proofs of both statements at the same time. \\
By $\frac{d}{dt_+}$ we will denote the right derivative and it obviously holds for $i\in \{1,...,N\}$ that
\begin{align*}
& \frac{d}{dt_+}\sup_{0\le s\le t}|[^1\Psi^N_{s,0}(X)]_i-{^1\varphi^N_{s,0}}(X_i)|\\ \le &  |[^2\Psi^N_{t,0}(X)]_i-{^2\varphi^N_{t,0}}(X_i)|\\
\le &|\int_{0}^t\frac{1}{N}\sum_{j\neq i}f^{N}([^1\Psi^N_{s,0}(X)]_i-[^1\Psi^N_{s,0}(X)]_j)-f^{N}*\widetilde{k}^N_s({^1\varphi^N_{s,0}}(X_i))ds|.
\end{align*}
Our primary aim in the following is to derive an appropriate upper bound for the last term. To this end, we need to distinguish if the considered particle belongs to the set of the `good' or the `bad'. The reason for this lies in our assumption on the closeness of a particle to its related `mean-field particle' which results in having less information on the positions of `bad' compared to `good' particles in the case $\alpha=2$. The proof of statement $(ii)$ (concerning $\alpha \in (1,\frac{4}{3}]$) will already be concluded in the current part because we will show in the end that typically no `bad' particle occurs in this case. For statement $(i)$ on the other hand further considerations will be necessary.\\ 
Thus, we start with the case that the observed particle $i$ is `good' which means that only configurations $X$ are considered where $i\in \mathcal{M}^N_g(X)$. 
\paragraph{Controlling the deviations of the `good' particles:}
First, it obviously holds for $i\in \mathcal{M}^N_g(X)$ and $0\le t_1\le t\le T$ that
{\allowdisplaybreaks \begin{align}
 & \big|\int_{t_1}^t\frac{1}{N}\sum_{j\neq i}f^{N}([^1\Psi^N_{s,0}(X)]_i-[^1\Psi^N_{s,0}(X)]_j)-f^{N}*\widetilde{k}^N_s({^1\varphi^N_{s,0}}(X_i))ds\big| \label{term00}\\
\le & \big|\int_{t_1}^t\frac{1}{N}\sum_{j\neq i}f^{N}([^1\Psi^N_{s,0}(X)]_i-[^1\Psi^N_{s,0}(X)]_j)\mathbf{1}_{(G^N(X_i))^C}(X_j)ds\big| \notag\\
& + \big|\int_{t_1}^t\Big(\frac{1}{N}\sum_{j\neq i}f^{N}([^1\Psi^N_{s,0}(X)]_i-[^1\Psi^N_{s,0}(X)]_j)\mathbf{1}_{G^N(X_i)}(X_j) \notag \\
&-f^{N}*\widetilde{k}^N_s({^1\varphi^N_{s,0}}(X_i))\Big)ds\big| .
\end{align}
Then we apply multiple times triangle inequality to show that the previous term is bounded by
\begin{align}
 & \big|\int_{t_1}^t\frac{1}{N}\sum_{j\neq i}f^{N}([^1\Psi^N_{s,0}(X)]_i-[^1\Psi^N_{s,0}(X)]_j)\mathbf{1}_{(G^N(X_i))^C}(X_j)ds\big| \label{rel.t1} \\
& + \big|\int_{t_1}^t\frac{1}{N}\sum_{j\neq i}\Big(  f^{N}([^1\Psi^N_{s,0}(X)]_i-[^1\Psi^N_{s,0}(X)]_j)\mathbf{1}_{G^N(X_i)}(X_j)  \notag \\
&-f^{N}(^1\varphi^N_{s,0}(X_i)-{^1\varphi^N_{s,0}}(X_j))\mathbf{1}_{G^N(X_i)}(X_j)\Big) ds\big| \label{rel.t3} \\
& +\big| \int_{t_1}^t \frac{1}{N}\sum_{j\neq i}f^{N}(^1\varphi^N_{s,0}(X_i)-{^1\varphi^N_{s,0}}(X_j))\mathbf{1}_{G^N(X_i)}(X_j)ds \notag \\
&-\int_{t_1}^t\int_{\mathbb{R}^6}f^N({^1\varphi^N_{s,0}}(X_i)-{^1\varphi^N_{s,0}}(Y))\mathbf{1}_{G^N(X_i)}(Y)k_0(Y)d^6Yds\big| \label{rel.t5}  \\
& +\big|\int_{t_1}^t\int_{\mathbb{R}^6}f^N({^1\varphi^N_{s,0}}(X_i)-{^1\varphi^N_{s,0}}(Y))\mathbf{1}_{G^N(X_i)}(Y)k_0(Y)d^6Yds\notag\\
&-\int_{t_1}^tf^{N}*\widetilde{k}^N_s(^1\varphi^N_{s,0}(X_i))ds\big| \label{rel.t7} 
\end{align}}
\noindent 
In the following we derive subsequently upper bounds for the four terms \eqref{rel.t1}, \eqref{rel.t3}, \eqref{rel.t5} and \eqref{rel.t7} starting with the simplest and concluding with the most complex.\\
{\bf Considerations for term \eqref{rel.t1}:} Since at the moment we only consider configurations where $i\in \mathcal{M}^N_g(X)$, it follows that $\sum_{j\neq i}\mathbf{1}_{(G^N(X_i))^C}(X_j)=0$. Thus, term \eqref{rel.t1} vanishes in this case and we only have to take the remaining terms into account.\\
{\bf Considerations for term \eqref{rel.t7}:} It is straightforward to derive an upper bound for term \eqref{rel.t7} since it holds due to \eqref{Liouville} that
\begin{align*}
& f^{N}*\widetilde{k}^N_s({^1\varphi^N_{s,0}}(X_i))\\
=& \int_{\mathbb{R}^6}f^{N}({^1\varphi^N_{s,0}}(X_i)-{^1Y})k^N_s(Y)d^6Y\\
=& \int_{ \mathbb{R}^6}f^{N}({^1\varphi^N_{s,0}}(X_i)-{^1\varphi^N_{s,0}}(Y))k_0(Y)d^6Y,
\end{align*}
which yields
\begin{align}
& \big|\int_{t_1}^t\int_{ \mathbb{R}^6}f^{N}(^1\varphi^N_{s,0}(X_i)-{^1\varphi^N_{s,0}}(Y))k_0(Y)\mathbf{1}_{G^N(X_i)}(Y)d^6Yds \notag \\
&-\int_{t_1}^tf^{N}*\widetilde{k}^N_s({^1\varphi^N_{s,0}}(X_i))ds\big| \notag \\
= & \big|\int_{t_1}^t\int_{ \mathbb{R}^6}f^{N}(^1\varphi^N_{s,0}(X_i)-{^1\varphi^N_{s,0}}(Y))k_0(Y)(\mathbf{1}_{G^N(X_i)}(Y)-1)d^6Yds\big| \notag \\
\le & T\|f^{N}\|_{\infty}\int_{ \mathbb{R}^6}\mathbf{1}_{(G^N(X_i))^C}(Y)k_0(Y)d^6Y \notag \\
\le & TN^{\alpha c}\mathbb{P}\big(Y \in \mathbb{R}^6:Y\notin G^N(X_i) \big). \label{bound1.0}
\end{align}
As mentioned the `good' set $G^N(Z)$ will be defined differently depending if $\alpha\in (1,\frac{4}{3}]$ or $\alpha=2$. Hence, we need two distinguish between these two options for estimating \eqref{bound1.0}.\\
{\bf Estimates for term \eqref{bound1.0} if $\alpha=2$:} In this case we identify for $Z\in \mathbb{R}^6$
\begin{align}
 G^N(Z):=\big(M^N_{6N^{-\frac{2}{9}-\sigma },N^{-\frac{2}{9}}}(Z)\big)^C \label{Def G.alpha=2}.
\end{align}
where the reason for this choice will become clear later in the proof. Then it holds due to item $(i)$ of Lemma \ref{lem3} that
\begin{align}
& \mathbb{P}\big(Y\in \mathbb{R}^6:Y\notin G^N(X_i) \big)\notag \\
= & \mathbb{P}\big(Y\in \mathbb{R}^6:Y\in M^N_{6N^{-\frac{2}{9}-\sigma },N^{-\frac{2}{9}}}(X_i) \big) \notag \\
 \le & C(N^{-\frac{2}{9}-\sigma})^2(N^{-\frac{2}{9}})^4 \notag \\
\le &  CN^{-\frac{4}{3}-2 \sigma} \label{prob G^N, alpha=2} 
\end{align}
This yields for term \eqref{bound1.0} that
\begin{align}
 & TN^{\alpha c}\mathbb{P}\big(Y \in \mathbb{R}^6:Y\notin G^N(X_i) \big)\le C N^{2(\frac{7}{18}-\sigma)}N^{-\frac{4}{3}-2 \sigma}\le CN^{-\frac{5}{9}} \label{bound1, alpha=2}
 \end{align}
{\bf Estimates for term \eqref{bound1.0} if $\alpha\in (1,\frac{4}{3}]$:} Here, the definition of the set $G^N(Z)$ is slightly more complex:
\begin{align} 
G^N(Z):=\bigcap_{\substack{k\in \mathbb{N}:\\
\frac{3}{4}-{\sigma} \le k {\sigma} \le 1}}\Big(M^N_{6N^{-\frac{{k\sigma}}{2} },N^{-\frac{1}{2}+ \frac{k\sigma}{6}+\frac{\sigma}{2}}}(Z) \cup  M^N_{ 6N^{-\frac{1}{2}+{\sigma}},N^{-\frac{5}{18}}}(Z)\Big)^C\label{Def G, alpha=4/3}
\end{align} 
Now it follows analogously to the previous case that
{\allowdisplaybreaks \begin{align}
& \mathbb{P}\big(Y \in \mathbb{R}^6:Y\notin  G^N(X_i)\big) \notag \\
\le &  \mathbb{P}\big( Y \in M^N_{ 6N^{-\frac{1}{2}+{\sigma}},N^{-\frac{5}{18}}}(X_i)\big) \notag \\
& + C\max_{\substack{k\in \mathbb{N}:\\
\frac{3}{4}-{\sigma} \le k {\sigma} \le 1}}\mathbb{P}\big(Y \in M^N_{6N^{-\frac{k\sigma}{2} },N^{-\frac{1}{2}+ \frac{k\sigma}{6}+\frac{\sigma}{2}}}(X_i)\big)\notag \\
\le & C (N^{-\frac{1}{2}+{\sigma}})^2(N^{-\frac{5}{18}})^{4} \notag \\
&+ C\max_{\substack{k\in \mathbb{N}:\\
\frac{3}{4}-{\sigma} \le k \sigma \le 1}}\big((N^{-\frac{k\sigma }{2} })^2 (N^{-\frac{1}{2}+ \frac{k\sigma}{6}+\frac{\sigma}{2}})^4\big) \notag \\
\le &  C (N^{-\frac{19}{9}+2{\sigma}}+\underbrace{\max_{\substack{k\in \mathbb{N}\\
\frac{3}{4}-{\sigma} \le k {\sigma} \le 1}}N^{-2+2\sigma-\frac{k\sigma}{3}}}_{\le CN^{-\frac{9}{4}+\frac{7}{3}\sigma}}) \label{prob G^N, alpha=4/3}
\end{align}} 
where the last term is bounded by $ C N^{-\frac{19}{9}+2{\sigma}}$ if $\sigma>0 $ is chosen sufficiently small (which in the current situations means $\sigma\le \frac{5}{12}$ and as mentioned at the beginning of the proof can be assumed without restriction). Thus, in this case term \eqref{bound1.0} is bounded by 
\begin{align}
 & TN^{\alpha c}\mathbb{P}\big(Y \in \mathbb{R}^6:Y\notin G^N(X_i) \big)\le T N^{\frac{4}{3}\frac{2}{3}}(CN^{-\frac{19}{9}+2 \sigma})\le CN^{-\frac{11}{9}+2\sigma} \label{bound1, alpha=4/3}.
 \end{align}
which concludes the estimates for this term.\\\\
\noindent For the remaining terms we need a version of the law of large numbers.
\begin{lem}\label{largenumberslem}
Let $\delta,C_0>0$, $N{\in \mathbb{N}} $ and let $(X_{k})_{k\in \mathbb{N}}$ be a sequence of i.i.d. random variables $X_k:\Omega \to \mathbb{R}^6$ distributed with respect to a probability density $k\in \mathcal{L}^1(\mathbb{R}^6)$. Moreover, let $(M^N_i)_{i\in I}$ be a family of (possibly $N$-dependent) sets $M^N_i\subseteq \mathbb{R}^6$ fulfilling $\bigcup_{i \in I}M^N_i=\mathbb{R}^6$ where $|I|<C_0$ and $h_N\colon \mathbb{R}^6\to \mathbb{R}$ measurable functions which fulfill on the one hand $\|h_N\|_{\infty}\le C_0N^{1-\delta}$ and on the other hand
\[\max_{i \in I}\int_{M^N_i}h_N(X)^2k(X)d^6X\le C_0N^{1-\delta}.\]
Then for any $\gamma>0$ there exists a constant $C_{1}>0$ such that for all $N\in \mathbb{N}$ it holds that
\begin{align*}
& \mathbb{P}\big(|\frac{1}{N}\sum_{k=1}^Nh_N(X_k)-\int_{\mathbb{R}^6}h_N(Z)k(Z)d^6Z|\geq 1\big)
\le  C_{1}N^{-\gamma}.
\end{align*}
\end{lem}
\vspace{0,4cm}
\noindent One might be confused by the rather inconvenient way of essentially claiming that $\int_{\mathbb{R}^6}h_N(X)^2k(X)d^6X\le CN^{1-\delta}$. However, in subsequent applications we will always apply such families $(M^N_i)_{i\in I}$ for verifying this constraint and thus we decided to adapt the formulation of the lemma to this circumstance. 
\begin{proof} 
We prove the lemma by applying Markov inequality. Let for this purpose be $M \in\mathbb{N}$, then it holds for all $N\in \mathbb{N}$ that
\begin{align*}
& \mathbb{P}\big(|\frac{1}{N}\sum_{k=1}^Nh_N(X_k)-\int_{\mathbb{R}^6}h_N(Z)k(Z)d^6Z|\geq 1\big)\\
\le  & \mathbb{E}[\frac{1}{N^{2M}}\Big(\sum_{k=1}^N\big( h_N(X_k)-\int_{\mathbb{R}^6}h_N(Z)k(Z)d^6Z\big)\Big)^{2M}]
\end{align*}
where $\mathbb{E}[\cdot]$ shall denote the expectation with respect to the $N$-fold product of $k$.\\
Let moreover 
$$ \mathcal{M}:=\big\{(\gamma_1,...,\gamma_{N})\in \mathbb{N}_0^{N} : \sum_{i=1}^{N}\gamma_i=2M\big\} $$ be a set of multi-indices then it follows that
\begin{align*}
& \frac{1}{N^{2M}}\mathbb{E}[\Big(\sum_{k=1}^N\big( h_N(X_k)-\int_{\mathbb{R}^6}h_N(Z)k(Z)d^6Z\big)\Big)^{2M}]\\
= & \frac{1}{N^{2M}}\sum_{(\gamma_1,...,\gamma_{N}) \in \mathcal{M}}\mathbb{E}[\prod_{k=1}^N \big(h_N(X_k)-\int_{\mathbb{R}^6}h_N(Z)k(Z)d^6Z\big)^{\gamma_k}].
\end{align*}
We abbreviate for $\gamma:=(\gamma_1,...,\gamma_N)$
$$ G_{\gamma} (X):=\prod_{k=1}^N\big(h_N(X_k)-\int_{\mathbb{R}^6}h_N(Z)k(Z)d^6Z\big)^{\gamma_k}.$$
If there exists an index $i\in \{1,...,N\}$ such that $\gamma_i=1$, then integration over the $i$-th variable first shows that in this case $\mathbb{E}[G_{\gamma}(X)]=0.$\\
Moreover, it holds on the one hand that
\begin{align}
&|\big(h_N(X_k)-\int_{\mathbb{R}^6}h_N(Z)k(Z)d^6Z\big)^{\gamma_k}| \notag\\
\le & 2^{\gamma_k}\big(|h_N(X_k)|^{\gamma_k}+|\int_{\mathbb{R}^6}h_N(Z)k(Z)d^6Z|^{\gamma_k}\big) \label{est.l.o.l.n.}
\end{align}
and since $\|h_N\|_{\infty}\le C_0N^{1-\delta}$, it follows on the other hand due to the remaining assumptions on the maps $h_N$ that for every natural number $n\geq 2$:
\begin{align*}
&\int_{\mathbb{R}^6}|h_N(X)|^nk(X)d^6X\\
\le & C_0\max_{i \in I}\int_{M^N_i}|h_N(X)|^nk(X)d^6X\\
\le & C_0 \|h_N\|_{\infty}^{n-2}\max_{i \in I}\int_{M^N_i}h_N(X)^2k(X)d^6X\\
\le & C_0\big(C_0^{n-2}  N^{(n-2)(1-\delta)}\big)\big(C_0N^{1-\delta}\big)
\end{align*}
Furthermore, if we identify $R:=\Big( \int_{\mathbb{R}^6} h_N^2(Z)k(Z)d^6Z\Big)^{\frac{1}{2}}$, then it holds that
\begin{align*}
& \int_{\mathbb{R}^6}|h_N(Z)|k(Z)d^6Z\\
\le & \frac{1}{R}\underbrace{\int_{\mathbb{R}^6}h_N^2(Z)k(Z)d^6Z}_{=R^2}+\underbrace{\int_{\mathbb{R}^6}|h_N(Z)|\mathbf{1}_{[0,R]}(|h_N(Z)|)k(Z)d^6Z}_{\le R}\\
\le &2\Big(C \max_{i\in I} \int_{M^N_i}h_N^2(Z)k(Z)d^6Z\Big)^{\frac{1}{2}}\\
\le & CN^{\frac{1}{2}(1-\delta)}.
\end{align*}
Since the constraints on the maps $h_N$ are more restrictive the larger the value of $\delta$ is chosen, we can limit the considered values to (for example) $(0,1]$. If we identify additionally $|\gamma|:=|\{i\in\{1,...,N\}: \gamma_i\neq 0\}|$ and recall that only tuples matter where $\gamma_i \neq 1\ \forall i\in \{1,...,N\}$ as well as $\sum_{i=1}^N\gamma_i=2M$, then application of these estimates and relation \eqref{est.l.o.l.n.} yields that
\begin{align*}
 \mathbb{E}[G_{\gamma}(X)]
\le \prod_{1\le i\le N: \gamma_i\geq 2} (C^{\gamma_i}N^{(\gamma_i-2)(1-\delta)}N^{1-\delta})
\le  C^{2M}N^{2M(1-\delta)}N^{|\gamma|(-1+\delta)}.
\end{align*}
We can determine an upper bound for the number of multi-indices $\gamma \in \mathcal{M}$ related to a certain value $|\gamma|$ by simple combinatorics: Any such $\gamma$ can be identified by first choosing the set of indices which shall fulfill $\gamma_i\neq 0$ and then assigning each of the elements belonging to this set a number in $\{1,...,2M\}$ such that the sum over these numbers equals $2M$. Since $\mathbb{E}[G_{\gamma}(X)]$ vanishes if there is at least one index where $\gamma_i=1$, we only have to take into account terms $G_{\gamma}(X)$ where $|\gamma|\le M$. Consequently, we get an upper bound for the number of different tuples $\gamma$ where 
$|\gamma|=k\le M$ by
$$\sum_{\substack{\gamma \in \mathcal{M}\\|\gamma|=k}}1\le \binom{N}{k}(2M)^k\le N^k(2M)^M.$$
Altogether we get
\begin{align*}
&\frac{1}{N^{2M}}\sum_{\gamma \in \mathcal{M}}\mathbb{E}[G_{\gamma}(X)] \\ \le & \frac{1}{N^{2M}}N^{2M(1-\delta)}\sum_{\gamma \in \mathcal{M}}C^{M}N^{|\gamma|(-1+\delta)} \\ \le & C^M N^{-2M\delta}\sum_{k=1}^{M}N^{k}(2M)^MN^{k(-1+\delta)}\\
\le & (CM)^M N^{-M\delta }
\end{align*}
which proves the lemma since $M$ can be chosen arbitrarily large.
\end{proof}
\noindent
{\bf Considerations for term \eqref{rel.t5}:} We want to apply the law of large numbers to show that term \eqref{rel.t5} stays sufficiently small for typical initial data. To this end, we define for an arbitrary $Y\in \mathbb{R}^6$ the function
\begin{align}
h^{t}_N(Y,\cdot):\mathbb{R}^6\to \mathbb{R}^3, \ Z\mapsto N^{\beta}\int_0^t f^{N}(^1\varphi^N_{s,0}(Y)-{^1\varphi^N_{s,0}}(Z))ds\mathbf{1}_{G^N(Y)}(Z) \label{def.h_N}
\end{align}
where $0<\beta\le \frac{1}{2}(1-\sigma)$ is a parameter which will later be chosen differently depending whether we consider the case $\alpha=2$ or $\alpha\in (1,\frac{4}{3}]$. Thus, we will fix it later when the general considerations are concluded respectively when we start to distinguish between these two options. The notation $h^t_N(Y,\cdot)$ shall emphasize the correspondence to the function applied in Lemma \ref{largenumberslem}. \\
Though $h^t_N(Y,\cdot)$ does not map to $\mathbb{R}$ as claimed in the assumptions of Lemma \ref{largenumberslem}, it can still be applied on each component separately. If for each of the three components the difference to its expectation stays typically small, then the same holds for the related vector-valued map. The subsequent considerations will show that each component of this map fulfills the conditions for the application of Lemma \ref{largenumberslem}.\\ 
In accordance with the assumptions of the lemma, we begin by providing a suitable cover of $\mathbb{R}^6$ to check the constraints on the considered map. The subsequent list shows an appropriate family of `collision classes' (see \eqref{def.coll.cl.1}) yielding such a cover where $k,l\in \mathbb{Z}$, $N\in \mathbb{N}\setminus \{1\}$, $\delta>0$ and $0\le r,v \le 1$:
\begin{enumerate}
\item[(i)]  $M^N_{(0,r),(0,v)}(Y)$ 
\item[(ii)]$M^N_{(0,r),(N^{l\delta}v,N^{(l+1)\delta}v)}(Y)$,\ $0\le l \le \lfloor  \frac{\ln(\frac{1}{v})}{\delta\ln(N)}\rfloor $\\
\item[(iii)]  $M^N_{(0,r),(1,\infty)}(Y)$\\
\item[(iv)] $M^N_{(N^{k\delta}r,N^{(k+1)\delta}r),(0,v)}(Y)$, $0\le k \le \lfloor  \frac{\ln(\frac{1}{r})}{\delta\ln(N)}\rfloor $  \\
\item[(v)] $M^N_{(N^{k\delta}r,N^{(k+1)\delta}r),(N^{l\delta}v,N^{(l+1)\delta} v)}(Y)$, $0\le k \le \lfloor \frac{\ln(\frac{1}{r})}{\delta\ln(N)}\rfloor ,\ 0\le l \le \lfloor \frac{\ln(\frac{1}{v})}{\delta\ln(N)}\rfloor $  \\
\item[(vi)]  $M^N_{(N^{k\delta}r,N^{(k+1)\delta}r),(1,\infty)}(Y)$, $0\le k \le \lfloor  \frac{\ln(\frac{1}{r})}{\delta\ln(N)}\rfloor$ \\
\item[(vii)]$M^N_{(N^{-\delta},\infty),(0,\infty)}(Y)$  
\end{enumerate} 
Sketch \eqref{partition} might simplify the comprehension of the basic form of the stated cover.
\\ The axes correspond to the parameters which characterize a `collision class'. Though this sketch might suggest that the sets related to items (i) to (vii) are disjoint, there is partly actually some overlap. While the dashed line shall outline the respective areas belonging to these items, the dotted line shall describe the additional intersection in smaller `collision classes' within these items. Obviously, the `collision classes' are chosen `finer' as the related `collision strength' becomes larger. If the particles keep a distance of almost order 1 to each other (see item (vii)), even no splitting is necessary. If we choose $r=v:=N^{-c}$, then the number of sets belonging to this list is some integer $I_{\delta}$ (independent of $N$) and if we label the related parameters of these sets consecutively by $r_i,R_i,v_i$ and $V_i$, then $(M^N_{(r_i,R_i),(v_i,V_i)}(Y))_{i\in I_{\delta}}$ provides us the desired family of `collision classes'.\\  
It remains to check that for each set of this family the assumptions of Lemma \ref{largenumberslem} are fulfilled.\\ 
To this end, we abbreviate $\widetilde{r}:=\max(r,N^{-c})$ for $r\geq 0$.
Then we obtain by Corollary \ref{cor1} and Lemma \ref{lem3} that for $0\le v\le V ,\ 0\le r\le R $ the following holds:
{\allowdisplaybreaks
\begin{align}
&\int_{M^N_{(r,R),(v,V)}(Y)}\big(\int_{0}^t |f^{N}(^1\varphi^N_{s,0}(Z)-{^1\varphi^N_{s,0}}(Y))|ds\big)^2k_0(Z) d^6Z \notag \\
\le & C\big(\min(\frac{1}{\widetilde{r}^{\alpha}},\frac{1}{\widetilde{r}^{\alpha-1}v})\big)^{2}\int_{M^N_{(r,R),(v,V)}(Y)}k_0(Z) d^6Z\notag \\
\le &C\min \big(\frac{1}{\widetilde{r}^{2\alpha}},\frac{1}{\widetilde{r}^{2(\alpha-1)}v^2}\big)\min \big(1,R^2,R^2V^4+R^3\max(V^3,R^3)\big)\notag \\
\le &C\min\big(\frac{1}{\widetilde{r}^{2(\alpha-1)}v^2},\frac{R^2}{\widetilde{r}^{2(\alpha-1)}v^2},\frac{R^2V^4}{\widetilde{r}^{2(\alpha-1)}\max(\widetilde{r},v)^2}+\frac{R^6}{\widetilde{r}^{2\alpha}}\big) \notag\\
\le &C\min\big(\frac{1}{\widetilde{r}^2v^2},\frac{R^2}{\widetilde{r}^{2}v^2},\frac{R^2V^4}{\widetilde{r}^{2}\max(\widetilde{r},v)^2}+\frac{R^6}{\widetilde{r}^4}\big) \label{up.bound.}.
\end{align}}
where we applied in the last step that $\alpha\le 2.$\\
By application of these estimates we can determine upper bounds for each set of the `collision class'-family. The notation for the terms on the left used in the following list is chosen such that it should become obvious which of the upper limits derived in \eqref{up.bound.} is applied in the respective case. If the third of the different upper bounds given in \eqref{up.bound.} is used, only the larger addend is stated. For a compact notation we drop the constant $C$ for this list and reintroduce it afterwards.
\begin{itemize}
\item[(i)]  $\frac{(N^{-c})^6}{(N^{-c})^4}=N^{-2c}$
\item[(ii)] $\frac{(N^{-c})^2(N^{ (k+1)\delta}N^{-c})^4}{(N^{-c})^{2}(N^{k\delta }N^{-c})^{2}}=N^{-2c+2(k+2)\delta}$,\ $0 \le k\le \lfloor \frac{{c}}{\delta} \rfloor $\\
\item[(iii)] $\frac{(N^{-c})^2}{(N^{-c})^{2}}=1$\\
\item[(iv)] $\frac{(N^{(k+1)\delta}N^{-c})^{6}}{(N^{k\delta}N^{-c})^4}= N^{-2c+2k\delta +6\delta},$ $0 \le k\le \lfloor \frac{c}{\delta} \rfloor $  \\
\item[(v)] $\frac{(N^{(k+1)\delta}N^{-c})^2(N^{(l+1)\delta}N^{-c})^4}{(N^{k\delta}N^{-c})^{2}(N^{l\delta}N^{-c})^{2}}+ \frac{(N^{(k+1)\delta}N^{-c})^6}{(N^{k\delta}N^{-c})^4}$\\
$\ = N^{ -2c+2l\delta+6\delta}+N^{-2c+2k\delta +6\delta}$, $0 \le k,l \le \lfloor  \frac{c}{\delta} \rfloor $\\
\item[(vi)]   $\frac{(N^{(k+1)\delta}N^{-c})^2}{(N^{k\delta}N^{-c})^2}= N^{2\delta}$, $0 \le k \le \lfloor  \frac{c}{\delta} \rfloor $\\
\item[(vii)]$\frac{1}{(N^{-\delta})^4}=N^{4\delta}$ 
\end{itemize}
After regarding the possible values for $k$ and $l$ related to the respective items it is straightforward to see that all these terms are bounded by $N^{6\delta}$ (where the expressions stated in (iv) and (v) determine this choice). Thus, it follows that for all $i\in I_\delta $
\begin{align*}
& \int_{M^N_{(r_i,R_i),(v_i,V_i)}(Y)}h^t_N(Y,Z)^2k_0(Z)d^{6}Z
\le (N^\beta)^2\big(CN^{6 \delta}\big)\le CN^{2 (3\delta+\beta)} .
\end{align*} 
If we choose $\delta>0$ small enough, then the related constraint of Lemma \ref{largenumberslem} is fulfilled because in this case $\beta \in (0,\frac{1}{2}(1-\sigma)]$ (which we assumed initially) implies the relation $2(3\delta+\beta)<1$. Now we want to check if the same is true for the assumption concerning $\|h^t_N(Y,\cdot)\|_{\infty}$. For this purpose $t_{min}$ shall once again denote (one of) the point(s) in time where the considered `mean-field particles' are closest to each other in space. Then it holds according to Corollary \ref{cor1} that
\begin{align}
& \int_{0}^t|f^{N}(^1\varphi^N_{s,0}(Y)-{^1\varphi^N_{s,0}}(Z))|\mathbf{1}_{G^N(Z)}(Y)ds \notag \\
\le & \min\Big(\frac{Ct}{|^1\varphi^N_{t_{min},0}(Y)-{^1\varphi^N_{t_{min},0}}(Z)|^\alpha},\frac{CN^{(\alpha-1)c}}{|^2\varphi^N_{t_{min},0}(Y)-{^2\varphi^N_{t_{min},0}}(Z)|},\notag\\
&\frac{C}{|^1\varphi^N_{t_{min},0}(Y)-{^1\varphi^N_{t_{min},0}}(Z)|^{\alpha-1}|^2\varphi^N_{t_{min},0}(Y)-{^2\varphi^N_{t_{min},0}}(Z)|}\Big)\mathbf{1}_{G^N(Z)}(Y) \label{est.ass2.lln0}
\end{align} 
To estimate the last term we need to distinguish between the two cases $\alpha=2$ and $\alpha \in (1,\frac{4}{3}]$ since the definition of $G^N(\cdot)$ and (as mentioned previously) also the choice of $\beta\in (0,\frac{1}{2}(1-\sigma)]$ are different for these two options. We start with the slightly more elaborate case.\\
{\bf Estimates for \eqref{est.ass2.lln0} if $\alpha\in (1,\frac{4}{3}]$:} Now we finally fix $\beta$ and choose for the current case $\beta:=\frac{1}{2}(1-\sigma)$. The set $G^N(Y)$ was basically constructed such that \eqref{est.ass2.lln0} keeps sufficiently small and we recall that for $\alpha \in (1,\frac{4}{3}]$:
\begin{align*}
G^N(Y):=\bigcap_{\substack{k\in \mathbb{N}:\\
\frac{3}{4}-{\sigma} \le k {\sigma} \le 1}}\Big(M^N_{6N^{-\frac{{k\sigma}}{2} },N^{-\frac{1}{2}+ \frac{k \sigma }{6}+\frac{\sigma}{2}}}(Y) \cup  M^N_{ 6N^{-\frac{1}{2}+{\sigma}},N^{-\frac{5}{18}}}(Y)\Big)^C
\end{align*}
Hence, we obtain different upper bounds for the relative velocity between `mean-field particles' during `collisions' where their minimal spatial distance lies below certain values. To see this one has to regard that for configurations $Y\in G^N(Z)$ and $k\in \mathbb{N}$ where $\frac{3}{4}-\sigma \le  k\sigma\le 1 $ the following implication holds
\begin{align*}
6N^{-\frac{(k+1)\sigma}{2}}\le  x_{min}&:=|^1\varphi^N_{t_{min},0}(Y)-{^1\varphi^N_{t_{min},0}}(Z)|\le 6N^{-\frac{k\sigma}{2}}\\
\Rightarrow v_{min}&:=|^2\varphi^N_{t_{min},0}(Y)-{^2\varphi^N_{t_{min},0}}(Z)|\geq N^{-\frac{1}{2}+ \frac{k \sigma }{6}+\frac{\sigma}{2}}
\end{align*}
and thus in this case \eqref{est.ass2.lln0} is bounded by $$C(N^{\frac{(k+1)\sigma}{2} })^{\alpha-1}N^{\frac{1}{2}-\frac{k \sigma }{6}-\frac{\sigma}{2}}\le CN^{\frac{1}{2}-\frac{\sigma}{3}}$$ since $\alpha\in (1,\frac{4}{3}]$.\\ If on the other hand $x_{min}\le 6N^{-\frac{1}{2}+\sigma}$, then it holds due to the definition of $G^N(Z)$ that
$$|^2\varphi^N_{t_{min},0}(Y)-{^2\varphi^N_{t_{min},0}}(Z)|\geq N^{-\frac{5}{18}}$$ and thus this time  \eqref{est.ass2.lln0} is bounded by 
$$CN^{c(\alpha-1)}N^{\frac{5}{18}}\le CN^{\frac{2}{3}\frac{1}{3}+\frac{5}{18}}=CN^{\frac{1}{2}}.$$ 
If none of the previous conditions is fulfilled, then $x_{min}\geq 6N^{-\frac{1}{2} \frac{3}{4}}=6N^{-\frac{3}{8}}$ is the remaining option which in turn yields:
$$\frac{1}{x_{min}^{\alpha}}\le CN^{\frac{1}{2}}$$ 
This eventually shows that in total $CN^{\frac{1}{2}}$ determines an upper bound for term \eqref{est.ass2.lln0} and thus $$\|h^t_N(Y,\cdot)\|_{\infty}\le N^{\beta}CN^{\frac{1}{2}}\le CN^{\frac{1}{2}(1-\sigma)}N^{\frac{1}{2}}\le CN^{1-\frac{\sigma}{2}} .$$
Hence, also the second assumption of Lemma \ref{largenumberslem} is fulfilled in this case and it remains to consider $\alpha =2$.\\
\noindent{\bf Estimates for term \eqref{est.ass2.lln0} if $\alpha=2 $:} For this setting we choose $\beta=c+\sigma=\frac{7}{18}$ and recall that for $\alpha=2$: 
$$ G^N(Y):=\big(M^N_{6N^{-\frac{2}{9}-\sigma },N^{-\frac{2}{9}}}(Y)\big)^C$$
Then we can once again apply the reasoning of the previous case to conclude that this time term \eqref{est.ass2.lln0} is bounded by
\begin{align*}
& \frac{Ct}{N^{\alpha(-\frac{2}{9}-\sigma)}}+\frac{C}{N^{(1-\alpha)c}N^{-\frac{2}{9}}}\\
\le &  CN^{\frac{4}{9}+2\sigma}+CN^{c+\frac{2}{9}}.
\end{align*}
The second upper bound `controls' constellations where $x_{min}\le 6N^{-\frac{2}{9}-\sigma}$ and the first all remaining.\\
This, however, yields for small enough $\sigma>0$ that
\begin{align*}
\|h^t_N(Y,\cdot)\|_{\infty}\le &  N^{\beta}C\big(N^{\frac{4}{9}+2\sigma}+N^{c+\frac{2}{9}}\big)\le   CN^{1-\sigma}
\end{align*} 
where we regarded that $\beta=c+\sigma$ and $c=\frac{7}{18}-\sigma$. Consequently, also the second assumption of the law of large numbers is satisfied in this case.\\\\ 
Before we are able to show that term \eqref{rel.t5} stays typically sufficiently small, we need to introduce the sets $\mathcal{B}^{N,\sigma}_{1,i}\subseteq \mathbb{R}^{6N}$, $i\in \{1,...,N\}$:
\begin{align}
\begin{split}
& X\in \mathcal{B}_{1,i}^{N,\sigma}\subseteq \mathbb{R}^{6N}\\ \label{def.B_1}
\Leftrightarrow & \exists t_1,t_2\in [0,T]:\\
& \Big|\frac{1}{N}\sum_{j\neq i}\int_{t_1}^{t_2}f^{N}(^1\varphi^N_{s,0}(X_i)-{^1\varphi^N_{s,0}}(X_j))\mathbf{1}_{G^N(X_i)}(X_j)ds\\
&  -\int_{\mathbb{R}^6}\int_{t_1}^{t_2} f^{N}(\varphi^N_{s,0}(X_i)-\varphi^N_{s,0}(Y))\mathbf{1}_{G^N(X_i)}(Y)dsk_0(Y)d^6Y\Big| > N^{-\beta} .
\end{split}
\end{align}
where in accordance with the choice of the parameter $\beta$ in the previous estimates we identify
 \begin{align}
\beta:=\begin{cases}\frac{1}{2}(1-\sigma)\ ,\  &\alpha\in (1,\frac{4}{3}]\\
\frac{7}{18}\ ,\  & \alpha=2  \end{cases} . \label{Def.beta}
\end{align}
Our next aim is to show that configurations belonging to this set are untypical. Unfortunately, for the considered functions $h^t_N(X_i,\cdot) $ the version of the law of large numbers only makes statements about the probability of fluctuations at a certain point in time. But on the other hand, it tells us that at the considered moment large fluctuations are extremely unlikely. Furthermore, on very short time intervals fluctuations can not change significantly since $\|f^N\|_{\infty}\le N^{\alpha c}$. Thus, this problem can be easily solved.\\ 
For $\delta_N >0$ it holds that
\begin{align}
& X\in \mathcal{B}_{1,i}^{N,\sigma} \notag  \\ 
\Rightarrow & \exists k\in \{0,...,\lfloor \frac{T}{\delta_N } \rfloor \}:  \notag \\
&  \Big(\big|\int_{0}^{k \delta_N }\Big(\frac{1}{N}\sum_{j\neq i}f^{N}(^1\varphi^N_{s,0}(X_i)-{^1\varphi^N_{s,0}}(X_j))\mathbf{1}_{G^N(X_i)}(X_j)  \notag \\
& -\int_{\mathbb{R}^6} f^{N}(^1\varphi^N_{s,0}(X_i)-{^1\varphi^N_{s,0}}(Y))\mathbf{1}_{G^N(X_i)}(Y)k_0(Y)
d^6Y\Big) ds\big| \geq \frac{N^{-\beta}}{4}\Big) \ \vee  \notag \\
& \Big(\int_{k \delta_N }^{(k+1) \delta_N }\Big(\big|\frac{1}{N}\sum_{j\neq i}f^{N}(^1\varphi^N_{s,0}(X_i)-{^1\varphi^N_{s,0}}(X_j))\mathbf{1}_{G^N(X_i)}(Y)
\big| \notag \\
& +\big|\int_{\mathbb{R}^6} f^{N}(^1\varphi^N_{s,0}(X_i)-{^1\varphi^N_{s,0}}(Y))\mathbf{1}_{G^N(X_i)}(Y)k_0(Y)
d^6Y\big|\Big) ds \geq \frac{N^{-\beta}}{4}\Big)
\label{s.scale.dev.}.
\end{align} 
This follows easily by the definition of the set $\mathcal{B}_{1,i}^{N,\sigma}$ if one takes into account that for any continuous map $a:\mathbb{R}\to \mathbb{R}^m, \ m\in \mathbb{N},\  t_1,t_2\in [0,T]$ it holds that
\begin{align}
& \big|\int_{t_1}^{t_2 }a(s)ds\big| \notag \\
= &  \big|\int_{0}^{t_2 }a(s)ds-\int_{0}^{t_1 }a(s)ds\big| \notag \\
\le & \big|\int_{0}^{\lfloor \frac{t_2}{\delta_N}\rfloor \delta_N }a(s)ds\big|+\int_{\lfloor \frac{t_2}{\delta_N}\rfloor \delta_N }^{t_2}|a(s)|ds +  \big|\int_{0 }^{\lfloor \frac{t_1}{\delta_N}\rfloor \delta_N}a(s)ds\big|+\int_{\lfloor \frac{t_1}{\delta_N}\rfloor \delta_N }^{t_1}|a(s)|ds \notag \\
\le & 2\max_{k\in \{0,...,\lfloor \frac{T}{\delta_N}\rfloor \}}\Big(\big|\int_{0 }^{k \delta_N}a(s)ds\big| +\int_{k \delta_N }^{(k+1) \delta_N }|a(s)|ds\Big).
\end{align}
Since $\|f^{N}\|_{\infty}\le N^{c\alpha}$, the second constraint of assumption \eqref{s.scale.dev.} is true for all configurations if we choose $\delta_N:=\frac{N^{-\beta}}{8\|f^N\|_{\infty}}\le CN^{-\beta-c\alpha}$. After some more detailed analysis one could easily show that this inequality is still true with extremely high probability if $\delta_N $ is of much larger order because for the current estimate we essentially assumed the worst case scenario that all particles form a single cluster. But on the other hand this does not lead to a relevant improvement of the result. According to the previous reasoning for at least one $k\in \{0,....,\lfloor \frac{T}{\delta_N}\rfloor\}$ the event related to the first constraint of \eqref{s.scale.dev.} must occur if $X\in \mathcal{B}_{1,i}^{N,\sigma}$. However, the law of large numbers yields that for any of these events and any $\gamma>0$ there exists $C_{\gamma}>0$ such that its probability is smaller than $C_\gamma N^{-\gamma}$ . By regarding that $c=\frac{2}{3}$ if $\alpha\in (1,\frac{4}{3}]$ and $c=\frac{7}{18}-\sigma $ for $\alpha=2$ as well as $\beta< \frac{1}{2} $ it follows that the number of such events is bounded by $\lfloor \frac{T}{\delta_N}\rfloor+1\le CN^{\frac{1}{2}+\frac{7}{18}2}=CN^{\frac{23}{18}}$ for $\alpha=2$ and by $CN^{\frac{1}{2}+\frac{2}{3}\frac{4}{3}}=CN^{\frac{25}{18}}$ if $\alpha\in (1,\frac{4}{3}]$ and thus it holds for all $N\in \mathbb{N}$ that
\begin{align}
& \mathbb{P}\big(\exists i \in \{1,...,N\}:X\in \mathcal{B}_{1,i}^{N,\sigma}\big) \notag \\
\le & N\mathbb{P}\big(X\in \mathcal{B}_{1,i}^{N,\sigma}\big)\le N\big(CN^{\frac{25}{18}}(C_{\gamma+\frac{43}{18}}N^{-(\gamma+\frac{43}{18})})\big)\le \widetilde{C}_\gamma N^{-\gamma} \label{prob.b.1}
\end{align} 
where $\widetilde{C}_\gamma$ is simply an adjusted $\gamma$-dependent constant. Eventually, this yields that for typical initial data and large enough $N\in \mathbb{N}$ term \eqref{rel.t5} stays indeed smaller than $$N^{-\beta}=\begin{cases}N^{-\frac{1}{2}(1-\sigma)},\ & \alpha\in(1,\frac{4}{3}]\\
N^{-\frac{7}{18}},\ &  \alpha=2 \end{cases}.$$
We finally arrived at the estimates for the last remaining term.\\
{\bf Considerations for term \eqref{rel.t3}:} For the subsequent part we abbreviate: 
\begin{align}
 \Delta^N_g(t,X)&:=\max_{j\in \mathcal{M}^N_g(X)}\sup_{0\le s\le t}|[^1\Psi^N_{s,0}(X)]_j-{^1\varphi^N_{s,0}}(X_j)| \notag \\
 \Delta^N_b(t,X)&:=\max_{j\in \mathcal{M}^N_b(X)}\sup_{0\le s\le t}|[^1\Psi^N_{s,0}(X)]_j-{^1\varphi^N_{s,0}}(X_j)| \notag \\
 \widetilde{G}^N(\cdot)&:=G^N(\cdot)\cap \big(M^N_{3N^{-\frac{1}{2}+\sigma},\infty}(\cdot)\big)^C \label{abbrev.thm1} 
\end{align}
Hence, $ \Delta^N_g(t,X)$ describes the largest spatial deviation of the `good' particles and $\Delta^N_b(t,X)$ the corresponding value for the `bad' ones.\\
By definition of $\widetilde{G}^N(\cdot)$ (applied for the first inequality) and the stopping time $\tau^N(X)$ (see \eqref{Def.stopping}) it holds for $X_j\in \widetilde{G}^N(X_i)$ and times $s\in [0,\tau^N(X)]$ that
\begin{align*}
& \max\big( 2N^{-c},\frac{2}3|{^1\varphi^N_{s,0}}(X_j)-{^1\varphi^N_{s,0}}(X_i)|\big)\geq  \max \big( 2N^{-c},2N^{-\frac{1}{2}+\sigma}\big) \geq 2\Delta^N_g(t,X).
\end{align*}
Furthermore, the map $g^N$ was defined such that $|f^N(q+\delta)-f^N(q)|\le g^N(q)|\delta|$ for $q,\delta \in \mathbb{R}^3$ where $ \max\big(2 N^{-c},\frac{2}{3}|q|\big)\geq |\delta|$ (see definition \eqref{Def.g^N}) and thus the subsequent estimates are fulfilled for all $0\le t_1\le t\le \tau^N(X)$:
\begin{align}
& \big|\int_{t_1}^t\Big(\frac{1}{N}\sum_{j\neq i}\Big(f^{N}([^1\Psi^N_{s,0}(X)]_j-[^1\Psi^N_{s,0}(X)]_i)  \notag  \\
&-f^{N}(^1\varphi^N_{s,0}(X_j)-{^1\varphi^N_{s,0}}(X_i))\Big)\mathbf{1}_{G^N(X_i)}(X_j)\Big) ds\big| \label{term0} \\
\le & \int_{0}^t\Big(\frac{1}{N}\sum_{\substack{j\neq i\\ j\in \mathcal{M}^N_b(X)}}\Big(\big|f^{N}([^1\Psi^N_{s,0}(X)]_j-[^1\Psi^N_{s,0}(X)]_i) \notag  \\
&- f^{N}(^1\varphi^N_{s,0}(X_j)-{^1\varphi^N_{s,0}}(X_i))\big|\Big)\mathbf{1}_{G^N(X_i)}(X_j)\Big) ds\label{term1}\\
&+\int_{0}^t\Big(\frac{1}{N}\sum_{\substack{j\neq i\\ j\in \mathcal{M}^N_g(X)}}\Big(\big|f^{N}([^1\Psi^N_{s,0}(X)]_j-[^1\Psi^N_{s,0}(X)]_i)\big| \notag  \\
&+ \big|f^{N}(^1\varphi^N_{s,0}(X_j)-{^1\varphi^N_{s,0}}(X_i))\big|\Big)\mathbf{1}_{G^N(X_i)\cap M^N_{3N^{-\frac{1}{2}+\sigma},\infty}(X_i)}(X_j)\Big) ds \label{term1,5}\\
 & + \int_{0}^t\frac{2}{N}\sum_{\substack{j\neq i\\ j\in \mathcal{M}^N_g(X)}}g^{N}(^1\varphi^N_{s,0}(X_j)-{^1\varphi^N_{s,0}}(X_i))\Delta^N_g(s,X)\mathbf{1}_{\widetilde{G}^N(X_i)}(X_j) ds \label{term2}
\end{align}
where we just divided the addends into three groups and applied for the last arising term the previous considerations.\\
We continue by defining a set which  in the end will turn out to be crucial for the estimates related to term \eqref{term2}: 
\begin{align}
\begin{split}
& X\in \mathcal{B}_{2,i}^{N,\sigma}\subseteq \mathbb{R}^{6N} \label{def.B_2}\\
\Leftrightarrow & \exists t_1,t_2\in [0,T]:\\
& \Big|\frac{1}{N}\sum_{j\neq i}\int_{t_1}^{t_2}g^{N}({^1\varphi^N_{s,0}}(X_j)-{^1\varphi^N_{s,0}}(X_i))\mathbf{1}_{\widetilde{G}^N(X_i)}(X_j)ds\\
&  -\int_{\mathbb{R}^6}\int_{t_1}^{t_2} g^{N}(^1\varphi^N_{s,0}(Y)-{^1\varphi^N_{s,0}}(X_i))\mathbf{1}_{\widetilde{G}^N(X_i)}(Y)dsk_0(Y)
d^6Y\Big| > 1
\end{split} 
\end{align}\
For $Y,Z\in \mathbb{R}^6$ (and $\alpha\in (1,\frac{4}{3}]\cup \{2\}$) it holds that
\begin{align}
& \int_{0}^t g^{N}(^1\varphi^N_{s,0}(Y)-{^1\varphi^N_{s,0}}(Z)) \mathbf{1}_{\widetilde{G}^N(Z)}(Y)ds \notag \\
\le  & C\begin{cases}N^{\frac{1}{2}-\sigma}\int_{0}^t|f^{N}(^1\varphi^N_{s,0}(Y)-{^1\varphi^N_{s,0}}(Z))|\mathbf{1}_{G^N(Z)}(Y)ds,\  \ &\text{if }\alpha\in (1, \frac{4}{3}] \\ 
N^{\frac{7}{18}-\sigma}\int_{0}^t|f^{N}(^1\varphi^N_{s,0}(Y)-{^1\varphi^N_{s,0}}(Z))|\mathbf{1}_{G^N(Z)}(Y)ds,\  \ &\text{if }\alpha=2  \end{cases} \label{check.ass.l.o.l.n}
\end{align}
where we regarded that according to the definition of $\widetilde{G}^N(\cdot)$ (see \eqref{abbrev.thm1}) the `mean-field particles' which are relevant for this term keep at least a distance of order $N^{-\frac{1}{2}+\sigma}$ to each other as well as the definition of $g^N$ (see \eqref{Def.g^N}). We remark that instead of checking if $|h^t_N(Y,\cdot)|$ (see \eqref{def.h_N}) fulfills the assumptions of the law of large numbers we even verified them for a map which has the form of the term on the right-hand side of \eqref{check.ass.l.o.l.n} (and where the prefactor is even of slightly larger order with respect to $N$). Consequently, the same reasoning as used previously for the map $h^t_N(Y,\cdot)$ works to show that for an arbitrary $\gamma>0$ there exists $C_{\gamma}>0$ such that for all $N\in \mathbb{N}$:
\begin{align}
\mathbb{P}\big(\exists i\in \{1,...,N\}:X\in \mathcal{B}_{2,i}^{N,\sigma}\big)\le C_{\gamma} N^{-\gamma} \label{prob.b.2}
\end{align}
It remains to determine an upper bound for term \eqref{term1,5} and to show that the `bad' particles do typically not `infect' the `good' ones which corresponds to deriving a suitable bound for term \eqref{term1}. Since the allowed maximal value for $\Delta^N_b(t,X)$ (resp. for the largest deviation of a `bad' particle) is distinctly larger than the corresponding value for $\Delta^N_g(t,X)$ (at least if $\alpha=2$), problems could arise if the number of `bad' particles coming close to a `good' one exceeds a certain value. In the subsequent part we want to show that the probability of such an event gets vanishingly small as $N$ increases. It will turn out in the end that for $\alpha\in (1,\frac{4}{3}]$ typically no `bad' particle occurs at all and thus we restrict the estimates for the following term to the relevant case $\alpha=2$.  \\ 
{\bf Estimates for term \eqref{term1} if $\alpha=2$:}\\
After introducing $h^t_N(Y,\cdot)$ (see\eqref{def.h_N}) we also implemented a family of `collision classes' $\big(M^N_{(r_i,R_i),(v_i,V_i)}(Y)\big)_{i \in I_\delta}$ yielding a cover of $\mathbb{R}^6$  and checked if $h^t_N(Y,\cdot)$ in combination with this cover fulfills the assumptions of the law of large numbers (resp. Lemma \ref{largenumberslem}). Let $\big(M^N_{(r_i,R_i),(v_i,V_i)}(Y)\big)_{i \in I_{\delta}}$ denote again the family related to the list stated there but this time for the parameters $r=v:=6N^{-\frac{2}{9}-\sigma}$ (instead of $r=v:=N^{-c}$) as well as $\delta:=\sigma $ and we define for $i\in  \{1,...,N\}$ the set $ \mathcal{B}_{3,i}^{N,\sigma}\subseteq \mathbb{R}^{6N}$ as follows:
\begin{align}
\begin{split}
& X\in \mathcal{B}_{3,i}^{N,\sigma} \subseteq \mathbb{R}^{6N} \\
\Leftrightarrow & \exists l\in I_{\sigma}: \Big(R_l\neq \infty\ \land\\ &\sum_{j \in \mathcal{M}^N_b(X)}\mathbf{1}_{M^N_{(r_l,R_l),(v_l,V_l)}(X_i)}(X_j) \geq  N^{\frac{2\sigma}{3}}\big\lceil N^{\frac{2}{3}} R_l^2\min\big(\max(V_l,R_l),1\big)^4\big\rceil\Big) \ \vee \label{def.B_3}\\
& \sum_{j \in \mathcal{M}^N_b(X)}1=| \mathcal{M}^N_b(X)|\geq N^{\frac{2}{3}(1+\sigma)}
\end{split}
\end{align} 
In a first step we derive a suitable upper bound for term \eqref{term1} under the condition that $X\in \big(\mathcal{B}_{3,i}^{N,\sigma}\big)^C$ and prove in a second step that $\mathbb{P}\big( X\in \mathcal{B}_{3,i}^{N,\sigma}\big)$ gets vanishingly small as $N$ increases.\\ 
We start with some general estimates and apply them for the relevant sets afterwards. To this end, we abbreviate for $0\le r\le R $ and $0\le v \le V$:
$$ \widetilde{M}^N_{(r,R),(v,V)}(X_i):=G^N(X_i)\cap M^N_{(r,R),(v,V)}(X_i)$$
However, we only consider values of $r$ and $R$ fulfilling the constraint
\begin{align}
\big(r=0 \land R=6\delta^N_b=6N^{-\frac{2}{9}-\sigma}\big)\vee \big( r\geq 6\delta^N_b\land R=N^{\sigma}r\big) \label{cond.para.r,R} 
\end{align}
because for the family $\big(M^N_{(r_i,R_i),(v_i,V_i)}(Y)\big)_{i \in I_{\sigma}}$ essentially only values matter where one of these relations is fulfilled. Finally, we recall that $$\sup_{0\le s \le t}|\Psi^N_{s,0}(X)-\Phi^N_{s,0}(X)|_{\infty}\le N^{-\frac{2}{9}-\sigma}=\delta^N_b$$ for $\alpha=2$ and times before the stopping time is `triggered'. Thus, we obtain that for $0\le t\le \tau^N(X)$:
{\allowdisplaybreaks \begin{align}
 & \int_{0}^t\frac{1}{N}\sum_{\substack{j\neq i\\ j\in \mathcal{M}^N_b(X)}}\Big(\big|f^{N}([^1\Psi^N_{s,0}(X)]_j-[^1\Psi^N_{s,0}(X)]_i)\notag  \\
&- f^{N}(^1\varphi^N_{s,0}(X_j)-{^1\varphi^N_{s,0}}(X_i))\big|\Big) \mathbf{1}_{\widetilde{M}^N_{(r,R),(v,V)}(X_i)}(X_j) ds \label{t.imp.b.part.1}\\
\le & \int_{0}^t\frac{1}{N}\sum_{\substack{j\neq i\\ j\in \mathcal{M}^N_b(X)}} \Big(\big|f^{N}([^1\Psi^N_{s,0}(X)]_j-[^1\Psi^N_{s,0}(X)]_i)\big|\notag  \\
&+ \big|f^{N}(^1\varphi^N_{s,0}(X_j)-{^1\varphi^N_{s,0}}(X_i))\big|\Big)\mathbf{1}_{\widetilde{M}^N_{(r,R),(v,V)}(X_i)}(X_j) ds \mathbf{1}_{[0,6\delta^N_b]}(r)\notag \\
& +\frac{2}{N}\Delta^N_b(t,X)\sup_{Y\in \widetilde{M}^N_{(r,R),(v,V)}(X_i)}\int_0^t g^{N}(^1\varphi^N_{s,0}(Y)-{^1\varphi^N_{s,0}}(X_i))ds  \notag \\ 
& \cdot  \sum_{\substack{j\neq i\\ j\in \mathcal{M}^N_b(X)}} \mathbf{1}_{\widetilde{M}^N_{(r,R),(v,V)}(X_i)}(X_j)\mathbf{1}_{[6\delta^N_b,\infty)}(r) 
\end{align}
where we regarded once again that $|f^N(q+\delta)-f^N(q)|\le g^N(q)|\delta|$ for $q,\delta \in \mathbb{R}^3$ provided that $ \max\big(2 N^{-c},\frac{2}{3}|q|\big)\geq |\delta|$ (see \eqref{Def.g^N}). Now application of Corollary \ref{cor1} (i)+(ii) yields that the previous term is bounded by
\begin{align}
& \frac{C}{N}\frac{1}{N^{- c}v}\sum_{\substack{j\neq i\\ j\in \mathcal{M}^N_b(X)}} \mathbf{1}_{\widetilde{M}^N_{(r,R),(v,V)}(X_i)}(X_j)\mathbf{1}_{[0,6\delta^N_b]}(r)\notag \\
&+ \frac{C}{N}\frac{\Delta_b^N(t,X)}{r^{2}\max(r,v)}\sum_{\substack{j\neq i\\ j\in \mathcal{M}^N_b(X)}} \mathbf{1}_{\widetilde{M}^N_{(r,R),(v,V)}(X_i)}(X_j)\mathbf{1}_{[6\delta^N_b,\infty)}(r) . \label{bound4.0}
\end{align}}
We remark that the assumptions of Corollary \ref{cor1} (ii) (which we needed to estimate the first of these terms) are indeed fulfilled in the current situation since according to the constraints on the possible parameters (see \eqref{cond.para.r,R}) $r\in [0,6\delta^N_b]$ implies $R=\delta^N_b$ and $r=0$. Hence, by regarding the definition of $ G^N(X_i)$ (see \eqref{Def G.alpha=2}) it follows that
$$\widetilde{M}^N_{(0,6\delta^N_b),(v,V)}(X_i)=M^N_{(0,6\delta^N_b),(v,V)}(X_i)\cap G^N(X_i)\subseteq \big(M^N_{6\delta^N_b,N^{-\frac{2}{9}}}(X_i)\big)^C$$ which in turn provides us the necessary implication:
\begin{align}
& X_j\in \widetilde{M}^N_{(0,6\delta^N_b),(v,V)}(X_i) \notag\\
\Rightarrow
& |^2\varphi^N_{t_{min},0}(X_j)-{^2\varphi^N_{t_{min},0}(X_i)}|\geq N^{-\frac{2}{9}}=N^{\sigma}\delta_b^N \notag \\
&  \geq \begin{cases}  N^{\sigma} \sup_{0\le s \le \tau^N(X)}|\Psi^N_{s,0}(X)-\Phi^N_{s,0}(X)|_{\infty}\\
\frac{N^{\sigma}}{6} |^1\varphi^N_{t_{min},0}(X_j)-{^1\varphi^N_{t_{min},0}(X_i)}| \end{cases} \label{rel.values.v}
\end{align}  
where as usual $t_{min}$ shall denote a point in time where $|^1\varphi^N_{\cdot,0}(X_j)-{^1\varphi^N_{\cdot,0}(X_i)}|$ takes its minimum on $[0,T]$.\\ Now we want to derive an upper bound for term \eqref{bound4.0} under the condition that
\begin{align*}
& \sum_{j \in \mathcal{M}_b^N(X)}\mathbf{1}_{M^N_{(r,R),(v,V)}(X_i)}(X_j) \le   N^{\frac{2\sigma}{3}}\big\lceil N^{\frac{2}{3}} R^2\min\big(\max(V,R),1\big)^4\big\rceil.
\end{align*}
For a clearer presentation we deal with the addends related to $\mathbf{1}_{[0,6\delta^N_b]}(r)$ and $\mathbf{1}_{ [6\delta^N_b, \infty)}(r)$ separately. For the first of them we already discussed that $r=0$ and $R=6\delta^N_b$ due to condition \eqref{cond.para.r,R}. By regarding additionally that according to \eqref{rel.values.v} we only have to consider values $v>N^{-\frac{2}9}$ for estimating this term we obtain that
\begin{align}
&\frac{C}{N}\frac{1}{N^{-c}v}\sum_{\substack{j\neq i\\ j\in \mathcal{M}^N_b(X)}} \mathbf{1}_{\widetilde{M}^N_{(r,R),(v,V)}(X_i)}(X_j)\notag \\
\le & \frac{C}{N}\Big(\frac{ N^{\frac{2}{3}
(1+\sigma)}R^2\min(V,1)^4}{N^{-c}\max(N^{-\frac{2}{9}},v)} + \frac{N^{\frac{2\sigma}{3}}}{N^{-c}\max(N^{-\frac{2}{9}},v)} \Big) \notag \\
\le & C\Big(\frac{R^2\min(V,1)^4}{\max(N^{-\frac{2}{9}},v)} N^{\frac{1}{18}}+ N^{-\frac{7}{18}}\Big)\notag \\
\le & C\Big(\frac{\min(V,1)^4}{\max(N^{-\frac{2}{9}},v)} N^{-\frac{7}{18}-2\sigma}+ N^{-\frac{7}{18}}\Big)  \label{bound4.1}
\end{align} 
where we applied $c=\frac{7}{18}-\sigma$ if $\alpha=2$ and $R=\delta^N_b=N^{-\frac{2}{9}-\sigma}$.\\
Taking additionally into account that $\Delta^N_b(t,X)\le N^{-\frac{2}{9}-\sigma}= \delta^N_b$ as well as $R=N^{\sigma}r$ for $r\geq 6\delta^N_b$ (see \eqref{cond.para.r,R}) it follows for the second term of \eqref{bound4.0} that
\begin{align}
& \frac{C}{N}\frac{\Delta_b^N(t,X)}{r^{2}\max(r,v)}\sum_{\substack{j\neq i\\ j\in \mathcal{M}^N_b(X)}} \mathbf{1}_{\widetilde{M}^N_{(r,R),(v,V)}(X_i)}(X_j) \notag \\
\le & \frac{C}{N}\Big( \frac{N^{\frac{2}{3}(1+\sigma)}R^2\min\big(\max(V,R),1\big)^4}{r^2 \max(r,v)}  +\frac{N^{\frac{2\sigma}{3}}}{r^2 \max(r,v)}\Big) N^{-\frac{2}{9}-\sigma} \notag \\
\le & C\Big( \frac{\min\big(\max(V,R),1\big)^4}{\max(r,v)} N^{-\frac{5}{9}+2\sigma}+N^{-\frac{5}{9}+3\sigma}\Big). \label{bound4.2}
\end{align} 
The sum of terms \eqref{bound4.1} and \eqref{bound4.2} forms an upper bound for term \eqref{bound4.0} under the current assumption. Furthermore, all sets which belong to the family $\big(M^N_{(r_i,R_i),(v_i,V_i)}(Y)\big)_{i \in I_\sigma}$ except for $M^N_{(N^{-\sigma},\infty),(0,\infty)}(Y)$ are contained in a `collision class' which takes one of the subsequent forms for suitable parameter $r,v\in [0,1]$ (see the list previous to sketch \ref{partition}):
\begin{multicols}{2}
\begin{itemize}
\item[(i)] $ M^N_{(0,6\delta^N_b),(0,6\delta^N_b)}(Y)$
\item[(ii)]$M^N_{(0,6\delta^N_b),(v,N^{\sigma}v)}(Y) $
\item[(iii)]  $M^N_{(0,6\delta^N_b),(1,\infty)}(Y)$
\item[(iv)] $M^N_{(r,N^{\sigma}r),(0,6\delta^N_b)}(Y)$
 \item[(v)] $M^N_{(r,N^\sigma r),(v,N^\sigma v)}(Y)$
\item[(vi)] $  M^N_{(r,N^\sigma r),(1,\infty)}(Y)$
\end{itemize}
\end{multicols}
\noindent 
By comparing the possible values for $r,R,v,V$ appearing in this list with estimates \eqref{bound4.1} and \eqref{bound4.2} it is straightforward to conclude that for the considered terms a set of kind (ii), (iv) or (v) with $v=N^{-\sigma}$ or $r=N^{-\sigma}$ yields the `worst case option' and thus in total term \eqref{term1} is bounded by
\begin{align}
 &  C\Big( N^{-\frac{7}{18}}+  N^{-\frac{5}{9}+3\sigma}\Big) 
\le  CN^{-\frac{7}{18}} \label{bound4}
\end{align}
if $X\in \big(\mathcal{B}^{N,\sigma}_{3,i}\big)^C$ and $\sigma>0$ is chosen small enough. To this end, we regarded additionally that the number of `collision classes' belonging to the applied cover $|I_{\sigma}|$ is bounded (independent of $N$). Moreover, we already used that for the only class where the previous general considerations can not be applied (which is $M^N_{(N^{-\sigma},\infty),(0,\infty)}(Y)$) the following holds if $X\in \big(\mathcal{B}^{N,\sigma}_{3,i}\big)^C$ (see \eqref{def.B_3} for the definition) and $t\le \tau^N(X)$:
\begin{align}
& \int_{0}^t\frac{1}{N}\sum_{\substack{j\neq i\\ j\in \mathcal{M}^N_b(X)}}\Big(\big|f^{N}([^1\Psi^N_{s,0}(X)]_j-[^1\Psi^N_{s,0}(X)]_i)\notag  \\
&- f^{N}(^1\varphi^N_{s,0}(X_j)-{^1\varphi^N_{s,0}}(X_i))\big|\Big) \mathbf{1}_{M^N_{(N^{-\sigma},\infty),(0,\infty)}(X_i)}(X_j) ds \notag \\
\le & \frac{2}{N}\sup_{Y\in M^N_{(N^{-\sigma},\infty),(0,\infty)}(X_i)}\int_0^t g^{N}(^1\varphi^N_{s,0}(Y)-{^1\varphi^N_{s,0}}(X_i))ds  \notag   \sum_{\substack{j\neq i\\ j\in \mathcal{M}^N_b(X)}} \underbrace{\Delta^N_b(t,X)}_{\le N^{-\frac{2}{9}-\sigma}}\notag \\
\le & \frac{2}{N} \big(T\frac{C}{(N^{-\sigma})^3}\big)N^{-\frac{2}{9}-\sigma}\underbrace{|\mathcal{M}^N_b(X)|}_{\le  N^{\frac{2}{3}(1+\sigma)}} \notag \\
\le & CN^{-\frac{5}{9}+\frac{8}{3}\sigma}
\end{align}
which is distinctly smaller than necessary (for small enough $\sigma>0$).\\ This concludes the estimates for term \eqref{term1} and it remains to show that the probability related to the set $\mathcal{B}^{N,\sigma}_{3,i}$ is indeed small enough. First, we bring the assumptions of this set into a form which is easier to handle. To this end, we recall that $j\in \mathcal{M}^N_b(X)$ implies that $X_k\in \big(G^N(X_j)\big)^C$ for some $k\in \{1,...,N\}\setminus \{j\}$ and thus it holds for $R,V>0$ and $$M:=\big\lceil N^{\frac{2}{3}\sigma}\lceil N^{\frac{2}{3}} R^2\min\big(\max(V,R),1\big)^4\rceil \big\rceil $$ that
\begin{align}
& \sum_{j \in \mathcal{M}_b^N(X)}\mathbf{1}_{M^N_{R,V}(X_i)}(X_j)\geq M \label{cond.prob.est.1}\\
\Rightarrow & \big(\ \exists j \in \{1,...,N\}: \sum_{k=1}^N \mathbf{1}_{(G^N(X_j))^C}(X_k)\geq\lceil  \frac{N^{\frac{\sigma}{3}}}{2}\rceil  \big)\ \vee \label{ass.bad.inf.} \\
& \Big(\exists \mathcal{S}\subseteq \{1,...,N\}^2\setminus \bigcup_{n=1}^N\{(n,n)\}: \notag \\
& \ \text{(i)}\ \ \ |\mathcal{S}|= \lceil  \frac{N^{-\frac{\sigma}{3}}M}{2}\rceil  \notag \\ & \  \text{(ii)} \ \ \forall (j,k)\in \mathcal{S}:X_j\in(G^N(X_k))^C\cap M^N_{R,V}(X_i)   \notag \\
&\ \text{(iii)} \ (j_1,k_1),(j_2,k_2)\in \mathcal{S}\Rightarrow  \{j_1,k_1\}\cap \{j_2,k_2\}=\emptyset\Big)\label{ass.dis.coll.1}
\end{align}
For the explanation why this relationship holds we will name for convenience the event $X_m\in  M^N_{R,V}(X_n)$ by the phrase `\textit{collision} between particles $m,n$' and the phrase `\textit{hard collision} between particles $m,n$' will be applied synonymous to the event $X_m \in (G(X_n))^C$. \\
If assumption \eqref{ass.bad.inf.} does not hold, then an arbitrary `bad' particle can `infect' at most $\lceil \frac{N^{\frac{\sigma}{3}}}{2}\rceil $ further particles to belong to the set $\mathcal{M}_b^N(X)$ (or in our language it can have at most $\lceil \frac{N^{\frac{\sigma}{3}}}{2}\rceil $ \emph{hard collisions} with different particles). For the following considerations we assume that this is the case (respectively that the event related to \eqref{ass.bad.inf.} does not occur) and we argue that under this constraint the relation $$\sum_{j \in \mathcal{M}_b^N(X)}\mathbf{1}_{M^N_{R,V}(X_i)}(X_j)\geq M=\big\lceil N^{\frac{2}{3}\sigma}\lceil N^{\frac{2}{3}} R^2\min\big(\max(V,R),1\big)^4\rceil \big\rceil $$ implies that the event related to \eqref{ass.dis.coll.1} is indeed fulfilled. This can be seen as follows: If \eqref{cond.prob.est.1} is fulfilled, then there exists a set $\mathcal{C}_0\subseteq \mathcal{M}^N_b(X),\ |\mathcal{C}_0|\geq M$ of `bad' particles which all have a \textit{collision} with the particle which belongs to label $i$. If additionally the event related to \eqref{ass.bad.inf.} does not occur, then there exist at most $\lfloor \frac{N^{\frac{\sigma}{3}}}{2}\rfloor $ particles having a \textit{hard collision} with particle $i$ and we `remove' all of those which are (possibly) contained in $\mathcal{C}_0$ from this set. Hence, we obtain a new set which we call $\mathcal{C}_1\subseteq \mathcal{C}_0$ and it obviously holds that $|\mathcal{C}_1|\geq M -\lfloor \frac{N^{\frac{\sigma}{3}}}{2}\rfloor \geq 1 $. Take one of these remaining `bad' particles $j_1$ out of $\mathcal{C}_1$. Since $j_1\in \mathcal{C}_1\subseteq \mathcal{C}_0\subseteq \mathcal{M}^N_b(X)$, there must be at least one further particle having a \textit{hard collisions} with $j_1$ (which by construction of $\mathcal{C}_1$, however, can not be $i$). Let $k_1$ be one of them and we get our first tuple $(j_1,k_1)$ fulfilling condition (ii) of the set $\mathcal{S}$ appearing in \eqref{ass.dis.coll.1}. Now `remove' $j_1$ and $k_1$ as well as all of their at most $(2\lfloor \frac{N^{\frac{\sigma}{3}}}{2}\rfloor -2)$ (possibly existing) remaining `\textit{hard collision} partners' from $\mathcal{C}_1$ to obtain a new set $\mathcal{C}_2\subseteq \mathcal{C}_1$. Finally, the procedure can be repeated (provided that $\mathcal{C}_2\neq \emptyset$) by choosing the next particle (label) $j_2$ out of $\mathcal{C}_2$ and afterwards an arbitrary one of its \textit{hard collision} partners $k_2$ (which by construction of $\mathcal{C}_2$ must be unequal to $i,j_1$ and $k_1$). Then the next round can start after having removed $j_2$ and $k_2$ as well as their (possibly existing) remaining `\textit{hard collision} partners' from $\mathcal{C}_2$ to obtain $\mathcal{C}_3\subseteq \mathcal{C}_2$. Since after each round at most $2\lfloor \frac{N^{\frac{\sigma}{3}}}{2}\rfloor $ `particle labels' are removed from the set $\mathcal{C}_k$ to obtain $\mathcal{C}_{k+1}$, this procedure can be repeated at least $ \lceil \frac{M -\lfloor \frac{N^{\frac{\sigma}{3}}}{2}\rfloor }{N^{\frac{\sigma}{3}}}\rceil\geq \lceil \frac{N^{-\frac{\sigma}{3}}M}{2}\rceil $ times (where we additionally regarded that $M\geq N^{\frac{2\sigma}{3}}$) and thus provides us a set $\mathcal{S}$ consisting of tuples $(j_i,k_i)$ like claimed in \eqref{ass.dis.coll.1}. The respective removal of the remaining `\textit{hard collision} partners' of $(j_i,k_i)$ after the related round ensures that also condition (iii) is fulfilled.  \\
After these considerations we can easily determine an upper bound for the probability $\mathbb{P}(X\in \mathcal{B}_{3,i}^{N,\sigma})$. Let to this end be $R, V>0$. We start with assumption \eqref{ass.dis.coll.1} and abbreviate for the moment 
$$M_1:=\lceil \frac{N^{-\frac{\sigma}{3}}M}{2}\rceil .$$ 
First, we remark that there obviously exist less than $\binom{N^2}{K}$ different possibilities to choose $K$ `disjoint' pairs $(j,k)$ belonging to $\{1,...,N\}^2\setminus \bigcup_{n=1}^N\{(n,n)\}$ (where by `disjoint' we mean that condition (iii) of \eqref{ass.dis.coll.1} is fulfilled). Application of this in the first step, subsequently Lemma \ref{lem3} in the third step and finally
$$\sup_{Y\in \mathbb{R}^6}\mathbb{P}\big(X_1\in (G^N(Y))^C\big)\le C N^{-\frac{4}{3}-2\sigma}$$ (which was shown in \eqref{prob G^N, alpha=2}) yields that
{\allowdisplaybreaks \begin{align}
&\mathbb{P}\Big(\exists \mathcal{S}\subseteq \{1,...,N\}^2\setminus \bigcup_{n=1}^N\{(n,n)\}:|\mathcal{S}|=M_1 \ \land \notag \\ & \hspace{0,6cm} \big(\forall (j,k)\in \mathcal{S}:X_j\in(G^N(X_k))^C\cap M^N_{R,V}(X_i)  \big) \ \land \notag \\
& \hspace{0,6cm} \big((j_1,k_1),(j_2,k_2)\in \mathcal{S}\Rightarrow  \{j_1,k_1\}\cap \{j_2,k_2\}=\emptyset\big)\Big) \notag  \\
\le & \binom{N^2}{M_1}\mathbb{P}\Big( \forall (j,k)\in \{(2,3),(4,5),...,(2M_1,2M_1+1)\}: \notag \\
&\hspace{1,5cm} X_j\in(G^N(X_k))^C\cap M^N_{R,V}(X_1) \Big) \notag  \\
\le &\frac{N^{2M_1}}{M_1!}\Big(\sup_{Y\in \mathbb{R}^6}\mathbb{P}\big(X\in (G^N(Y))^C\big)
\sup_{Z\in \mathbb{R}^6}\mathbb{P}\big(X\in M^N_{R,V}(Z)\big)\Big)^{M_1} \notag \\
\le & C^{M_1}\frac{N^{2M_1}}{M_1^{M_1}} \big(N^{-\frac{4}{3}-2\sigma}\big)^{M_1}\Big(R^2\min\big(\max(V,R),1\big)^4\Big)^{M_1}\notag \\
\le &  (CN^{-\frac{7\sigma}{3}} )^{\frac{N^{\frac{\sigma}{3}} }{2}} 
\end{align} }
where the last step follows after regarding that
\begin{align*}
M_1=\lceil \frac{N^{-\frac{\sigma}{3}} }{2}M\rceil \text{ and } M=\big\lceil N^{\frac{2}{3}\sigma}\lceil N^{\frac{2}{3}} R^2\min\big(\max(V,R),1\big)^4\rceil \big\rceil
\end{align*} which in particular implies that $M_1\geq \frac{N^{\frac{\sigma}{3}}}{2} $. Consequently, this probability decays distinctly faster than necessary for any class which appears in $\big(M^N_{(r_i,R_i),(v_i,V_i)}(Y)\big)_{i \in I_{\delta}}$ where $R_l\neq \infty$ and now it suffices to derive a suitable upper bound for the probability of the event related to assumption \eqref{ass.bad.inf.}. But previous to this we recall that we also have to show that typically $\sum_{k\in \mathcal{M}^N_b(X)}1\le N^{\frac{2}{3}(1+\sigma)}$. However, for this purpose the preceding reasoning can be applied as well if the collision class parameters $R,V$ are both set to infinity so that we obtain the trivial event $\mathbf{1}_{M^N_{\infty,\infty}(X_i)}(X_j)=1$. Everything stays the same except for the slight difference that this time $M_1:=\lceil \frac{N^{\frac{2}{3}+\frac{\sigma}{3}}}{2}\rceil$ and $\mathbb{P}\big(X_1\in M^N_{R,V}(Y)\big)=1$. Inserting this into the previous estimates shows that
$$\mathbb{P}\big(\sum_{k\in \mathcal{M}^N_b(X)}1\le N^{\frac{2}{3}(1+\sigma)}\big)\le CN^{-\sigma N^{\frac{2}{3}}} $$ 
which a fortiori is small enough.\\ 
Now we continue with the considerations for assumption \eqref{ass.bad.inf.}. To this end, we abbreviate $M_2:=\lceil  \frac{N^{\frac{\sigma}{3}}}{2}\rceil $ and it holds according to estimates \eqref{prob G^N, alpha=2} that
{\allowdisplaybreaks
\begin{align}
& \mathbb{P}\Big(X\in \mathbb{R}^{6N}: \big(\exists j \in \{1,...,N\}: \sum_{k\neq j} \mathbf{1}_{(G^N(X_j))^C}(X_k)\geq  M_2\big)\Big)\notag \\
\le & N\mathbb{P}\Big(X\in \mathbb{R}^{6N}: \sum_{k=2}^N \mathbf{1}_{(G^N(X_1))^C}(X_k)\geq M_2\Big) \notag \\
\le & N\binom{N}{M_2}\sup_{Y\in \mathbb{R}^6}\mathbb{P}\big(Z\in \mathbb{R}^{6}: Z\in (G^N(Y))^C  \big)^{M_2} \notag \\
\le & N \frac{N^{M_2}}{M_2!}  \big(CN^{-\frac{4}{3}-2\sigma}\big)^{M_2} \notag \\
\le & CN^{-\frac{1}{3}\lceil  \frac{N^{\frac{\sigma}{3}}}{2}\rceil }. \label{prob.est.} 
\end{align}}
In total we obtain
\begin{align}
& \mathbb{P}\big(X\in \mathcal{B}^{N,\sigma}_{3,i} \big) \notag \\
\le & |I_{\sigma}|\sup_{\substack{R,V>0}}\mathbb{P}\Big(\sum_{j \in \mathcal{M}_b^N(X)}\mathbf{1}_{M^N_{R,V}(X_i)}(X_j) \geq  N^{\frac{2\sigma}{3}}\big\lceil N^{\frac{2}{3}} R^2\min\big(\max(R,V),1\big)^4\big\rceil \Big) \notag \\
& + \mathbb{P}\big(\sum_{k\in \mathcal{M}^N_b(X)}1\geq N^{\frac{2}{3}(1+\sigma)}\big)  \notag\\
\le &(CN^{-\frac{7\sigma}{3}} )^{\frac{N^{\frac{\sigma}{3}}}{2} } 
 \label{prob.b.3}
\end{align}
which concludes the reasoning for term \eqref{term1}.\\\\ While for the last term we only had to consider the singularity parameter $\alpha= 2$ it will be necessary to take again both options $\alpha= 2$ respectively $\alpha\in (1,\frac{4}{3}]$ into account for the remaining term \eqref{term1,5} before we finally arrive at the concluding part.\\ We identify $$ v^N_{min}:=\begin{cases}N^{-\frac{5}{18}} &\text{ if }\alpha\in(1,\frac{4}{3}]\\
N^{-\frac{2}{9}}&\text{ if }\alpha=2 \end{cases}.$$ Analogous to the estimates for term \eqref{term1} it follows by application of Corollary \ref{cor1} that
{\allowdisplaybreaks
\begin{align}
& \int_{0}^t\Big(\frac{1}{N}\sum_{\substack{j\neq i\\ j\in \mathcal{M}^N_g(X)}}\Big(\big|f^{N}([^1\Psi^N_{s,0}(X)]_j-[^1\Psi^N_{s,0}(X)]_i)\big| \notag  \\
&+ \big|f^{N}(^1\varphi^N_{s,0}(X_j)-{^1\varphi^N_{s,0}}(X_i))\big|\Big)\mathbf{1}_{G^N(X_i)\cap M^N_{3N^{-\frac{1}{2}+\sigma},\infty}(X_i)}(X_j)\Big) ds \notag \\
\le & \frac{C}{N}\frac{1}{N^{-(\alpha-1)c} v^N_{min}} \sum_{j\neq i} \mathbf{1}_{G^N(X_i)\cap M^N_{3N^{-\frac{1}{2}+\sigma},N^{-\frac{1}{9}+3\sigma}}(X_i)}(X_j) \notag \\
&+  \frac{C}{N}\frac{1}{N^{-(\alpha-1)c}N^{-\frac{1}{9}+3\sigma}} \sum_{j\neq i} \mathbf{1}_{ G^N(X_i)\cap  M^N_{3N^{-\frac{1}{2}+\sigma},\infty}(X_i)}(X_j) \label{term1,5+}
\end{align}}
where the first term takes into account collisions where the relative velocity is below order $N^{-\frac{1}{9}+3\sigma}$ (but still larger than order $v_{min}$ if $X_j\in G^N(X_i)$) while the second deals with the rest. Corollary \ref{cor1} (ii) is applicable since the relative velocity values for the considered `collision classes' are of distinctly larger order than the deviation between corresponding particle trajectories of the microscopic and the auxiliary system. More precisely, we applied that $G^N(X_i)\subseteq M^N_{ 6\delta^N_b, v^N_{min}}(X_i)$ where $\delta_b^N=N^{-\frac{2}{9}-\sigma}$ if $\alpha=2$ and $\delta_b^N=N^{-\frac{1}{2}+\sigma}$ if $\alpha\in (1,\frac{4}{3}]$ (see \eqref{Def G.alpha=2} and \eqref{Def G, alpha=4/3}) as well as 
\begin{align*}
&\max_{i\in \mathcal{M}^N_g(X)}\sup_{0\le s \le \tau^N(X)}|[\Psi^N_{s,0}(X)]_i-\varphi^N_{s,0}(X_i)|\\
\le &\begin{cases}  N^{-\frac{1}{2}+\sigma}=N^{-\frac{2}{9}+\sigma}  v^N_{min} ,& \text{ if }\alpha\in (1,\frac{4}{3}]\\
N^{-\frac{7}{18}+\sigma}=N^{-\frac{1}{6}+\sigma} v^N_{min} ,& \text{ if }\alpha=2 \end{cases}
\end{align*}
where we recall that $v^N_{min}=N^{-\frac{5}{18}}$ if $\alpha\in(1,\frac{4}{3}]$ and $v^N_{min}=N^{-\frac{2}{9}}$ if $\alpha=2$. This shows that the assumptions of Corollary \ref{cor1} (ii) are fulfilled because as discussed at the beginning of the proof $\sigma>0$ can be chosen `small' (and in particular smaller than $\frac{1}{6}$).\\
Now we define a fourth set of `inappropriate' initial data as follows:
\begin{align}
\begin{split}
& X\in \mathcal{B}_{4,i}^{N,\sigma}\subseteq \mathbb{R}^{6N} \\
\Leftrightarrow &  \sum_{j \neq i}\mathbf{1}_{M^N_{6N^{-\frac{1}{2}+\sigma},N^{-\frac{1}{9}+3\sigma}}(X_i)}(X_j)\geq  N^{\frac{\sigma}{2}}\ \land \\
& \sum_{j \neq i}\mathbf{1}_{M^N_{6N^{-\frac{1}{2}+\sigma},\infty}(X_i)}(X_j)\geq  N^{3\sigma}
\end{split} \label{def.B_4}
\end{align}
The choice $N^{-\frac{1}{9}+3\sigma}$ for the velocity parameter is more or less random at the moment but will turn out to be reasonable during subsequent proofs where this set will be applied again. Due to our estimates it holds for configurations belonging to the complement of this set that term \eqref{term1,5+} (and thereby \eqref{term1,5}) is bounded by
\begin{align}
&\frac{C}{N}\frac{1}{N^{-(\alpha-1)c}v^N_{min}} N^{\frac{\sigma}{2}} +  \frac{C}{N}\frac{1}{N^{-(\alpha-1)c}N^{-\frac{1}{9}+3\sigma}} N^{3\sigma}\notag \\
\le &\begin{cases} CN^{-\frac{1}{2}(1-\sigma)} +CN^{-\frac{2}{3}}\le  CN^{-\frac{1}{2}(1-\sigma)}\ , & \ \alpha \in (1,\frac{4}{3}]\\
CN^{-\frac{7}{18}-\frac{\sigma}{2}} +CN^{-\frac{1}{2}}\ , & \ \alpha =2 \end{cases} \label{bound5}
\end{align}
where we regarded that $c=\frac{2}{3}$ and $v^N_{min}=N^{-\frac{5}{18}}$ if $\alpha\in (1,\frac{4}{3}]$ as well as $c=\frac{7}{18}-\sigma$ and $v^N_{min}=N^{-\frac{2}{9}}$ if $\alpha=2$.\\
We abbreviate this time $M_1:=\lceil N^{\frac{\sigma}{2}} \rceil$, $M_2:=\lceil N^{3\sigma} \rceil$ and by essentially the same reasoning as applied in \eqref{prob.est.} and application of Lemma \ref{lem3} we obtain that
\begin{align}
& \mathbb{P}\big(X\in  \mathcal{B}^{N,\sigma}_{4,i}\big) \notag \\
\le & \frac{N^{M_1}}{M_1!}\sup_{Y\in \mathbb{R}^6}\mathbb{P}\big(X_i\in M^N_{6N^{-\frac{1}{2}+\sigma},N^{-\frac{1}{9}+3\sigma}}(Y)\big)^{M_1} \notag \\
& + \frac{N^{M_2}}{M_2!}\sup_{Y\in \mathbb{R}^6}\mathbb{P}\big(X_i\in M^N_{6N^{-\frac{1}{2}+\sigma},\infty}(Y)\big)^{M_2} \notag \\
\le & \frac{(CN)^{M_1}}{M_1!}\big(N^{2(-\frac{1}{2}+\sigma)}\big)^{M_1}\big(N^{4(-\frac{1}{9}+3\sigma)}\big)^{M_1}  + C^{M_2}\frac{N^{M_2}}{(N^{3\sigma})^{M_2}}\big(N^{2(-\frac{1}{2}+\sigma)}\big)^{M_2}  \notag \\
\le &C\big(N^{-\frac{4}{9}+14\sigma} \big)^{N^{\frac{\sigma}{2}}} +\big( CN^{-\sigma}\big)^{N^{3\sigma}} \label{prob.b.4}
\end{align} 
which at least for small enough $\sigma>0$ decreases fast enough.\\  
Now we can finally conclude the proof for the first case where we have to distinguish between the two possible options $\alpha=2$ and $\alpha \in (1,\frac{4}{3}]$.\\
{ \bf Concluding estimates for $\alpha=2$:} Due to the previous probability estimates (see \eqref{prob.b.1}, \eqref{prob.b.2},\eqref{prob.b.3} and \eqref{prob.b.4}) it easily follows that for small enough $\sigma>0$ and an arbitrary $\gamma>0$ there exists a constant $C>0$ such that $$\mathbb{P}\big(\bigcup_{j\in \{1,2,3,4\}}\bigcup_{i=1}^N \mathcal{B}^{N,\sigma}_{j,i}\big)\le CN^{-\gamma}.$$ 
Initially we stated a sum of four terms which determines an upper limit for the term of our interest (see \eqref{term00}):
$$  \big|\int_{t_1}^t\frac{1}{N}\sum_{j\neq i}f^{N}([^1\Psi^N_{s,0}(X)]_i-[^1\Psi^N_{s,0}(X)]_j)-f^{N}*\widetilde{k}^N_s({^1\varphi^N_{s,0}}(X_i))ds\big|.$$ 
We point out that for $X\in\Big(\bigcup_{j\in \{1,2,3,4\}}\bigcup_{i=1}^N \mathcal{B}^{N,\sigma}_{j,i}\Big)^C$ the respective upper bounds which we derived in the previous part hold for any $t_1,t\in [0,\tau^N(X)]$. For the remaining part we restrict ourselves to these `good' configurations and it remains to merge the different bounds (see \eqref{term2}, \eqref{bound4}, \eqref{bound5}, \eqref{bound1, alpha=2} as well as definition \eqref{def.B_1}). First, we remark that $CN^{-\frac{7}{18}}$ dominates all of these upper bounds (for a suitable constant $C>0$) except for term \eqref{term2}. However, this term is exactly the reason why we introduced the set $\mathcal{B}^{N,\sigma}_{2,i}$ (see \eqref{def.B_2}) because for configurations $X\in \bigcap_{i=1}^N\big(\mathcal{B}^{N,\sigma}_{2,i}\big)^C$ it holds for any $i\in \{1,...,N\}$ and $t_1,t\in [0,T]$ that 
\begin{align*}
& \big|\int_{t_1}^t\Big(\frac{1}{N}\sum_{j=1}^N g^{N}(^1\varphi^N_{s,0}(X_j)-{^1\varphi^N_{s,0}}(X_i))\mathbf{1}_{G^N(X_i)}(X_j)\\
&  -\int_{\mathbb{R}^6} g^{N}(^1\varphi^N_{s,0}(Y)-{^1\varphi^N_{s,0}}(X_i))\mathbf{1}_{G^N(X_i)}(Y)k_0(Y)
d^6Y\Big)ds\Big|\le 1
\end{align*}
and thus for $N>1$ and $t_1\le t$
\begin{align}
& \int_{t_1}^t\frac{1}{N}\sum_{j=1}^N g^{N}(^1\varphi^N_{s,0}(X_j)-{^1\varphi^N_{s,0}}(X_i))\mathbf{1}_{G^N(X_i)}(X_j)ds \notag \\
\le & 1+\int_{t_1}^t\int_{\mathbb{R}^6} g^{N}(^1\varphi^N_{s,0}(Y)-{^1\varphi^N_{s,0}}(X_i))\mathbf{1}_{G^N(X_i)}(Y)k_0(Y)d^6Yds \notag \\
\le &1+ C\ln(N)(t-t_1) \label{est.sum.g}
\end{align}
where we regarded that (for $N>1$)
\begin{align*}
& \sup_{t_1\le s\le t}\int_{\mathbb{R}^6} g^{N}({^1\varphi^N_{s,0}}(Y) -{^1\varphi^N_{s,0}}(X_i))\mathbf{1}_{G^N(X_i)}(Y)k_0(Y)d^6Y  \\
\le &  C\sup_{t_1\le s\le t}\int_{\mathbb{R}^3}\min\big(N^{3c},\frac{1}{|Y-{^1\varphi^N_{s,0}}(X_i)|^3}\big)\widetilde{k}^N_s(Y)d^3Y  \\
\le & C\ln(N). 
\end{align*}
 If we abbreviate 
\begin{align}
\delta^N_g(t,X):=\max_{i\in \mathcal{M}^N_g(X)}|[^2\Psi^{N}_{t,0}(X)]_i-{^2\varphi^{N}_{t,0}}(X_i)| \label{abbrev.delta}
\end{align}
and recall the previously introduced abbreviation
$$\Delta^N_g(t,X):=\max_{i\in \mathcal{M}^N_g(X)}\sup_{0\le s\le t}|[^1\Psi^N_{s,0}(X)]_i-{^1\varphi^N_{s,0}}(X_i)|,$$ then it holds in particular for $t_1\le t$ that 
$$\Delta^N_g(t,X)\le \Delta^N_g(t_1,X)+\int_{t_1}^t\delta^N_g(s,X)ds .$$
If we choose for some constant $C_1>0$ the subsequent sequence of time steps 
\begin{align*}
& t_n=n\frac{C_1}{\sqrt{\ln(N)}} \text{ for }n\in \{0,...,\lceil \frac{\sqrt{\ln(N)}}{C_1}\tau^N(X)\rceil-1\},\\
& t_{\lceil \frac{\sqrt{\ln(N)}}{C_1}\tau^N(X)\rceil}=\tau^N(X)
\end{align*}
and abbreviate $t^*:=t_{n+1}-t_n= \frac{C_1}{\sqrt{\ln(N)}}$, then the previous relation easily implies that for $ t_n\le t\le \tau^N(X)$ 
\begin{align}
\Delta^N_g(t,X)\le \sum_{k=1}^n\sup_{0\le s \le t_k}\delta^N_g(s,X)t^*+\int_{t_n}^t\delta^N_g(s,X)ds
\label{rel.v.q.bound}
\end{align}
By merging the preceding considerations and by regarding in particular relation \eqref{est.sum.g} in the third step it follows that for any `good' particle $i\in \mathcal{M}^N_g(X)$, the considered configurations and for all times $t\in [t_n,t_{n+1}]$ where $n\in \{0,...,\lceil \frac{\sqrt{\ln(N)}}{C_1}\tau^N(X)\rceil-1\}$ the following inequality holds:
\begin{align}
&\delta^N_g(t,X)\notag \\
\le & \delta^N_g(t_n,X)+\max_{i\in \mathcal{M}^N_g(X)} \big|\int_{t_n}^t\Big(\frac{1}{N}\sum_{j\neq i}f^{N}([^1\Psi^N_{s,0}(X)]_i-[^1\Psi^N_{s,0}(X)]_j) \notag \\
&   -f^{N}*\widetilde{k}^N_s({^1\varphi^N_{s,0}}(X_i))\Big)ds\big| \notag  \\
\le &\max_{i\in \{1,...,N\}}\int_{t_n}^t \frac{2}{N}\sum_{j=1}^N g^{N}(^1\varphi^N_{s,0}(X_j)-{^1\varphi^N_{s,0}}(X_i))\mathbf{1}_{G^N(X_i)}(X_j)\underbrace{\Delta^N_g(s,X)}_{\le \Delta^N_g(t,X)}ds  \notag \\
&+\delta^N_g(t_n,X) +CN^{-\frac{7}{18}} \notag \\
\le &\big(1+ C\ln(N)\underbrace{(t-t_n)}_{\le t_{n+1}-t_n=t^*}\big)\big( \sum_{k=1}^n\sup_{0\le s \le t_k}\delta^N_g(s,X)t^*+\int_{t_n}^t\delta^N_g(r,X)dr\big)\notag \\
& +\delta^N_g(t_n,X)+CN^{-\frac{7}{18}} \notag \\
\le &\big(1+ C\ln(N)t^*\big)\int_{t_n}^t\delta^N_g(r,X) dr\notag \\
&  + \big(2+ C\ln(N)(t^*)^2\big)\sum_{k=1}^n\sup_{0\le s \le t_k}\delta^N_g(s,X) +CN^{-\frac{7}{18}}. 
\end{align} 
Application of Gronwall`s Lemma implies that for $t\in [t_n,t_{n+1}]$
\begin{align}
 & \delta^N_g(t,X)\notag \\
\le &  \Big(\big(2+ C\ln(N)(t^*)^2\big)\sum_{k=1}^n\sup_{0\le s \le t_k}\delta^N_g(s,X) +CN^{-\frac{7}{18}}\Big) e^{t^*+ C\ln(N){(t^*)}^2}.
\end{align}
Since this relation holds for all $t\in [t_n,t_{n+1}]$ and a fortiori for $t\in [0,t_n]$, we can `replace' the left-hand side by its supremum over $[0,t_{n+1}]$. Moreover, if we choose $C_1:=\min\big(\frac{1}{\sqrt{C}},1\big)$ for the constant appearing in the definition of $t^*=\frac{C_1}{\sqrt{\ln(N)}}$, then the previous relation implies 
\begin{align}
 & \sup_{0\le s \le t_{n+1}}\delta^N_g(s,X)
\le   3e^2\sum_{k=1}^n\sup_{0\le s \le t_k}\delta^N_g(s,X) +Ce^2N^{-\frac{7}{18}}. \label{Gron.velocity}
\end{align}
Due to this relation it follows for $n\in \{1,...,\lceil \frac{\sqrt{\ln(N)}}{C_1}\tau^N(X)\rceil\}$ that
\begin{align}
& \sup_{0\le s \le t_n}\delta^N_g(s,X)
\le  Ce^2N^{-\frac{7}{18}}(3e^2+1)^{n-1}. \label{upp.bound.v.thm1}
\end{align}
For $n=1$ the relation is obvious due to \eqref{Gron.velocity} and if it holds for $k\in \{1,...,n\}$, $n\in \mathbb{N}$ (where we fix for these estimates the constant $C$), then we obtain that
\begin{align*}
& \sup_{0\le s \le t_{n+1}}\delta^N_g(s,X)\\
\le &  3e^2\sum_{k=1}^n \underbrace{\sup_{0\le s \le t_k}\delta^N_g(s,X) }_{\le   Ce^2N^{-\frac{7}{18}}(3e^2+1)^{k-1}}+Ce^2N^{-\frac{7}{18}} \\
\le & 3e^2\big(  Ce^2N^{-\frac{7}{18}}\frac{(3e^2+1)^n-1}{(3e^2+1)-1}\big)+Ce^2N^{-\frac{7}{18}}\\
=&  Ce^2N^{-\frac{7}{18}}(3e^2+1)^n
\end{align*}
which confirms the claim. Hence, it follows (for large enough $N\in \mathbb{N}$) that 
\begin{align}
\sup_{0\le s \le \tau^N(X)}\delta^N_g(s,X)
\le &  Ce^2N^{-\frac{7}{18}}(3e^2+1)^{\lceil \frac{\sqrt{\ln(N)}}{C_1}\tau^N(X)\rceil-1} \notag \\
\le & Ce^2N^{-\frac{7}{18}}N^{ \frac{\ln(3e^2+1)}{\ln(N)}\frac{\sqrt{\ln(N)}}{C_1}T} \notag \\
\le & CN^{-\frac{7}{18}+\frac{\sigma}{2}}.
\end{align}
This concludes the estimates because the received upper limit for the velocity deviation easily implies that
\begin{align}
& \max_{i\in \mathcal{M}^N_g(X)}\sup_{0\le s \le \tau^N(X)}|[\Psi^N_{s,0}(X)]_i-\varphi^N_{s,0}(X_i)|  
\le  CN^{-\frac{7}{18}+\frac{\sigma}{2}}
\end{align}
which is smaller than the allowed deviation $N^{-\frac{7}{18}+\sigma}$ for large enough values of $N$. \\
Finally, we can focus again on the case $\alpha\in (1,\frac{4}{3}]$ and prove in the following that typically no particle `triggers' the stopping time in this setting. \\
{ \bf Concluding estimates for $\alpha\in (1,\frac{4}{3}]$:} Large parts of the reasoning are very similar (or simpler) in the current situation but first we have to verify our claim that in the less singular case typically no `bad' particle occurs. Fortunately, we already implemented the necessary estimates and if we define
\begin{align}
& X\in \mathcal{B}^{N,\sigma}_5  \notag \\
\Leftrightarrow & X\in \mathbb{R}^{6N}:\big(\exists (i,j)\in \{1,...,N\}^2:i\neq j \land X_j\notin  G^N(X_i)\big), \label{def.B_5}
\end{align} 
then it holds according to estimate \eqref{prob G^N, alpha=4/3} that
\begin{align*}
\mathbb{P}\big(X\in \mathcal{B}^{N,\sigma}_5\big)
\le \binom{N}{2}\mathbb{P}\big(X_2 \notin G^N(X_1)\big)\le  N^2(C N^{-\frac{19}{9}+2{\sigma}})\le C N^{-\frac{1}{9}+2{\sigma}}.
\end{align*} 
In the current case we restrict our initial data even further than in the previous one and identify to this end 
\begin{align}
 \mathcal{G}^{N,\sigma}_1:= \Big(\mathcal{B}^{N,\sigma}_5\cup \bigcup_{j\in \{1,2,3,4\}}\bigcup_{i=1}^N \mathcal{B}^{N,\sigma}_{j,i}\Big)^C \label{def.G_1} 
\end{align}
which leads to a slower decay of the probability of excluded configurations:
$$\mathbb{P}\Big(X\in \big(\mathcal{G}^{N,\sigma}_1\big)^C\Big)\le C N^{-\frac{1}{9}+2\sigma}$$ 
However, this decay is still sufficiently fast for verifying the Theorem since according to the discussion at the beginning of the proof $\sigma>0$ can be chosen arbitrarily small and thus in particular smaller than $\frac{\epsilon}{2}$ for any given $\epsilon>0$. Hence, it remains to show that for such configurations the stopping time does not get triggered. At this point we already remark that the set $\mathcal{G}^{N,\sigma}_1$ will play also a crucial role in all proofs concerning Theorem \ref{thm2}.\\
By restricting the initial data further the advantage arises that the contribution of term \eqref{term1} disappears since as claimed no `bad' particle exists for such configurations. One easily comprehends that all upper limits which arise again by terms \eqref{term2}, \eqref{bound5}, \eqref{bound1, alpha=4/3} as well as definition \eqref{def.B_1} are bounded by $CN^{-\frac{1}{2}(1-\sigma)}$ this time except for \eqref{term2}. However, the same reasoning as in the first case applies with the slight (but important) difference that for $\alpha\in (1,\frac{4}{3}]$:
\begin{align*}
& \int_{\mathbb{R}^6} g^{N}({^1\varphi^N_{s,0}}(Y) -{^1\varphi^N_{s,0}}(X_i))\mathbf{1}_{G^N(X_i)}(Y)k_0(Y)d^6Y  \\
\le &  C\int_{\mathbb{R}^3}\min\big(N^{(\alpha+1)c},\frac{1}{|Y-{^1\varphi^N_{s,0}}(X_i)|^{\alpha+1}}\big)\widetilde{k}^N_s(Y)d^3Y  \\
\le & C
\end{align*}
and thus 
\begin{align*}
& \int_{0}^t\frac{1}{N}\sum_{j=1}^N g^{N}(^1\varphi^N_{s,0}(X_j)-{^1\varphi^N_{s,0}}(X_i))\mathbf{1}_{G^N(X_i)}(X_j)ds
\le  C
\end{align*} for the considered configurations and $t\in [0,T]$. Hence, it follows by basically copying the estimates of the previous case that $\forall t\in [0, \tau^N(X)]$ and the relevant initial data:
\begin{align}
&\delta^N_g(t,X)\notag \\
\le & \max_{i\in \mathcal{M}^N_g(X)} \big|\int_{0}^t\frac{1}{N}\sum_{j\neq i}f^{N}([^1\Psi^N_{s,0}(X)]_i-[^1\Psi^N_{s,0}(X)]_j) -f^{N}*\widetilde{k}^N_s({^1\varphi^N_{s,0}}(X_i))ds\big| \notag  \\
\le &\max_{i\in \{1,...,N\}}\int_0^t \frac{1}{N}\sum_{j=1}^N g^{N}(^1\varphi^N_{s,0}(X_j)-{^1\varphi^N_{s,0}}(X_i))\mathbf{1}_{G^N(X_i)}(X_j)ds \Delta^N_g(t,X) \notag \\
& +CN^{-\frac{1}{2}(1-\sigma)} \notag \\
\le &C \int_0^t\delta^N_g(s,X)ds +CN^{-\frac{1}{2}(1-\sigma)} . 
\end{align} 
where we recall that $\delta^N_g(t,X)$ was defined in \eqref{abbrev.delta}.\\
Application of Gronwall`s Lemma and taking the supremum over $[0,t]$ subsequently yields that 
\begin{align}
 & \sup_{0\le s \le t}\delta^N_g(t,X)
\le   CN^{-\frac{1}{2}(1-\sigma)}e^{Ct} \label{Gron-case 4/3} .
\end{align} 
and thus also
\begin{align}
\max_{i\in \mathcal{M}^N_g(X)}\sup_{0\le s\le t}|[^1\Psi^N_{s,0}(X)]_i-{^1\varphi^N_{s,0}}(X_i)|\le tCN^{-\frac{1}{2}(1-\sigma)}e^{Ct}.
\end{align}
By taking into account that $ \mathcal{M}^N_g(X)=\{1,...,N\}$ for the considered configurations we can finally conclude that
\begin{align}
\sup_{0\le s \le \tau^N(X)}|{\Psi^{N}_{s,0}(X)}-{\Phi^{N}_{s,0}}(X)|_{\infty}\le CN^{-\frac{1}{2}+\frac{\sigma}{2}}. \label{Gron-case 4/3+}
\end{align}
which for large $N$ keeps sufficiently small so that the stopping time is not `triggered`.\\
Consequently, the proof for statement (ii) of the theorem is already completed and we can go on with the estimates for the `bad' particles if $\alpha=2$.
\paragraph{Controlling the deviations of the `bad' particles:}
For the whole section we only consider the case $\alpha=2$ and thus $c=\frac{7}{18}-\sigma$. Most estimates for the second part are similar. The only new problem is that we have less control on the position of the observed `bad' particle. While the maximal distance of a `good' particle to its `mean-field particle partner' is of the same order as the cut-off radius, the situation is different in this case. The strategy to handle this problem will be as follows: Since the vast majority of particles is typically `good', we have at least sufficient control on the positions of most potential `collision partners'. Furthermore, we know that the considered `bad' particle moves in a ball of radius $N^{-\frac{2}{9}-\sigma}$ around its related `mean-field particle'. In a small area containing this ball at the initial time we will introduce a homogeneously distributed cloud of auxiliary `mean-field particles' having respectively a distance slightly smaller than the cut-off size to their nearest neighbor (and the properties of the mean-field dynamics ensure that this homogeneity propagates in time to a certain extend). `Hard' collisions might cause that the observed particle departs too far from its initially corresponding `mean-field particle'. However, in this case there always exists another auxiliary particle belonging to this cloud which is closer to the observed one than the cut-off radius. If we exchange the currently applied auxiliary particle by such a closer one as soon as its distance to the related particle of the microscopic system becomes to large, then it will turn out that most of the estimates can simply be copied from case 1. At first sight it might seem strange to introduce a whole `cloud' of such auxiliary particles instead of introducing every time a single new one at the position of the observed particle when it departs to far from its current `mean-field particle partner'. The arising problem is that in this case the introduced auxiliary particle would be correlated with the remaining particles since its position in phase space would depend on the whole configuration. This issue does not occur if the mentioned `cloud' is applied because for a certain particle the initial positions of the related auxiliary particles are chosen independent of the remaining configuration (but at the price that we need to introduce many of them). In case 1 we showed that for typical initial data the related `mean-field particles' fulfill a number of properties which made it possible to prove that the effective and the microscopic dynamics are usually close. This time we show that all of the auxiliary particles which belong to the small `cloud' fulfill corresponding demands with high probability and thus we will end up in a very similar situation as in case 1.     \\\\  
To implement this idea we first define 
\begin{align}
J_N := & \{-\lceil N^{\frac{1}{6}} \rceil ,...,-1,0,1,..., \lceil  N^{\frac{1}{6}} \rceil\}^6  \label{J_N}
\end{align}
and for $(k_1,...,k_6)\in J_N$ the positions $X^i_{k_1,...,k_6}:=X_i+\sum_{j=1}^6 k_jN^{-\frac{7}{18}+\frac{\sigma}{2}}{e}_j$ (where ${e}_j$, $j\in \{1,...,6\}$ shall denote the canonical basis vectors of $\mathbb{R}^6$). These configurations can be understood as the initial data of the auxiliary particles described in the preliminary considerations. According to Lemma \ref{lem1} (which is applied in the first step) and the condition on the distance between corresponding `real' and `mean-field particle' for times before the stopping time is `triggered' (used for the second inequality), it holds for arbitrary $t_1\in [0,\tau^N(X)]$ and large enough $N$ that 
\begin{align*}
|\varphi^N_{0,t_1}([\Psi^N_{t_1,0}(X)]_i)-X_i|\le C|[\Psi^N_{t_1,0}(X)]_i-\varphi^N_{t_1,0}(X_i)|<CN^{-\frac{2}{9}-\sigma}\le N^{-\frac{2}{9}}.
\end{align*}
Thus, this distance is of smaller order with respect to $N$ than the diameter of the auxiliary `particle cloud' around $X_i$ and thereby it is always possible to find a tuple $(k_1,...,k_6)\in J_N$ such that 
\begin{align}
|\varphi^N_{0,t_1}([\Psi^N_{t_1,0}(X)]_i)-X^i_{k_1,...,k_6}|\le \frac{\sqrt{6}}{2} N^{-\frac{7}{18}+\frac{\sigma}{2}} 
\end{align}
if $N$ is sufficiently large. Lemma \ref{lem1} implies in turn that
\begin{align}
|[\Psi^N_{t_1,0}(X)]_i-\varphi^N_{t_1,0}(X^i_{k_1,...,k_6})|\le C  N^{-\frac{7}{18}+\frac{\sigma}{2}}. \label{t_1,t_2}
\end{align} 
Let $N\in \mathbb{N}$ be large enough such that $C  N^{-\frac{7}{18}+\frac{\sigma}{2}}< \frac{1}{2}  N^{-\frac{7}{18}+\sigma}$ then there exists a further point in time $t_2 \in (t_1, T]$ such that
$$\sup_{s\in [t_1,t_2]}|[^1\Psi^N_{s,0}(X)]_i-{^1\varphi^N_{s,0}(X^i_{k_1,...,k_6})}|\le  N^{-\frac{7}{18}+\sigma}$$
holds as well as the following bound for the velocity deviation if $\sigma>0$ is sufficiently small:
$$\sup_{s\in [t_1,t_2]}|[^2\Psi^N_{s,0}(X)]_i-{^2\varphi^N_{s,0}(X^i_{k_1,...,k_6})}|\le   N^{-\frac{2}{9}-\sigma}$$
For this time span the auxiliary particle related to the initial data $X^i_{k_1,...,k_6}$ provides a sufficiently good approximation for the trajectory of the corresponding `real' one and we want to apply it in the following to show that $\sup_{t_1\le s \le t}|[\Psi^N_{s,0}(X)]_i-{\varphi^N_{s,0}}(X_i)|$ grows slow enough on this interval. More precisely, we utilize that this variable is bounded by
\begin{align}
\sup_{t_1\le s \le t}|[\Psi^N_{s,0}(X)]_i-{\varphi^N_{s,0}}(X^i_{k_1,...,k_6})|+\sup_{t_1\le s \le t}|{\varphi^N_{s,0}}(X^i_{k_1,...,k_6})-{\varphi^N_{s,0}}(X_i)| \label{double.upp.bound}
\end{align}
and derive upper bounds for the growth of these deviations instead. The reason for this is based on the fact that the considerations for the first of these two terms is mostly analogous to the estimates of case 1 because the spatial distance between the considered auxiliary particle and the `real' particle is bounded from above by $N^{-\frac{7}{18}+\sigma}$ (which is also the largest allowed deviation for a `good' particle).\\
For the remaining part of the proof we will always assume that for an arbitrary point in time $t_1\in [0,\tau^N(X))$ and $X\in \mathbb{R}^{6N}$ the initial position of the auxiliary particle $X^i_{k_1,...,k_6}$ and $t_2\in (t_1,\tau^N(X)]$ are chosen such that the previously introduced demands are fulfilled on $[t_1,t_2]$. Obviously $t_2$ and the choice of $(k_1,...,k_6)\in J_N$ depend on $i,t_1$ and $X$ but for a clear notation we omit to make this explicit. For the same reason we simply abbreviate $\widetilde{X}_i:=X^i_{k_1,...,k_6}$.\\
We start with the second of these two terms because controlling its growth is a simple application of Lemma \ref{lem1}. Due to this lemma it follows for arbitrary $t\in [t_1,t_2]$ that
\begin{align}
& |{\varphi^N_{t,0}}(\widetilde{X}_i)-{\varphi^N_{t,0}}(X_i)| \notag \\
\le & e^{C(t-t_1)}|{\varphi^N_{t_1,0}}(\widetilde{X}_i)-{\varphi^N_{t_1,0}}(X_i)| \notag \\
\le & e^{C(t-t_1)}\big(\big|\varphi^N_{t_1,0}({X}_i)-[\Psi^N_{t_1,0}(X)]_i\big|+\big|[\Psi^N_{t_1,0}(X)]_i-\varphi^N_{t_1,0}(\widetilde{X}_i)\big|\big)\notag  \\
\le & e^{C(t-t_1)}\big(\big|\varphi^N_{t_1,0}({X}_i)-[\Psi^N_{t_1,0}(X)]_i\big|+N^{-\frac{7}{18}+\sigma}\big). \label{est.eff.bad.term}
\end{align}
where we regarded the allowed upper bound for $\big|[\Psi^N_{t_1,0}(X)]_i-\varphi^N_{t_1,0}(\widetilde{X}_i)\big|$ according to the choice of $\widetilde{X}_i$. This already concludes the estimates for this term and we return to it later.\\
For the second term we first remark that  
\begin{align}
&  |[^2\Psi^N_{t,0}(X)]_i-{^2\varphi^N_{t,0}}(\widetilde{X}_i)|  \notag \\
\le  & |[^2\Psi^N_{t_1,0}(X)]_i-{^2\varphi^N_{t_1,0}}(\widetilde{X}_i)|  \notag \\
& + \big|\int_{t_1}^{t}\Big(\frac{1}{N}\sum_{j\neq i}f^{N}([^1\Psi^N_{s,0}(X)]_i-[^1\Psi^N_{s,0}(X)]_j)-f^{N}*\widetilde{k}_s(^1\varphi^N_{s,0}(\widetilde{X}_i))\Big)ds\big|. 
\end{align}
Now we want to derive an upper bound for the force term and as noted the situation is almost the same as in the previous case. Hence, it is not surprising that we apply again multiple times triangle inequality to obtain essentially the four terms of case 1:
{\allowdisplaybreaks
\begin{align}
& \big|\int_{t_1}^{t}\frac{1}{N}\sum_{j\neq i}f^{N}([^1\Psi^N_{s,0}(X)]_i-[^1\Psi^N_{s,0}(X)]_j)-f^{N}*\widetilde{k}_s(^1\varphi^N_{s,0}(\widetilde{X}_i))ds\big| \label{force.bad case}\\
\le &  \big|\int_{t_1}^{t}\frac{1}{N}\sum_{j\neq i}f^{N}([^1\Psi^N_{s,0}(X)]_i-[^1\Psi^N_{s,0}(X)]_j)\mathbf{1}_{(G^N(\widetilde{X}_i))^C}(X_j)ds\big| \label{bad.term} \\
&+  \big|\int_{t_1}^{t}\frac{1}{N}\sum_{j\neq i}\Big(
f^{N}([^1\Psi^N_{s,0}(X)]_i-[^1\Psi^N_{s,0}(X)]_j)\mathbf{1}_{G^N(\widetilde{X}_i)}(X_j) \notag  \\
&-f^{N}(^1\varphi^N_{s,0}(\widetilde{X}_i)-{^1\varphi^N_{s,0}}(X_j))\mathbf{1}_{G^N(\widetilde{X}_i)}(X_j)\Big) ds\big| \label{c_2t3} \\
& +\big| \int_{t_1}^{t} \frac{1}{N}\sum_{j\neq i}f^{N}(^1\varphi^N_{s,0}(\widetilde{X}_i)-{^1\varphi^N_{s,0}}(X_j))\mathbf{1}_{G^N(\widetilde{X}_i)}(X_j)ds  \notag\\
&-\int_{t_1}^{t}\int_{\mathbb{R}^6}f^N(^1\varphi^N_{s,0}(\widetilde{X}_i)-{^1\varphi^N_{s,0}}(Y))\mathbf{1}_{G^N(\widetilde{X}_i)}(Y)k_0(Y)d^6Yds\big| \label{c_2t5}  \\
& +\big|\int_{t_1}^{t}\Big(\int_{\mathbb{R}^6}f^N(^1\varphi_{s,0}^N(\widetilde{X}_i)-{^1\varphi^N_{s,0}}(Y))\mathbf{1}_{G^N(\widetilde{X}_i)}(Y)k_0(Y)d^6Y  \notag\\
&-f^{N}*\widetilde{k}^N_s(^1\varphi^N_{s,0}(\widetilde{X}_i))\Big)ds\big| \label{c_2t6,2}
\end{align} }
A suitable upper bound for term \eqref{c_2t6,2} can be derived analogously to the estimates leading to term \eqref{bound1, alpha=2} and thus is given by $ CN^{-\frac{5}{9}}$. Next, we care for term \eqref{c_2t3} and at the same time for \eqref{c_2t5}. Since according to the choice of $t_1, t_2$ and $\widetilde{X}_i$ it holds that $\sup_{t_1\le s \le t_2}|[^1\Psi^N_{s,0}(X)]_i-{^1\varphi^N_{s,0}}(\widetilde{X}_i)|\le N^{-\frac{7}{18}+\sigma}$ (which as stated is the same value as the upper bound for the allowed deviations of `good' particles), it follows with the help of the map $g^N$ (see \eqref{Def.g^N}) that
\begin{align}
& \big|\int_{t_1}^{t}\frac{1}{N}\sum_{j\neq i}\Big(f^{N}([^1\Psi^N_{s,0}(X)]_i-[^1\Psi^N_{s,0}(X)]_j)\mathbf{1}_{G^N(\widetilde{X}_i)}(X_j) \notag  \\
&-f^{N}(^1\varphi^N_{s,0}(\widetilde{X}_i)-{^1\varphi^N_{s,0}}(X_j))\mathbf{1}_{G^N(\widetilde{X}_i)}(X_j)\Big) ds\big| \label{good force,bad case} \\
\le  &  \big|\int_{t_1}^{t}\frac{1}{N}\sum_{j\in \mathcal{M}^N_b(X)\setminus \{i\}}\Big(f^{N}([^1\Psi^N_{s,0}(X)]_i-[^1\Psi^N_{s,0}(X)]_j)\mathbf{1}_{G^N(\widetilde{X}_i)}(X_j) \notag   \\
&-f^{N}(^1\varphi^N_{s,0}(\widetilde{X}_i)-{^1\varphi^N_{s,0}}(X_j))\mathbf{1}_{G^N(\widetilde{X}_i)}(X_j)\Big) ds\big| \notag  \\
& +\int_{t_1}^{t}\frac{1}{N}\sum_{j\in \mathcal{M}^N_g(X)\setminus \{i\}}\Big( g^{N}(^1\varphi^N_{s,0}(\widetilde{X}_i)-{^1\varphi^N_{s,0}}(X_j))\mathbf{1}_{G^N(\widetilde{X}_i)}(X_j) \notag  \\
& \cdot \big(|[^1\Psi^N_{s,0}(X)]_i-{^1\varphi^N_{s,0}}(\widetilde{X}_i)|+|[^1\Psi^N_{s,0}(X)]_j-{^1\varphi^N_{s,0}}({X}_j)| \big)\Big) ds. \label{est.b2}
\end{align}
Obviously, all these terms have basically the same structure as in case 1. We just have to adjust the definitions of the sets $\mathcal{B}^{N,\sigma}_{i,j}$ from the previous to the current situation. More precisely, we recall that $\widetilde{X}_i$ was only an abbreviation for $X^i_{k_1,...k_6}$ and define for $(k_1,...,k_6)\in J_N$:
\begin{align}
& X\in \mathcal{B}_{1,i,(k_1,..,k_6)}^{N,\sigma}\subseteq \mathbb{R}^{6N} \notag\\
\Leftrightarrow & \exists t_1',t_2'\in [0,T]:\notag \\
& \Big|\int_{t_1'}^{t_2'}\Big(\frac{1}{N}\sum_{j\neq i}f^{N}(^1\varphi^N_{s,0}(X^i_{k_1,...,k_6})-{^1\varphi^N_{s,0}}(X_j))\mathbf{1}_{G^N(X^i_{k_1,...k_6})}(X_j) \notag \\
&  -\int_{\mathbb{R}^6} f^{N}(^1\varphi^N_{s,0}(X^i_{k_1,...,k_6})-{^1\varphi^N_{s,0}}(Y)) \notag \\
& \hspace{3,5cm} \cdot \mathbf{1}_{G^N(X^i_{k_1,...k_6})}(Y)k_0(Y)
d^6Y\Big)ds\Big| > N^{-\frac{7}{18}}\ \vee \label{Def.B-bad.2} \\
& \Big|\int_{t_1'}^{t_2'}\Big(\frac{1}{N}\sum_{j\neq i}g^{N}(^1\varphi^N_{s,0}(X^i_{k_1,...,k_6})-{^1\varphi^N_{s,0}}(X_j))\mathbf{1}_{G^N(X^i_{k_1,...k_6})}(X_j) \notag \\
&  -\int_{\mathbb{R}^6}  g^N(^1\varphi^N_{s,0}(X^i_{k_1,...,k_6})-{^1\varphi^N_{s,0}}(Y)) \notag \\
& \hspace{3,5cm}\cdot \mathbf{1}_{G^N(X^i_{k_1,...k_6})}(Y)k_0(Y)
d^6Y\Big)ds\Big| > 1\ \label{Def.B-bad.1} 
\end{align}
Hence, statement \eqref{Def.B-bad.2} is the same as applied in the definition of the set $\mathcal{B}^{N,\sigma}_{1,i}$ (see \eqref{def.B_1}) except for the irrelevant difference that $X_i$ is replaced by the initial data of another auxiliary particle $X^i_{k_1,...,k_6}:=X_i+\sum_{j=1}^6k_j N^{-\frac{7}{18}+\frac{\sigma}{2}}e_j$. For statement \eqref{Def.B-bad.1}, on the other hand, a corresponding relationship holds, however, with respect to $\mathcal{B}^{N,\sigma}_{2,i}$. Consequently, it follows analogous to the reasoning applied for the sets $\mathcal{B}^{N,\sigma}_{j,i}, \ j\in \{1,2\}$ that for any $\gamma>0$ there exists $C_\gamma>0$ such that for all $N\in \mathbb{N}$
$$
\mathbb{P}\big(X\in \mathcal{B}^{N,\sigma}_{1,i,(k_1,...,k_6)}\big)\le C_\gamma N^{-\gamma}
.$$ Like in case 1 restricting the initial data to this set is already enough to handle term \eqref{c_2t5} and the second term of \eqref{est.b2}. Thus, we can start with the considerations for the first term of \eqref{est.b2} (and after that only term \eqref{bad.term} remains). In the proof of case 1 the set $\mathcal{B}_{3,i}^{N,\sigma}$ (see \eqref{def.B_3}) was introduced to deal with the corresponding term and since the situation is basically the same we just have to modify the definition such that it applies for $X^i_{k_1,...,k_6}$ for $(k_1,...,k_6)\in J_N$:
\begin{align}
& X\in \mathcal{B}_{2,i,(k_1,...,k_6)}^{N,\sigma} \subseteq \mathbb{R}^{6N} \notag  \\
\Leftrightarrow & \exists l\in I_{\sigma}: \Big(R_l\neq \infty\ \land \notag  \\ &\sum_{j \in \mathcal{M}^N_b(X)\setminus \{i\}}\mathbf{1}_{M^N_{(r_l,R_l),(v_l,V_l)}(X^i_{k_1,...,k_6})}(X_j) \geq  N^{\frac{2\sigma}{3}}\big\lceil N^{\frac{2}{3}} R_l^2\min\big(\max(V_l,R_l),1\big)^4\big\rceil\Big) \ \vee \notag  \\
& \sum_{j \in \mathcal{M}^N_b(X)\setminus \{i\}}1\geq N^{\frac{2}{3}(1+\sigma)}  \label{Def.B_3.bad}
\end{align}
For $X\in  \big(\mathcal{B}_{2,i,(k_1,...,k_6)}^{N,\sigma} \big)^C$ and $t\in [t_1,t_2]$ the term 
\begin{align}
& \big|\int_{t_1}^{t}\frac{1}{N}\sum_{j\in \mathcal{M}^N_b(X)\setminus \{i\}}\Big(f^{N}([^1\Psi^N_{s,0}(X)]_j-[^1\Psi^N_{s,0}(X)]_i)\mathbf{1}_{G^N(\widetilde{X}_i)}(X_j) \notag   \\
&-f^{N}(^1\varphi^N_{s,0}(X_j)-{^1\varphi^N_{s,0}}(\widetilde{X}_i))\mathbf{1}_{G^N(\widetilde{X}_i)}(X_j)\Big) ds\big|  \label{imp.bad.part.bad.case}
\end{align}
can be handled by the same estimates as applied in case 1. For this purpose, one has to take into account the choice of the interval $[t_1,t_2]$ because for this time span it holds that
\begin{align*}
& \sup_{t\in [t_1,t_2]}|[^1\Psi^N_{s,0}(X)]_j-{^1\varphi^N_{s,0}}(\widetilde{X}_i)|\le N^{-\frac{7}{18}+\sigma}\ \land \\
& \sup_{t\in [t_1,t_2]}|[^2\Psi^N_{s,0}(X)]_j-{^2\varphi^N_{s,0}}(\widetilde{X}_i)|\le N^{-\frac{2}{9}-\sigma}.
\end{align*}
On can easily comprehend by the considerations starting after \eqref{def.B_3} that these estimates can be copied in the current situation and hence also the previously derived upper bound $CN^{-\frac{7}{18}}$ can be applied (see \eqref{bound4}).\\
This concludes the considerations for term \eqref{c_2t3} respectively \eqref{good force,bad case} and we record that due to definition \eqref{Def.B-bad.1} and the subsequent reasoning it holds for configurations $X\in \big( \mathcal{B}^{N,\sigma}_{1,i,(k_1,...,k_6)}\cup  \mathcal{B}^{N,\sigma}_{2,i,(k_1,...,k_6)}\big)^C$ and $t\in [t_1,t_2]$ that
{\allowdisplaybreaks
\begin{align}
& \big|\int_{t_1}^{t}\frac{1}{N}\sum_{j\in \mathcal{M}^N_b(X)\setminus \{i\}}\Big(f^{N}([^1\Psi^N_{s,0}(X)]_i-[^1\Psi^N_{s,0}(X)]_j)\mathbf{1}_{G^N(\widetilde{X}_i)}(X_j) \notag   \\
&-f^{N}(^1\varphi^N_{s,0}(\widetilde{X}_i)-{^1\varphi^N_{s,0}}(X_j))\mathbf{1}_{G^N(\widetilde{X}_i)}(X_j)\Big) ds\big| \notag  \\
& +\int_{t_1}^{t}\frac{1}{N}\sum_{j\in \mathcal{M}^N_g(X)\setminus \{i\}}\Big( g^{N}(^1\varphi^N_{s,0}(\widetilde{X}_i)-{^1\varphi^N_{s,0}}(X_j))\mathbf{1}_{G^N(\widetilde{X}_i)}(X_j)\notag  \\
& \cdot \big(|[^1\Psi^N_{s,0}(X)]_i-{^1\varphi^N_{s,0}}(\widetilde{X}_i)|+|[^1\Psi^N_{s,0}(X)]_j-{^1\varphi^N_{s,0}}({X}_j)| \big)\Big) ds \notag \\
\le &  CN^{-\frac{7}{18}} \notag  \\
& +\Big(1+\int_{t_1}^{t}\int_{\mathbb{R}^6}  g^{N}(^1\varphi^N_{s,0}(\widetilde{X}_i)-{^1\varphi^N_{s,0}}(Y))k_0(Y)
d^6Yds\Big) \notag\\
& \cdot \sup_{s\in [t_1,t]}\Big(|[^1\Psi^N_{s,0}(X)]_i-{^1\varphi^N_{s,0}}(\widetilde{X}_i)|+\max_{j\in \mathcal{M}_g^N(X) }|[^1\Psi^N_{s,0}(X)]_j-{^1\varphi^N_{s,0}}({X}_j)| \Big) \notag \\
\le & CN^{-\frac{7}{18}} +C\big(1+(t-t_1)\ln(N)\big)N^{-c} \label{est.b1}
\end{align}}
The derivation of the upper bound for the first term was already discussed previously. For the upper bound of the second term we regarded that $0\le g^N(q)\le C\min(N^{3c},\frac{1}{|q|^3})$ which leads to the factor $C\ln(N)$ after the integration as well as 
$$|[^1\Psi^N_{s,0}(X)]_i-{^1\varphi^N_{s,0}}(\widetilde{X}_i)|+\max_{j\in \mathcal{M}_g^N(X) }|[^1\Psi^N_{s,0}(X)]_j-{^1\varphi^N_{s,0}}({X}_j)|\le 2N^{-c}$$ for $s\in [t_1,t]$ due to $t\in [t_1,t_2]\subseteq [t_1,\tau^N(X)]$, the constraints on $t_2$ and the definition of the stopping time (see \eqref{Def.stopping}).\\
We finally arrived at the last remaining term which is \eqref{bad.term}:
\begin{align*}
 \big|\int_{t_1}^{t}\frac{1}{N}\sum_{j\neq i}f^{N}([^1\Psi^N_{s,0}(X)]_i-[^1\Psi^N_{s,0}(X)]_j)\mathbf{1}_{(G^N(\widetilde{X}_i))^C}(X_j)ds\big|
\end{align*} This term takes into account the impact of the `hard' collisions which were excluded for the `good' particles. Thus, it constitutes basically the first significant modification in contrast to the considerations for the `good' particles. Fortunately, the estimates for this remaining term are straightforward but first we need to define for the last time in this proof a set of inappropriate initial data for $(k_1,...,k_6)\in J_N$ and $i\in \{1,...,N\}$:
\begin{align}
\begin{split}
& X\in \mathcal{B}_{3,i,(k_1,...,k_6)}^{N,\sigma}\subseteq \mathbb{R}^{6N} \\
\Leftrightarrow &  \sum_{j \neq i}\mathbf{1}_{M^N_{6N^{-\frac{2}{9}-\sigma},N^{-\frac{2}{9}}}(X^i_{k_1,...,k_6})}(X_j)\geq  N^{\frac{\sigma}{2}}
\end{split} \label{def.B_3.bad}
\end{align} 
After recalling that $(G^N(Z))^C=M^N_{6N^{-\frac{2}{9}-\sigma},N^{-\frac{2}{9}}}(Z)$ (if $\alpha=2$), it follows for configurations $X \notin  \mathcal{B}_{3,i,(k_1,...,k_6)}^{N,\sigma}$ that
this last remaining term is bounded by
\begin{align} 
 CN^{\frac{\sigma}{2}-1}\|f^N\|_{\infty}|t-t_1| & \le  CN^{\frac{\sigma}{2}-1}\big(N^{\frac{7}{18}-\sigma}\big)^2|t-t_1| \notag \\
 & \le CN^{-\frac{2}{9}-\frac{3\sigma}{2}}|t-t_1|. \label{upp.bound.bad.term}
 \end{align}
 Moreover, by taking into account that $\mathbb{P}\big(Y\in \mathbb{R}^6:Y\notin G^N(X_i) \big)
\le   CN^{-\frac{4}{3}-2 \sigma}$ (see \eqref{prob G^N, alpha=2}) it follows that
 \begin{align}
& \mathbb{P}\big( X\in \mathcal{B}^{N,\sigma}_{3,i,(k_1,...,k_6)}\big)\
\le  \binom{N}{\lceil N^{\frac{\sigma}{2}}\rceil}\big(  CN^{-\frac{4}{3}-2 \sigma}\big)^{\lceil N^{\frac{\sigma}{2}}\rceil}\le CN^{-\frac{1}{3}\lceil N^{\frac{\sigma}{2}}\rceil}
 \end{align}
which obviously drops sufficiently fast.\\
 We arrived at the point where we have to merge all upper bounds which we derived in the previous part. However, first we restrict our initial data to those configurations where all applied estimates work for arbitrary $t_1,t_2$ fulfilling the initially introduced demands. 
For this purpose, we consider for the remaining part only configurations 
$$X \in \Big(\bigcup_{j\in \{1,2,3\}}\bigcup_{i=1}^N\bigcup_{(k_1,...,k_6)\in J_N}\mathcal{B}^{N,\sigma}_{j,i,(k_1,...,k_6)}\Big)^C.$$
We already discussed that for any $\gamma>0$ there exists a constant $C_\gamma>0$ such that $
\mathbb{P}\big(X\in \mathcal{B}^{N,\sigma}_{1,i,(k_1,...,k_6)}\big)
\le C_{\gamma}N^{-\gamma}$ and according to the proof of the first case it holds that $\mathbb{P}\big(X\in \mathcal{B}^{N,\sigma}_{2,i,(k_1,...,k_6)}\big) \le (CN^{-\frac{7\sigma}{3}} )^{\frac{N^{\frac{\sigma}{3}}}{2}}$ (see \eqref{prob.b.3}). Since $|J_N|\le (3\lceil N^\frac{1}{6}\rceil )^6 \le CN $ (see \eqref{J_N}), it is again possible to choose the constant $C_{\gamma}>0$ such that 
\begin{align*}
& \mathbb{P}\Big(\bigcup_{j\in \{1,2,3\}}\bigcup_{i=1}^N\bigcup_{(k_1,...,k_6)\in J_N}\mathcal{B}^{N,\sigma}_{j,i,(k_1,...,k_6)} \Big) \le C_{\gamma}N^{-\gamma}
\end{align*}
holds for a given $\gamma>0$ and all $N\in \mathbb{N}$. \\
For configurations
$$X\in \Big(\bigcup_{j\in \{1,2,3\}}\bigcup_{i=1}^N\bigcup_{(k_1,...,k_6)\in J_N}\mathcal{B}^{N,\sigma}_{j,i,(k_1,...,k_6)} \Big)^C$$ all derived upper bounds are fulfilled for arbitrary `triples' $t_1,t_2$ and $\widetilde{X}_i$ provided they are chosen according to the introduced constraints on them (which we will review shortly). Under these conditions we obtain that \eqref{c_2t3} is bounded by $C(1+(t-t_1)\ln(N))N^{-\frac{7}{18}}$ (see \eqref{est.b1}), while the upper limit for term \eqref{c_2t5} is $N^{-\frac{7}{18}}$ (see definition \eqref{Def.B-bad.2}), the bound for \eqref{c_2t6,2} (which is $CN^{-\frac{5}{9}}$) was already derived in case 1 (see \eqref{bound1, alpha=2}) and $CN^{-\frac{2}{9}-\frac{3\sigma}{2}}(t-t_1)$ constitutes an upper bound for \eqref{bad.term} (see \eqref{upp.bound.bad.term}. Hence, it follows that for $t\in [t_1,t_2]$ (and for small enough $\sigma>0$) the force term \eqref{force.bad case} is dominated by
$$C\big(N^{-\frac{2}{9}-\frac{3\sigma}{2}}(t-t_1)+N^{-\frac{7}{18}} \big).$$
This brings us to the concluding estimates. By regarding this upper bound we obtain that for any $i\in \{1,...,N\}$ and for all times $t\in [t_1,t_2]$ the following inequality holds for the considered configurations:
\begin{align}
&|[^2\Psi^{N}_{t,0}(X)]_i-{^2\varphi^{N}_{t,0}}(\widetilde{X}_i)|\notag \\
\le & |[^2\Psi^{N}_{t_1,0}(X)]_i-{^2\varphi^{N}_{t_1,0}}(\widetilde{X}_i)|\notag \\
&+\big|\int_{t_1}^t\Big(\frac{1}{N}\sum_{j\neq i}f^{N}([^1\Psi^N_{s,0}(X)]_i-[^1\Psi^N_{s,0}(X)]_j) -f^{N}*\widetilde{k}^N_s({^1\varphi^N_{s,0}}(\widetilde{X}_i))\Big)ds\big| \notag  \\
\le & \underbrace{|[^2\Psi^{N}_{t_1,0}(X)]_i-{^2\varphi^{N}_{t_1,0}}(\widetilde{X}_i)|}_{\le \frac{N^{-\frac{7}{18}+\sigma}}{2}} +C\big(N^{-\frac{2}{9}-\frac{3\sigma}{2}}(t-t_1)+N^{-\frac{7}{18}} \big)\label{best.t_2,v}
 \end{align} 
 where we regarded the condition on the distance between the applied auxiliary particle of the `cloud' and the related particle of the microscopic system at the starting time of the observation period $t_1$.
Now it is straightforward to indicate an upper bound for the spatial deviation for $t\in [t_1,t_2]$:
\begin{align}
& |[^1\Psi^{N}_{t,0}(X)]_i-{^1\varphi^{N}_{t,0}}(\widetilde{X}_i)|\notag \\
\le & \underbrace{|[^1\Psi^{N}_{t,0}(X)]_i-{^1\varphi^{N}_{t,0}}(\widetilde{X}_i)|}_{ \le \frac{N^{-\frac{7}{18}+\sigma}}{2}}+\int_{t_1}^t|[^2\Psi^{N}_{s,0}(X)]_i-{^2\varphi^{N}_{s,0}}(\widetilde{X}_i)|ds \notag \\
\le & \frac{N^{-\frac{7}{18}+\sigma}}{2}+  C\big(N^{-\frac{2}{9}-\frac{3\sigma}{2}}(t-t_1)^2+N^{-\frac{7}{18}}(t-t_1) \big)\label{best.t_2,x}
\end{align}
Eventually, we recall the conditions on $t_1,t_2$ and $\widetilde{X}_i$ which we introduced previous to the estimates for the `bad' particles so that we can discuss what we achieved so far. $t_1$ denotes basically an arbitrary moment in $[0,\tau^N(X))$ and we argued that it is always possible to find an auxiliary particle of the introduced `cloud' which is closer (in phase space) to the observed `real' particle than $\frac{N^{-c}}{2}=\frac{N^{-\frac{7}{18}+\sigma }}{2}$ (at least for large enough $N$) at this point in time. $\widetilde{X}_i$, on the other hand, was simply an abbreviation for the initial position of the respectively considered auxiliary particle $X^i_{k_1,...,k_6}=X_i+\sum_{j=1}^6k_jN^{-\frac{7}{18}+\frac{\sigma}{2}}e_j$ where $(k_1,...,k_6)\in J_N$. Finally, $t_2$ was defined as a point in time of $ (t_1,\tau^N(X)]$ where the distance in (physical) space between this auxiliary particle and the `real' one still fulfills 
$$\sup_{t_1\le t \le t_2}|[^1\Psi^{N}_{t,0}(X)]_i-{^1\varphi^{N}_{t,0}}(\widetilde{X}_i)|\le N^{-\frac{7}{18}+\sigma}$$ while for the velocity deviation the much larger upper bound $$\sup_{t_1\le t \le t_2}|[^2\Psi^{N}_{t,0}(X)]_i-{^2\varphi^{N}_{t,0}}(\widetilde{X}_i)|\le N^{-\frac{2}{9}-\sigma}$$ was allowed. After the time $t_2$ (possibly) a new auxiliary particle of the `cloud' which is closer to the observed `real' particle must be chosen for further estimates. However, relations \eqref{best.t_2,v} and \eqref{best.t_2,x} provide us the opportunity to determine a lower bound for the possible length of such an interval $[t_1,t_2]$ where in any case the same auxiliary particle can be applied. For large enough $N\in \mathbb{N}$ and small enough $\sigma>0$ the subsequent implication holds
\begin{align*}
&t-t_1\le N^{-\frac{1}{12}}
\Rightarrow \begin{cases}\frac{N^{-\frac{7}{18}+\sigma}}{2}+C\big(N^{-\frac{2}{9}-\frac{3\sigma}{2}}(t-t_1)^2+N^{-\frac{7}{18}}(t-t_1) \big)\le N^{-\frac{7}{18}+\sigma} \\
\frac{N^{-\frac{7}{18}+\sigma}}{2}+C\big(N^{-\frac{2}{9}-\frac{3\sigma}{2}}(t-t_1)+N^{-\frac{7}{18}} \big)\le CN^{-\frac{11}{36}-\frac{3\sigma}{2}}\le N^{-\frac{2}{9}-\sigma} \end{cases}
\end{align*}
and thus according to relations \eqref{best.t_2,v} and \eqref{best.t_2,x} $t_2:=t_1+N^{-\frac{1}{12}}$ is a possible option such that the constraints on $t_2$ are fulfilled. Hence, \eqref{best.t_2,v} and \eqref{best.t_2,x} yield for this choice of $t_2$ (and small enough $\sigma>0$) that
\begin{align*}
\sup_{t_1\le s \le t_2}|[\Psi^{N}_{t,0}(X)]_i-{\varphi^{N}_{t,0}}(\widetilde{X}_i)|\le CN^{-\frac{2}{9}-\frac{3\sigma}{2}}(t_2-t_1)=CN^{-\frac{11}{36}-\frac{3\sigma}{2}}.
\end{align*}
Eventually, we can return to term \eqref{double.upp.bound} and by regarding additionally estimate \eqref{est.eff.bad.term} we obtain for $t\in [t_1,t_1+N^{-\frac{1}{12}}]$, the considered configurations, large enough $N$ and sufficiently small $\sigma>0$ that
\begin{align}
& \sup_{t_1\le s \le t}|[\Psi^N_{s,0}(X)]_i-{\varphi^N_{s,0}}(X_i)|\notag \\
\le &\sup_{t_1\le s \le t}|[\Psi^N_{s,0}(X)]_i-{\varphi^N_{s,0}}(X^i_{k_1,...,k_6})|+\sup_{t_1\le s \le t}|{\varphi^N_{s,0}}(X^i_{k_1,...,k_6})-{\varphi^N_{s,0}}(X_i)|  \notag \\
 \le & CN^{-\frac{11}{36}-\frac{3\sigma}{2}}+ e^{C(t-t_1)}\big|[\Psi^N_{t_1,0}(X)]_i-\varphi^N_{t_1,0}({X}_i)\big|. 
\end{align}
Since $t_1\in [0,\tau^N(X))$ was chosen arbitrarily, we can define a sequence of time steps 
$$t_n:=nN^{-\frac{1}{12}}\text{ for }n\in \{0,...,\lceil \tau^N(X)N^{\frac{1}{12}}\rceil-1\} \text{ and }t_{\lceil \tau^N(X)N^{\frac{1}{12}}\rceil}:=\tau^N(X)$$
and receive a corresponding sequence of inequalities 
 \begin{align*}
& \sup_{t_n\le s \le t_{n+1}}|[\Psi^N_{s,0}(X)]_i-{\varphi^N_{s,0}}(X_i)|\notag \\
\le & CN^{-\frac{11}{36}-\frac{3\sigma}{2}}+ e^{CN^{-\frac{1}{12}}}\big|[\Psi^N_{t_n,0}(X)]_i-\varphi^N_{t_n,0}({X}_i)\big|.
\end{align*}
Now it is straightforward to derive inductively that 
\begin{align*}
\sup_{0\le s \le t_n}|[\Psi^N_{s,0}(X)]_i-{\varphi^N_{s,0}}(X_i)|
\le CN^{-\frac{11}{36}-\frac{3\sigma}{2}} \sum_{k=0}^{n-1} e^{2CN^{-\frac{1}{12}}k} 
\end{align*}
which after regarding that $\lceil T N^{\frac{1}{12}}\rceil$ constitutes an upper bound for the possible values of $n$ yields that
\begin{align*}
\sup_{0\le s \le \tau^N(X)}|[\Psi^N_{s,0}(X)]_i-{\varphi^N_{s,0}}(X_i)|\le CN^{-\frac{2}{9}-\frac{3}{2}\sigma}.
\end{align*}
This value stays smaller than $N^{-\frac{2}{9}-\sigma}$ for sufficiently large $N$ which shows that also the `bad' particles do typically not `trigger' the stopping time for the relevant $N$ and $\sigma$. Hence, the main part of the proof is finally completed.\\\\
Like mentioned in the discussion of the strategy, we conclude the proof of Theorem \ref{thm1} by showing that for $N>1$
\begin{align}
\sup\limits_{x\in \mathbb{R}^6}\sup_{0\le s \le T}|^1\varphi_{s,0}^N(x)-{^1\varphi^{\infty}_{s,0}}(x)|\le e^{C\sqrt{\ln(N)}}N^{-2c} \label{dist.effect.flow.limit}
\end{align}
which obviously is a considerably smaller bound than necessary for verifying the statement. \\
In the following we identify $\Delta_N(t):=\sup\limits_{x\in \mathbb{R}^6}\sup_{0\le s \le t}|^1\varphi_{s,0}^N(x)-{^1\varphi^{\infty}_{s,0}}(x)|$. Let $t\in [0,T]$ be such that still $\Delta_N(t)\le N^{-c}$, then it holds for $x\in \mathbb{R}^6$ and $N\in \mathbb{N}\setminus \{1\}$ that
{\allowdisplaybreaks 
\begin{align*}
 & |^2\varphi_{t,0}^N(x)-{^2\varphi^{\infty}_{t,0}}(x)|\\
\le &\big|\int_0^t\int_{\mathbb{R}^6}\big(f^N(^1\varphi^N_{s,0}(x)-{^1\varphi^N_{s,0}}(y))-f^{\infty}({^1\varphi^{\infty}_{s,0}}(x)-{^1\varphi^{\infty}_{s,0}}(y))\big)k_0(y)d^6yds\big|\\
\le &\big|\int_0^t\int_{\mathbb{R}^6}\big(f^N(^1\varphi^N_{s,0}(x)-{^1\varphi^N_{s,0}}(y))-f^N({^1\varphi^{\infty}_{s,0}}(x)-{^1\varphi^{\infty}_{s,0}}(y))\big)k_0(y)d^6yds\big|\\
 & +\big|\int_0^t\int_{\mathbb{R}^6}\big(f^N(^1\varphi^{\infty}_{s,0}(x)-{^1\varphi^{\infty}_{s,0}}(y))-f^{\infty}({^1\varphi^{\infty}_{s,0}}(x)-{^1\varphi^{\infty}_{s,0}}(y))\big)k_0(y)d^6yds\big|\\
\le & \int_{0}^t2\Delta_N(s)\int_{\mathbb{R}^6} g^N({^1\varphi^N_{s,0}}(x)-{^1\varphi^N_{s,0}}(y))k_0(y)d^6yds\\
 & +\big|\int_0^t\int_{\mathbb{R}^6}\big(f^N({^1\varphi^{\infty}_{s,0}}(x)-{^1y})-f^{\infty}({^1\varphi^{\infty}_{s,0}}(x)-{^1y})\big)k^{\infty}_s(y)d^6yds\big|\\
\le & C\ln(N)\int_{0}^t\Delta(s) ds+
\big|\int_0^t\int_{\mathbb{R}^6} \frac{^1y}{|^1y|^{\alpha+1}}\mathbf{1}_{(0,N^{-c}]}(|^1y|) k^{\infty}_s(y+{\varphi}^{\infty}_{s,0}(x))d^6yds\big|\\
&+ \big|\int_0^t\int_{\mathbb{R}^6} {^1y} N^{c(\alpha+1)}\mathbf{1}_{(0,N^{-c}]}(|^1y|) k^{\infty}_s(y+{\varphi}^{\infty}_{s,0}(x))d^6yds\big|
\end{align*}}
where we remark that for the stated application of the map $g^N$ (see \eqref{Def.g^N} for the definition) in the second step the assumption $\Delta_N(t)\le N^{-c}$ was applied. Moreover, since $g^N(q)\le C\min\big( N^{(\alpha+1)c},\frac{1}{|q|^{\alpha+1}}\big)$ for all $q\in \mathbb{R}^3$ the factor $\ln(N)$ only arises if $\alpha=2$ and is not necessary for the remaining smaller values of $\alpha$ which are considered.\\
It remains to take a closer look at the last two terms. However, we consider in the following only the first of these terms since the second can be treated analogously. We use again the notation $x=(^1x,{^2x})\in \mathbb{R}^6$ where the first component shall describe the position in space and the second the velocity. We will show that due to the slowly varying mass or charge density cancellations arise such that this term keeps small enough.
{\allowdisplaybreaks
\begin{align}
&\big|\int_0^t\int_{\mathbb{R}^6} \frac{^1y}{|^1y|^{\alpha+1}}\mathbf{1}_{(0,N^{-c}]}(|^1y|) k^{\infty}_s(y+{\varphi}^{\infty}_{s,0}(x))d^6yds\big|\notag \\
= &\big|\int_0^t\int_{\mathbb{R}^6} \frac{^1y}{|^1y|^{\alpha+1}}\mathbf{1}_{(0,N^{-c}]}(|^1y|) \Big(\big(k^{\infty}_s(y+{\varphi}^{\infty}_{s,0}(x))-k^{\infty}_s((0,{^2y})+{\varphi}^{\infty}_{s,0}(x))\big) \notag \\
&+k^{\infty}_s((0,{^2y})+{\varphi}^{\infty}_{s,0}(x))\Big)d^6yds\big| \notag \\
\le  & \int_0^t\int_{\mathbb{R}^6} \frac{1}{|^1y|^\alpha}\mathbf{1}_{(0,N^{-c}]}(|^1y|)\Big(\big|k^{\infty}_s(y+{\varphi}^{\infty}_{s,0}(x))-k^{\infty}_s((0,{^2y})+{\varphi}^{\infty}_{s,0}(x))\big|\Big)d^6yds \label{concl.term}
\end{align}}
where the last step follows since 
\begin{align*}
& \big|\int_0^t\int_{\mathbb{R}^6} \frac{^1y}{|^1y|^{\alpha+1}}\mathbf{1}_{(0,N^{-c}]}(|^1y|)k^{\infty}_s((0,{^2y})+{\varphi_{s,0}^{\infty}}(x))d^6yds\big|\\
=&\big|\int_0^t\widetilde{k}^{\infty}_s({^1\varphi_{s,0}^{\infty}}(x))\int_{\mathbb{R}^3} \frac{q}{|q|^{\alpha+1}}\mathbf{1}_{(0,N^{-c}]}(|q|)d^3qds\big|= 0
\end{align*}
due to the symmetry properties of the force kernel.\\ By application of the condition on our initial density $|\nabla k_0(x)|\le \frac{C}{(1+|x|)^{3+\delta}}$ as well as Lemma \ref{lem1} (in the second last step) it follows for arbitrary $z\in \mathbb{R}^6$ and $s\in [0,T]$ that
\begin{align}
&\big|k^{\infty}_s(y+z)-k^{\infty}_s((0,{^2y})+z)\big|\mathbf{1}_{(0,N^{-c}]}(|^1y|) \notag \\
=& \big|k_0(\varphi^{\infty}_{0,s}(y+z))-k_0(\varphi^{\infty}_{0,s}((0,{^2y})+z))\big| \mathbf{1}_{(0,N^{-c}]}(|^1y|) \notag  \\
\le & \sup_{z'\in \overline{\varphi^{\infty}_{0,s}(y+z)\varphi^{\infty}_{0,s}((0,{^2y})+z)}}|\nabla k_0(z')| \notag  \\
&\cdot \mathbf{1}_{(0,N^{-c}]}(|^1y|)\Big( \big|\varphi^{\infty}_{0,s}(y+z)-\varphi^{\infty}_{0,s}((0,{^2y})+z)\big|\Big) \notag \\
 \le &   \sup_{z'\in \overline{\varphi^{\infty}_{0,s}(y+z)\varphi^{\infty}_{0,s}((0,{^2y})+z)}}\frac{C}{(1+|z'|)^{3+\delta}} \notag \\
&  \cdot \mathbf{1}_{(0,N^{-c}]}(|^1y|)\Big(C\big|(y+z)-\big((0,{^2y})+z \big) \big|\Big) \notag \\
\le & \sup_{y'\in \mathbb{R}^3:|y'|\le N^{-c}}\sup_{z'\in \overline{\varphi^{\infty}_{0,s}((y',{^2y})+z)\varphi^{\infty}_{0,s}((0,{^2y})+z)}}\frac{CN^{-c}}{(1+|z'|)^{3+\delta}} \label{t.decay.grad.dens.}
\end{align}
where $\overline{xy}:=\{(1-\lambda)x+\lambda y\in \mathbb{R}^6: \lambda\in [0,1]\}$ for $x,y\in \mathbb{R}^6$.\\
At this point we only note that if the value of $|^2y|$ (appearing in this expression) is chosen large enough, then all configurations of the set over which the supremum is taken have a velocity value of this order due to the bounded mean-field force. Hence, term \eqref{t.decay.grad.dens.} drops like $\frac{CN^{-c}}{(1+|^2y|)^{3+\delta}}$ as $|^2y|$ increases. For a more rigorous argumentation see the reasoning utilized in the proof of Lemma \ref{lem1} (starting after \eqref{est.grad.dens.}) which is essentially analogous.\\ 
Now we can apply these considerations to estimate term \eqref{concl.term}. It follows that for arbitrary $z\in \mathbb{R}^6$ (and in particular $z:={\varphi}^{\infty}_{s,0}(x)$) and $\alpha\in (1,2]$ 
 \begin{align*}
&\big|\int_{\mathbb{R}^6} \frac{^1y}{|^1y|^{\alpha+1}}\mathbf{1}_{(0,N^{c}]}(|^1y|) k^{\infty}_s(y+z)d^6y\big|\\
\le  &\int_{\mathbb{R}^3} \frac{1}{|^1y|^\alpha}\mathbf{1}_{(0,N^{-c}]}(|^1y|)d^3(^1y) \\
& \cdot \int_{\mathbb{R}^3} \sup_{y'\in \mathbb{R}^3:|y'|\le N^{-c}}\sup_{z'\in \overline{\varphi^{\infty}_{0,s}((y',{^2y})+z)\varphi^{\infty}_{0,s}((0,{^2y})+z)}}\frac{CN^{-c}}{(1+|z'|)^{3+\delta}}d^3(^2y)\\
\le & CN^{-2c}.
\end{align*}
Consequently, it holds that for any $x\in \mathbb{R}^6$ that
\begin{align}
&\sup_{0\le s \le t}|^1\varphi_{s,0}^N(x)-{^1\varphi^{\infty}_{s,0}}(x)|\notag\\
\le & \int_0^t|^2\varphi_{s,0}^N(x)-{^2\varphi^{\infty}_{s,0}}(x)| ds\notag \\
\le & C\ln(N)\int_{0}^t\int_0^s\Delta_N(r)drds+CN^{-2c}t. \label{last.Gron.}
\end{align}
By means of this inequality one easily derives by Gronwall lemma \ref{Gron.lem.} that $$ \Delta_N(t)=\sup_{x\in \mathbb{R}^6}\sup_{0\le s \le t}|{^1\varphi_{s,0}^N}(x)-{^1\varphi^{\infty}_{s,0}}(x)|\le CN^{-2c}te^{\sqrt{C\ln(N)}t}$$ 
which shows that the initial assumption $\Delta_N(t)\le N^{-c}=N^{-\frac{7}{18}+\sigma}$ stays true for arbitrarily large times $t$ provided that $N\in \mathbb{
N}$ is large enough.\\
Applying this bound additionally on the relation $$|^2\varphi_{t,0}^N(x)-{^2\varphi^{\infty}_{t,0}}(x)| \notag 
\le C\ln(N)\int_{0}^t\Delta_N(s)ds+CN^{-2c}$$ yields the claimed result 
\begin{align}
\sup\limits_{x\in \mathbb{R}^6}\sup_{0\le s \le T}|\varphi_{s,0}^N(x)-{\varphi^{\infty}_{s,0}}(x)|\le e^{C\sqrt{\ln(N)}}N^{-2c} \label{dist.mean-field.flow}
\end{align}
for the relevant $N$ which concludes the proof of Theorem \ref{thm1}.

\newpage
\section{Discussion of the first main result}
\noindent Up to now, we showed that for typical initial conditions all particles keep very close to their related `mean-field particles' for the considered systems. The result related to item (ii) is basically only an interim result that will be extended in the following part of the work. However, actually the focus of our interest concerns the Vlasov-Poisson system which is considered in item (i). Unfortunately, for this system it is not possible to extend the outcome by the approach which will be applied for the `less singular systems' of item (ii) where the reasons for this will be discussed later in the work. First, it makes sense to summarize what we have achieved so far for the Vlasov-Poisson system and to discuss the still existing shortcomings of the result. As mentioned initially, our main intention was to reduce the cut-off size to an order below the mean inter-particle distance which is $N^{-\frac{1}{3}}$ in $3$-dimensional space. Since the cut-off size which we consider is $N^{-\frac{7}{18}+\sigma}$ (where $\sigma>0$ can be chosen arbitrarily small), this aim is any case reached. In fact, it is still possible to improve the result by basically the same method where a finer subdivision in `particle classes' is applied than just the distinction between `bad' and `good'. However, one can imagine after comprehending the current proof that each additional class is connected with a significant increase in estimates. Moreover, the value of the attainable improvement is questionable since the method fails in any case at a cut-off order above $N^{-\frac{1}{2}}$. The reason why the approach has to fail at latest at this point lies in the circumstance that the law of large numbers does not yield better control than $N^{-\frac{1}{2}}$-fluctuations around the expectation. This is a problem because for the Vlasov-Poisson system the method (in the presented form) only works if up to a small number of exceptions all particles are as close or closer than the cut-off order to their related `mean-field particle'. The reason for this arises by the circumstance that we have to assume that the interacting particles take the `worst trajectories' possible within the bounds determined by their respective stopping times. If for the vast majority of particles the deviations are smaller than the cut-off size, then even in this worst-case scenario the number of `encounters' between `mean-field particles' where the inter-particle distance falls blow the cut-off size is typically of the same order as the corresponding number in the system of interacting particles. The situation, however, is completely different if for a significant number of particles the allowed deviations are of larger order than the cut-off size. This is also the main reason why simply permitting bigger deviations between corresponding particles does not improve the results that can be achieved. Hence, what actually can be seen as one of the strengths of the approach introduced initially by Pickl and Boers \cite{Peter} resp. Pickl and Lazarovici \cite{PeterDustin} (and thereby of the approach applied in the proof of Theorem \ref{thm1}) determines also its limit: All information about the particle distribution is obtained by their related `mean-field particles'. This allows in many situations to prove quite strong results with comparatively low expense (where this becomes particularly apparent in the previous works and less in the current). However, strong closeness assumptions between related particles are necessary which additionally get stronger as the considered systems get `more involved' (for example by regarding a smaller cut-off size). Thus, at a certain point the method has to fail without arguing additionally that the assumed `worst-case scenarios' for the interacting system are not typical but exceptions. It will be discussed in more detail during the introduction to the second main result how such a reasoning might look like.\\  
Nevertheless, some aspects of the received result can be seen as relevant improvements. Of course, the most obvious progress is that at an arbitrary point in time the force acting on the majority of the particles does (typically) not change if the cut-off is removed. If, however, a given particle is observed for a longer time period, then this particle will very likely `run into' the cut-off of some other particle. Hence, this is still no rigorous justification that Propagation of Chaos holds for the Vlasov-Poisson system. Although heuristically no surprises are expected in this regard, proving that typical trajectories do not change significantly as the cut-off is removed seems to be quite a hard problem for the Vlasov-Poisson system (while for the less singular systems of item (ii), showing this will be exactly our aim in the next chapter).   \\
One further interesting question which is studied for instance in the book of Spohn \cite{SpohnBook} for different systems is how the local particle distribution looks like and if the solution to the effective equation also provides information about this distribution. By `local' we mean that we consider for $x\in \mathbb{R}^6$ some region of the form $$\Delta^N_{r_1,r_2}(x):= B_{N^{-\frac{1}{3}}r_1}(^1x)\times B_{r_2}(^2x) \subseteq \mathbb{R}^6,\ r_1,r_2>0$$ in which typically an $N$-independent number of particles is located where as usual $B_r(x):=\{x'\in \mathbb{R}^3:|x'-x|<r\}$. More precisely, for a given $N$ the expected number of `mean-field particles' in such a region is
\begin{align*}
\lambda^N_{r_1,r_2}(x):=N\int_{\Delta^N_{r_1,r_2}(x)}k_t(x')d^6x'
\end{align*}
and it holds that
\begin{align*}
\lim_{N\to \infty}\lambda^N_{r_1,r_2}(x)&= \lim_{N\to \infty}N\int_{\Delta^N_{r_1,r_2}(x)}k_t(x')d^6x'\\
& = \frac{4}{3}\pi r_1^3\int_{B_{r_2}({^2x})}k_t({^1x},v')d^3(v')=:\lambda_{r_1,r_2}(x).
\end{align*}
 Moreover, since the `mean-field particles' are i.i.d. with density $k_0$ the distribution of the particle number in such a region converges to a Poisson distribution with parameter $\lambda_{r_1,r_2}(x)$. However, actually we want to obtain information about the local distribution of the interacting particles. To this end, we abbreviate
$$p^N_1:=\mathbb{P}\big(\exists i\in \mathcal{M}^N_b(X):{^1\varphi_{t,0}}(X_i)\in B_{2N^{-\frac{2}{9}-\sigma}}(^1x)\big)$$ (where we recall that $\mathcal{M}^N_b(X)$ denotes the set of `bad' particles) and 
\begin{align*}
p^N_2:= \mathbb{P}\Big(&\exists i\in \{1,...,N\}:\big(N^{-\frac{1}{3}}r_1-N^{-\frac{7}{18}+\sigma}\le |{^1x}-{^1\varphi_{t,0}}(X_i)|\le N^{-\frac{1}{3}}r_1+N^{-\frac{7}{18}+\sigma}\big)\vee\\
& \big({^1\varphi_{t,0}}(X_i)\in B_{N^{-\frac{1}{3}}r_1}(^1x)\land r_2-N^{-\frac{7}{18}+\sigma}\le  |{^2x}-{^2\varphi_{t,0}}(X_i)|\le r_2+N^{-\frac{7}{18}+\sigma}\big)\Big),
\end{align*} 
then it holds according to the results which we derived in the proof of Theorem \ref{thm1} that for $t\in [0,T]$, $k\in \mathbb{N}$, $x\in \mathbb{R}^6$, arbitrary $\gamma>0$ and for large enough $N\in \mathbb{N}$:
\begin{align*}
 & \mathbb{P}\Big(X\in \mathbb{R}^{6N}: \sum_{i=0}^N\mathbf{1}_{\Delta^N_{r_1,r_2}(x)}([\Psi^N_{t,0}(X)]_i)\neq \sum_{i=0}^N\mathbf{1}_{\Delta^N_{r_1,r_2}(x)}(\varphi_{t,0}(X_i))\Big)\\
\le & p^N_1+p^N_2+C_{\gamma}N^{-\gamma}
\end{align*} 
To this end, we regarded that according to the proof of Theorem \ref{thm1} the deviation between related `real' and `mean-field particles' does typically (resp. with a probability larger than $1-C_{\gamma}N^{-\gamma}$) stay below order $N^{-\frac{7}{18}+\sigma}$ in the case of `good' particles and $N^{-\frac{2}{9}-\sigma}$ in the general case if $N$ is large and $\sigma>0$ small enough. Thus, if the event `related to' $p_2^N$ does not occur, then we obtain that typically no `mean-field particle' is located in the border region of $\Delta^N_{r_1,r_2}(x)$ so that the numbers of `good' interacting and `mean-field particles' inside this volume coincide. If there is additionally no `bad mean-field particle' in the spatial region $B_{2N^{-\frac{2}{9}-\sigma},r_2}({^1x})$, then obviously even the total number of particles in $\Delta^N_{r_1,r_2}(x)$ is the same for the two systems.\\ 
By regarding $\sup_{0\le s \le T}\|\widetilde{k}_s\|_{\infty}<C$ it follows that 
\begin{align*}
p_2^N\le N \big(CN^{-\frac{7}{18}+\sigma}N^{2(-\frac{1}{3})}\big)\le CN^{-\frac{1}{18}+\sigma}
\end{align*} 
 and 
\begin{align*}
&p_1^N\\
\le & \mathbb{P}\big(\exists i \in \{1,...,N\}:\\
& \hspace*{0,32cm}\underbrace{\exists j\in \{1,...,N\}\setminus\{i\}: X_i\in M^N_{6N^{-\frac{2}{9}-\sigma},N^{-\frac{2}{9}}}(X_j)}_{\Leftrightarrow i \in \mathcal{M}^N_b(X)} \land {^1\varphi_{t,0}(X_i)}\in B_{2N^{-\frac{2}{9}-\sigma}}(^1x)\big)\\
\le & N^2 \sup_{Y\in \mathbb{R}^6}\mathbb{P}\big( Z\in M^N_{6N^{-\frac{2}{9}-\sigma},N^{-\frac{2}{9}}}(Y) \big) \mathbb{P}\big(Z\in \mathbb{R}^6:  {^1\varphi_{t,0}(Z)}\in B_{2N^{-\frac{2}{9}-\sigma}}(^1x)\big)\\
\le & N^{2}\big( CN^{2(-\frac{2}{9}-\sigma)}N^{4(-\frac{2}{9})}\big)\big( CN^{3(-\frac{2}{9}-\sigma)}\big)\\
\le & CN^{-5\sigma}
\end{align*}
where we applied Lemma \ref{lem3} in the second last step. Since $\sigma>0$ can be chosen arbitrarily small, both probabilities vanish as $N$ goes to infinity which yields that the property to be locally Poisson distributed transfers from the `mean-field particle' system to the interacting system. This concludes the discussion of the first result.\\
However, the previous analysis already provides first indications on how an extension of the result might work which will be considered in more detail in the subsequent introduction to the second main result.
\chapter{Vlasov equation as the mean-field limit of particle systems with singular potentials \label{sec.main2}}
\section{Introduction to the desired goals and heuristic proceeding} Our aim is to show that for the systems considered in item (ii) of Theorem \ref{thm1} reducing the cut-off size has very little influence on the dynamics. However, the cut-off will never be removed completely but only made arbitrarily small. Hence, the interaction is always Lipschitz continuous, so that we do not have to care about the solution theory which at least in the attractive case is non-trivial without regularization. On the other hand, the proof includes showing that for very small cut-off sizes and typical initial configurations the inter-particle distance will never fall below the cut-off diameter for any particle pair on the considered time span. Thus, in this case corresponding trajectories of the regularized and of the non-regularized systems coincide. This yields additionally to the actually aspired aim that for typical initial data (with respect to the considered measure) the $N$-particle dynamics are well-defined also without cut-off. A similar reasoning was also applied in \cite{Hauray2013}. For achieving this aim it will be necessary to derive some additional information about the distribution of the interacting particles which is not provided by the closeness to their related `mean-field particles' alone. The idea is to apply a similar approach as in the proof of the first main result. In that case we introduced a system of `mean-field particles' for which we have plenty of information about their dynamics and their distribution and exploited this to show that their trajectories keep close to the corresponding trajectories related to the microscopic system with high probability. The crucial point was that as long as these trajectories are close, the information about the `mean-field particles' can to a wide extend be transferred on the interacting particles. At the current point we already derived much information about the considered $N$-particle systems with cut-off parameter $c=\frac{2}{3}$ and in the next step we will extend this even further. Subsequently, we will apply the related particles in the same way as the `mean-field particles' were applied previously: More precisely, this time we want to `transfer' the information which we have about the trajectories of the system with (comparatively) large cut-off on the corresponding trajectories of systems where the cut-off size might be arbitrarily small.\\ Before we are able to indicate the precise statement we first need to introduce the following set for $\delta>0$ and $N\in \mathbb{N}$:
\begin{align}
\begin{split}
\mathcal{L}^N_{\delta}:=\Big\{Y\in \mathbb{R}^6 \ | \ \forall Z_1,Z_2\in \mathbb{R}^6:& \big( \max_{i\in \{1,2\}}|Z_i-Y|\le N^{-\frac{1}{3}}\ \land Z_1\neq Z_2\big) \\ & \ \ \Rightarrow  \frac{|k_0(Z_1)-k_0(Z_2)|}{|Z_1-Z_2|}\le N^{\frac{\delta}{2}}k_0(Z_1)\Big\} 
\end{split}\label{Def.L}
\end{align}
This sets contains the configurations where the initial density fulfills a special local Lipschitz property where the reasons leading to the definition will become clear during subsequent proofs. Of course, one could in principle replace the $N^{\frac{\delta}{2}}$ factor appearing in the condition by $N^{\delta}$ but it will turn out to be slightly beneficial for the notation to define it this way. Furthermore, the appearing value $N^{-\frac{1}{3}}$ is only one of arbitrarily many possible choices for this variable.  \\ 
The second main theorem reads as follows:
\begin{thm} \label{thm2}
Let $T>0$ and $k_0\in\mathcal{L}^{1}(\mathbb{R}^{6})$ a probability density fulfilling the assumptions of Theorem \ref{thm1}. Moreover, let $(\Psi_{t,s}^{N,c})_{s,t \in \mathbb{R}}$ be the microscopic flow defined in \eqref{Def.micro.sys.} for $\alpha\in (1,\frac{4}{3}]$. If $ c_2\geq c_1: =\frac{2}{3}$, then for any $\sigma,\epsilon>0$ there exist $C_1,\sigma'>0$ such that for all $N\in \mathbb{N}$
\begin{align}
& \mathbb{P}\big(X\in \mathbb{R}^{6N}:\sup_{0\le s \le T}|\Psi_{s,0}^{N,c_1}(X)-\Psi_{s,0}^{N,c_2}(X)|_{\infty}> N^{-\frac{1}{2}+\sigma} \big) \notag \\
\le  &C_{1}N^{-\frac{1}{9}+\epsilon}+\mathbb{P}\big(X\in \mathbb{R}^{6N}:\exists i\in \{1,...,N\}:X_i\notin \mathcal{L}^N_{\sigma'}\big)\label{result+}.
\end{align}
\end{thm} 
\vspace{0,4cm} 
\noindent If in inequality \eqref{result+} the second addend of the right side vanishes sufficiently fast, then we can conclude by application of Theorem \ref{thm1} that
\begin{align*}
\sup_{0\le s \le T}|\Phi_{s,0}^{\infty}(X)-\Psi_{s,0}^{N,c}(X)|_{\infty}\le 2N^{-\frac{1}{2}+\sigma}
\end{align*} holds for typical initial data and arbitrary $c\geq \frac{2}{3}$ if $N\in \mathbb{N}$ is large enough. The upper bound 
\begin{align}
\mathbb{P}\big(X\in \mathbb{R}^{6N}:\exists i\in \{1,...,N\}:X_i\notin \mathcal{L}^N_{\sigma'}\big) \label{thm2.upp.bound.prob.}
\end{align} 
can in principle be improved in the sense that this Lipschitz constraint does not necessarily need to hold for all particles. Actually, it would be sufficient if configurations $X\in \mathbb{R}^{6N}$ which fulfill a constraint of the form   
\begin{align*}
\exists Z\in \mathbb{R}^{6N}:|Z-X|_{\infty}\le N^{-\frac{1}{3}}\land \Big(\prod_{i=1}^Nk_0(X_i)> Ce^{N^{\frac{\sigma'}{2}}|Z-X|_1} \prod_{i=1}^Nk_0(Z_i)\Big)
\end{align*}
(instead of $\exists i\in \{1,...,N\}:X_i\notin \mathcal{L}^N_{\sigma'}$)
are untypical where $C>1$ shall be some arbitrary constant. However, since for a considerable class of initial densities the probability stated in the Theorem vanishes anyway sufficiently fast, there seem to be at least equally serious shortcomings of the current result and thus we are content with the stated version which perhaps is slightly more tangible. On the other hand, it would of course be desirable to further generalize the possible initial data in the long term.\\
Furthermore, it is also possible to give some comments on the specific choice of $\sigma'>0$ which appears in \eqref{thm2.upp.bound.prob.}. Usually one can simply choose $\sigma'=\sigma$. However, it will turn out that the estimates implemented in the proof only work for small enough values of $\sigma>0$. This is a consequence of the circumstance that for a given $\sigma>0$ we will restrict the initial data to a set which depends on this parameter. We will see that this set is not typical for large values of $\sigma$ and additionally it is not suited for implementing appropriate estimates in this case. On the other hand, it suffices to show that the event $\sup_{0\le s \le T}|\Psi_{s,0}^{N,c_1}(X)-\Psi_{s,0}^{N,c_2}(X)|_{\infty}> N^{-\frac{1}{2}+\sigma} $ is untypical for small values of $\sigma$ to conclude that the same is true for larger values. But as a result we have to choose for instance $\sigma'=\min(\sigma,\sigma^*)$ for some sufficiently small $\sigma^*>0$ such that starting from a certain value of $\sigma $ the upper bound for the probability determined by term \eqref{thm2.upp.bound.prob.} does not improve anymore.\\ 
It is quite obvious that proving such a statement includes showing that the interacting particles typically keep a certain minimal distance to each other which in turn yields that a corresponding statement is true even for the non-regularized system. To this end, one has to consider that the parameter $c_2>0$ can simply be chosen large enough such that the cut-off radius $N^{-c_2}$ falls short of this minimal distance.\\ 
Additional to the results which we already received in the previous part of the work we will need further properties of the microscopic dynamics. The next step will be to show that with very high probability `shifting' the positions of arbitrary particles a little bit does not affect the evolution of the remaining particles in manner relevant to us if the cut-off radius is $N^{-\frac{2}{3}}$.\\
For this purpose we recall the set 
of `good' initial data applied for proving the first main result:
\begin{align}
\mathcal{G}_{1,T}^{N,\sigma} :=\big( \mathcal{B}^{N,\sigma}_5\cup\bigcup_{j\in \{1,2,3,4\}}\bigcup_{i=1}^N \mathcal{B}^{N,\sigma}_{j,i} \big)^C 
\end{align}
We note that previously we dropped the $T$-dependence of this set for a clearer notation. Obviously, the sets $\mathcal{B}^{N,\sigma}_{j,i}$ and $\mathcal{B}^{N,\sigma}_5$ also depend on this parameter but we will continue to ignore this dependency in the notation, as it will always be clear from the context. The definitions of these sets can be found in \eqref{def.B_1},\eqref{def.B_2}, \eqref{def.B_3}, \eqref{def.B_4} and \eqref{def.B_5}. The configurations belonging to $\mathcal{G}_{1,T}^{N,\sigma}$ have many good properties which will be very important during the subsequent proofs. In particular for such configurations all `real' particles keep closer to their corresponding `mean-field particle' than order $CN^{-\frac{1}{2}+\frac{\sigma}{2}}$ which was derived in \eqref{Gron-case 4/3+}. Since this is such a crucial property for the remaining part, we introduce an own Corollary for this statement.
\begin{cor} \label{cor.shift.}
Let $T>0$ and $k_0$ be a probability density fulfilling the assumptions of Theorem \ref{thm1}. Moreover, let $(\Psi_{t,s}^{N,c})_{s,t \in \mathbb{R}}$ be the flow defined in \eqref{Def.micro.sys.} and $(\Phi_{t,s}^{N,c})_{s,t \in \mathbb{R}}$ the `lifted' effective flow related to system \eqref{def.flow} with initial data $k_0$ for $\alpha\in (1,\frac{4}{3}]$ and $c=\frac{2}{3}$. If $\sigma>0$ is sufficiently small, then there exists $C_1>0$ such that for all $N\in \mathbb{N}$ and $X\in \mathcal{G}^{N,\sigma}_{1,T}$ it holds that
\begin{align*}
\sup_{0\le s \le T}|\Psi^{N,c}_{s,0}(X)-\Phi_{s,0}^{N,c}(X)|_{\infty}\le C_1N^{-\frac{1}{2}+\frac{1}{2}\sigma} . 
\end{align*}
\end{cor}
\vspace{0,4cm}
\noindent
We point out that it might appear strange that $\sigma>0$ needs to be sufficiently small such that the previous relation is fulfilled. This is caused by the $\sigma$-dependence of the `good' set $\mathcal{G}^{N,\sigma}_{1,T}$ which for large values of this parameter becomes more or less worthless for the estimates. \\ In the following we will only consider configurations of this `good' set $\mathcal{G}^{N,\sigma}_{1,T}$ and do not care for the remaining untypical initial data. In fact, the plan is to restrict the initial data steadily further until we have reached a set for which the estimates necessary to prove Theorem \ref{thm2} can be successfully implemented. However, we point out that the `strongest' restriction already took place by excluding the configuration of the set $\mathcal{B}^{N,\sigma}_5$. Compared to the upper bound which we derived for the probability of this set the probability related to the set of all remaining excluded configurations will turn out to be negligibly small.
\newpage

\section{Proof of the second main result}
\subsection{Preliminary studies}
Theorem \ref{thm1} already yields us very much information about the positions of the `real particles' in phase space by their related `mean-field particles'. However, though the uncertainty about their positions is only of order $N^{-\frac{1}{2}+\sigma}$, we still have to expect them to behave in the worst case possible within these constraints at the moment. Thus, if the (spatial) distance between two `mean-field particles' falls below order $N^{-\frac{1}{2}+\sigma}$, then we have to assume that their corresponding `real' particles get as close to each other as possible (which usually leads to a large deviation to the mean-field dynamics if the cut-off size is chosen very small). If the $N$-particle density $F_t^N$ typically does not change too fast in the neighborhood of a considered configuration, then heuristically it should be possible to argue that there is no tendency of the interacting particles to run into the (removed) singularity. The aim of the subsequent lemma is to show a property of the (regularized) $N$-particle dynamics which is very helpful to prove such a statement.
\begin{lem} \label{shift-lem} 
Let $k_0$ be a probability density fulfilling the assumptions of Theorem \ref{thm1} and let $(\Psi_{t,s}^{N,c})_{s,t \in \mathbb{R}}$ be the $N$-particle flow defined in \eqref{Def.micro.sys.} for $1<\alpha\le \frac{4}{3}$ and $c=\frac{2}{3}$. Let additionally for ${\sigma}>0,\ N\in \mathbb{N}$, $t_{1},t_{2}\in [0,T]$ and $C_0>0$ the set $\mathcal{G}^{N,\sigma}_{2,(t_1,t_2)}\subseteq \mathbb{R}^{6N}$ be defined as follows:
\begin{align*}
&X\in \mathcal{G}^{N,\sigma}_{2,(t_1,t_2)}\subseteq \mathbb{R}^{6N}\\
\Leftrightarrow & \forall Y_1,Y_2\in \mathbb{R}^{6N}:  \\
& \max_{i\in \{1,2\}}\min_{0\le s \le T}|{\Psi^{N,c}_{s,0}}(Y_i)-{\Psi^{N,c}_{s,0}}(X)|_{\infty}> \frac{N^{-\frac{1}{2}+\sigma}}{C_0}\ \vee \\
&  \bigg(  \max_{t_1 \le s \le t_2}|{\Psi^{N,c}_{s,0}}(Y_1)-{\Psi^{N,c}_{s,0}}(Y_2)|_1 \\
 & \le   N^{-\sigma}(t_2-t_1)+C_0\min_{t_1\le s \le t_2}|{\Psi^{N,c}_{s,0}}(Y_1)-{\Psi^{N,c}_{s,0}}(Y_2)|_1  \ \land\\
& \ \   \max_{0 \le s \le T}|{\Psi^{N,c}_{s,0}}(Y_1)-{\Psi^{N,c}_{s,0}}(X)|_{\infty}\\
 & \le  C_0 \big( N^{-\frac{1}{2}+\frac{\sigma}{2}}+\min_{0\le s \le T}|{\Psi^{N,c}_{s,0}}(Y_1)-{\Psi^{N,c}_{s,0}}(X)|_{\infty}\big)\bigg) .
\end{align*} 
If the constant appearing in this definition $C_0>0$ is sufficiently large and $\sigma>0$ small enough, then there exist $C_1>0$ and $\epsilon>0$ such that for all $N\in \mathbb{N}$ and $t_1,t_2\in [0,T]$ which fulfill $t_{2}-t_{1}\geq N^{-\frac{1}{3}}$ it holds that 
\[\mathbb{P}(X \in \mathcal{G}^{N,\sigma}_{1,T}\cap (\mathcal{G}_{2,(t_1,t_2)}^{N,\sigma})^C)\le C_1N^{- N^{\epsilon}} .\]
\end{lem}
\vspace{0,4cm}
\noindent 
Before we begin with the proof, we note that the dependence of the set $\mathcal{G}^{N,\sigma}_{2,(t_1,t_2)}$ on the constant $C_0>0$ is not made explicit to avoid an even more overloaded notation. Furthermore, since the lemma will be crucial on many occasions throughout the subsequent proofs we outline its essential statement in words: If a trajectory belonging to the `good' initial data $X\in\mathcal{G}^{N,\sigma}_{1,T}$ is close with respect to $|\cdot|_{\infty}$ to another trajectory at an arbitrary point in time belonging to $[0,T]$, then they are typically close (in this sense) for all times in $[0,T]$. If on the other hand two trajectories (having initial data $Y_1,Y_2 \in \mathbb{R}^{6N}$) are respectively close to such a `good' trajectory (with respect to $|\cdot |_\infty$) at a certain point in time, then for their distance with respect to $|\cdot |_1$ one of the following two options is fulfilled with extremely high probability: Their distance keeps almost of the same order on $[t_1,t_2]\subseteq [0,T]$ where $t_1+N^{-\frac{1}{3}}\le t_2$ or it does at least not exceed order $N^{-{\sigma}}(t_2-t_1)$ if $N\in \mathbb{N}$ is large enough.\\
After the short introduction we start with the proof of this important lemma.
\begin{proof} 
The estimates applied in the proof only need to be fulfilled for large enough $N\in \mathbb{N}$ and sufficiently small $\sigma>0$ so that the statement of the lemma holds. To avoid too much redundant formulations we apply as a general assumption that for the respective estimates this is indeed the case and mention it only partly explicitly. Since the cut-off parameter $c=\frac{2}{3}$ is again fixed, we will drop the related indices in the notation.\\
First, we prove the statement that such trajectories keep close with respect to $|\cdot |_{\infty}$ (which is pretty obvious after having proved Theorem \ref{thm1}) and show afterwards the more interesting statement about their distance with respect to $|\cdot |_1$.\\
Let $C_0>0$, $X\in \mathcal{G}^{N,\sigma}_{1,T}$ and $X'\in \mathbb{R}^{6N}$ a configuration for which there exists $t_0\in [0,T]$ such that $|\Psi^N_{t_0,0}(X')-\Psi^N_{t_0,0}(X)|_{\infty}\le \frac{N^{-\frac{1}{2}+\sigma}}{C_0}$. To keep the notation as compact as possible we define $Y:=\Psi_{t_0,0}^N(X)$ as well as $Z:=\Psi_{t_0,0}^N(X')$. 
The estimates will be confined on showing that the trajectories keep close until time $T$. Proving the closeness for times in $[0,t_0]$ works analogously due to the time-symmetry of the dynamics.\\ If (for example) $C_0\geq 4$, then it holds for any point in time $t \in [0,T-t_0]$ where the condition $\sup_{0\le s  \le  t}|\Psi^N_{s,0}(Y)-\Psi^N_{s,0}(Z)|_{\infty}\le  \frac{1}{2}
N^{-\frac{1}{2}+\sigma}$ is still fulfilled that
{\allowdisplaybreaks
\begin{align}
& \frac{d}{dt_+}\sup_{0 \le s \le t}|^1{\Psi}^N_{s,0}(Y)-{^1\Psi^N_{s,0}}(Z)|_{\infty} \\
\le & |^2{\Psi}^N_{t,0}(Y)-{^2\Psi^N_{t,0}}(Z)|_{\infty}\\
\le & |^2Y-{^2Z}|_{\infty} \notag \\
& + \frac{1}{N}\max_{i\in\{1,...,N\}}\big|\int_{0}^t\sum_{i\neq j}\Big( f^N([^1\Psi^N_{s,0}(Y)]_j-[^1\Psi^N_{s,0}(Y)]_i) \notag \\
&-f^N([^1\Psi^N_{s,0}(Z)]_j-[^1\Psi^N_{s,0}(Z)]_i)\Big)ds\big|\\
\le & |^2Y-{^2Z}|_{\infty}\\
& + \frac{1}{N}\max_{i\in\{1,...,N\}}\big|\int_{0}^t\sum_{i\neq j}\Big( f^N([^1\Psi^N_{s,0}(Y)]_j-[^1\Psi^N_{s,0}(Y)]_i) \notag \\
&-f^N({^1\varphi_{s+t_0,0}^N}(X_j)-{^1\varphi_{s+t_0,0}^N}(X_i))\Big)ds\big| \label{shift:t2}\\
& +  \frac{1}{N}\max_{i\in\{1,...,N\}}\big|\int_{0}^t\sum_{i\neq j}\Big( f^N({^1\varphi_{s+t_0,0}^N}(X_j)-{^1\varphi_{s+t_0,0}^N}(X_i))\notag \\
&-f^N([^1\Psi^N_{s,0}(Z)]_j-[^1\Psi^N_{s,0}(Z)]_i)\Big)ds\big| \label{shift:t4}\\
\le & |^2Y-{^2Z}|_{\infty} \notag\\
& + C\sup_{0\le s \le t}|\Psi^N_{s,0}(Y)-\Phi^N_{s+t_0,0}(X)|_{\infty}+CN^{-\frac{1}{2}+\frac{1}{2}\sigma} \notag  \\
& + C\sup_{0\le s \le t}|\Psi^N_{s,0}(Z)-\Phi^N_{s+t_0,0}(X)|_{\infty}+CN^{-\frac{1}{2}+\frac{1}{2}\sigma}.  \label{shift:t5}
\end{align}}
For the explanation why the last relation holds we recall that $X\in \mathcal{G}^{N,\sigma}_{1,T}$ implies in particular $X\in \big(\mathcal{B}^{N,\sigma}_5\big)^C$ which yields $\forall i\neq j:X_i\in G^N(X_j)$ (see \eqref{def.B_5}) and thus the upper bound for term \eqref{shift:t2} was already derived in the proof of Theorem \ref{thm1} by the estimates for term \eqref{rel.t3}. The upper bound for term \eqref{shift:t4} follows on the one hand due to the choice of the time $t$ and on the other hand due to Corollary \ref{cor.shift.} which yields that for $s\in [0,t]$
\begin{align*}
& |\Psi_{s,0}^N(Z)-\Phi_{s+t_0,0}^N(X)|_{\infty}\\
\le & |\Psi_{s,0}^N(Z)-\Psi_{s,0}^N(Y)|_{\infty}+|\underbrace{\Psi_{s,0}^N(Y)}_{=\Psi_{s+t_0,0}^N(X)}-\Phi_{s+t_0,0}^N(X)|_{\infty}\\
\le & \frac{1}{2}
N^{-\frac{1}{2}+\sigma}+CN^{-\frac{1}{2}+\frac{\sigma}{2}}.
\end{align*}
At least for large enough $N$ this bound is smaller than $N^{-\frac{1}{2}+\sigma}$ and thus in particular smaller than the maximal value for the deviation $$|\Psi_{s,0}^N(Y)-\Phi_{s+t_0,0}^N(X)|=|\Psi_{s+t_0,0}^N(X)-\Phi_{s+t_0,0}^N(X)|$$
which was allowed according to the stopping time introduced in the proof of Theorem \ref{thm1}. We point out that in the proof of this Theorem we only applied this information (namely that the distance with respect to $|\cdot|_{\infty}$ between the $N$-particle trajectory and the applied auxiliary trajectory $\Phi^N_{\cdot,0}(X)$ is still below $N^{-\frac{1}{2}+\sigma}$) and $X \in \mathcal{G}^{N,\sigma}_{1,T}$ to implement the different estimates. It was, on the other hand, not important that they start at the same initial data. Hence, it is straightforward to see that term \eqref{shift:t4} can be handled by the same reasoning as applied for term \eqref{rel.t3} which leads to a corresponding result.\\ Since $\sup_{0\le s \le t}|\Psi^N_{s,0}(Y)-\Phi^N_{s+t_0,0}(X)\big|_{\infty}\le CN^{-\frac{1}{2}+\frac{\sigma}{2}}$ holds due to Corollary \ref{cor.shift.}, it follows by triangle inequality that 
\begin{align*}
& \frac{d}{dt_+}\sup_{0 \le s \le t}|^1{\Psi}^N_{s,0}(Y)-{^1\Psi^N_{s,0}}(Z)|_{\infty}\\
\le & |^2{\Psi}^N_{t,0}(Y)-{^2\Psi^N_{t,0}}(Z)|_{\infty}\\
\le & C\sup_{0 \le s \le t}|^1{\Psi}^N_{s,0}(Y)-{^1\Psi^N_{s,0}}(Z)|_{\infty}+CN^{-\frac{1}{2}+\frac{\sigma}{2}}+|^2Y-{^2Z}|_{\infty}.
\end{align*}
With the help of Gronwall's lemma and a subsequent application of the inequality limiting the velocity deviation which appears between the second and the third line of the previous relation one easily concludes that 
\begin{align}
 &\sup_{0 \le s \le t}|{\Psi}^N_{s,0}(Y)-{\Psi^N_{s,0}}(Z)|_{\infty} \notag \\
 \le & \big(|Y-Z|_{\infty}+CN^{-\frac{1}{2}+\frac{\sigma}{2}} \big)e^{Ct} \le \big(\frac{N^{-\frac{1}{2}+\sigma}}{C_0}+CN^{-\frac{1}{2}+\frac{\sigma}{2}} \big)e^{Ct}  \label{upp.b.shift.lem}
\end{align}
By choosing $C_0$ large enough we can gather that the assumption
$$\sup_{0 \le s \le t}|{\Psi}^N_{s,0}(Y)-{\Psi^N_{s,0}}(Z)|_{\infty}\le  \frac{N^{-\frac{1}{2}+\sigma}}{2} $$
which we applied for the estimates keeps in fact valid for all $t\in [0,T-t_0]$ if $N$ is sufficiently large. After having noticed this, the first inequality appearing in \eqref{upp.b.shift.lem} concludes the proof regarding the deviation with respect to $|\p|_{\infty}$. \\
Proving the second statement is bit more complex but still large parts of the previous ideas and estimates can be recycled. We will focus on the arguments which are new and keep the familiar part a little shorter.  \\ 
By application of the sets $M^{N,(t_1,t_2)}_{r,v}(X)$ for $0\le t_1<t_2\le T$ (defined in \eqref{def.coll.cl.1}) we identify 
\begin{align}
&  \mathcal{C}^{i,N}_{1,(t_1,t_2)}(X):=\big\{j\in \{1,...,N\}\setminus\{i\}: X_j \in M^{N,(t_1,t_2)}_{6N^{-\frac{1}{2}+{\sigma}},N^{-\frac{1}{9}+3{\sigma}}}(X_i) \big\} \notag \\ 
& \label{sets C_1,C_2} \mathcal{C}^{i,N}_{2,(t_1,t_2)}(X):=\big\{j\in \{1,...,N\}\setminus\{i\}: X_j \in M^{N,(t_1,t_2)}_{6N^{-\frac{1}{2}+{\sigma}},\infty}(X_i) \big\}\setminus \mathcal{C}^{i,N}_{1,(t_1,t_2)}(X) 
\end{align} 
and finally $\mathcal{C}^{i,N}_{(t_1,t_2)}(X)=\mathcal{C}^{i,N}_{1,(t_1,t_2)}(X)\cup \mathcal{C}^{i,N}_{2,(t_1,t_2)}(X)$. Thus, these sets contain the labels of the `mean-field particles' which come `close' (in physical space) to the $i$-th `mean-field particle' at some moment in $[t_1,t_2]$ and in case of $\mathcal{C}^{i,N}_{1,(t_1,t_2)}(X)$ have additionally a low relative velocity value (or more precisely lower than $N^{-\frac{1}{9}+3\sigma}$). \\
As before, let $X\in \mathcal{G}^{N,\sigma}_{1,T}$ and additionally $Y',Z' \in \mathbb{R}^{6N}$ such that 
\[\max_{\widetilde{X}\in \{Y',Z'\}}\inf_{0\le s\le T}|\Psi^N_{s,0}(\widetilde{X})-\Psi^N_{s,0}(X)|_{\infty}\le \frac{N^{-\frac{1}{2}+\sigma}}{C_0}.\]
Then, as we previously showed it holds for large enough $C_0,N$ that
\begin{align}
\max_{\widetilde{X}\in \{Y',Z'\}}\sup_{0\le s \le T}|\Psi^N_{s,0}(\widetilde{X})-\Psi^N_{s,0}(X)|_{\infty}\le \frac{N^{-\frac{1}{2}+\sigma}}{2}\label{dist.inf.shift.lem}.
\end{align}
Let this be the case and let the times $t_1,t_2\in [0,T]$ be chosen such that the condition $t_2-t_1\geq N^{-\frac{1}{3}}$ mentioned in the assumptions of the lemma is fulfilled. Moreover, we assume that $t_0\in [t_1,t_2]$ is a point in time where the distance $|\Psi^N_{\p,0}(Y')-\Psi^N_{\p,0}(Z')|_{1}$ attains its minimal value on this compact interval and identify again for ease of notation $Y:=\Psi_{t_0,0}^N(Y')$ as well as $Z:=\Psi_{t_0,0}^N(Z')$. Like in the preceding case we will only show explicitly that in positive time direction (respectively for times in $[t_0,t_2]$) the considered distance between the related trajectories stays typically sufficiently small and remark that for times in $[t_1,t_0]$ an analogous reasoning can be applied.\\ 
Finally, we abbreviate for $i\in \{1,...,N\}$
\begin{align}
{^1\Delta^N_i}(Y,Z,t):=\sup_{0\le s \le t}|[^1\Psi^N_{s,0}(Y)]_i-[^1\Psi^N_{s,0}(Z)]_i| \label{def.Delta^1} 
\end{align}
and note that due to a `mean value argument' it holds for $q,q'\in \mathbb{R}^3$ that 
\begin{align}
|f^N(q) -f^N(q')|
\le & C\big(\frac{1}{(|q|+N^{-{c}})^{\alpha+1}}+\frac{1}{(|q'|+N^{-{c}})^{\alpha+1}} \big)|q-q'|.\label{lip.coll.}
\end{align} 
We apply this relation in the third step of the subsequent estimates to obtain term \eqref{problem.t.shift.}. Moreover, we make use of the sets $\mathcal{C}^{i,N}_{(t_1,t_2)}(X)$ as well as $\mathcal{C}^{i,N}_{n,(t_1,t_2)}(X)$ for $n=1,2$ (see\eqref{sets C_1,C_2}) to conclude that for $t\in [0,t_2-t_0]$ the following holds:
{\allowdisplaybreaks \begin{align}
&\frac{d}{dt_+}\sum_{i=1}^N {^1\Delta^N_i}(Y,Z,t)\label{shift.t.1-norm}\\
\le & \sum_{i=1}^N \sup_{0\le s \le t}|[^2\Psi^N_{s,0}(Y)]_i-[^2\Psi^N_{s,0}(Z)]_i|  \label{t.v.} \\
\le & |^2Y-{^2Z}|_1 \notag\\
&+ \frac{1}{N}\int_{0}^{t}\sum_{i=1}^N\sum_{j\neq i}\Big(\big| f^N([^1\Psi^N_{s,0}(Y)]_j-[^1\Psi^N_{s,0}(Y)]_i) \notag \\
&  -f^N([^1\Psi^N_{s,0}(Z)]_j-[^1\Psi^N_{s,0}(Z)]_i)\big|\Big)ds  \\
\le & |^2Y-{^2Z}|_1 \notag \\
&+   \frac{C}{N}  \sum_{i=1}^N\sum_{ j\in \big(\mathcal{C}^{i,N}_{(t_1,t_2)}(X)\big)^C}\int_0^{t} g^N(^1\varphi^N_{s+t_0,0}(X_j)-{^1\varphi^N_{s+t_0,0}(X_i)})\notag \\
& \cdot \big(^1\Delta^N_i(Y,Z,s)+{^1\Delta^N_j}(Y,Z,s) \big)ds \label{t_31} \\
& +\frac{C}{N}\sum_{i=1}^N\sum_{ j\in \mathcal{C}^{i,N}_{2,(t_1,t_2)}(X)}\int_0^t\Big(\frac{{^1\Delta^N_i}(Y,Z,s)+{^1\Delta^N_j}(Y,Z,s) }{(|[^1\Psi^N_{s,0}(Y)]_j-[{^1\Psi^N_{s,0}(Y)}]_i|+N^{-c})^{\alpha+1}}\notag \\
& +\frac{{^1\Delta^N_i}(Y,Z,s)+{^1\Delta^N_j}(Y,Z,s)  }{(|[^1\Psi^N_{s,0}(Z)]_j-[{^1\Psi^N_{s,0}(Z)}]_i|+N^{-c})^{\alpha+1}}\Big) ds\label{problem.t.shift.}\\
& +  \frac{1}{N}\sum_{i=1}^N\sum_{ j\in \mathcal{C}^{i,N}_{1,(t_1,t_2)}(X)}\int_0^{t}\Big(|f^N([^1\Psi^N_{s,0}(Y)]_j-[^1\Psi^N_{s,0}(Y)]_i)| \notag \\
& +|f^N([^1\Psi^N_{s,0}(Z)]_j-[^1\Psi^N_{s,0}(Z)]_i)|\Big)ds  \label{t_32}.
\end{align}}
\noindent For receiving term \eqref{t_31} we applied the properties of $g^N$ (see \eqref{ineq.g^N+}) together with Corollary \ref{cor.shift.} and estimate \eqref{dist.inf.shift.lem} which yield the following implication for large enough $N$ and $\widetilde{X}\in Y,Z$:
\begin{align*}
& j\in  \big(\mathcal{C}^{i,N}_{(t_1,t_2)}(X)\big)^C \\
\Rightarrow &\forall s\in [0,t_2-t_0]:\\
& |[^1\Psi^N_{s,0}(\widetilde{X})]_j-[{^1\Psi^N_{s,0}(\widetilde{X})}]_i|\geq |[^1\Psi^N_{s+t_0,0}(X)]_j-[{^1\Psi^N_{s+t_0,0}(X)}]_i|-N^{-\frac{1}{2}+\sigma}\\
&\hspace*{4,4cm} \geq   \underbrace{|^1\varphi^N_{s+t_0,0}(X_j)-{^1\varphi^N_{s+t_0,0}}(X_i)|}_{\geq 6N^{-\frac{1}{2}+\sigma}}-2N^{-\frac{1}{2}+\sigma}\\
&\hspace*{4,4cm} > \frac{|^1\varphi^N_{s+t_0,0}(X_j)-{^1\varphi^N_{s+t_0,0}(X_i)|}}{2}
\end{align*}
A first step for simplifying these estimates is to notice that 
$$j\in \mathcal{C}^{i,N}_{n,(t_1,t_2)}(X)\Leftrightarrow i\in \mathcal{C}^{j,N}_{n,(t_1,t_2)}(X)$$ for $n=1,2$ and thus it holds for the index sets of the sums that
$$ \{(i,j)\in \{1,...,N\}^2:j\in \mathcal{C}^{i,N}_{n,(t_1,t_2)}(X)\}= \{(i,j)\in \{1,...,N\}^2: i\in   \mathcal{C}^{j,N}_{n,(t_1,t_2)}(X)\}.$$
By these considerations we can simplify terms \eqref{t_31} to \eqref{t_32} as follows:
\begin{align}
& |^2Y-{^2Z}|_1 \notag \\
&+  \max_{\substack{k\in \{1,...,N\}\\ \widetilde{X}\in \{Y,Z\} }}\bigg(\sum_{ j\in \big(\mathcal{C}_{t_1,t_2}^{k,N}(X)\big)^C}\int_0^t g^N(^1\varphi^N_{s+t_0,0}(X_j)-{^1\varphi^N_{s+t_0,0}}(X_k))ds \notag \\
& +\sum_{ j\in \mathcal{C}^{k,N}_{2,(t_1,t_2)}(X)}\int_0^t\frac{1}{(|[^1\Psi^N_{s,0}(\widetilde{X})]_j-[{^1\Psi^N_{s,0} (\widetilde{X})}]_k|+N^{-c})^{\alpha+1}}ds\bigg) \frac{C}{N}\sum_{i=1}^N{^1\Delta^N_i}(t,Y,Z)\notag \\
& + \frac{2}{N}\max\limits_{\widetilde{X}\in \{Y,Z\}}\sum_{i=1}^N\sum_{ j\in \mathcal{C}^{i,N}_{1,(t_1,t_2)}(X)}\int_0^{t}|f^N([^1\Psi^N_{s,0}(\widetilde{X})]_j-[^1\Psi^N_{s,0}(\widetilde{X})]_i)|ds. \label{shift.t}
\end{align}
Application of $X\in \mathcal{G}^{N,\sigma}_{1,T}$ yields $X\in (\mathcal{B}^{N,\sigma}_{5})^C$ and thus $X_j\in G^N(X_i)$ for all $ j\neq i$ (see \eqref{def.B_5}) as well as $X\in  \bigcap_{i=1}^N (\mathcal{B}^{N,\sigma}_{2,i})^C$ (see \eqref{def.B_2}) which implies in total that for all $i\in \{1,...,N\}$
\begin{align}
& \frac{1}{N}\sum_{j\in \big(\mathcal{C}_{t_1,t_2}^{i,\sigma}(X)\big)^C}\int_0^T g^N(^1\varphi^N_{s,0}(X_j)-{^1\varphi^N_{s,0}}(X_i))ds \notag \\
\le & \frac{1}{N}\sum_{j\neq i }\int_0^T g^N(^1\varphi^N_{s,0}(X_j)-{^1\varphi^N_{s,0}}(X_i))\mathbf{1}_{G^N(X_i)\cap (M^N_{3N^{-\frac{1}{2}+\sigma},\infty}(X_i))^C}(X_j) ds \notag \\
\le & 1+\int_0^T\int_{\mathbb{R}^6}g^N(^1\varphi^N_{s,0}(Y)-{^1\varphi^N_{s,0}}(X_i))k_0(Y)d^6Y ds \notag \\
\le &  C.  \label{est.sum.g+}
\end{align}
Furthermore, due to $X \in  \mathcal{G}^{N,\sigma}_{1,T} \subseteq \bigcap_{i=1}^N (\mathcal{B}^{N,\sigma}_{4,i})^C$ (see \eqref{def.B_4}) it holds that
$$\big|\mathcal{C}^{i,N}_{2,(t_1,t_2)}(X)\big| \le \sum_{k\neq i} \mathbf{1}_{M^N_{6N^{-\frac{1}{2}+\sigma},\infty}(X_i)}(X_k)\le N^{3\sigma}.$$ It follows by Corollary \ref{cor1} (ii) that for any $i\in \{1,...,N\}$ and $\widetilde{X}\in \{Y,Z\}$
\begin{align}
& \frac{C}{N}\sum_{ j\in \mathcal{C}^{i,N}_{2,(t_1,t_2)}(X)}\int_0^t\frac{1}{(|[^1\Psi^N_{s,0}(\widetilde{X})]_j-[{^1\Psi^N_{s,0} (\widetilde{X})}]_i|+N^{-c})^{\alpha+1}}ds \notag \\
\le & \frac{C}{N}\frac{1}{N^{-c\alpha}N^{-\frac{1}{9}+3\sigma}}\big|\mathcal{C}^{i,N}_{2,(t_1,t_2)}(X)\big|\notag \\
\le & \frac{C}{N}\frac{1}{N^{-c\alpha}N^{-\frac{1}{9}+3\sigma}}N^{3\sigma} \notag \\
\le &  C. \label{expl.appl.cor1}
\end{align}
In the last step we took into account that $c=\frac{2}{3}$ respectively $\alpha\in (1,\frac{4}{3}]$. Moreover, we remark that the assumptions of Corollary \ref{cor1} (ii) are fulfilled since $ j\in \mathcal{C}^{i,N}_{2,(t_1,t_2)}(X) $ implies that the value of relative velocity between the $i$-th and $j$-th `mean-field particle' at the time of their closest encounter is at least of order $N^{-\frac{1}{9}+3\sigma}$ (see \eqref{sets C_1,C_2}) and thus of much larger order than the deviation
\begin{align*}
&\max\limits_{\widetilde{X}\in \{Y,Z\}}\sup_{0\le s \le T-t_0}|\Psi^N_{s,0}(\widetilde{X})-\Phi^N_{t_0+s,0}(X)|_{\infty}\\
\le & \max\limits_{\widetilde{X}\in \{Y,Z\}}\sup_{0\le s \le T-t_0}|\Psi^N_{s,0}(\widetilde{X})-\Psi^N_{t_0+s,0}(X)|_{\infty} +\sup_{0\le s \le T}|\Psi^N_{s,0}(X)-\Phi^N_{s,0}(X)|_{\infty}\\ \le &  N^{-\frac{1}{2}+\sigma}+CN^{-\frac{1}{2}+\frac{\sigma}{2}}
\end{align*}
 where we applied relation \eqref{dist.inf.shift.lem}, $Y=\Psi^N_{t_0,0}(Y')$, $Z=\Psi^N_{t_0,0}(Z')$ and Corollary \ref{cor.shift.}. Hence, it follows for large enough $N\in \mathbb{N}$ and $t\in [0,t_2-t_0]$ that
\begin{align}
&\frac{d}{dt_+}\sum_{i=1}^N {^1\Delta^N_i}(Y,Z,t) \notag \\
\le & \sum_{i=1}^N \sup_{0\le s \le t}|[^2\Psi^N_{s,0}(Y)]_i-[^2\Psi^N_{s,0}(Z)]_i|  \\
\le & C\sum_{i=1}^N {^1\Delta^N_i}(Y,Z,t) +|^2Y-{^2Z}|_1\notag \\
&+ \frac{2}{N}\max\limits_{\widetilde{X}\in \{Y,Z\}}\sum_{i=1}^N\sum_{ j\in \mathcal{C}^{i,N}_{1,(t_1,t_2)}(X)}\int_0^{T-t_0}|f^N([^1\Psi^N_{s,0}(\widetilde{X})]_j-[^1\Psi^N_{s,0}(\widetilde{X})]_i)|ds. \label{Gron.shift}
\end{align}
Until now all estimates can be implemented for arbitrary $X\in \mathcal{G}^{N,\sigma}_{1,T}$ and it remains to show that the last term keeps typically sufficiently small (i.e. smaller than $N^{-\sigma}(t_2-t_1)$) before we are able to conclude the proof.\\ 
First, we recall that $j\in \mathcal{C}^{i,N}_{1,(t_1,t_2)}(X)$ implies that $X_j\in M^{N,(t_1,t_2)}_{6N^{-\frac{1}{2}+{\sigma}},N^{-\frac{1}{9}+3{\sigma}}}(X_i)$. Consequently, it suffices to cover this set by a certain number of finer subdivided `collision classes' and to show that the `impact' related to each class stays typically small enough. \\ 
For $k\in \mathbb{Z}$ one possibility for such a cover is given by
\begin{align}
 &\text{(i)}&   C^N_1(X_i)&:=M^{N,(t_1,t_2)}_{(0,6N^{-\frac{1}{2}+{\sigma}}),(0,N^{-\frac{1}{3}})}(X_i) \notag \\
&\text{(ii)}& C^N_{2,k}(X_i)&:=M^{N,(t_1,t_2)}_{(0,6N^{-\frac{1}{2}+{\sigma}}),(N^{-(k+1){\sigma}},N^{-k{\sigma}})}(X_i),\ \frac{1}{9}-3\sigma \le k{\sigma} \le  \frac{1}{3} \notag \\
&\text{(iii)}& C^N_3(X_i)&:=M^{N,(t_1,t_2)}_{(0,6N^{-\frac{1}{2}+{\sigma}}),(N^{-\lceil\frac{1}{9\sigma}-3 \rceil \sigma   },N^{- \frac{1}{9}+3\sigma} )}(X_i) \label{Def.C_k}
\end{align}
because it holds that
\[ M^{N,(t_1,t_2)}_{6N^{-\frac{1}{2}+{\sigma} },N^{-\frac{1}{9}+3{\sigma} }}(X_i)\subseteq C^N_1(X_i)  \cup C^N_3(X_i)\cup \bigcup_{\substack{k\in \mathbb{N}:\\
 \frac{1}{9}-3\sigma    \le k{\sigma}  \le \frac{1}{3}}}C^N_{2,k}(X_i).\]  
Since we only regard configurations belonging to $ \mathcal{G}_{1,T}^{N,\sigma}$, no `mean-field particle pair' fulfills the conditions of classes where the relative velocity values are very low (which in particular is the case for $C_1(X_i)$) and thus these classes can be neglected for the subsequent estimates.\\
Let $N\in \mathbb{N},\ k\in \{\lceil \frac{1}{9\sigma}-3 \rceil -1,...,\lfloor \frac{1}{3\sigma} \rfloor \}$, $r:=6N^{-\frac{1}{2}+\sigma}$ and 
$$v_k:= \begin{cases}N^{-k\sigma} & \text{, if} \  \frac{1}{9}-3\sigma  \le k\sigma \le \frac{1}{3}\\
N^{-\frac{1}{9}+3\sigma} &\text{, if} \     k=\lceil \frac{1}{9\sigma}-3 \rceil -1\end{cases}.$$
Once again we see that the different values for $v_k$ are of distinctly larger order with respect to $N$ than the deviation 
\begin{align*}
&\max\limits_{\widetilde{X}\in \{Y,Z\}}\sup_{-t_0\le s \le T-t_0}|\Psi^N_{s,0}(\widetilde{X})-\Phi^N_{t_0+s,0}(X)|_{\infty}\le N^{-\frac{1}{2}+\sigma}+CN^{-\frac{1}{2}+\frac{\sigma}{2}}
\end{align*} 
(if $\sigma>0$ is chosen small enough). Hence, also in the current situation we can apply our collision estimates described in Corollary \ref{cor1} to conclude that for $\widetilde{X}\in \{Y,Z\}$
\begin{align}
& \frac{2}{N}\sum_{i=1}^N\sum_{ j\neq i}\int_0^{T-t_0}|f^N([^1\Psi^N_{s,0}(\widetilde{X})]_j-[^1\Psi^N_{s,0}(\widetilde{X})]_i)|\mathbf{1}_{M^{N,(t_1,t_2)}_{(0,r),(v_{k+1},v_k)}(X_i)}(X_j) ds \notag \\
\le & \frac{C}{N}\frac{1}{N^{-c(\alpha-1)}v_{k+1}}\sum_{i=1}^N\sum_{ j\neq i}\mathbf{1}_{M^{N,(t_1,t_2)}_{(0,r),(v_{k+1},v_k)}(X_i)}(X_j) .
 \label{t-shift:impact}
\end{align}
In the next step we show that for $\epsilon\geq \sigma$ typically 
$$\sum_{i=1}^N\sum_{ j\neq i}\mathbf{1}_{M^{N,(t_1,t_2)}_{(0,r),(v_{k+1},v_k)}(X_i)}(X_j) \le N^{2\epsilon}\big\lceil N^2r^2v_k^3\big(r+v_k(t_2-t_1)\big) \big\rceil.$$ Very similar to the reasoning applied previously to handle assumption \eqref{cond.prob.est.1} we obtain the subsequent relationship:
\begin{align}
& \sum_{i=1}^N\sum_{ j\neq i}\mathbf{1}_{M^{N,(t_1,t_2)}_{(0,r),(v_{k+1},v_k)}(X_i)}(X_j) \geq N^{2\epsilon}\big\lceil N^2r^2v_k^3\big(r+v_k(t_2-t_1)\big) \big\rceil \label{cond.shift}\\
\Rightarrow & \big(\exists i\in \{1,...,N\}: \sum_{ j\neq i}\mathbf{1}_{M^{N,(t_1,t_2)}_{(0,r),(0,v_k)}(X_i)}(X_j)\geq \lfloor N^{\epsilon} \rfloor \big) \ \vee \label{cond.shift1}\\
&\Big(\exists \mathcal{S}\subseteq \{1,...,N\}^2\setminus \bigcup_{n=1}^N\{(n,n)\}: \notag \\
& \ \text{(i)}\ \  |\mathcal{S}|\geq \frac{N^{\epsilon}}{4}\big\lceil N^2r^2v_k^3\big(r+v_k(t_2-t_1)\big) \big\rceil \notag \\ & \  \text{(ii)} \ \ \forall (i,j)\in \mathcal{S}:X_j\in M^{N,(t_1,t_2)}_{(0,r),(0,v_k)}(X_i)  \notag \\
&\ \text{(iii)} \ (i_1,j_1),(i_2,j_2)\in \mathcal{S}\Rightarrow  \{i_1,j_1\}\cap \{i_2,j_2\}=\emptyset\Big)\label{cond.shift2}
\end{align}
This can be seen as follows:\\ If condition \eqref{cond.shift} is fulfilled, then there exists a set $\mathcal{S'}\subseteq \{1,...,N\}^2\setminus \bigcup_{n=1}^N\{(n,n)\}$ where $$|\mathcal{S}'|\geq N^{2\epsilon}\big\lceil N^2r^2v_k^3\big(r+v_k(t_2-t_1)\big) \big\rceil\ \land  \ \forall (i,j)\in \mathcal{S}':X_j \in M^{N,(t_1,t_2)}_{(0,r),(0,v_k)}(X_i).$$ 
Let us assume that condition \eqref{cond.shift1} is not fulfilled respectively there exists no particle having a collision of the considered kind with at least $\lfloor N^{\epsilon} \rfloor $ different particles. Then we argue that it is possible to find a set $\mathcal{S}\subseteq \mathcal{S}'$ containing $ \frac{N^{\epsilon}}{4}\big\lceil N^2r^2v_k^3\big(r+v_k(t_2-t_1)\big) \big\rceil$ particle pairs having such a collision with each other (which corresponds to item (i)+(ii) of condition \eqref{cond.shift2}) but with no further particle of the remaining pairs belonging to $\mathcal{S}$ (which corresponds to item (iii) of condition \eqref{cond.shift2}): Just choose an arbitrary pair of $(i_1,j_1)\in\mathcal{S}'$. Since by assumption $$\forall i\in \{1,...,N\}:\sum_{ j\neq i}\mathbf{1}_{M^{N,(t_1,t_2)}_{(0,r),(0,v_k)}(X_i)}(X_j)\le \lfloor N^{\epsilon}\rfloor-1, $$ it follows that there exist respectively at most $(\lfloor N^{\epsilon}\rfloor-2)$ further particles which have such a collision with a particle related to labels $i_1$ or $j_1$ which in turn corresponds to at most $4(\lfloor N^{\epsilon}\rfloor-2)$ further tuples $(i,j)$ that are contained in $\mathcal{S}'$ where $i_1\in \{i,j\}$ or $j_1\in \{i,j\}$ except for $(i_1,j_1)$ and $(j_1,i_1)$. Now `remove' the at most $4\lfloor N^{\epsilon}\rfloor-6$ tuples of $\mathcal{S}'$ where $i_1$ or $j_1$ are `contained' and choose the next pair $(i_2,j_2)$ out of the remaining ones for the second round of the approach. Since at least $$\big\lceil \frac{N^{2\epsilon}\lceil N^2r^2v_k^3\big(r+v_k(t_2-t_1)\big) \rceil}{ 4N^{\epsilon}} \big\rceil \geq  \frac{N^{\epsilon}}{4}\big\lceil N^2r^2v_k^3\big(r+v_k(t_2-t_1)\big) \big\rceil $$ rounds of this routine are possible, it provides us a set $\mathcal{S}$ like desired. The `removal' of the tuples after each round ensures that item (iii) of condition \eqref{cond.shift2} is fulfilled. \\
Fortunately, the event related to assumption \eqref{cond.shift1} can not occur for the considered configurations $X\in \mathcal{G}^{N,\sigma}_{1,T}\subseteq \bigcap_{i=1}^N \big(\mathcal{B}^{N,\sigma}_{4,i}\big)^C$ because by definition of this set (see \eqref{def.B_4}) it holds for $r=6N^{-\frac{1}{2}+\sigma}$ and the possible values for $v_k$ that
$$\sum_{ j\neq i}\mathbf{1}_{M^{N,(t_1,t_2)}_{(0,r),(0,v_k)}(X_i)}(X_j)\le  \sum_{j \neq i}\mathbf{1}_{M^N_{6N^{-\frac{1}{2}+\sigma},N^{-\frac{1}{9}+3\sigma}}(X_i)}(X_j)< N^{\frac{\sigma}{2}}\le N^\epsilon $$
due to the constraint $\epsilon\geq \sigma$. Now we apply Lemma \ref{lem3} and the property that the `mean-field particles' are i.i.d. to derive an upper bound for the probability of the `event' described by assumption \eqref{cond.shift2}. For this purpose, we abbreviate $$M:= \big\lceil \frac{N^{\epsilon}}{4} \big\lceil N^2r^2v_k^3\big(r+v_k(t_2-t_1)\big) \big\rceil\big\rceil$$ and note that the number of possibilities for choosing $M$ `disjoint collision pairings' (where the expression `disjoint' refers to item (iii) of condition \eqref{cond.shift2}) is bounded by $\binom{N^2}{M}$. Thus, it holds that 
{\allowdisplaybreaks 
\begin{align*}
& \mathbb{P}\Big(\exists \mathcal{S}\subseteq \{1,...,N\}^2\setminus \bigcup_{n=1}^N\{(n,n)\}:\\
& \hspace{0,5cm} |\mathcal{S}|\geq  M \ \land \ \forall(i,j)\in \mathcal{S}:X_j\in M^{N,(t_1,t_2)}_{(0,r),(0,v_k)}(X_i)\ \land \\
& \hspace{0,5cm} (i_1,j_1),(i_2,j_2)\in \mathcal{S}\Rightarrow  \{i_1,j_1\}\cap \{i_2,j_2\}=\emptyset \Big)\\
\le & \binom{N^2}{M}\mathbb{P}\big(\forall (i,j)\in \{(1,2),(2,3),...,(2M-1,2M)\}: X_j \in  M^{N,(t_1,t_2)}_{(0,r),(0,v_k)}(X_i) \big)\\
\le & C^M\frac{N^{2M}}{M^M}\big(r^2v_k^4 (t_2-t_1)+r^3v_k^3 \big)^M\\
\le & \big(CN^{-\epsilon})^{ \frac{N^\epsilon}{4}}
\end{align*} }
 In the first step we regarded that we only have to care for `disjoint' tuples according to the assumption $\{i_1,j_1\}\cap \{i_2,j_2\}=\emptyset$ and in the second step we simply applied the probability estimates of Lemma \ref{lem3}. Finally, the last step follows by regarding the choice of $M$ and in particular $M\geq \frac{N^\epsilon}{4}$. It is straightforward to see that this upper bound for the probability fulfills the `decay condition' claimed in the assumptions of the current lemma and thus we can assume for the remaining part that the configurations considered by us indeed fulfill
$$\sum_{i=1}^N\sum_{ j\neq i}\mathbf{1}_{M^{N,(t_1,t_2)}_{(0,r),(v_{k+1},v_k)}(X_i)}(X_j)\le N^{2\epsilon}\big\lceil N^2r^2v_k^3\big(r+v_k(t_2-t_1)\big) \big\rceil$$ for each set of the previously defined cover. For the last conclusion we remark additionally that the number of classes belonging to the cover is bounded by some constant (which only depends on $\sigma$) and thus the probability that for any of these classes the stated upper bound is violated fulfills the same `decay condition'.\\ It remains to verify that under this assumption term \eqref{t-shift:impact} stays sufficiently small for all of these classes:
{\allowdisplaybreaks \begin{align*}
&\frac{C}{N}\frac{1}{N^{-c(\alpha-1)}v_{k+1}}\sum_{i=1}^N\sum_{ j\neq i}\mathbf{1}_{M^{N,(t_1,t_2)}_{(0,r),(v_{k+1},v_k)}(X_i)}(X_j)\\
\le &  \frac{C}{N}\frac{1}{N^{-c(\alpha-1)}v_{k+1}} N^{2\epsilon}\big\lceil N^2r^2v_k^3\big(r+v_k(t_2-t_1)\big) \big\rceil\\
\le & \frac{C}{N}\frac{1}{N^{-\frac{2}{9}} v_{k+1}}\big( N^{2\epsilon}+ N^{2+2\epsilon} r^2v_k^3\big(r+v_k(t_2-t_1)\big) \big)\\
\le & C\Big(\frac{N^{-\frac{7}{9}+2\epsilon+\sigma}}{v_k(t_2-t_1)}+\frac{N^{-\frac{5}{18}+2\epsilon+4\sigma}v_k^2}{t_2-t_1}+N^{\frac{2}{9}+2\epsilon+3\sigma}v_k^3\Big)(t_2-t_1)
\end{align*}}
where we applied that $v_{k+1}=N^{-\sigma}v_k$, $r=6N^{-\frac{1}{2}+\sigma}$, $c=\frac{2}{3}$ and $\alpha\in (1,\frac{4}{3}]$. Thus, the interesting values for $v_k$ are the largest and the smallest possible, respectively $N^{-\frac{1}{9}+3\sigma}$ and $N^{-\frac{1}{3}}$ (where as mentioned before the class $C^N_1(X_i)$ can be neglected since $X\in \mathcal{G}^{N,\sigma}_{1,T}$). By regarding this and additionally $t_2-t_1\geq N^{-\frac{1}{3}}$ as well as $\epsilon\geq \sigma$ we finally obtain that the previous expression is bounded by
\begin{align*}
 & C\big(N^{-\frac{1}{9}+3\epsilon}+ N^{-\frac{1}{6}+12\epsilon}+N^{-\frac{1}{9}+14\epsilon}\big)(t_2-t_1).
\end{align*}
Application of this estimate on inequality \eqref{Gron.shift} yields that for all such configurations there exists $C>0$ such that for $t\in [0,t_2-t_0]$:
\begin{align}
&\sum_{i=1}^N {^1\Delta^N_i}(Y,Z,t) \notag \\
\le &|^1Y-{^1Z}|_1+\int_0^t \sum_{i=1}^N \sup_{0\le r\le s}|[^2\Psi^N_{r,0}(Y)]_i-[^2\Psi^N_{r,0}(Z)]_i|ds \notag  \\
\le &\int_0^t C\sum_{i=1}^N {^1\Delta^N_i}(Y,Z,s)ds+|^1Y-{^1Z}|_1 \notag \\
&+|^2Y-{^2Z}|_1(t_2-t_1)+CN^{-\frac{1}{9}+14\epsilon}(t_2-t_1)^2 \label{ineq.chain (shift-lem)}.
\end{align}  
Hence, we can apply Gronwall's lemma and obtain that for $t\in [0,t_2-t_0]$: 
\begin{align}
& \sum_{i=1}^N {^1\Delta^N_i(Y,Z,t)
}\notag \\
\le & \Big(|^1Y-{^1Z}|_1+|^2Y-{^2Z}|_1(t_2-t_1)+CN^{-\frac{1}{9}+14\epsilon}(t_2-t_1)^2\Big)e^{Ct} \label{ineq.shift.lem.r}
\end{align} 
Moreover, due to the previous estimates it holds for the considered configurations that
\begin{align}
&\sum_{i=1}^N \sup_{0\le s \le t}|\big([^2\Psi^N_{s,0}(Y)]_i-[^2\Psi^N_{s,0}(Z)]_i-({^2Y_i}-{^2Z_i)}|   \notag\\
\le & \frac{1}{N}\int_{0}^{t}\sum_{i=1}^N\sum_{j\neq i}\Big(\big| f^N([^1\Psi^N_{s,0}(Y)]_j-[^1\Psi^N_{s,0}(Y)]_i) \notag \\
&  -f^N([^1\Psi^N_{s,0}(Z)]_j-[^1\Psi^N_{s,0}(Z)]_i)\big|\Big)ds \notag  \\
\le &  C\sum_{i=1}^N {^1\Delta^N_i}(Y,Z,t)+CN^{-\frac{1}{9}+14\epsilon}(t_2-t_1) \label{ineq.shift.lem.v0}\\
\le &C\Big(|^1Y-{^1Z}|_1+|^2Y-{^2Z}|_1(t_2-t_1)+N^{-\frac{1}{9}+14\epsilon}(t_2-t_1)\Big) \label{ineq.shift.lem.v}
\end{align}
where we point out that the inequality related to \eqref{ineq.shift.lem.v0} will become important for the subsequent Corollary.\\
Since the value of $\sigma>0$ and thereby of $\epsilon>0$ can be chosen arbitrarily small, it follows in particular that they can be chosen such that for large enough $N\in \mathbb{N}$ and $t\in [0,t_2-t_0]$
\begin{align}
& \sup_{0\le s \le t}|\Psi^N_{s,0}(Y)-\Psi^N_{s,0}(Z)|_1 \notag \\ 
\le &  C|X-Z|_1+CN^{-\frac{1}{9}+14\epsilon}(t_2-t_1) \le  C|X-Z|_1+ N^{-\sigma}(t_1-t_2)
\label{ineq.shift.lem.tot.}
\end{align} which eventually completes the proof.
\end{proof}
\vspace*{0,5cm}
\noindent Although the proof is completed, we directly continue to examine the inequality related to term \eqref{ineq.shift.lem.v0} closer. In the proof of Theorem \ref{thm2} we will derive upper bounds for the growth of deviations (with respect to $|\cdot|_1$) between different trajectories on short time intervals which in turn will be applied to determine an upper bound for their deviation on longer time spans. At first glance, inequality \eqref{ineq.shift.lem.v0} appears inappropriate for this purpose because an already existing spatial deviation would be translated in a velocity deviation of (possibly) multiple value independent of the length of the considered time interval. Extrapolated to longer times the order of the deviation would be greatly overestimated without further reasoning. One could get rid of this issue by deriving a better suited estimate. On the other hand, the described problem looses significance the longer the considered time interval $[t_1,t_2]$ is and by applying additionally the second order nature of the dynamics, it is possible to extend the estimates step by step to longer time spans. Very broadly speaking, if the velocity deviation is of distinctly larger order than the deviation in position space, then the spatial deviation must first `catch up' before it contributes in a relevant way to the further (relative) growth of the velocity deviations.\\ As an exception to the common proceeding, we implement the estimates first this time and summarize the derived results in a Corollary afterwards because the situation which we consider is exactly the same as in the previous proof.\\
As mentioned, we want to extend the relation provided by inequality \eqref{ineq.shift.lem.v0}. For the following we choose a sufficiently big $N\in \mathbb{N}$ and $\epsilon,\epsilon_0>0$ small enough such that the relation $N^{-\frac{1}{9}+14\epsilon} \le N^{-\epsilon_0}$ is fulfilled for the corresponding addend appearing in term \eqref{ineq.shift.lem.v0}. Then the related inequality takes the form
\begin{align}
&\sum_{i=1}^N \sup_{t_1\le s \le t}\Big(|[^2\Psi^N_{s,0}(Y')]_i-[^2\Psi^N_{s,0}(Z')]_i-([^2\Psi^N_{t_1,0}(Y')]_i-[^2\Psi^N_{t_1,0}(Z')]_i)|\Big)   \notag \\
\le &  C\sum_{i=1}^N \sup_{t_1\le s \le t}|[^1\Psi^N_{s,0}(Y')]_i-[^1\Psi^N_{s,0}(Z')]_i|+CN^{-\epsilon_0}(t_2-t_1). \label{ineq.shift.lem.v2}
\end{align} 
for $t\in [t_1,t_2]$ and according to the previous proof we know that it holds for typical initial data. In particular, the inequality is satisfied for configurations $Y',Z'\in \mathbb{R}^{6N}$ which fulfill $$\max_{X'\in \{Y',Z'\}}\min_{0\le s \le T}|{\Psi^N_{s,0}}(X')-{\Psi^N_{s,0}}(X)|_{\infty}\le  N^{-\frac{1}{2}+\frac{4}{5}}$$ for some $X\in\widetilde{\mathcal{G}}^{N,\sigma,\epsilon_0}_{2,(t_1,t_2)}\cap \mathcal{G}^{N,\sigma}_{1,T}$ where $\widetilde{\mathcal{G}}^{N,\sigma,\epsilon_0}_{2,(t_1,t_2)}\subseteq \mathbb{R}^{6N}$ shall be defined as follows for $t_1,t_2\in [0,T]$ (however, without making the $T$-dependence explicit in the notation):
\begin{align}
\begin{split}
& X \in \widetilde{\mathcal{G}}^{N,\sigma,\epsilon_0}_{2,(t_1,t_2)}\subseteq \mathbb{R}^{6N}   \\
\Leftrightarrow &  \forall Y \in \mathbb{R}^{6N}: \Big(\min_{0\le s \le T}|{\Psi^N_{s,0}}(Y)-{\Psi^N_{s,0}}(X)|_{\infty}\le  N^{-\frac{1}{2}+\frac{4}{5}\sigma} \\
&\Rightarrow  \frac{1}{N}\sum_{i=1}^N\sum_{ j\neq i}\int_{0}^T|f^N([^1\Psi^N_{s,0}(Y)]_j-[^1\Psi^N_{s,0}(Y)]_i)|  \\
& \hspace*{3,25cm}\cdot \mathbf{1}_{ M^{N,(t_1,t_2)}_{6N^{-\frac{1}{2}+{\sigma}},N^{-\frac{1}{9}+3{\sigma}}}(X_i)}(X_j) ds\le N^{-\epsilon_0}(t_2-t_1) \Big) \label{Def G 2 tilde}
\end{split}
\end{align}
To see this, one has to regard that the force term in this definition corresponds to term \eqref{Gron.shift}. The probability estimates which we implemented in the previous proof were only necessary to show that this term keeps typically small enough for trajectories which at some moment in $[0,T]$ are `close' to a trajectory of the `good' set $\mathcal{G}^{N,\sigma}_{1,T}$. If this force term fulfills the stated bound, then according to the reasoning of the previous proof the assumption $X\in \mathcal{G}^{N,\sigma}_{1,T}$ deals with the rest so that all estimates which lead to inequality \eqref{ineq.shift.lem.v2} can be carried out for such configurations. Moreover, the probability estimates of the proof of Lemma \ref{shift-lem} imply that for sufficiently small $\sigma,\epsilon_0>0$ there exist $C>0$ and $\epsilon' >0$ such that for all $N\in \mathbb{N}$ and $t_1,t_2\in T$ where $t_2\geq t_1+N^{-\frac{1}{3}}$ it holds that
$$\mathbb{P}\big(X\in \mathcal{G}^{N,\sigma}_{1,T}\cap \big(\widetilde{\mathcal{G}}^{N,\sigma,\epsilon_0}_{2,(t_1,t_2)}\big)^C\big)<CN^{-N^{\epsilon'}}.$$
This can be comprehended by going through the considerations which start after term \eqref{Gron.shift}. As a last remark we point out that despite their similar notation this set should not be confused with the (for later considerations distinctly more important) set $\mathcal{G}^{N,\sigma}_{2,(t_1,t_2)}$ which is defined in the assumptions of Lemma \ref{shift-lem}. \\
 Let $ N^{-\epsilon_0}\le N^{-\epsilon_1}a$ for $\epsilon_1>0$ and a variable $a>0$ as well as $N^{-\frac{1}{3}}\le\delta_t<\frac{1}{K_1^2}$ where $K_1>1$ is a constant which will be important in the following. Finally, we assume that
 \begin{align}
 X\in \mathcal{G}^{N,\sigma}_{1,T}\cap \mathcal{G}^{N,\sigma}_{2,(0,T)}  \bigcap_{k=0}^{\lceil \frac{T}{\delta_t}\rceil-1}\widetilde{\mathcal{G}}^{N,\sigma,\epsilon_0}_{2,(k\delta_t,(k+1)\delta_t)} \label{ass.X.cor}.
 \end{align}
Here, $\delta_t$ corresponds to the length of the short time span mentioned in the introductory explanations where we want to compare the trajectories in the proof of Theorem \ref{thm2}. The meaning of the variable $a$ will become clear shortly. Next, we introduce a differentiable map $h:\mathbb{R}\to \mathbb{R}^6$ fulfilling $\frac{d}{ds}{^1h(s)}={^2h(s)}$ for all $s\in \mathbb{R}$ where like in the remaining work the notation $h(s)=(^1h(s),{^2h(s)})\in \mathbb{R}^{6}$ shall distinguish between `position' and `velocity components'. Moreover, for $t_k=k\delta_t$, $k\in \{0,...,\lceil \frac{T}{\delta_t}\rceil-1\}$ the map shall fulfill the following constraints:
\begin{itemize}
\item[(i)] $\displaystyle \sum_{i=1}^N\sup\limits_{t_k\le s\le  t_{k+1}}|[^1\Psi^N_{s,t_k}(h(t_k))]_i-[{^1h(s)}]_i|\le aK_1 \delta_t^2 $
\item[(ii)] $\displaystyle \sum_{i=1}^N\sup\limits_{t_k\le s\le  t_{k+1}}|[^2\Psi^N_{s,t_k}(h(t_k))]_i-[{^2h(s)}]_i|\le aK_1  \delta_t $
\item[(iii)] $\sup\limits_{0\le s \le T}|\Psi^N_{s,0}(X)-h(s)|_\infty\le N^{-\frac{1}{2}+\frac{3}{5}\sigma} $
\end{itemize}
Conditions (i) and (ii) should be interpreted as follows: While $(\Psi^N_{r,s})_{r,s\in \mathbb{R}}$ is the $N$-particle flow we studied in Lemma \ref{shift-lem}, the map $h$ shall describe a trajectory which arises by (possibly) different dynamics. If we `follow' the trajectory $h$ and start at time $t_k$ to observe how fast the distance between $h$ and the corresponding trajectory related to the $(\Psi^N_{r,s})_{r,s\in \mathbb{R}}$-dynamics grows, then constraints (i) and (ii) shall determine upper bounds for the allowed deviations on short time spans of length $\delta_t$. Only by application of these assumptions we will in the subsequent part derive suitable upper bounds for the long-term deviations $$\sup\limits_{0\le s\le  T} |{^1\Psi^N_{s,0}(h(0))}-{^1h(s)}|_1\text{  and  }\sup\limits_{0\le s\le  T} |{^2\Psi^N_{s,0}(h(0))}-{^2h(s)}|_1.$$ 
We already point out that in the proof of Theorem \ref{thm2} we will show that such constraints are typically fulfilled if in place of the map $h$ a trajectory of the $N$-particle dynamics $\Psi^{N,c}_{\p,0}(X)$ is considered where the cut-off parameter  $c\geq \frac{2}{3}$ may be chosen arbitrarily large (which corresponds to an arbitrarily small cut-off). Application of this Corollary will then take care of the rest.\\
First, we remark that the condition related to item (iii) ensures that for large enough $N$ and for $t_k=k\delta_t$, $k\in \{0,...,\lceil \frac{T}{\delta_t}\rceil-1\}$
\begin{align}
|\Psi^N_{t_k,0}(X)-h(t_k)|_\infty\le \sup\limits_{0\le s \le T}|\Psi^N_{s,0}(X)-h(s)|_\infty\le N^{-\frac{1}{2}+\frac{3}{5}\sigma}\label{cond iii}
\end{align} and since $X\in \mathcal{G}^{N,\sigma}_{1,T}\cap  \widetilde{\mathcal{G}}^{N,\sigma,\epsilon_0}_{2,(t_k,t_{k+1})}$, it follows that the conditions for the subsequent application of the inequality related to \eqref{ineq.shift.lem.v0} (respectively its current form \eqref{ineq.shift.lem.v2}) are satisfied (where we note that the conditions are discussed after \eqref{ineq.shift.lem.v2}):
\begin{align}
&\sum_{i=1}^N \sup_{t_k\le s \le t_{k+1} }\Big(|[^2\Psi^N_{s,0}(h(0)]_i-[^2\Psi^N_{s,t_k}(h(t_k))]_i-([^2\Psi^N_{t_k,0}(h(0))]_i-{^2h(t_k)})| \Big)  \notag\\
\le &  C\sum_{i=1}^N\sup\limits_{t_k\le s\le  t_{k+1}}|[^1\Psi^N_{s,0}(h(0))]_i-[^1\Psi^N_{s,t_k}(h(t_k))]_i|+CN^{-\epsilon_0}\delta_t  \label{dev.rek.}
\end{align}
We point out that the constant $C>0$ will exceptionally be kept fixed for the whole estimates. Moreover, if $f_1,f_2:\mathbb{R}\to \mathbb{R}^{6N}$ are functions which have the same `structure' as $\Psi^N_{\cdot,t}(X)$ or the map $h$, then we will abbreviate in the following for convenience
\begin{align}
|f_1-f_2|^{s,t}_1:=\sum_{i=1}^N\sup\limits_{s\le r\le  t}|[f_1(r)]_i-[{f_2(r)}]_i| \label{abbrev.norm}
\end{align}
and introduce a corresponding abbreviation for such maps restricted to the `position or velocity components':
\begin{align}
|{^lf_1}-{^lf_2}|^{s,t}_1:=\sum_{i=1}^N\sup\limits_{s\le r\le  t}|[^lf_1(r)]_i-[{^lf_2(r)}]_i| \text{  for } l=1,2 
\end{align}
Let $n\in \mathbb{N}_0$ be such that 
\begin{align}
&|^1\Psi^N_{\p,0}(h(0))-{^1h(\p)}|_1^{0,t_n}\notag \\
=&\sum_{i=1}^N\sup\limits_{0\le s\le  t_n}|[^1\Psi^N_{s,0}(h(0))]_i-[{^1h(s)}]_i|\le a\delta_t^{\frac{3}{2}}\label{cond.n}
\end{align}
and due to constraint (i) on the map $h$ and $\delta_t<\frac{1}{K_1^2}$ it follows that in any case $n$ can be chosen larger or equal to $1$.\\
 We obtain by application of relation \eqref{dev.rek.} and $N^{-\epsilon_0}\le N^{-\epsilon_1}a$ that for all $s\in [t_k,{t_{k+1}}]$ and $k\le n-1$
\begin{align}
&|^2\Psi^N_{\p,0}(h(0))-{^2h(\p)}|_1^{t_k,s} \notag \\
\le &|^2\Psi^N_{\p,0}(h(0))-{^2\Psi^N_{\p,t_k}(h(t_k))}|_1^{t_k,s}+|{^2\Psi^N_{\p,t_k}(h(t_k))}-{^2h(\p)}|_1^{t_k,s} \notag \\  
\le &  |^2\Psi^N_{\p,0}(h(0))-{^2\Psi^N_{\p,t_k}}(h(t_k))-({^2\Psi^N_{t_k,0}(h(0))}-{^2h(t_k))}|_1^{t_k,s}\notag \\ 
& +\underbrace{|{^2\Psi^N_{t_k,0}(h(0))}-{^2h(t_k)}|_1^{t_k,s}}_{=|{^2\Psi^N_{t_k,0}(h(0))}-{^2h(t_k)}|_1}+\underbrace{|^2\Psi^N_{\p,t_k}(h(t_k))-{^2h(\p)}|_1^{t_k,s}}_{\le K_1a \delta_t} \notag \\
\le & \big(C|^1\Psi^N_{\p,0}(h(0))-{^1h(\p)}|^{t_k,t_{k+1}}_1+C\underbrace{N^{-\epsilon_0}}_{\le N^{-\epsilon_1}a}\delta_t   \big) \notag \\
&+|^2\Psi^N_{t_{k},0}(h(0))-{^2h(t_k)}|_1+K_1 a\delta_t \notag \\
\le &Ca\delta_t^{\frac{3}{2}}+ |^2\Psi^N_{t_{k},0}(h(0))-{^2h(t_k)}|_1+(CN^{-\epsilon_1}+K_1)a\delta_t \notag \\
\le & |^2\Psi^N_{t_{k},0}(h(0))-{^2h(t_k)}|_1+\big(C(N^{-\epsilon_1}+\delta_t^{\frac{1}{2}})+K_1\big)a\delta_t \label{dev.v.est.}
\end{align}
where in the third step we applied relation \eqref{dev.rek.} and item (ii) of the constraints on the map $h$ while the second last step follows due to condition \eqref{cond.n}. By application of the recursive relation determined by this inequality (and the circumstance that $s\in [t_k,t_{k+1}]$ was given arbitrarily) we obtain the following estimate for $k\le n$:
\begin{align}
 &|^2\Psi^N_{\p,0}(h(0))-{^2h(\p)}|_1^{0,t_k} \notag \\
 =&\sum_{i=1}^N\sup\limits_{0\le s\le  t_k}|[^2\Psi^N_{s,0}(h(0))]_i-[{^2h(s)}]_i|\le k\big(C(N^{-\epsilon_1}+\delta_t^{\frac{1}{2}})+K_1\big)a\delta_t \label{res.rec.v.}.
\end{align} Moreover, it holds for $k\le n-1$ and all $s\in [t_k,t_{k+1}]=[k\delta_t,(k+1)\delta_t]$ that 
\begin{align*}
& |^1\Psi^N_{\p,0}(h(0))-{^1h(\p)}|_1^{t_k,s} \\
\le &|^1\Psi^N_{t_{k},0}(h(0))-{^1h(t_k)}|_1\\
&+ \sum_{i=1}^N \sup_{t_k\le s \le t_{k+1} }|[^2\Psi^N_{s,0}(h(0))]_i-[{^2h(s)}]_i|\underbrace{(t_{k+1}-t_k)}_{=\delta_t}
\end{align*}
where we regarded for this conclusion that $\frac{d}{ds}{^1h(s)}={^2}h(s)$.\\
Application of relation \eqref{res.rec.v.} yields for $k\le n-1$ (see \eqref{cond.n}) that
\begin{align}
& |^1\Psi^N_{\p,0}(h(0))-{^1h(\p)}|_1^{t_k,t_{k+1}} \notag \\
\le & |^1\Psi^N_{t_{k},0}(h(0))-{^1h(t_k)}|_1+(k+1)\big(C(N^{-\epsilon_1}+\delta_t^{\frac{1}{2}})+K_1\big)a\delta_t^2\label{res.rec.r.0}
\end{align}
We assume for the rest of the considerations that $K_1\geq 4C$. By taking into account that $\delta_t\le 1$ this recursive relation implies for $k\le n$:  
\begin{align}
 |^1\Psi^N_{\cdot,0}(h(0))-{^1h(\cdot)}|_1^{0,t_k}& \le \big(C(N^{-\epsilon_1}+\delta_t^{\frac{1}{2}})+K_1\big)a\delta_t^2\sum_{i=0}^{k}i\notag \\
 &  \le K_1\big(\underbrace{\frac{C}{K_1}}_{\le \frac{1}{4}}(N^{-\epsilon_1}+\delta_t^{\frac{1}{2}})+1\big)a\delta_t^2\frac{k(k+1)}{2}\notag \\
& \le  k(k+1)\frac{3}{4}K_1a\delta_t^2 \label{res.rec.r.}
\end{align}
By application of this we can finally determine a lower bound such that the condition on $n$ (respectively $|^1\Psi^N_{\p,0}(h(0))-{^1h(\p)}|_1^{0,t_n}\le a\delta_t^{\frac{3}{2}}$) is fulfilled in dependence on the remaining variables. Due to \eqref{res.rec.r.} the following implication holds:  
\begin{align*}
 k(k+1)\frac{3}{4}K_1a\delta_t^2\le a \delta_t^{\frac{3}{2}} \Rightarrow
|^1\Psi^N_{\p,0}(h(0))-{^1h(\p)}|_1^{0,t_k}\le  a \delta_t^{\frac{3}{2}}
\end{align*}
Thus, relation \eqref{cond.n} holds if $k(k+1)\le \frac{4}{3K_1}\delta_t^{-\frac{1}{2}}$. Hence, we can choose $$n:=\lfloor\frac{2}{\sqrt{3K_1}}\delta_t^{-\frac{1}{4}}\rfloor-1,$$  and identify
\begin{align}
&  \Delta_t:=n\delta_t=\big(\lfloor\frac{2}{\sqrt{3K_1}}\delta_t^{-\frac{1}{4}}\rfloor-1\big)\delta_t. \label{Def.Delta_t} 
\end{align} 
After regarding that $n=\frac{\Delta_t}{\delta_t}$, relation \eqref{res.rec.r.} and \eqref{res.rec.v.} imply that
\begin{align}
& |^1\Psi^N_{\cdot,0}(h(0))-{^1h(\cdot)}|_1^{0,\Delta_t}& \notag\\
 \le &  \frac{\Delta_t}{\delta_t}(\frac{\Delta_t}{\delta_t}+1)\frac{3}{4}K_1a\delta_t^2=(1+\frac{\delta_t}{\Delta_t})\frac{3}{4}K_1a\Delta_t^2 \label{new.in.con.x0} 
\end{align}
as well as
\begin{align}
 &|^2\Psi^N_{\cdot,0}(h(0))-{^2h(\cdot)}|_1^{0,\Delta_t}\notag \\
  \le &\frac{\Delta_t}{\delta_t}\big(C(N^{-\epsilon_1}+\delta_t^{\frac{1}{2}})+K_1\big)a\delta_t
  \le  \big(1+\underbrace{\frac{C}{K_1}}_{\le \frac{1}{4}}(N^{-\epsilon_1}+\delta_t^{\frac{1}{2}})\big)K_1a\Delta_t. \label{new.in.con.v} 
\end{align}
So far we only assumed that $\delta_t<\frac{1}{K_1^2}\le 1$ as well as $K_1\geq 4C$. Now we assume additionally that $\delta_t\le \frac{1}{K_2}$ where $K_2>0$ shall be a sufficiently large constant such that 
\begin{align}
\Big(\delta_t\le \frac{1}{K_2}\Rightarrow \frac{\delta_t}{\Delta_t}\le \frac{1}{3}\Big) \land  \frac{1}{K_1^2}> \frac{1}{K_2}  \label{ass.K_2}
\end{align}
which is obviously possible due to the definition of $\Delta_t$ (see \eqref{Def.Delta_t}). In this case the first of the previous two inequalities implies that
\begin{align}
 |^1\Psi^N_{\cdot,0}(h(0))-{^1h(\cdot)}|_1^{0,\Delta_t} \le   K_1a\Delta_t^2
  \label{new.in.con.x}
\end{align}
This concludes the main part of the estimates and we can start to discuss the current result. First, we remark that due to the symmetry of the situation we could just as well have chosen another `starting point' $h(t_{k'})$ instead of $h(0)$ and would have obtained
\begin{align}
 |^1\Psi^N_{\cdot,t_{k'}}(h(t_{k'}))-{^1h(\cdot)}|_1^{t_{k'},t_{k'}+\Delta_t} \le & K_1a\Delta_t^2 \label{k-shift.x}
\end{align}
instead of \eqref{new.in.con.x} and 
\begin{align}
 &  |^2\Psi^N_{\cdot,t_{k'}}(h(t_{k'}))-{^2h(\cdot)}|_1^{t_{k'},t_{k'}+\Delta_t}    \le  \big(1+\frac{N^{-\epsilon_1}+\delta_t^{\frac{1}{2}}}{4}\big)K_1a\Delta_t  \label{k-shift.v} 
\end{align} instead of \eqref{new.in.con.v} (provided that $ t_{k'}+\Delta_t\le T$). In the introduction we mentioned that we want to extend the estimates step by step to longer time spans. This is exactly the plan we want to implement now. By application of the previously derived inequalities it is possible to find a new sequence of time steps such that conditions corresponding to those of items (i),(ii) and (iii) are fulfilled but for a new set of parameters. More precisely, after replacing $\delta_t=:\delta_{t_1}$ by $\Delta_t=:\delta_{t_2}$ and $a=:a_1$ by $\big(1+\frac{N^{-\epsilon_1}+\delta_t^{\frac{1}{2}}}{4}\big)a=:a_2$ it holds that:
\begin{itemize}
\item[(i)] $|^1\Psi^N_{\cdot,k\Delta_t}(h(k\Delta_t))-{^1h(\cdot)}|_1^{k\Delta_t,\min((k+1)\Delta_t,T)} \le a_2 K_1 \Delta_t^2 $
\item[(ii)] $|^2\Psi^N_{\cdot,k\Delta_t}(h(k\Delta_t))-{^2h(\cdot)}|_1^{k\Delta_t,\min((k+1)\Delta_t,T)}\le a_2 K_1  \Delta_t $
\item[(iii)] $\sup\limits_{0\le s \le T}|\Psi^N_{s,0}(X)-h(s)|_\infty\le N^{-\frac{1}{2}+\frac{3}{5}\sigma} $
\end{itemize}
While the constraint related to item (iii) is the same as in the previous case, the adjusted items (i) and (ii) follow directly by inequalities \eqref{k-shift.x} and \eqref{k-shift.v} applied for $$t_{k'}=k\Delta_t=k(n\delta_t)=t_{kn}.$$ If additionally $$X\in \bigcap_{k=0}^{\lceil \frac{T}{\Delta_t}\rceil-1}\widetilde{\mathcal{G}}^{N,\sigma,\epsilon_0}_{2,(k\Delta_t,(k+1)\Delta_t)}$$ is fulfilled, then we end up in exactly the same initial situation as in the first case but with respect to the adjusted values $a_2$ and $\delta_{t_2}$. For the implementation of the estimates we only needed that $\delta_t\le \frac{1}{K_2}$ (where the constraints on the constant $K_2$ are stated in \eqref{ass.K_2}). Thus, if also $\delta_{t_2}=\Delta_t\le  \frac{1}{K_2}$, then we can apply the same estimates as in the first case which finally provided inequalities \eqref{new.in.con.x0} and  \eqref{new.in.con.x} so that we obtain corresponding relations (but, of course, for the new set of parameters this time).\\
More generally, if we identify $a_{n+1}:=\big(1+\frac{N^{-\epsilon_1}+\delta_{t_{n}}^{\frac{1}{2}}}{4}\big)a_n$ where $a_1:=a$,
\begin{align}
\delta_{t_{n+1}}:=\max\Big(1,\big(\lfloor\frac{2}{\sqrt{3K_1}}\delta_{t_n}^{-\frac{1}{4}}\rfloor-1\big)\Big)\delta_{t_n},\ \ \delta_{t_1}=\delta_t \label{rek.rel.t}
\end{align}
(in correspondence to definition \eqref{Def.Delta_t}), then the same estimates work as long as \begin{align} 
 X\in \bigcap_{k=0}^{\lceil \frac{T}{\delta_{t_n}}\rceil-1}\widetilde{\mathcal{G}}^{N,\sigma,\epsilon_0}_{2,(k\delta_{t_n},(k+1)\delta_{t_n})} \label{Def.G_delta}
 \end{align} and the condition $\delta_{t_n}\le\frac{1}{K_2}$ is still fulfilled. Hence, this sequence of estimates does not (need to) end before $\delta_{t_n}>\frac{1}{K_2}$ and thereby the derived inequalities yield us upper bounds for the deviations even for a time span larger than some constant $\frac{1}{K_2}>0$ (which in particular is independent of the starting variables $a$ and $\delta_t$). What remains is to show that the arising upper bounds keep sufficiently small. According to the recursive relations (i) and (ii) adjusted to the case of general $n\in \mathbb{N}$ it follows (by regarding the recursion for $a_n$) that for $n\geq 1$ where $\delta_{t_n}\le \frac{1}{K_2}$ and $k\in \{0,...,\lceil \frac{T}{\delta_{t_n}}\rceil-1\}$:
 \begin{align}
 & |^1\Psi^N_{\cdot,k\delta_{t_n}}(h(k\delta_{t_n}))-{^1h(\cdot)}|_1^{k\delta_{t_n},\min((k+1)\delta_{t_n},T)} \notag \\
    \le &  K_1a_{n}\delta_{t_n}^2 
= K_1a\delta_{t_n}^2\prod_{i=1}^{n-1}\big(1+\frac{N^{-\epsilon_1}+\delta_{t_i}^{\frac{1}{2}}}{4}\big) \label{gen.rel.dis.x}
\end{align}
and correspondingly
 \begin{align}
&  |^2\Psi^N_{\cdot,k\delta_{t_n}}(h(k\delta_{t_n}))-{^2h(\cdot)}|_1^{k\delta_{t_n},\min((k+1)\delta_{t_n},T)} \notag \\
    \le &  K_1a_{n}\delta_{t_n} 
= K_1a\delta_{t_n}\prod_{i=1}^{n-1}\big(1+\frac{N^{-\epsilon_1}+\delta_{t_i}^{\frac{1}{2}}}{4}\big). \label{gen.rel.dis.v}
\end{align}
As stated, the recursion is in any case applicable if $\delta_{t_n}$ is still smaller than or equal to $\frac{1}{K_2}$. Let $n_{max}\in \mathbb{N}$ be maximal with this property, then the condition $\frac{\delta_{t_n}}{\delta_{t_{n+1}}}\le \frac{1}{3}$ is fulfilled for $n\le n_{max}$ according to the choice of $K_2$ and yields us that
\begin{align} 
\delta_{t_{n_{max}-k}}\le  \big(\frac{1}{3}\big)^k \delta_{t_{n_{max}}}\le \frac{1}{3^k} \frac{1}{K_2}  \label{rec.t_n}
\end{align} 
Moreover, since $K_2>1$ and $\delta_t\geq N^{-\frac{1}{3}}$ as well as
$ 3^{\ln(N)}=N^{\ln(3)} \geq N^{\frac{1}{3}} $ it follows that the number of factors $n_{max}$ is bounded by $\lceil \ln(N)\rceil$. Hence, we obtain that for the `large' $N$ which we consider the following relationship holds:
\begin{align*}
&\prod_{i=1}^{n_{max}}\big(1+\frac{N^{-\epsilon_1}+\delta_{t_i}^{\frac{1}{2}}}{4}\big) \\
\le & \prod_{i=1}^{n_{max}}\big(1+\frac{\delta_{t_i}^{\frac{1}{2}}}{2}\big)\prod_{i=1}^{n_{max}}\big(1+\frac{N^{-\epsilon_1}}{2}\big)
\le  \exp\big(\sum_{i=1}^{n_{max}} \frac{\delta_{t_i}^{\frac{1}{2}}}{2}\big) \big(1+\frac{N^{-\epsilon_1}}{2}\big)^{\lceil \ln(N)\rceil} \\
\le  & \exp\Big(\frac{1}{2}\sum_{k=0}^{\infty} \big(\frac{1}{3^k} \frac{1}{K_2}  \big)^{\frac{1}{2}}\Big) 2 \le   2\exp\Big(\frac{1}{2\sqrt{K_2}}\frac{\sqrt{3}}{\sqrt{3}-1}\Big)
\end{align*}
where we used relation \eqref{rec.t_n} in the second last step.\\
For convenience we abbreviate in the following $t^*:=\delta_{t_{n_{max}+1}}$ and assume additionally that $X\in \bigcap_{i=0}^{\lceil\frac{T}{t^*}\rceil-1}\widetilde{\mathcal{G}}^{N,\sigma,\epsilon_0}_{2,(it^*,(i+1)t^*)} $. Application of the previous estimate on inequalities \eqref{gen.rel.dis.x} and \eqref{gen.rel.dis.v} yields that there exists a further constant $K_3:= 2K_1\exp\Big(\frac{1}{2\sqrt{K_2}}\frac{\sqrt{3}}{\sqrt{3}-1}\Big)$ such that for $k\in \{0,...,\lceil \frac{T}{t^*}\rceil-1\}$
 \begin{align}
 & |^1\Psi^N_{\cdot,kt^*}(h(kt^*))-{^1h(\cdot)}|_1^{kt^*,\min((k+1)t^*,T)}\le   K_3a(t^*)^2 
 \label{end.rel.dis.x}
\end{align}
as well as
 \begin{align}
&  |^2\Psi^N_{\cdot,kt^*}(h(kt^*))-{^2h(\cdot)}|_1^{kt^*,\min((k+1)t^*,T)} \le   K_3at^*. \label{end.rel.dis.v}
\end{align}
Since $t^*>\frac{1}{K_2}$, these relations are finally suited to implement the concluding step of the estimates. On the other hand, the recursive definition \eqref{rek.rel.t} and $K_1\geq 1$ imply that $t^*=\delta_{t_{n_{max}+1}}\le 1$ and thus the stated inequalities yield for $s\in[kt^*,\min((k+1)t^*,T)]$ where $k\in \{0,...,\lceil \frac{T}{t^*}\rceil-1\}$ that 
 \begin{align}
&|\Psi^N_{s,0}(h(0))-{h(s)}|_1 \notag  \\
 \le & |\Psi^N_{s,0}(h(0))-\Psi^N_{s,kt^*}(h(kt^*))|_1+\underbrace{|\Psi^N_{s,kt^*}(h(kt^*))-{h(s)}|_1}_{\le 2K_3a t^*}\notag \\
\le & C\big(|\Psi^N_{kt^*,0}(h(0))-h(kt^*)|_1+\underbrace{N^{-\epsilon_0}t^*}_{\le N^{-\epsilon_1 }at^*}\big)+2K_3a t^*   .
\end{align}
The estimate for the first addend arises by the same reasoning as utilized for inequalities \eqref{ineq.chain (shift-lem)} and \eqref{ineq.shift.lem.v} but where $Y$ is replaced by $\Psi^N_{kt^*,0}(h(0))$ and $Z$ by $h(kt^*)$. To this end, one has to regard that $X\in \mathcal{G}^{N,\sigma}_{1,T}\cap\widetilde{\mathcal{G}}^{N,\sigma,\epsilon_0}_{2,(kt^*,(k+1)t^*)}$ which together with the constraint provided by item (iii) $\sup\limits_{0\le s \le T}|\Psi^N_{s,0}(X)-h(s)|_\infty\le N^{-\frac{1}{2}+\frac{3}{5}\sigma}$ ensures that the requirements for the argumentation leading to these inequalities is fulfilled, with the slight difference that for the current configurations the factor $N^{-\frac{1}{9}+14\epsilon}$ appearing in \eqref{ineq.chain (shift-lem)} and \eqref{ineq.shift.lem.v} needs to be replaced by $N^{-\epsilon_0}$ (which then leads to the stated result). \\
Since the relation holds for arbitrary $s\in[kt^*,\min((k+1)t^*,T)]$, it follows inductively that for the relevant $k$, $N$ and $C>1$
\begin{align}
&\sup_{0\le s \le kt^*}|\Psi^N_{s,0}(h(0))-{h(s)}|_1  \notag \\
\le & 3K_3a t^*   
\sum_{i=0}^{k-1}C^i\le 3K_3a t^*\big(\frac{C^{k}-1}{C-1}\big). \notag 
\end{align}
Hence, it follows that
\begin{align}
&\sup_{0\le s \le T}|\Psi^N_{s,0}(h(0))-{h(s)}|_1\le 3K_3a t^*    \big(\frac{C^{\lceil \frac{T}{t^*}\rceil}-1}{C-1}\big).
\end{align}
which due to $ \frac{1}{K_2}\le t^*\le 1$ shows that $$\sup_{0\le s \le T}|\Psi^N_{s,0}(h(0))-{h(s)}|_1\le 3K_3    \big(\frac{C^{\lceil K_2T\rceil}-1}{C-1}\big) a.$$ 
Before we formulate a Corollary which records the derived results, we state the conditions on $X$ which we applied for the estimates again in a more transparent form. However, we point out that this set will only be important for the formulation of the Corollary so that (in contrast to other sets of `good' initial data) the details of the definition have no direct relevance for later proofs. We needed in particular that $$X\in \bigcap_{i=0}^{\lceil\frac{T}{\delta_{t_n}}\rceil-1}\widetilde{\mathcal{G}}^{N,\sigma,\epsilon_0}_{2,(i\delta_{t_n},(i+1)\delta_{t_n})} $$ for all $n\in \mathbb{N}$ where $\delta_{t_{n}}\le \frac{1}{K_2}$ and also for the next larger natural number, after which this relation is no longer satisfied (previously called $n_{max}+1$). $\delta_{t_n}$ is defined by the recursion \eqref{rek.rel.t} and $K_2$ is the constant which we applied in the considerations. The set $\widetilde{\mathcal{G}}^{N,\sigma,\epsilon_0}_{2,(t_1,t_2)}$ was defined in \eqref{Def G 2 tilde}
and by regarding the condition appearing in its definition
\begin{align*}
 &\forall Y\in \mathbb{R}^{6N}:\Big( \min_{0\le s \le T}|{\Psi^N_{s,0}}(Y)-{\Psi^N_{s,0}}(X)|_{\infty}\le  N^{-\frac{1}{2}+\frac{4}{5}\sigma} \\
\Rightarrow & \frac{1}{N}\sum_{i=1}^N\sum_{ j\neq i}\int_{0}^T|f^N([^1\Psi^N_{s,0}(Y)]_j-[^1\Psi^N_{s,0}(Y)]_i)|\\
& \hspace{2,7cm} \cdot  \mathbf{1}_{ M^{N,(t_1,t_2)}_{6N^{-\frac{1}{2}+{\sigma}},N^{-\frac{1}{9}+3{\sigma}}}(X_i)}(X_j) ds\le N^{-\epsilon_0}(t_2-t_1) \Big)
\end{align*} 
together with the fact that respectively $\delta_{t_{n+1}}=k_n\delta_{t_n}$ for some $k_n\in \mathbb{N}$ as well as $\delta_{t_n}\le 1$ if $\delta_{t_1}\le  1$, it is straightforward to see that for $\delta_{t_1}\le  1$
\begin{align*}
X\in \bigcap_{i=0}^{\lceil\frac{T+1}{\delta_{t_1}}\rceil-1}\widetilde{\mathcal{G}}^{N,\sigma,\epsilon_0}_{2,(i\delta_{t_1},(i+1)\delta_{t_1})}\Rightarrow \forall n\geq 1: X\in \bigcap_{i=0}^{\lceil\frac{T}{\delta_{t_n}}\rceil-1}\widetilde{\mathcal{G}}^{N,\sigma,\epsilon_0}_{2,(i\delta_{t_n},(i+1)\delta_{t_n})} .
\end{align*}
Hence, a set of good initial data unifying all necessary properties can be defined as follows for $N^{-\frac{1}{3}}\le \delta_{t}\le 1$:
\begin{align}
\begin{split}
&X\in \mathcal{G}^{N,\sigma,\epsilon_0}_{\delta_t,T}\subseteq \mathbb{R}^{6N} \\
\Leftrightarrow & 
X\in \mathcal{G}^{N,\sigma}_{1,T} \cap \mathcal{G}^{N,\sigma}_{2,[0,T]}  \  \land  X\in \bigcap_{i=0}^{\lceil\frac{T+1}{\delta_t}\rceil-1}\widetilde{\mathcal{G}}^{N,\sigma,\epsilon_0}_{2,(i\delta_t,(i+1)\delta_t)} \label{Def.good set}
\end{split}
\end{align} 
Moreover, as we mentioned previously, it holds according to the proof of Lemma \ref{shift-lem} that for small enough $\sigma,\epsilon_0>0$ there exist $C>0$ and $\epsilon'>0$ such that for all $N\in \mathbb{N}$ and $t_1,t_2\in [0,T]$ where $t_2\geq t_1+N^{-\frac{1}{3}}$:
\begin{align*}
\mathbb{P}\big(X\in \mathcal{G}^{N,\sigma}_{1,T}\cap \big(\widetilde{\mathcal{G}}^{N,\sigma,\epsilon_0}_{2,(t_1,t_2)}\big)^C\big)\le CN^{-N^{\epsilon'}}.
\end{align*}
More precisely, according to reasoning starting after term \eqref{Gron.shift} and leading to relation \eqref{ineq.shift.lem.v0} the probability estimates are applicable if $N^{-\frac{1}{9}+14\epsilon}\le N^{-\epsilon_0}$. However, the only constraint on $\epsilon$ was that $\epsilon\geq \sigma$ and $\sigma>0$ can be chosen arbitrarily small. Hence, if $\epsilon_0<\frac{1}{9}$, then $\sigma>0$, the constant $C>0$ and $\epsilon'>0$ can in principle be chosen such that the stated upper bound for the probability is valid. Furthermore, it holds according to the reasoning stated after definition \eqref{def.G_1} that for a given $\epsilon>0$ and for sufficiently small $\sigma>0$ it holds that 
\begin{align*}
\mathbb{P}\big(X\in (\mathcal{G}^{N,\sigma}_{1,T})^C \big)\le CN^{-\frac{1}{9}+\epsilon}.
\end{align*}
This implies in total that for a given $\epsilon>0$ and small enough parameters $\sigma,\epsilon_0>0$ there exist $\epsilon'>0$ and $C>0$ such that
\begin{align}
& \mathbb{P}\big(X\in \mathcal{G}^{N,\sigma,\epsilon_0}_{\delta_t,T} \big)\notag \\
\geq & 1-\mathbb{P } \big(X\in \big(\mathcal{G}^{N,\sigma}_{1,T} \big)^C\big)-\mathbb{P } \big(X\in \mathcal{G}^{N,\sigma}_{1,T}\cap \big( \mathcal{G}^{N,\sigma}_{2,(0,T)} \big)^C \big) \notag \\
& -C\underbrace{\delta_t^{-1}}_{\le N^{\frac{1}{3}}}\max_{i\in \mathbb{N}_0: i\delta_{t}\le T+1}\underbrace{\mathbb{P } \big(X\in \mathcal{G}^{N,\sigma}_{1,T}\cap \big(\widetilde{\mathcal{G}}^{N,\sigma,\epsilon_0}_{2,(i\delta_{t},(i+1)\delta_{t})}\big)^C \big)}_{\le CN^{-N^{\epsilon'}}}\notag \\\
\geq & 1-CN^{-\frac{1}{9}+\epsilon}-\mathbb{P } \big(X\in \mathcal{G}^{N,\sigma}_{1,T}\cap \big( \mathcal{G}^{N,\sigma}_{2,(0,T)} \big)^C \big)  \label{prob.good set}
\end{align}
where we applied that $\delta_{t_1}=\delta_t\geq N^{-\frac{1}{3}}$.\\
The stated probability estimates will become important in the proof of Theorem \ref{thm2}, however, now we conclude by summarizing the essential results in the subsequent Corollary.
\begin{samepage}
\begin{cor} \label{Cor.shift2} 
Let $T,C_1,\sigma>0$, $N\in \mathbb{N}$, $N^{-\frac{1}{3}}\le \delta_t\le 1$ as well as $\epsilon_0, \epsilon_1,a>0$ such that $N^{-\epsilon_0}\le N^{-\epsilon_1}a$. Moreover, let $(\Psi_{t,s}^{N,c})_{s,t \in \mathbb{R}}$ be the $N$-particle flow defined in \eqref{Def.micro.sys.} for $1<\alpha\le \frac{4}{3}$, $c=\frac{2}{3}$ and $h:\mathbb{R}\to \mathbb{R}^{6N}$ a differentiable map where the constraints \begin{itemize}
\item[(i)] $\displaystyle \sum_{i=1}^N\sup\limits_{k\delta_t \le s\le  (k+1)\delta_t}|[^1\Psi^{N,c}_{s,k\delta_t}(h(k\delta_t))]_i-[{^1h(s)}]_i|\le aC_1 \delta_t^2 $
\item[(ii)] $\displaystyle \sum_{i=1}^N\sup\limits_{k\delta_t \le s\le  (k+1)\delta_t}|[^2\Psi^{N,c}_{s,k\delta_t}(h(k\delta_t))]_i-[{^2h(s)}]_i|\le aC_1  \delta_t $
\item[(iii)] $\sup\limits_{0\le s \le T}|\Psi^{N,c}_{s,0}(X)-h(s)|_\infty\le N^{-\frac{1}{2}+\frac{3}{5}\sigma} $
\item[(iv)]  $\forall s\in [0,T]:\frac{d}{ds}{^1h(s)}={^2h(s)}$
\end{itemize}
are fulfilled for a configuration $X\in \mathcal{G}^{N,\sigma,\epsilon_0}_{\delta_t,T}$ (defined in \eqref{Def.good set}) and all $k\in \{0,...,\lceil \frac{T}{\delta_t} \rceil-1\}$. If $\sigma>0$ is chosen small enough, then there exist $N_0\in \mathbb{N}$ and $C_2>0$ such that for all $N\geq N_0$ and all $a>0$ which fulfill the introduced conditions, the stated constraints on the map $h$ imply that 
\begin{align}
&\sup_{0\le s \le T}|\Psi^{N,c}_{s,0}(h(0))-{h(s)}|_1
\le  C_2a.
\end{align}
\end{cor}
\end{samepage}
\vspace{0,4cm}
\noindent 
Regarding the assumptions of the Corollary, it is important to note that we only know that configurations $X\in\mathcal{G}^{N,\sigma,\epsilon_0}_{\delta_t,T}$ are typical if $\sigma>0$ and $\epsilon_0>0$ are sufficiently small. This is discussed in more detail previous to estimate \eqref{prob.good set}. However, for the single application of the Corollary (which will be in the proof of Theorem \ref{thm2}) we will only consider the case $a= N^{-\sigma}$ (for small $\sigma>0$) and thus it is obvious that for example the choice $\epsilon_0=2\sigma,\epsilon_1=\sigma$ fulfills this demand as well as the constraint $N^{-\epsilon_0}\le N^{-\epsilon_1}a$ stated in the assumptions of the Corollary.\\ If the part previous to the Corollary has been skipped, then we point out to the reader that the proof starts shortly before relation \eqref{ineq.shift.lem.v2} and continue now with further preliminaries for the proof of Theorem \ref{thm2}.\\ 
The analysis concerning the dynamical properties is now concluded and we continue by implementing a lemma which improves our capabilities to make probability estimates for the interacting particles.\\
Our final aim is to show that increasing the cut-off parameter $c$ (which corresponds to `shrinking' the cut-off diameter for fixed $N\in\mathbb{N}$) will barely change the trajectories of the $N$-particle dynamics if $c\geq \frac{2}{3}$. Heuristically one would expect such a property because for a system of particles moving independently from each other it is easy to show that with high probability only a tiny fraction of them gets close enough such that the cut-off `comes into play' (provided that its size is very small). However, it is a priori unclear if this is also true for a system of $N$ interacting particles. If the particles tend to run into the (regularized) singularity systematically, then the deviation between dynamics with small but different-sized cut-offs could be significant. In the following we want to show that under certain assumptions the particles of the systems considered by us behave `good' (meaning that there is typically not an unexpectedly large number of them coming very close to each other). Our first step will be to show that this is indeed true for the previously considered system where the cut-off diameter is of order $N^{-\frac{2}{3}}$. Since the particles of the microscopic system interact with each other, the initial product structure of the probability density $F^{N,c}_t$ gets lost as time passes. However, the next lemma will provide us a tool which enables us to determine suitable upper bounds for the probability of arbitrary (Borel-measurable) events in this system also at later times.\\ 
Before starting with the lemma we have to introduce a further set of `good' initial data:
\begin{align}
\begin{split}
\mathcal{G}_{3,T}^{N,\sigma}:=\big\{X\in \mathbb{R}^{6N} \ |  &\ \exists Y\in \mathcal{G}_{1,T}^{N,\sigma}\cap \mathcal{G}_{2,(0,T)}^{N,\sigma}:|X-Y|_{\infty}\le N^{-\frac{1}{2}+\frac{3\sigma}{4}}\land \\
& \  \forall i \in \{1,...,N\}:Y_i\in \mathcal{L}_{\sigma}^N\big\}. \label{def.G_3}
\end{split}
\end{align}
Because of its importance for the subsequent Lemma we recall the definition of the set $\mathcal{L}^N_{\delta}$ which was introduced previous to Theorem \ref{thm2}:
\begin{align}
\begin{split}
\mathcal{L}^N_{\delta}:=\Big\{Y\in \mathbb{R}^6 \ | \ \forall Z_1,Z_2\in \mathbb{R}^6:& \big( \max_{i\in \{1,2\}}|Z_i-Y|\le N^{-\frac{1}{3}}\ \land Z_1\neq Z_2\big) \\ & \ \ \Rightarrow  \frac{|k_0(Z_1)-k_0(Z_2)|}{|Z_1-Z_2|}\le N^{\frac{\delta}{2}}k_0(Z_1)\Big\} 
\end{split}
\end{align}
where $\delta>0$ and $N\in \mathbb{N}$.

\begin{lem} \label{prod-Lem}
Let $N\in \mathbb{N},\ T>0$ and $k_0\in \mathcal{L}^1(\mathbb{R}^6)$ be a probability density fulfilling the assumptions of Theorem \ref{thm1}. Moreover, let $(\Psi_{t,s}^{N,c})_{s,t \in \mathbb{R}}$ be the $N$-particle flow defined in \eqref{Def.micro.sys.} for $1<\alpha\le \frac{4}{3}$ and $c=\frac{2}{3}$ as well as $k^{N,c}_t:=k_0(\varphi^{N,c}_{0,t}(\cdot))$ where $(\varphi^{N,c}_{s,t})_{s,t \in \mathbb{R}}$ is the effective flow defined in \eqref{def.flow}. If $\sigma>0$ is sufficiently small, then there exist $C_1>0$, $N_0\in\mathbb{N}$ such that for all $N\in \mathbb{N}$ where $N\geq N_0$, $M\in \{1,...,N\}$, $t\in [0,T]$ and $\mathcal{S}=\mathcal{S}'\times \mathbb{R}^{6(N-M)}\subseteq  \mathbb{R}^{6N}$ where $\mathcal{S}'\subseteq  \mathbb{R}^{6M}$ shall be Borel-measurable the following holds:
\begin{align*}
& \mathbb{P}\big(X\in \mathbb{R}^{6N}:X\in \mathcal{G}^{N,\sigma}_{3,T} \land \Psi^{N,c}_{t,0}(X) \in \mathcal{S} \big)\\
= & \int_{\mathbb{R}^{6N}}\mathbf{1}_{\Psi^{N,c}_{t,0}(\mathcal{G}^{N,\sigma}_{3,T})}(X)\mathbf{1}_{\mathcal{S}'}(X_1,...,X_M)F^{N,c}_t(X)d^{6N}X \\
\le &  C_1^M\int_{\mathbb{R}^{6M}}\mathbf{1}_{\mathcal{S}'}(X_1,...,X_M)\prod_{i=1}^Mk^N_t(X_i)d^{6M}X
\end{align*}
\end{lem}
\vspace{0,4cm}
\begin{proof}
\noindent The intention for restricting the initial data to $ \mathcal{G}^{N,\sigma}_{3,T}$ might seem strange at the moment but will resolve itself during the proof of Theorem \ref{thm2}. Since the cut-off parameter is fixed to $c=\frac{2}{3}$, we will once again omit to make the related indices explicit in the notation during the proof.
The applied estimates and statements of the proof only need to be fulfilled if $N\in \mathbb{N}$ is large enough and $\sigma>0$ sufficiently small which will be important on several occasions. For convenience and for avoiding redundant formulations we will often omit to mention this explicitly. \\
Now let $t\in [0,T],\ M\in \mathbb{N}$ and $\mathcal{S}=\mathcal{S}'\times \mathbb{R}^{6(N-M)}\subseteq  \mathbb{R}^{6N}$ where $\mathcal{S}'\subseteq  \mathbb{R}^{6M}$ shall be Borel-measurable. Moreover, let $X\in \Psi^N_{t,0}(\mathcal{G}^{N,\sigma}_{3,T})$, then by definition there exists $Y\in \mathcal{G}^{N,\sigma}_{1,T}\cap \mathcal{G}^{N,\sigma}_{2,(0,T)}$ where $|\Psi^N_{0,t}(X)-Y|_{\infty}\le N^{-\frac{1}{2}+\frac{3\sigma}{4}}$ and $Y_i\in \mathcal{L}_{\sigma}^N$ for all $i\in \{1,...,N\}$. It follows due to Lemma \ref{shift-lem} that for large enough $N\in \mathbb{N}$ and $Z\in \mathbb{R}^{6N}$ the subsequent implication holds: 
\begin{align}
 |Z-X|_{\infty} & \le N^{-\frac{1}{2}+\frac{3\sigma}{4}} \notag \\
\Rightarrow \big(|Z-\Psi^N_{t,0}(Y)|_{\infty} & \le |Z-X|_{\infty} + |X-\Psi^N_{t,0}(Y)|_{\infty} \notag \\
&\le   N^{-\frac{1}{2}+\frac{3\sigma}{4}}+C(N^{-\frac{1}{2}+\frac{\sigma}{2}}+|\Psi^N_{0,t}(X)-Y|_{\infty})\notag \\
& \le CN^{-\frac{1}{2}+\frac{3\sigma}{4}}\big). \label{est.prod.lem.}
\end{align}
If we keep the assumption $|Z-X|_{\infty} \le N^{-\frac{1}{2}+\frac{3\sigma}{4}}$, then a further application of Lemma \ref{shift-lem} implies for the relevant $N,\sigma$ that:
\begin{align*}
&|\Psi^N_{0,t}(Z)-Y|_{\infty}\\ 
\le &C(N^{-\frac{1}{2}+\frac{\sigma}{2}}+|Z-\Psi^N_{t,0}(Y)|_{\infty})   \le CN^{-\frac{1}{2}+\frac{3\sigma}{4}}<N^{-\frac{1}{3}}
\end{align*}  After regarding additionally that $Y_i\in \mathcal{L}_{\sigma}^N\ \forall  i\in \{1,...,N\}$ we can apply the definition of $\mathcal{L}^N_\sigma$ in the third step (see \eqref{Def.L}) to obtain that in this case:
 {\allowdisplaybreaks
\begin{align}
 & F^N_t(X) \notag \\ 
=& \prod_{i=1}^N k_0([\Psi^N_{0,t}(X)]_i) \notag\\
\le & \prod_{i=1}^N \Big(\big|k_0([\Psi^N_{0,t}(X)]_i)-k_0([\Psi^N_{0,t}(Z)]_i)\big|+k_0([\Psi^N_{0,t}(Z)]_i) \Big)\notag\\
\le &   \prod_{i=1}^N \Big(\big(1+N^{\frac{\sigma}{2}}\big|[\Psi^N_{0,t}(X)]_i-[\Psi^N_{0,t}(Z)]_i\big|\big)k_0([\Psi^N_{0,t}(Z)]_i)\Big) \notag\\
\le & e^{N^{\frac{\sigma}{2}}|\Psi^N_{0,t}(X)-\Psi^N_{0,t}(Z)|_1}F^N_t(Z) \notag \\
\le & e^{N^{\frac{\sigma}{2}}\max(N^{-\sigma},C|X-Z|_1)}F^N_{t}(Z)\label{est.dens.0} 
\end{align}}
In the second last step we used that $\prod_{i=1}^N(1+|\delta_i|)\le e^{|\delta|_1}$ for $\delta\in \mathbb{R}^{6N}$ and in the last step that according to Lemma \ref{shift-lem}
$$ |\Psi^N_{0,t}(X)-\Psi^N_{0,t}(Z)|_1\le C\max\big(N^{-\sigma},|X-Z|_1\big)$$
where we regarded that $Y\in \mathcal{G}^{N,\sigma}_{1,T}\cap \mathcal{G}^{N,\sigma}_{2,(0,T)}$ as well as our previous estimates which yield:
$$\max_{\widetilde{X}\in \{X,Z\}}|\Psi^N_{t,0}(\widetilde{X})-Y|\le CN^{-\frac{1}{2}+\frac{3\sigma}{4}}$$ 
Now we can apply these estimates for the first crucial step of the proof. To his end, we abbreviate $r_N:=N^{-\frac{1}{2}+\frac{3\sigma}{4}}$, $X^M:=(X_1,...,X_M)$ as well as $\mu\big(B_{r}(0)\big):=\int_{\mathbb{R}^6}\mathbf{1}_{B_{r}(0)}(Y)d^6Y$ where $B_{r}(Y):=\{Y'\in \mathbb{R}^{6}:|Y'-Y|<r\}$. Then it holds that
{\allowdisplaybreaks
\begin{align}
&\int_{\Psi^N_{t,0}(\mathcal{G}^{N,\sigma}_{3,T})}\mathbf{1}_{\mathcal{S}'}(X^M)F^N_t(X)d^{6N}X \notag \\ 
= & \int_{\Psi^N_{t,0}(\mathcal{G}^{N,\sigma}_{3,T})}\mathbf{1}_{\mathcal{S}'}(X^M)    
\Big(\int_{\mathbb{R}^{6M}}\frac{F^N_t(X)}{\mu(B_{r_N}(0))^M} 
\prod_{i=1}^M\mathbf{1}_{B_{r_N}(X_{i})}(Z_i)d^{6M}Z \Big) d^{6N}X \notag  \\
\le & e^{N^{\frac{\sigma}{2}}\max(N^{-\sigma},CMr_N)}\int_{\Psi^N_{t,0}(\mathcal{G}^{N,\sigma}_{3,T})}\mathbf{1}_{\mathcal{S}'}(X^M) \notag\\
&\cdot  \Big(\int_{\mathbb{R}^{6M}}\frac{F^N_t(Z_1,...,Z_M,X_{M+1},...,X_N)}{\mu(B_{r_N}(0))^M}  \prod_{i=1}^M\mathbf{1}_{B_{r_N}(X_{i})}(Z_i)d^{6M}Z   \Big)d^{6N}X. \label{cruc.terms prod-lem} 
\end{align}} 
where we applied estimates \eqref{est.dens.0} and regarded that due to the appearing indicator functions it holds that
$$ e^{N^{\frac{\sigma}{2}}\max(N^{-\sigma},C|X-(Z_1,...,Z_M,X_{M+1},...,X_N)|_1)}\le  e^{N^{\frac{\sigma}{2}}\max(N^{-\sigma},CMr_N)}.$$
We will drop this factor for the subsequent steps and reintroduce it in the end. Moreover, for convenience we abbreviate in the first line $\overline{Z}:=(Z_1,...,Z_M,X_{M+1},...,X_N)$ and `rearrange' the previous integral as follows:
\begin{align}
& \int_{\Psi^N_{t,0}(\mathcal{G}^{N,\sigma}_{3,T})}\mathbf{1}_{\mathcal{S}'}(X^M) \int_{\mathbb{R}^{6M}}\frac{F^N_t(\overline{Z})}{\mu(B_{r_N}(0))^M}  \prod_{i=1}^M\mathbf{1}_{B_{r_N}(X_{i})}(Z_i)d^{6M}Z d^{6N}X \notag \\
= & \int_{\mathbb{R}^{6M}}\Big( \int_{\mathbb{R}^{6N}}\mathbf{1}_{\Psi^N_{t,0}(\mathcal{G}^{N,\sigma}_{3,T})}(X_1,...,X_M,\overline{Z}_{M+1},...,\overline{Z}_N)F^N_t(\overline{Z}) \prod_{i=1}^M\mathbf{1}_{B_{r_N}(X_{i})}(\overline{Z}_i)d^{6N}\overline{Z}\Big) \notag \\
& \cdot \frac{\mathbf{1}_{\mathcal{S}'}(X^M)  }{\mu(B_{r_N}(0))^M} d^{6M}(X_1,...,X_M)  \label{prod.lem.inn.int.}
\end{align} 
where we recall the abbreviation $X^M=(X_1,...,X_M)$ to emphasize why the factor $\mathbf{1}_{\mathcal{S}'}(X^M)$ can be `pulled out' of the integral.\\
We can focus on the inner integral for the further considerations. To visualize the basic ideas for estimating this integral, we first discuss heuristically the relevant steps by application of two sketches. The rigorous estimates where the details are stated will be implemented afterwards.\\
The initial situation is sketched in figure \eqref{di.p.s.}.
\begin{figure}[!h]
   \centering
   \def\svgscale{1.1}
   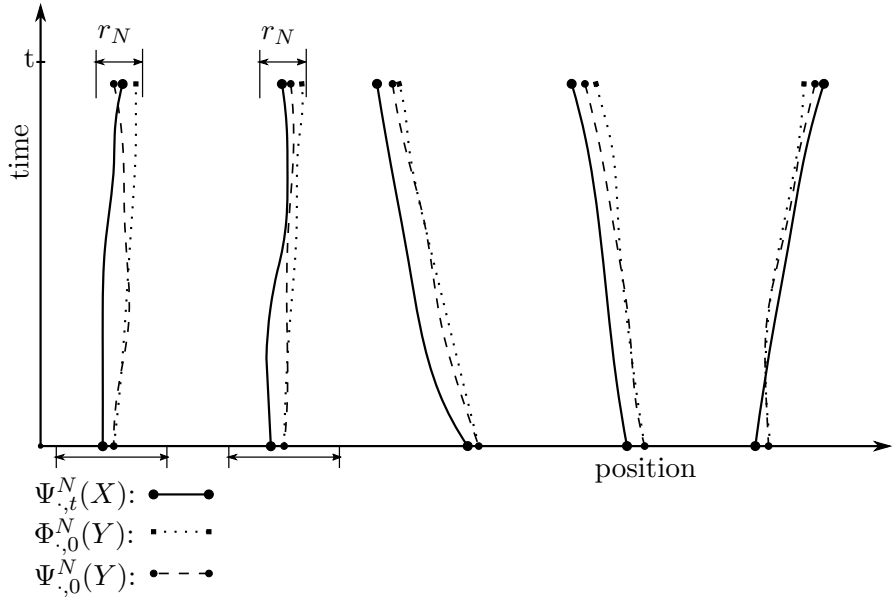
   \caption{Introduction to initial situation}
   \label{di.p.s.}
\end{figure} 
This drawing shows one of the trajectories $\Psi^N_{\cdot,0}(\Psi^N_{0,t}(X))=\Psi^N_{\cdot,t}(X)$ whose initial particle positions $[\Psi^N_{0,t}(X)]_i$ are distributed according to the $N$-fold product of $k_0$ restricted to the `good' set $\mathcal{G}^{N,\sigma}_{3,T}$ (respectively $\Psi^N_{0,t}(X)\in \mathcal{G}^{N,\sigma}_{3,T}$). The integration variables for the inner integral \eqref{prod.lem.inn.int.} are indicated by $\overline{Z}=(Z_1,...,Z_M,X_{M+1},...,X_N)$ but due to the appearing indicator functions we only have to take configurations $\overline{Z}\in \mathbb{R}^{6N}$ into account where $|X_i-Z_i|\le r_N$ for $i\in \{1,...,M\}$ which is sketched in \eqref{di.p.s.} by the measure lines around the particle positions on the left provided with $r_N$ (for the special case $M=2$). Hence, each relevant configuration of the integration set can be identified with a trajectory $\Psi^N_{\cdot,t}(\overline{Z})$ which at time $t$ is located at $\overline{Z}$ where $|X_i-Z_i|\le r_N$ for $i\in \{1,...,M\}$. In the following we want to derive constraints on the initial data $\Psi^N_{0,t}(\overline{Z})$ of such trajectories which are easy to handle. The drawing shows two further trajectories $\Psi^N_{\cdot,0}(Y)$ and $\Phi^N_{\cdot,0}(Y)$ which will be crucial for this purpose. More precisely, $\Psi^N_{0,t}(X)\in \mathcal{G}^{N,\sigma}_{3,T}$ implies that there exists $Y\in \mathcal{G}^{N,\sigma}_{1,T}\cap \mathcal{G}^{N,\sigma}_{2,(0,T)}$ such that $|Y-\Psi^N_{0,t}(X)|_{\infty}\le N^{-\frac{1}{2}+\frac{3\sigma}{4}}$ and $\forall i \in \{1,...,N\}:Y_i\in \mathcal{L}^N_{\sigma}$. According to Corollary \ref{cor.shift.} and Lemma \ref{shift-lem} not only $\Psi^N_{\p,0}(Y)$ and $\Phi^N_{\p,0}(Y)$ keep close on the whole time span but all the `good' properties which are connected with initial data $\mathcal{G}^{N,\sigma}_{1,T}\cap \mathcal{G}^{N,\sigma}_{2,(0,T)}$ can in large parts be `transferred' to those $N$-particle trajectories which are at some point in time $s\in [0,T]$ `close' (with respect to $|\p|_\infty$) to $\Psi^N_{\p,0}(Y)$. For the moment the most crucial property is that the particles related to such trajectories move up to a small deviation like `mean-field particles'. Thus, like implied in the first sketch the relation $|Y-\Psi^N_{0,t}(X)|_{\infty}\le N^{-\frac{1}{2}+\frac{3\sigma}{4}}$ directly yields that also $\Psi^N_{\p,t}(X)$ and $\Phi^N_{\p,0}(Y)$ are `close' on $[0,T]$ (and thereby in particular the configurations $\Phi^N_{t,0}(Y)$ and $X$). Moreover, the measure lines around the initial data $Y$ shall describe the small area where the special Lipschitz condition arising by $Y_i\in \mathcal{L}^N_{\sigma}$ is fulfilled. The reason why this property is important will become clear later.\\ By means of the second sketch we can now start with the essential heuristic considerations. Let to this end be $\Psi^N_{\cdot,0}(\overline{Z})$ one of the trajectories `related to' the relevant configurations of the integration set. Then, as mentioned before, we want to determine a suitable condition on the initial data $\Psi^N_{t,0}(\overline{Z})$. Simply `translating' the current condition $\max_{i\in \{1,...,M\}}|Z_i-X_i|\le r_N$ to the initial time is not very helpful since the arising constraint is hard to evaluate without further reasoning. By application of the following sketch (where we dropped the trajectories belonging to the `good' initial data $Y$ for a clearer presentation) we will discuss a more promising approach.
\begin{figure}[!h]
   \centering
   \def\svgscale{1.1}
   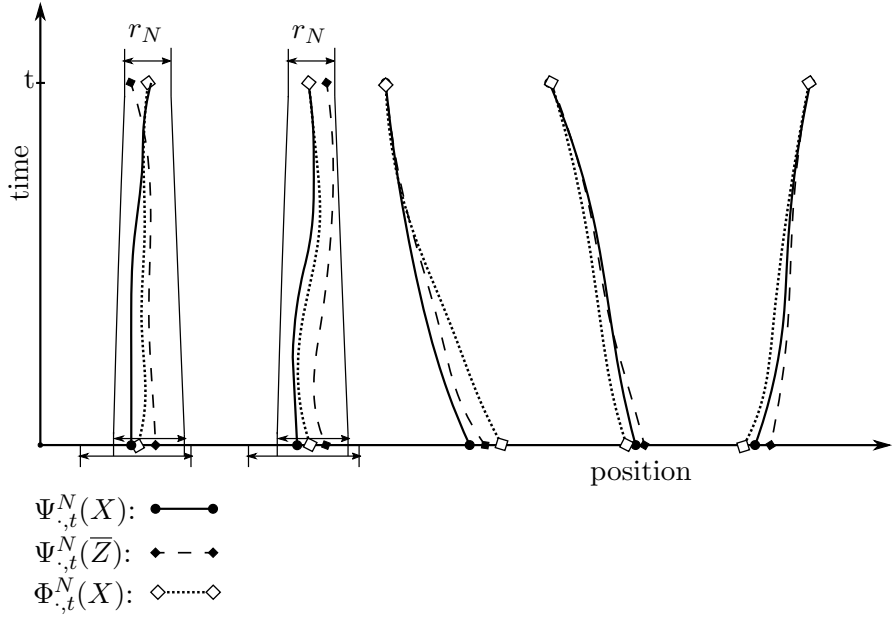
   \caption{Sketch of relevant trajectories}
   \label{di.p.s.1}
\end{figure} 
According to the assumptions, the distance (with respect to $|\cdot |_{\infty}$) between the relevant $\overline{Z}$ and $X$ is at most of order $r_N$ and by their closeness to  $\Psi^N_{t,0}(Y)$ resp. $\Phi^N_{t,0}(Y)$ it follows due to our previous considerations that these configurations evolve more or less like `mean-field particles' when they are `driven' by the (regularized) $N$-particle flow $(\Psi^N_{r,s})_{r,s\in \mathbb{R}}$. Hence, the distance between $\Psi^N_{s,t}(\overline{Z})$ and $\Phi^N_{s,t}(X)$ is bounded by $Cr_N$ for $s\in [0,t]$. This is sketched in the drawing by the newly introduced measure/border lines which describe the in past direction slowly growing upper bounds for the possible deviations. Consequently, it follows that for the considered integration set only configurations $\overline{Z}$ are relevant where the related trajectories $\Psi^N_{\p,t}(\overline{Z})$ start at initial positions which fulfill
\begin{align}
\max_{i\in \{1,...,N\}}|\varphi^N_{0,t}(X_i)-[\Psi^N_{0,t}(\overline{Z})]_i| \le Cr_N. \label{heu.cond.c1-lem}
\end{align} On the one hand, this implies that the distance between $[\Psi^N_{0,t}(\overline{Z})]_i$ and the i.i.d. positions $[\Psi^N_{0,t}(X)]_i$ is also bounded by $Cr_N$ for $i\in \{1,...,M\}$. Hence, if $\overline{Z}$ is one of the relevant configurations, then it holds that $$k_0([\Psi^N_{0,t}(\overline{Z})]_i)\approx k_0([\Psi^N_{0,t}(X)]_i)\text{ for } i\in \{1,...,M\}$$ where we regarded the `local Lipschitz property' of $k_0$ which holds in the neighborhood of $[\Psi^N_{0,t}(X)]_i$. However, at least equally important, for each $i\in \{1,...,M\}$ the constraint on the initial position $[\Psi^N_{0,t}(\overline{Z})]_i$ provided by \eqref{heu.cond.c1-lem} only depends on $X_i$ and not on the remaining configuration which is crucial for the estimates. \\\\ We will finally start to implement these preliminary consideration in a more rigorous way. The basic proceeding will be in line with the presented steps though this might seem a bit hidden due to the different notation which arises by substitutions of the integration variables. \\ 
Now we continue with the estimates for the inner integral of \eqref{prod.lem.inn.int.}:  
\begin{align}
& \int_{\mathbb{R}^{6N}}\mathbf{1}_{\Psi^N_{t,0}(\mathcal{G}^{N,\sigma}_{3,T})}(X_1,...,X_M,\overline{Z}_{M+1},...,\overline{Z}_N)F^N_t(\overline{Z}) \prod_{i=1}^M\mathbf{1}_{B_{r_N}(X_{i})}(\overline{Z}_i)d^{6N}\overline{Z}\notag\\
=& \int_{\mathbb{R}^{6N}}\mathbf{1}_{\Psi^N_{t,0}(\mathcal{G}^{N,\sigma}_{3,T})}( X_1,...,X_M,[\Psi^N_{t,0}(Z)]_{M+1},...,[\Psi^N_{t,0}(Z)]_N) \prod_{j=M+1}^N k_0(Z_j) \notag \\
& \cdot\prod_{i=1}^M\Big(\mathbf{1}_{B_{{r_N}}(X_{i})}([\Psi^N_{t,0}(Z)]_i)k_0(Z_i) \Big)d^{6N}Z \label{est.col.0}
\end{align}
This time we abbreviate 
\begin{align} 
\widetilde{Z}:=( X_1,...,X_M,[\Psi^N_{t,0}(Z)]_{M+1},...,[\Psi^N_{t,0}(Z)]_N). \label{widetilde(Z)}
\end{align}
If $\widetilde{Z}\in \Psi^N_{t,0}(\mathcal{G}^{N,\sigma}_{3,T})$, then by definition of $\mathcal{G}^{N,\sigma}_{3,T}$ there exists $Y\in \mathcal{G}^{N,\sigma}_{1,T}\cap\mathcal{G}^{N,\sigma}_{2,(0,T)} $ where $|Y-\Psi^N_{0,t}(\widetilde{Z})|_{\infty}\le N^{-\frac{1}{2}+\frac{3\sigma}{4}}$ and $Y_i\in \mathcal{L}^N_{\sigma}$ for all $i\in \{1,...,N\}$. Hence, Lemma \ref{shift-lem} implies that for the relevant values for $N,\sigma$
$$|\Psi^N_{t,0}(Y)-\widetilde{Z}|_{\infty}\le CN^{-\frac{1}{2}+\frac{3\sigma}{4}}.$$
Applying this time that $Y\in \mathcal{G}_{1,T}^{N,\sigma}$ we obtain by Corollary \ref{cor.shift.} that under these conditions also
$$|\Phi^N_{t,0}(Y)-\widetilde{Z}|_{\infty}\le|\Phi^N_{t,0}(Y)-\Psi^N_{t,0}(Y)|_{\infty}+|\Psi^N_{t,0}(Y)-\widetilde{Z}|_{\infty}\le CN^{-\frac{1}{2}+\frac{3\sigma}{4}}$$
which after recalling the definition of $\widetilde{Z}$ (see \eqref{widetilde(Z)}) implies in particular that
\begin{align}
|\varphi^N_{t,0}(Y_i)-X_i|\le CN^{-\frac{1}{2}+\frac{3\sigma}{4}} \text{ for } i\in \{1,...,M\}. \label{cond.loc.lip.}
\end{align} 
If $|[\Psi^N_{t,0}(Z)]_i-X_i|\le r_N=N^{-\frac{1}{2}+\frac{3\sigma}{4}}$ for $i\in \{1,...,M\}$, then it holds in turn that
\begin{align*}
|\Psi^N_{t,0}(Z)-\Psi^N_{t,0}(Y)|_{\infty}& \le\displaystyle |\Psi^N_{t,0}(Z)-\widetilde{Z}|_{\infty}+|\widetilde{Z}-\Psi^N_{t,0}(Y)|_{\infty}\\
& \le \max_{i\in \{1,...,M\}}|[\Psi^N_{t,0}(Z)]_i-X_i|+CN^{-\frac{1}{2}+\frac{3\sigma}{4}} \le  CN^{-\frac{1}{2}+\frac{3\sigma}{4}}
\end{align*}
where for this conclusion we applied that $\widetilde{Z}=( X_1,...,X_M,[\Psi^N_{t,0}(Z)]_{M+1},...,[\Psi^N_{t,0}(Z)]_N)$. After regarding once again that $Y\in \mathcal{G}^{N,\sigma}_{1,T}\cap \mathcal{G}^{N,\sigma}_{2,(0,T)}$, the last estimates imply due to Lemma \ref{lem1} and Lemma \ref{shift-lem} that for $i\in \{1,...,M\}$    
\begin{align}
& |\varphi^N_{0,t}(X_i)-Z_i|\notag \\
\le &|\varphi^N_{0,t}(X_i)-\underbrace{\varphi^N_{0,t}(\varphi^N_{t,0}(Y_i))}_{=Y_i}|+|[\Psi^N_{0,t}(\Psi^N_{t,0}(Y))]_i-[\Psi^N_{0,t}(\Psi^N_{t,0}(Z))]_i| \notag \\
 \le & C|X_i-\varphi^N_{t,0}(Y_i)|+C\big(N^{-\frac{1}{2}+\frac{\sigma}{2}}+|\Psi^N_{t,0}(Y)-\Psi^N_{t,0}(Z)|_{\infty} \big)\notag \\
 \le & CN^{-\frac{1}{2}+\frac{3\sigma}{4}}. \label{ineq.prod-lem}
\end{align}
In total we obtain the subsequent implication for $i\in \{1,...,M\}$, $\widetilde{Z}\in \Psi^N_{t,0}(\mathcal{G}^{N,\sigma}_{3,T})$ and a suitable constant $C>0$:
\begin{align*}
 |[\Psi^N_{t,0}(Z)]_i-X_i|\le r_N \Rightarrow |\varphi^N_{0,t}(X_i)-Z_i|\le Cr_N 
\end{align*} 
respectively
\[\prod_{i=1}^M\mathbf{1}_{B_{r_N}(X_{i})}([\Psi^N_{t,0}(Z)]_{i})\le \prod_{i=1}^M\mathbf{1}_{B_{ Cr_N}(\varphi^N_{0,t}(X_i))}(Z_i)\] 
Moreover, due to relation \eqref{cond.loc.lip.} and Lemma \ref{lem1} it holds that $$|\varphi^N_{0,t}(X_i)-Y_i|\le C|X_i-\varphi^N_{t,0}(Y_i)|\le  CN^{-\frac{1}{2}+\frac{3\sigma}{4}}$$ for $i\in \{1,...,M\}$ which together with $|\varphi^N_{0,t}(X_i)-Z_i|\le Cr_N $ implies for the relevant $N,\sigma$ that $|Y_i-Z_i|\le N^{-\frac{1}{3}}$. Since additionally $Y_i\in \mathcal{L}^N_{\sigma}$ for all $i\in \{1,...,N\}$, it follows that the special `Lipschitz property' related to $\mathcal{L}^N_{\sigma}$ (see \eqref{Def.L}) is applicable under these constraints on $Z_i$ and $\varphi^N_{0,t}(X_i)$. This yields for the relevant $ N,\sigma$ that in this case
\begin{align*} k_0(Z_i)\le & k_0(\varphi^N_{0,t}(X_i))+\big|k_0(\varphi^N_{0,t}(X_i))-k_0(Z_i)\big|\\
\le & k^N_t(X_i)+N^{\frac{\sigma}{2}}k_0(\varphi^N_{0,t}(X_i))|\varphi^N_{0,t}(X_i)-Z_i|\le 2k^N_t(X_i).
\end{align*}
Eventually, we can merge these considerations and obtain (once again for the relevant $N,\sigma$) that term \eqref{est.col.0} is bounded by
\begin{align*}
& \int_{\mathbb{R}^{6N}}\mathbf{1}_{\Psi^N_{t,0}(\mathcal{G}^{N,\sigma}_{3,T})}( X_1,...,X_M,[\Psi^N_{t,0}(Z)]_{M+1},...,[\Psi^N_{t,0}(Z)]_N) \prod_{j=M+1}^N k_0(Z_j)  \\
& \cdot\prod_{i=1}^M\Big(\mathbf{1}_{B_{ Cr_N}(\varphi^N_{0,t}(X_i))}(Z_i)k_0(Z_i) \Big)d^{6N}Z \\
\le  & \int_{\mathbb{R}^{6M}}\prod_{i=1}^M\Big(\mathbf{1}_{B_{ Cr_N}(\varphi^N_{0,t}(X_i))}(Z_i)2k_t(X_i) \Big)d^{6M}(Z_1,...,Z_M) \\
\le & 2^M\mu(B_{Cr_N}(0))^M\prod_{i=1}^Mk^N_t(X_i)
\end{align*}  
which completes the considerations for the inner integral.\\
Finally, it remains to `insert' these estimates in term \eqref{prod.lem.inn.int.} and to reintroduce the previously dropped prefactor. After recalling that we abbreviated $r_N:=N^{-\frac{1}{2}+\frac{3\sigma}{4}}$ and $X^M:=(X_1,...,X_M)$ we can conclude the proof as follows:
\begin{align*}
& \int_{\Psi^N_{t,0}(\mathcal{G}^{N,\sigma}_{3,T})}\mathbf{1}_{\mathcal{S}'}(X_1,...,X_M)F^N_t(X)d^{6N}X \\
\le & \frac{e^{N^{\frac{\sigma}{2}}\max(N^{-\sigma},CMr_n)}}{ \mu(B_{r_N}(0))^M }\int_{\mathbb{R}^{6M}}
 \Big(2\mu(B_{Cr_N}(0))\big)^M\prod_{i=1}^M k^N_t(X_i) \Big)\mathbf{1}_{\mathcal{S}'}(X^M)d^{6M}X\\
 \le &  C^M\int_{\mathbb{R}^{6M}}\mathbf{1}_{\mathcal{S}'}(X_1,...,X_M)\prod_{i=1}^Mk^N_t(X_i)d^{6M}X
\end{align*}
\end{proof}
\noindent We will apply the previous lemma to do probability estimates regarding the `collision numbers' for the microscopic system with a cut-off size of order $N^{-\frac{2}{3}}$. More precisely, we want to show that the number of particle pairs having a collision related to a particular `collision class' does typically not exceed a certain value. It will turn out to be simpler to divide $[0,T]$ in many short intervals and to argue that an adjusted version of the previous statement is true for each of them. The reason for this lies in the property that for short times we can apply the mean-field dynamics to approximate the trajectories of the interacting particles. As long as the related trajectories are sufficiently close, a certain `collision' between two `mean-field particles' corresponds to a similar collision in the system of interacting particles. Thus, in this case it suffices to count the number of `mean-field particle collisions' belonging to a certain `collision class' to receive the corresponding number for the microscopic system. The great technical advantage of this approach arises by the fact that the `mean-field particles' move independently from each other. Of course, their initial positions for each of these intervals (except for the first at time $t=0$) are correlated since they start at the position where the interacting particles are respectively located. However, our previously established lemma (very broadly speaking) tells us that still `enough  independence' is left to get rid of this issue.    \\ 
Before starting with the lemma we have to introduce some definitions. First, we define a set which contains tuples related to `mean field particle' pairs which have a collision characterized by a certain `collision class' with each other.\\ Let $X\in \mathbb{R}^{6N},\ r,v \in \mathbb{R}_{\geq 0}\cup \{\infty\}$ and $t_1,t_2 \in [0,T]$, then we identify
\begin{align}  
\begin{split}
\mathcal{M}^{N,(t_1,t_2)}_{r,v}(X):=\big\{&(i,j)\in  \{1,...,N\}^2 :i \neq j \land  X_j\in M^{N,(t_1,t_2)}_{r,v}(X_i)\big\}\label{def.M}
\end{split}
\end{align}
where $M^{N,(t_1,t_2)}_{r,v}(X_i)$ are the `collision classes' defined in \eqref{def.coll.cl.1} for the mean-field dynamics with cut-off parameter $c:=\frac{2}{3}$.\\
The second set is the corresponding set for the `real' particles (once again for $c:=\frac{2}{3}$)
\begin{align}
&\mathcal{R}^{N,(t_1,t_2)}_{r,v}(X):=\big\{(i,j)\in  \{1,...,N\}^2 :i\neq j \land \big( \exists t \in [t_1,t_2]: \notag \\
&  \min\limits_{t_1\le s \le t_2}|[^1\Psi^{N,c}_{s,0}(X)]_j-[{^1\Psi^{N,c}_{s,0}(X)}]_i|=|[^1\Psi^{N,c}_{t,0}(X)]_j-[{^1\Psi^{N,c}_{t,0}(X)}]_i|\le r\ \land  \notag \\
&\hspace{1,245cm} |[^2\Psi^{N,c}_{t,0}(X)]_j-[{^2\Psi^{N,c}_{t,0}(X)]_i}|\le v \big)\big\}. \label{def.R}
\end{align}
\vspace*{0,2cm}
\begin{lem} \label{c_1-Lem}
Let $N\in \mathbb{N}$ and $k_0\in \mathcal{L}^1(\mathbb{R}^6)$ be a probability density fulfilling the assumptions of Theorem \ref{thm1}. Moreover, let $t_1,t_2\in [0,T]$ where $t_2-t_1\geq N^{-\frac{1}{3}}$ and $r,v>0$ where $N^{-\frac{7}{9}}\le r\le \min(1,v)$. If $\sigma>0$ is small enough, then there exist $ C_1>0$ and $N_0\in \mathbb{N}$ such that for all $N\geq N_0$, $M\in \mathbb{N}$ as well as $r,v,t_1$ and $t_2$ fulfilling the previously mentioned constraints the following holds: 
\begin{align*}
&\mathbb{P}\Big( X\in \mathbb{R}^{6N} : \big|\mathcal{R}^{N,(t_1,t_2)}_{r,v}(X)\big|\geq M(t_2-t_1) \land X\in \mathcal{G}^{N,\sigma}_{3,T} \Big)\\
 \le &  N^{\frac{1}{3}} (t_2-t_1) \Big(\frac{C_1  N^{2+3{\sigma}}r^2\min(v,1)^3}{M}\big(\min(v,1)+N^{\frac{1}{3}}r \big)\Big)^{\frac{M}8N^{-\frac{1}{3}-3\sigma}}.
\end{align*}
\end{lem}
\vspace{0,4cm}
\begin{proof}
\noindent Once again the applied estimates and statements of the proof only need to hold if $N\in \mathbb{N}$ is large enough and $\sigma>0$ sufficiently small. Hence, for convenience and for avoiding redundant formulations we will partly omit to mention this explicitly. Since the cut-off parameter of the effective and the microscopic dynamics is once again fixed to $c:=\frac{2}{3}$ (regarding this recall that $\mathcal{R}^{N,(t_1,t_2)}_{r,v}$ was defined for the $N$-particle dynamics where the parameter $c$ takes this value), the related index will not appear in the notation. \\
We start by implementing the idea described previous to the lemma which is dividing the interval $[t_1,t_2]$ in many shorter intervals (of length $\Delta>0$) and to prove an adjusted statement for each of them.\\
Let $M,N\in \mathbb{N}$, $r,v>0$ where $\ N^{-\frac{7}{9}}\le r \le \min(1,v)$ and $t_1,t_2\in [0,T]$ such that $t_2-t_1\geq N^{-\frac{1}{3}}$ as well as $0\le \Delta \le t_2-t_1$ then it holds that
\begin{align}
& \big|\mathcal{R}^{N,(t_1,t_2)}_{r,v}(X)\big|\geq M(t_2-t_1) \notag \\
\Rightarrow & \exists k \in \{0,...,\lceil \frac{t_2-t_1}{\Delta}\rceil-1\}: \notag \\ 
&|\mathcal{R}^{N,(k\Delta,(k+1)\Delta)}_{r,v}(\Psi^N_{t_1,0}(X))|\geq \frac{M(t_2-t_1)}{\lceil \frac{t_2-t_1}{\Delta}\rceil}\geq \frac{M \Delta}{2} .\label{assu1}
\end{align}
where we regarded that according to definition \eqref{def.R} it holds that 
$$\mathcal{R}^{N,(\tau_2,\tau_3)}_{r,v}(\Psi^N_{\tau_1,0}(X))=\mathcal{R}^{N,(\tau_1+\tau_2,\tau_1+\tau_3)}_{r,v}(X).$$
Moreover, let $X\in \mathcal{G}^{N,\sigma}_{3,T}$ (see \eqref{def.G_3}) then there exists $Y\in \mathcal{G}^{N,\sigma}_{1,T}\cap \mathcal{G}^{N,\sigma}_{2,(0,T)} $ such that $|Y-X|_{\infty}\le N^{-\frac{1}{2}+\frac{3\sigma}{4}}$. According to Lemma \ref{shift-lem}, Corollary \ref{cor.shift.} and Lemma \ref{lem1} (applied in this order on the three addends appearing in the second line) this yields for $s\in [0,t]$ that (at least for the relevant values of $\sigma$ and $N$) 
\begin{align}
& |\Psi_{s,0}^N(X)-\Phi_{s,0}^N(X)|_{\infty} \notag \\
\le & |\Psi_{s,0}^N(X)-\Psi_{s,0}^N(Y)|_{\infty}+|\Psi_{s,0}^N(Y)-\Phi_{s,0}^N(Y)|_{\infty}+ |\Phi_{s,0}^N(Y)-\Phi_{s,0}^N(X)|_{\infty} \notag \\
\le &CN^{-\frac{1}{2}+\frac{3\sigma}{4}}+ CN^{-\frac{1}{2}+\frac{\sigma}{2}}+CN^{-\frac{1}{2}+\frac{3\sigma}{4}}. \label{close.c_1}
\end{align} 
Thus, the statement of the Lemma is obviously only interesting for values of $r$ which are smaller than order $N^{-\frac{1}{2}+\sigma}$ since otherwise we have sufficient `control' on the interacting particles by their `mean-field particle partners'. More precisely, for larger values of $r$ we only have to `count' the respective collisions of their related `mean-field particles' and show that this number typically remains sufficiently small (which we have basically already done for example during the probability estimates for \eqref{cond.shift}). Hence, we focus on the distinctly more interesting choices for $r>0$ where the information about the positions of the `real' particles provided by their related `mean-field particles' is in general not sufficient anymore to predict the order of their minimal spatial inter-particle distance.\\
Again we distinguish two cases: On the one hand we consider the case that there indeed exists a particle coming closer than $r\le N^{-\frac{1}{2}+\sigma}$ to at least $\lceil N^{3{\sigma}}\rceil$ particles. And in the second case we assume that all particles have at most $ \lfloor N^{3{\sigma}} \rfloor$ such `collision partners' but there exists a set $\mathcal{S}\subseteq \{1,...,N\}$ containing at least $\lceil (\frac{M\Delta}{2})(\frac{1}{4 N^{3{\sigma}}} )\rceil= \lceil \frac{N^{-3\sigma}M\Delta}{8} \rceil $ different pairs of particles $(i,j)$ having a collision with each other (respectively where $(i,j)\in \mathcal{R}^{k\Delta,(k+1)\Delta}_{r,v}(\Psi^{N,c_1}_{t_1,0}(X))$) but with no further particle of the pairs belonging to $\mathcal{S}$ (which we will shortly state more formally).\\
Analogous to the reasoning applied for estimating the probability of `event' \eqref{cond.shift} one easily sees that assumption \eqref{assu1} indeed implies that one of these two options must occur. The choice $ N^{3{\sigma}} $ seems to be random but on the one hand we know by definition of $\mathcal{G}^{N,\sigma}_{1,T}\subseteq \Big(\bigcup_{i=1}^N \mathcal{B}_{4,i}^{N,\sigma}\Big)^C$ that $\max_{i\in \{1,...,N\}}\sum_{j\neq i}\mathbf{1}_{M^N_{6N^{-\frac{1}{2}+\sigma},\infty}(Y_i)}(Y_j)< N^{3\sigma}$ (see \eqref{def.B_4}) and on the other hand estimates \eqref{close.c_1} yield for sufficiently large $N$ and $Y_j\in (M^N_{6N^{-\frac{1}{2}+\sigma},\infty}(Y_i))^C$ that
\begin{align*}
& |[^1\Psi^N_{t,0}(X)]_i-[^1\Psi^N_{t,0}(X)]_j|
\geq |^1\varphi^N_{t,0}(Y_i)-{^1\varphi^N_{t,0}(Y_j)}|-CN^{-\frac{1}{2}+\frac{3\sigma}{4}}\geq N^{-\frac{1}{2}+\sigma}.
\end{align*} 
In total this implies that for $X\in \mathcal{G}^{N,\sigma}_{3,T}$, sufficiently large $N$ and $r\le N^{-\frac{1}{2}+\sigma}$
\begin{align*}
& \max_{i\in \{1,...,N\}}\sum_{j\neq i}\mathbf{1}_{ \{Z\in \mathbb{R}^{6N}: \min\limits_{t_1\le s \le t_2}|[^1\Psi^N_{s,0}(Z)]_j-[{^1\Psi^N_{s,0}(Z)}]_i|\le r\}}(X)< N^{3\sigma}
\end{align*}
and thus the first case of the considered two options can not occur. Also smaller bounds than $ N^{3{\sigma}}$ would work in the current situation since the length of the time interval $\Delta$ will be chosen very short. But to avoid further estimates, we are content with the current one. \\
Consequently, we see that for $X\in \mathcal{G}^{N,\sigma}_{3,T}$, the relevant $N$ and $r\le N^{-\frac{1}{2}+\sigma}$ assumption \eqref{assu1} implies that the remaining of the previously described two options must be fulfilled which (stated a bit more formally) is:
\begin{align}
& \exists k \in \{0,...,\lceil \frac{t_2-t_1}{\Delta}\rceil-1\}:\notag\\
& \Big(\exists \mathcal{S}\subseteq \{1,...,N\}^2\setminus \bigcup_{n=1}^N \{(n,n)\}: \notag \\
& \ \text{(i)}\  \ |\mathcal{S}|\geq \lceil \frac{N^{-3\sigma}M\Delta}{8} \rceil \notag \\ 
& \  \text{(ii)} \ \  \mathcal{S}\subseteq \mathcal{R}^{N,(k\Delta,(k+1)\Delta)}_{r,v}(\Psi^N_{t_1,0}(X))\  \notag \\
&\ \text{(iii)} \ (i_1,j_1),(i_2,j_2)\in \mathcal{S}\Rightarrow  \{i_1,j_1\}\cap \{i_2,j_2\}=\emptyset\Big) \label{r.ass.}
\end{align}
Like described in the preliminary considerations we will apply the mean-field dynamics to show that the probability of such an event is very small. For a simpler notation we `adjust' the definition of the sets $M^{N,(t_1,t_2)}_{r,v}(Y)$ (see \eqref{def.coll.cl.1}) to the current situation:
\begin{align*}
& Z\in M^{N,(t_1,t_2),t}_{r,v}(Y)\subseteq  \mathbb{R}^6 \\
\Leftrightarrow &Z\neq Y\ \land\ \exists \tau \in [t_1,t_2]: \\ & \min_{t_1\le s \le t_2}|^1\varphi^N_{s+t,t}(Z)-{^1\varphi^N_{s+t,t}}(Y)|=|^1\varphi^{N}_{\tau+t,t}(Z)-{^1\varphi^{N}_{\tau+t,t}}(Y)|\le r \ \land  \\
& \hspace{5,87cm} |^2\varphi^{N}_{\tau+t,t}(Z)-{^2\varphi^{N}_{\tau+t,t}}(Y)|\le  v
\end{align*}
Consequently, everything stays the same except for the `time shift' of the dynamics given by the new parameter $t\geq 0$. This modification is only relevant for the present proof. \\ 
For short times the mean-field trajectories provide a very good approximation for the trajectories of the interacting particles. Thus, if $(i,j)\in \mathcal{R}^{N,(k\Delta,(k+1)\Delta)}_{r,v}(\Psi^N_{t_1,0}(X))$, then typically $[\Psi^N_{t_1+k\Delta,0}(X)]_i\in M^{N,(0,\Delta),t_1+k\Delta}_{2r,2v}([\Psi^N_{t_1+k\Delta,0}(X)]_j)$ is fulfilled if we choose $\Delta $ sufficiently small. To avoid such elongated expressions we abbreviate for the rest $$t'_k:=t_1+k\Delta.$$ By application of these considerations we split assumption \eqref{r.ass.} further. More precisely, \eqref{r.ass.} implies that:
{\allowdisplaybreaks
\begin{align} 
& \exists k \in \{0,...,\lceil \frac{t_2-t_1}{\Delta}\rceil-1\}:\notag \\
& \Big( \exists \mathcal{S}\subseteq \{1,...,N\}^2\setminus \bigcup_{n=1}^N\{(n,n)\}:  \notag \\
& \ \text{(i)}\  \ |\mathcal{S}|\geq \lceil \frac{N^{-3\sigma}M\Delta}{8} \rceil \notag \\ 
& \  \text{(ii)} \ \  \forall
(i,j)\in \mathcal{S}: [\Psi^N_{t'_k,0}(X)]_i\in M^{N,(0,\Delta),t'_k}_{2r,2v}([\Psi^N_{t'_k,0}(X)]_j)  \notag \\
&\ \text{(iii)} \ (i_1,j_1),(i_2,j_2)\in \mathcal{S}\Rightarrow  \{i_1,j_1\}\cap \{i_2,j_2\}=\emptyset\Big) \ \vee \\
&\Big( \sup\limits_{t'_k \le s \le t'_{k+1}}|^1\Psi^N_{s,0}(X)-{^1\Phi^N_{s,t'_k}}(\Psi^N_{t'_k,0}(X))|_{\infty}\geq \frac{r}{2}\notag\\
& \hspace{0,23cm} \sup\limits_{t'_k \le s \le t'_{k+1}}|^2\Psi^N_{s,0}(X)-{^2\Phi^N_{s,t'_k}}(\Psi^N_{t'_k,0}(X))|_{\infty}\geq \frac{v}{2}\Big)
  \label{pro.est.1}
\end{align} }
First, we determine a suitably short choice for the length $\Delta>0$ of the time interval so that the assumptions described in the last two lines of \eqref{pro.est.1} can not occur for the values of $r$ and $v$ which are relevant to us.  \\ 
To this end, we recall that $X\in \mathcal{G}^{N,\sigma}_{3,T}$ yields the existence of $Y\in \mathcal{G}_{1,T}^{N,\sigma}\cap \mathcal{G}_{2,(0,T)}^{N,\sigma} $ where $|Y-X|_{\infty}\le N^{-\frac{1}{2}+\frac{3\sigma}{4}}$ and it holds for $s\in [t'_k,t'_{k+1}]= [t_1+ k\Delta, t_1+ (k+1)\Delta]$ that
\begin{align*}
& |\Psi^N_{s,0}(Y)-{\Phi^N_{s,t'_k}}(\Psi^N_{t'_k,0}(Y))|_{\infty}\\
\le & |\Psi^N_{s,0}(Y)-{\Phi^N_{s,0}}(Y)|_{\infty}+|\Phi^N_{s,0}(Y)-{\Phi^N_{s,t'_k}}(\Psi^N_{t'_k,0}(Y))|_{\infty} \\
\le & CN^{-\frac{1}{2}+\frac{\sigma}{2}} +|\Phi^N_{t'_k,0}(Y)-\Psi^N_{t'_k,0}(Y)|_{\infty} e^{C(s-t'_k)}\\
\le & CN^{-\frac{1}{2}+\frac{\sigma}{2}}
\end{align*}
where we used Corollary \ref{cor.shift.} (in the second and the third step) as well as Lemma \ref{lem1} (in the second step).\\
By taking into account that $\sup_{0\le s \le T}|\Psi^N_{s,0}(X)-\Psi^N_{s,0}(Y)|_{\infty}\le CN^{-\frac{1}{2}+\frac{3\sigma}{4}}$ (which holds due to Lemma \ref{shift-lem}) this easily implies that
\begin{align*}
& |\Psi^N_{s,0}(X)-{\Phi^N_{s,t'_k}}(\Psi^N_{t'_k,0}(X))|_{\infty} \\
\le & |\Psi^N_{s,0}(X)-\Psi^N_{s,0}(Y)|_{\infty}+|\Psi^N_{s,0}(Y)-{\Phi}^N_{s,t'_k}(\Psi^N_{t'_k,0}(Y))|_{\infty}  \\ 
&+ |{\Phi^N_{s,t'_k}}(\Psi^N_{t'_k,0}(Y))-{\Phi^N_{s,t'_k}}(\Psi^N_{t'_k,0}(X))|_{\infty} \\
\le & CN^{-\frac{1}{2}+\frac{3\sigma}{4}}
\end{align*}
where in the last step we applied once again Lemma \ref{lem1} to estimate the term in the third line.\\
This basically already shows that the event related to the last line of \eqref{pro.est.1} can not occur for the values of $v$ which are relevant to us since fortunately we do not have to care about cases where $v$ is smaller than order $N^{-\frac{1}{2}+\sigma}$. We will explain the reasons for this in more detail shortly. However, first we notice that by the second order nature of the dynamics the last inequality applied for $\Delta:=N^{-\frac{1}{3}}$ yields the following upper bound for the spatial distance between these trajectories:
\begin{align*} 
& \sup\limits_{t_1+k\Delta \le s \le t_1+(k+1)\Delta}|^1\Psi^N_{s,0}(X)-{^1\Phi^N_{s,t_1+k\Delta}}(\Psi^N_{t_1+k\Delta,0}(X))|_{\infty}\\
\le & CN^{-\frac{1}{2}+\frac{3\sigma}{4}}\Delta 
\le  CN^{-\frac{5}{6}+\frac{3\sigma}{4}}
\end{align*}
Since the assumptions of the lemma claim $r\geq N^{-\frac{7}{9}}$, this implies that the event
$$\sup\limits_{t_1+k\Delta \le s \le t_1+(k+1)\Delta}|^1\Psi^N_{s,t_1}(X)-{^1\Phi^N_{s,t_1+k\Delta}}(\Psi^N_{t_1+k\Delta,t_1}(X))|_{\infty}\geq \frac{r}{2}$$ of constraint \eqref{pro.est.1} can not occur for $\Delta=N^{-\frac{1}{3}}$ if $\sigma>0$ is chosen sufficiently small and $N\in \mathbb{N}$ large enough (which we can assume without restriction). For eventually justifying that we indeed do not have to care for values $v\le N^{-\frac{1}{2}+\sigma}$ one must simply recall that $Y\in \mathcal{G}^{N,\sigma}_{1,T}\subseteq  \big(\mathcal{B}^{N,\sigma}_5\big)^C$ (see \eqref{def.B_5}) which yields that for all $i\neq j$ 
\begin{align*}
Y_j \notin M^N_{6N^{-\frac{1}{2}+\sigma} ,N^{-\frac{5}{18}}}(Y_i).
\end{align*} 
Together with the circumstance that we only consider choices of $r$ where $r\le N^{-\frac{1}{2}+\sigma}$ this tells us that the relative velocity between `mean-field particles' related to such an initial configuration $Y$ is at least of order $N^{-\frac{5}{18}}$ at times when their spatial distance falls below $6N^{-\frac{1}{2}+\sigma}$. Once again we recall that it holds according to Lemma \ref{shift-lem} and Corollary \ref{cor.shift.} that 
\begin{align*}
& \sup_{0\le s \le T}|\Psi_{s,0}^N(X)-\Phi_{s,0}^N(Y)|_\infty\\
\le &  \sup_{0\le s \le T}|\Psi_{s,0}^N(X)-\Psi_{s,0}^N(Y)|_\infty+\sup_{0\le s \le T}|\Psi_{s,0}^N(Y)-\Phi_{s,0}^N(Y)|_\infty\\
\le & CN^{-\frac{1}{2}+\frac{3\sigma}{4}}.
\end{align*}
Consequently, it is straightforward to see that a corresponding condition `transfers' on the relative velocity of the `real' particles related to initial configurations $X$. However, this implies that $\mathcal{R}_{r,v}^{N,(0,T)}(X)$ is empty if $v>0$ is of smaller order than $N^{-\frac{5}{18}}$ (and in particular for $v\le N^{-\frac{1}{2}+\sigma}$) if $r\le N^{-\frac{1}{2}+\sigma}$. Hence, for such choices of $r,v$ the statement of the Lemma is trivially fulfilled.\\
It remains to show that on a time interval of length $N^{-\frac{1}{3}}$ the number of considered `collisions' between the auxiliary particles which we apply to approximate the trajectories of the interacting particles does not exceed a certain value with sufficiently high probability. \\
To this end, we identify $\Delta:=N^{-\frac{1}{3}}$ for the rest of the proof and continue by implementing step by step the probability estimates of the related assumption described in \eqref{pro.est.1}:
{\allowdisplaybreaks
\begin{align}
& \exists k \in \{0,...,\lceil \frac{t_2-t_1}{\Delta} \rceil-1 \}\\
&\exists \mathcal{S}\subseteq \{1,...,N\}^2 \setminus \bigcup_{n=1}^N \{(n,n)\}:  \notag \\
& \ \text{(i)}\  \ |\mathcal{S}|= \lceil \frac{N^{-3\sigma}M\Delta}{8} \rceil \notag \\ 
& \  \text{(ii)} \ \   \forall
(i,j)\in \mathcal{S}: [\Psi^N_{t'_k,0}(X)]_i\in M^{N,(0,\Delta),t'_k}_{2r,2v}([\Psi^N_{t'_k,0}(X)]_j)  \notag \\
&\ \text{(iii)} \ (i_1,j_1),(i_2,j_2)\in \mathcal{S}\Rightarrow  \{i_1,j_1\}\cap \{i_2,j_2\}=\emptyset \label{fin.as.2} 
\end{align}}
These conditions are very strongly reminiscent of \eqref{cond.shift2}.\\ 
A rough upper bound for the number of possibilities choosing such a set consisting of $|\mathcal{S}|$ `disjoint' pairs is given by $\binom{N^2}{|\mathcal{S}|}\le (3\frac{N^2}{|\mathcal{S}|})^{|\mathcal{S}|}$. Due to the symmetry of the distribution (and item (iii)) it suffices to implement the probability estimates for the special choice 
\[\mathcal{S}_M:=\{(1,2),(3,4),...,(2\lceil\frac{N^{-3{\sigma}}M\Delta}{8}\rceil -1,2\lceil\frac{N^{-3{\sigma}}M\Delta}{8}\rceil)\}\]
and to multiply the result with the combinatorial factor.\\
More precisely, we identify for $k\in  \{0,...,\lceil \frac{t_2-t_1}{\Delta} \rceil-1\}$
\begin{align}
\begin{split}
\mathcal{D}_ {k}:=\{& X\in \mathcal{G}_{3,T}^{N,\sigma} \land  \forall
(i,j)\in \mathcal{S}_M: [\Psi^N_{t'_k,0}(X)]_i\in M^{N,(0,\Delta),t'_k}_{2r,2v}([\Psi^N_{t'_k,0}(X)]_j)  \} \label{def.D_k}
\end{split}
\end{align}
and by merging the previous considerations it follows that
\begin{align}
&\mathbb{P}\Big( Z\in \mathbb{R}^{6N}:  \big|\mathcal{R}^{N,(t_1,t_2)}_{r,v}(Z)\big|\geq M(t_2-t_1) \land Z\in \mathcal{G}^{N,\sigma}_{3,T} \Big) \notag \\
\le & \sum_{ k =0}^{\lceil \frac{t_2-t_1}{\Delta}\rceil-1 }\mathbb{P}\Big( Z\in \mathcal{G}^{N,\sigma}_{3,T} \land \big|\mathcal{R}^{N,(k\Delta,(k+1)\Delta)}_{r,v}(\Psi^N_{t_1,0}(Z))\big|\geq \frac{M\Delta}{2} \Big) \notag\\
\le &  C\lceil \frac{t_2-t_1}{\Delta} \rceil\big(3 \frac{N^{2}}{|\mathcal{S}_M|}\big)^{|\mathcal{S}_M|}\max_{k \in \{0,...,\lceil \frac{t_2-t_1}{\Delta} \rceil-1\}}\mathbb{P}\big( Z\in \mathcal{D}_k \big). \label{prob.c_1-lem}
\end{align}
It remains to determine a suitable upper bound for the probability of the `events' $\mathcal{D}_k$:
{\allowdisplaybreaks
\begin{align}
& \mathbb{P}\big(X\in \mathcal{D}_k)\notag \\
= & \int_{\mathbb{R}^{6N}}\mathbf{1}_{\mathcal{D}_k}
(X)\prod_{i=1}^Nk_0(X_i) d^{6N}X \notag \\ 
= & \int_{\mathbb{R}^{6N}}\mathbf{1}_{\{Z\in \mathcal{G}_{3,T}^{N,\sigma} :\ \forall
(i,j)\in \mathcal{S}_M: [\Psi^N_{t'_k,0}(Z)]_j\in M^{N,(0,\Delta),t'_k}_{2r,2v}([\Psi^N_{t'_k,0}(Z)]_i) \}}
(\Psi^N_{0,t'_k}(X))F^N_{t'_k}(X) d^{6N}X  \notag \\
= & \int_{\Psi^N_{t'_k,0}(\mathcal{G}^{N,\sigma}_{3,T})}\mathbf{1}_{\{Z\in \mathbb{R}^{6(2|\mathcal{S}_M|)} :\  \forall (i,j)\in \mathcal{S}_M:Z_j  \in M^{N,(0,\Delta),t'_k}_{2r,2v}(Z_i)\}}(X_1,...,X_{2|\mathcal{S}_M|})F^N_{t'_k}(X) d^{6N}X  \notag 
\end{align}
If additionally $N\in \mathbb{N}$ is large enough and $\sigma>0$ sufficiently small, then we can apply Lemma \ref{prod-Lem} and obtain that the previous term is bounded by
\begin{align}
&  C^{|\mathcal{S}_M|}\int_{\mathbb{R}^{6(2|\mathcal{S}_M|)}}\mathbf{1}_{\{Z\in \mathbb{R}^{6(2|\mathcal{S}_M|)}: \forall (i,j)\in \mathcal{S}_M:Z_j  \in M^{N,(0,\Delta),t'_k}_{2r,2v}(Z_i)\}}(X) \prod_{i=1}^{2|\mathcal{S}_M|} k^N_{t'_k}(X_i)d^{6(2|\mathcal{S}_M|)}X  \notag \\
= &  C^{|\mathcal{S}_M|}\mathbb{P}\big(X\in \mathbb{R}^{6(2|\mathcal{S}_M|)}:\ \forall (i,j)\in \mathcal{S}_M:  X_j\in M_{2r,2v}^{N,(0,\Delta),t'_k}(X_i)\big) \notag \\
\le &   C^{|\mathcal{S}_M|}\Big(r^2\min(v,1)^3\big(\min(v,1)\Delta+r\big)\Big)^{|\mathcal{S}_M|}
\end{align} }
where the last step follows by Lemma \ref{lem3}.\\
After recalling that $|\mathcal{S}_M|=\lceil \frac{N^{-3\sigma}M\Delta}{8} \rceil$ as well as $\Delta=N^{-\frac{1}{3}}$ we can finally apply these estimates on term \eqref{prob.c_1-lem} to conclude the proof:
 \begin{align*}
&\mathbb{P}\Big( Z\in \mathbb{R}^{6N}:  \big|\mathcal{R}^{N,(t_1,t_2)}_{r,v}(Z)\big|\geq M(t_2-t_1) \land Z\in \mathcal{G}^{N,\sigma}_{3,T} \Big) \notag \\
\le &  C\lceil \frac{t_2-t_1}{\Delta} \rceil\big(3 \frac{N^{2}}{|\mathcal{S}_M|}\big)^{|\mathcal{S}_M|} \Big(C^{|\mathcal{S}_M|}\big(r^2\min(v,1)^3(\min(v,1)\Delta+r)\big)^{|\mathcal{S}_M|}\Big)\\
\le &  \lceil N^{\frac{1}{3}} (t_2-t_1) \rceil\Big(\frac{CN^{2+3\sigma}r^{2}\min(v,1)^3}{M}\big(\min(v,1)+N^{\frac{1}{3}}r\big)\Big)^{|\mathcal{S}_M|}. 
\end{align*}
\end{proof}
\vspace{0,4cm}
\noindent After all these preliminary considerations we are finally able to prove Theorem \ref{thm2}, which states that trajectories which have the same initial data but belong to systems with different-sized cut-offs, $N^{-c_1}$ and $N^{-c_2}$, typically remain close with respect to $|\cdot|_{\infty}$ if $c_2\geq c_1=\frac{2}{3}$. We remark that in this proof it will be for the first time necessary to denote the cut-off parameter explicitly for objects or maps (like $f^N_c$ or $g^N_c$) which depend on it. 
 
\subsection{Implementation of the proof}
\begin{figure}[h]
   \centering
   \def\svgscale{1.4}
   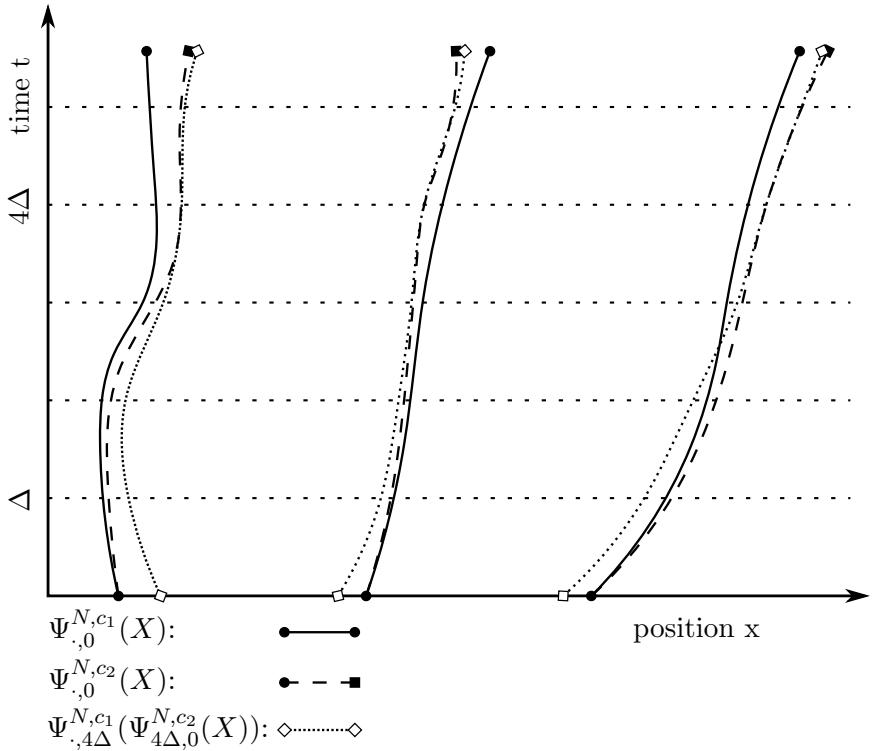
   \caption{Sketch of the essential proceeding}
   \label{fig.Thm2}
\end{figure}

Let $\sigma,\epsilon>0$ be given. We recall that according to the statement of the theorem we have to find a constant $C_1>0$ and a parameter $\sigma'>0$ such that 
\begin{align}
 & \mathbb{P}\big(X\in \mathbb{R}^{6N}:\sup_{0\le s \le T}|\Psi_{s,0}^{N,c_1}(X)-\Psi_{s,0}^{N,c_2}(X)|_{\infty}> N^{-\frac{1}{2}+\sigma} \big) \notag \\
\le & C_{1}N^{-\frac{1}{9}+\epsilon}+\mathbb{P}\big(X\in \mathbb{R}^{6N}:\exists i\in \{1,...,N\}:X_i\notin \mathcal{L}^N_{\sigma'}\big) \label{statement thm.2}
\end{align} holds for all $N\in \mathbb{N}$. If we are able to show that there exists $\sigma^*>0$ such that for any $\sigma_*\in (0, \sigma^*]$ it is possible to find a constant $C_1>0$ where $\forall N\in \mathbb{N}$
\begin{align}
& \mathbb{P}\big(X\in \mathbb{R}^{6N}:\sup_{0\le s \le T}|\Psi_{s,0}^{N,c_1}(X)-\Psi_{s,0}^{N,c_2}(X)|_{\infty}> N^{-\frac{1}{2}+\sigma_*} \big) \notag\\
\le &C_{1}N^{-\frac{1}{9}+\epsilon}+\mathbb{P}\big(X\in \mathbb{R}^{6N}:\exists i\in \{1,...,N\}:X_i\notin \mathcal{L}^N_{\sigma_*}\big) , \label{statement thm.2+}
\end{align}
then the statement of the theorem is proven: If the given $\sigma$ is smaller than or equal to $\sigma^*$, one just can identify $\sigma':=\sigma$ and the desired statement \eqref{statement thm.2} holds. If, on the other hand, $\sigma>\sigma^*$, then the choice $\sigma'=\sigma^*$ is possible which can be seen as follows: Under the condition that the inequality related to \eqref{statement thm.2+} holds for the choice $\sigma_*=\sigma^*$, then it holds a fortiori if the $\sigma^*$ appearing in the first line of the inequality is replaced by $\sigma$ where $\sigma>\sigma^*$ while the $\sigma^*$ in second line is kept fixed.\\
Hence, in the following we restrict ourselves to showing that the statement belonging to \eqref{statement thm.2+} indeed holds if $\sigma=\sigma_*>0$ is chosen small enough. Furthermore, like in previous proofs, applied estimates only need to hold for large enough $N\in \mathbb{N}$ in order that the statement of the theorem is fulfilled.
Thus, for several estimates and considerations we will assume that $N\in \mathbb{N}$ is chosen `large' and $\sigma>0$ `small'. However, for convenience and to avoid redundant formulations we often omit to mention this explicitly (or just call them the `relevant values').\\
Let $c_2\geq c_1:=\frac{2}{3}$. Now we identify for the last time a `good' set which unifies all the properties we will need for the proof:
\begin{align}
& X\in \mathcal{G}_T^{N,\sigma}\subseteq \mathbb{R}^{6N}  \notag \\ 
\Leftrightarrow &  X\in \mathcal{G}^{N,\sigma,2\sigma}_{N^{-\frac{1}{3}},T} \ \land \ \forall i \in \{1,...,N\}:X_i\in \mathcal{L}_{\sigma}^N \label{def.final.good.set}  
\end{align}
Sets of the form $\mathcal{G}^{N,\sigma,\epsilon}_{t,T}$ were defined in \eqref{Def.good set}. On the other hand, for all estimates which are explicitly implemented in the current proof (and which are not a consequence of a previous result) we basically only need that the considered configurations belong to $\mathcal{G}^{N,\sigma}_{1,T}\cap \mathcal{G}^{N,\sigma}_{2,(0,T)}\supseteq \mathcal{G}^{N,\sigma,2\sigma}_{N^{-\frac{1}{3}},T}$ and thus the details of definition \eqref{Def.good set} are not important for comprehending the proof.\\
Moreover, we recall that we omitted to make the dependence of the sets $\mathcal{G}^{N,\sigma}_{2,(t_1,t_2)}$ on the constant $C_0>0$ appearing in its definition (see the assumptions of Lemma \ref{shift-lem}) explicit to avoid an even more cluttered notation. For the current proof we assume that $C_0$ is chosen sufficiently large and $\sigma>0$ small enough such that the probability estimate $$\mathbb{P}(X \in \mathcal{G}^{N,\sigma}_{1,T}\cap (\mathcal{G}_{2,(t_1,t_2)}^{N,\sigma})^C)\le CN^{- N^{\epsilon'}} $$ which determines the essential statement of Lemma \ref{shift-lem} indeed holds for all $N\in \mathbb{N}$ and appropriate $\epsilon'>0$. Furthermore, we showed previously (see \eqref{prob.good set}) that for a given $\epsilon>0$ and small enough $\sigma>0$ it holds that
\begin{align}
& \mathbb{P}\big(X\in \mathcal{G}^{N,\sigma,2\sigma}_{N^{-\frac{1}{3}},T} \big)\notag \\
\geq & 1-CN^{-\frac{1}{9}+\epsilon}-\underbrace{\mathbb{P } \big(X\in \mathcal{G}^{N,\sigma}_{1,T}\cap \big( \mathcal{G}^{N,\sigma}_{2,(0,T)} \big)^C \big) }_{\le C N^{-N^{\epsilon'}}} .
\end{align}
Hence, it follows that in this case
\begin{align*}
& \mathbb{P}(X\in \mathcal{G}_T^{N,\sigma})\\
\geq  & 1-CN^{-\frac{1}{9}+\epsilon}- \mathbb{P}\big(X\in \mathbb{R}^{6N}: \exists i\in \{1,...,N\}:X_i\notin \mathcal{L}_{\sigma}^N\big).
\end{align*}
Comparing the derived lower bound for $\mathbb{P}(X\in \mathcal{G}_T^{N,\sigma})$ with \eqref{statement thm.2+} yields that it suffices to show  that for typical configurations belonging to $\mathcal{G}_T^{N,\sigma}$ the desired statement holds (for the relevant values of $\sigma$ and $N$) since 
\begin{align}
& \mathbb{P}\big(X\in \mathbb{R}^{6N}:\sup_{0\le s \le T}|\Psi_{s,0}^{N,c_1}(X)-\Psi_{s,0}^{N,c_2}(X)|_{\infty}>  N^{-\frac{1}{2}+\sigma} \big) \notag \\
\le & \mathbb{P}\big(X\in \mathcal{G}_T^{N,\sigma} \land \sup_{0\le s \le T}|\Psi_{s,0}^{N,c_1}(X)-\Psi_{s,0}^{N,c_2}(X)|_{\infty}>  N^{-\frac{1}{2}+\sigma} \big)  + \mathbb{P}\big(X\in \big(\mathcal{G}_T^{N,\sigma}\big)^C \big).
\end{align}
For the applied approach it will be necessary to control also the deviations with respect to $|\cdot|_1$ so that we actually show a stronger statement than claimed in the theorem. Moreover, it will turn out to be helpful to distinguish between configurations where collisions of a certain impact occur and such where this is not the case. For a clear presentation we introduce stopping times like in the proof of Theorem \ref{thm1} where the first two of them  shall control the deviations
\begin{align}
 \tau_{dev,1}^{N,\sigma}(X):=\sup &\Big\{t\in [0,T]:\sup\limits_{0\le s \le t}|\Psi^{N,c_1}_{s,0}(X)-\Psi^{N,c_2}_{s,0}(X)|_{\infty}<  N^{-\frac{1}{2}+\frac{3}{5}\sigma}\Big\}, \notag \\
 \tau_{dev,2}^{N,\sigma}(X):=\sup &\Big\{t\in [0,T]:\sup\limits_{0\le s \le t}|\Psi^{N,c_1}_{s,0}(X)-\Psi^{N,c_2}_{s,0}(X)|_{1}< N^{-\frac{\sigma}{2}}\Big\} \label{stopp.time dev2}
\end{align} 
and the third the upper bound for the impact of single collisions
\begin{align}
 \tau_{col}^{N,\sigma}(X):=\sup &\Big\{t\in [0,T]: \ \Big( \forall (i,j)\in \mathcal{M}^{N,(0,T)}_{6N^{-\frac{1}{2}+\sigma},\infty}(X)\ \forall l\in \{1,2\}: \notag \\
& \ \ \int_{0}^t|f^N_{c_l}([\Psi^{N,c_l}_{s,0}(X)]_i-[\Psi^{N,c_l}_{s,0}(X)]_j)|ds< N^{\frac{1}{2}-\frac{5}{2}{{\sigma}}} \Big)\Big\}.  \label{stopp.time col}
\end{align}
We note that the definition the set $\mathcal{M}^{N,(t_1,t_2)}_{r,v}(X)$ is given in \eqref{def.M}. Moreover, we identify: $$\tau^{N,\sigma}(X):=\min(\tau^{N,\sigma}_{dev,1}(X),\tau^{N,\sigma}_{dev,2}(X),\tau^{N,\sigma}_{col}(X)).$$
If we are able to show that for small enough values of $\sigma>0$ and large enough $N$ $\mathbb{P}(X\in \mathcal{G}_T^{N,\sigma} \land  \tau^{N,\sigma}(X)<T)$ is smaller than order $N^{-\frac{1}{9}}$, then according to our previous reasoning the statement of the theorem follows.\\
Obviously, it holds that 
\begin{align}
&  \tau^{N,\sigma}(X)<T  \notag \\
 \Rightarrow &   \tau^{N,\sigma}_{dev,1}(X) < \tau^{N,\sigma}_{col}(X)\ \vee \label{ev.1}  \\
 &\tau^{N,\sigma}_{dev,2}(X) < \tau^{N,\sigma}_{col}(X) \ \vee \label{ev.2} \\
 &  \Big(\tau^{N,\sigma}_{col}(X) \le  \min\big( \tau^{N,\sigma}_{dev,1}(X), \tau^{N,\sigma}_{dev,2}(X)\big) \  \land \ \tau^{N,\sigma}_{col}(X)<T \Big)  \label{ev.3}
. \end{align}  
Thus, instead of proving directly that configurations where $\tau^{N,\sigma}(X)<T$ are untypical we will subsequently derive upper bounds for the probability of the three events given in the last implication (starting with \eqref{ev.1} and concluding with \eqref{ev.3}).\\ 
If we abbreviate for the moment
$$a^N(t,X):=\sup\limits_{0\le s \le t}|\Psi^{N,c_1}_{s,0}(X)-\Psi^{N,c_2}_{s,0}(X)|_{\infty}$$ and
$$b^N(t,X):=\max_{l\in \{1,2\}}\max_{(i,j)\in \mathcal{M}^{N,(0,T)}_{6N^{-\frac{1}{2}+\sigma},\infty}}\int_{0}^t|f^N_{c_l}([\Psi^{N,c_l}_{s,0}(X)]_i-[\Psi^{N,c_l}_{s,0}(X)]_j)|ds, $$ then one can see for example by the following relation that event \eqref{ev.1} is Borel measurable:
\begin{align*}
&\big\{  X\in \mathbb{R}^{6N}:\tau^{N,\sigma}_{dev,1}(X) < \tau^{N,\sigma}_{col}(X)\big \}\\
=& \bigcup_{t\in [0,T)}\big\{X\in \mathbb{R}^{6N}:a^N(t,X)\geq  N^{-\frac{1}{2}+\frac{3}{5}\sigma}\ \land \ b^N(t,X)<N^{\frac{1}{2}-\frac{5}{2}\sigma}
\big\}\\
=& \bigcup_{t\in \mathbb{Q}\cap [0,T)}\big\{X\in \mathbb{R}^{6N}:a^N(t,X)\geq  N^{-\frac{1}{2}+\frac{3}{5}\sigma}\big\} \cap\big\{ b^N(t,X)<N^{\frac{1}{2}-\frac{5}{2}\sigma}
\big\}
\end{align*}
While the `$\supseteq$'-relation between second and last line is obvious, the `$\subseteq$'-relation can be seen as follows: If $X\in \mathbb{R}^{6N}$ fulfills the conditions of the set in the second line for some $t\in [0,T)$, then it follows by monotony of $a^N(\cdot, X)$ and continuity of $b^N(\cdot,X)$ that $a^N(t', X)$ and $b^N(t',X)$ fulfill the desired claims as well for some sufficiently small $t'\geq t$ where $t'\in  \mathbb{Q}\cap [0,T)$. For the event `$\tau^{N,\sigma}_{dev,2}(X) < \tau^{N,\sigma}_{col}(X)$` an analogous reasoning works while for the event related to \eqref{ev.3} one should take into account that a configuration fulfills \eqref{ev.3} if and only if it fulfills $\tau_{col}^{N,\sigma}(X)<T$ but none of the already discussed constraints related to \eqref{ev.1} or \eqref{ev.2}. \\
Large parts of the subsequent estimates are very similar to the reasoning applied in previous proof. It is helpful for the comprehension of the different steps to keep in mind that for times $t$ before $\tau_{dev,1}^{N,\sigma}(X)$ is `triggered' 
\begin{align}
\max_{l \in \{1,2\}}|\Psi^{N,c_l}_{t,0}(X)-\Phi^{N,c_1}_{t,0}(X)|\le N^{-\frac{1}{2}+\sigma} \label{thm2.dist.mean.real}
\end{align} 
holds for the relevant $N$ and $\sigma$ which can be seen by application of Corollary \eqref{cor.shift.} and the definition of the stopping time. Thus, both trajectories can be `controlled' by the mean-field trajectory in most situations. Only if two particles have a `close encounter' (below an inter-particle distance of order $N^{-\frac{1}{2}+\sigma}$) some further arguments are needed. \\ 
We start by showing that the deviations with respect to $|\cdot|_{\infty}$ keep sufficiently small for times before $\tau^{N,\sigma}_{col}(X)$. In fact, we show a slightly stronger statement which will turn out to be helpful later.\\ 
For $X\in \mathcal{G}_T^{N,\sigma}\subseteq \mathcal{G}_{1,T}^{N,\sigma}\subseteq \bigcap_{i=1}^N \big( \mathcal{B}^{N,\sigma}_{4,i}\big)^C$ (see \eqref{def.B_4}) it holds for all $i\in \{1,...,N\}$ that $$\sum_{j=1}^N \mathbf{1}_{\mathcal{M}^{N,(0,T)}_{6N^{-\frac{1}{2}+\sigma},\infty}(X)}(i,j)=\sum_{j=1}^N \mathbf{1}_{M^{N,(0,T)}_{6N^{-\frac{1}{2}+\sigma},\infty}(X_i)}(X_j) \le N^{3\sigma}.$$
By application of this, the definition of the stopping times and the good properties of the considered initial data it follows for $Y:=\Psi^{N,c_2}_{t_1,0}(X)$, the relevant values of $N$ and $\sigma$ as well as times $t$ where $0\le t_1\le t\le \min\big(\tau_{dev,1}^{N,\sigma}(X),\tau^{N,\sigma}_{col}(X)\big) $ that
{\allowdisplaybreaks 
\begin{align}
& \frac{d}{dt_+}\sup_{t_1\le s\le t}|{^1\Psi^{N,c_2}_{s,0}}(X)-{^1\Psi^{N,c_1}_{s,t_1}}(Y)|_{\infty} \notag \\
\le & \sup_{t_1\le s \le t}|{^2\Psi^{N,c_2}_{s,0}}(X)-{^2\Psi^{N,c_1}_{s,t_1}}(Y)|_{\infty} \notag \\
\le & \frac{1}{N}\max_{1\le i\le N}  \sum_{j\neq i} \int_{t_1}^{t}\Big(\big|f^N_{c_2}([{^1\Psi^{N,c_2}_{s,0}}(X)]_j-[{^1\Psi^{N,c_2}_{s,0}}(X)]_i) \notag \\
& - f^N_{c_1}([{^1\Psi^{N,c_1}_{s,t_1}}(Y)]_j-[{^1\Psi^{N,c_1}_{s,t_1}}(Y)]_i)\big| \Big) ds \notag \\
\le & \frac{1}{N}\max_{1\le i\le N}  \sum_{j\neq i}\Big( \int_{t_1}^{t}\Big(\big|f^N_{c_2}([{^1\Psi^{N,c_2}_{s,0}}(X)]_j-[{^1\Psi^{N,c_2}_{s,0}}(X)]_i) \notag \\
& - f^N_{c_1}([{^1\Psi^{N,c_1}_{s,t_1}}(Y)]_j-[{^1\Psi^{N,c_1}_{s,t_1}}(Y)]_i)\big|  \Big)ds\mathbf{1}_{(\mathcal{M}^{N,(0,T)}_{6N^{-\frac{1}{2}+\sigma},\infty}(X))^C}(i,j)\Big) \notag   \\
&+ \frac{1}{N}\max_{1\le i\le N}  \sum_{j\neq i}\Big( \int_{t_1}^{t}\Big(\big|f^N_{c_2}([{^1\Psi^{N,c_2}_{s,0}}(X)]_j-[{^1\Psi^{N,c_2}_{s,0}}(X)]_i) \notag \\
& - f^N_{c_1}([{^1\Psi^{N,c_1}_{s,t_1}}(Y)]_j-[{^1\Psi^{N,c_1}_{s,t_1}}(Y)]_i)\big|  \Big)ds\mathbf{1}_{\mathcal{M}^{N,(0,T)}_{6N^{-\frac{1}{2}+\sigma},\infty}(X)}(i,j)\Big)\label{est.infty.dist.thm2,0}
\end{align}
According to Lemma \ref{shift-lem} and Corollary \ref{cor.shift.} (both applied in the third step) it holds for the considered configurations, $(i,j)\in \big(\mathcal{M}^{N,(0,T)}_{6N^{-\frac{1}{2}+\sigma},\infty}(X)\big)^C$ (see \eqref{def.M}), the relevant $N\in \mathbb{N}$, $\sigma>0$ and $s\in  [t_1,\tau^{N,\sigma}(X)]$ that
\begin{align*}
&|[^1\Psi^{N,c_1}_{s,t_1}(Y)]_j-[{^1\Psi^{N,c_1}_{s,t_1}}(Y)]_i| \\
\geq &|{^1\varphi^{N,c_1}_{s,0}}(X_j)-{^1\varphi^{N,c_1}_{s,0}}(X_i)|-2|\Psi^{N,c_1}_{s,t_1}(Y)-\Phi^{N,c_1}_{s,0}(X)|_{\infty}\\
\geq &|{^1\varphi^{N,c_1}_{s,0}}(X_j)-{^1\varphi^{N,c_1}_{s,0}}(X_i)|-2|\underbrace{\Psi^{N,c_1}_{s,t_1}(Y)}_{  =\Psi^{N,c_1}_{s,t_1}(\Psi^{N,c_2}_{t_1,0}(X))}-\Psi^{N,c_1}_{s,0}(X)|_{\infty}\\
&-2|\Psi^{N,c_1}_{s,0}(X)-\Phi^{N,c_1}_{s,0}(X)|_{\infty}\\
\geq & |{^1\varphi^{N,c_1}_{s,0}}(X_j)-{^1\varphi^{N,c_1}_{s,0}}(X_i)|-C\big(N^{- \frac{1}{2}+\frac{\sigma}{2}}+|\Psi^{N,c_2}_{t_1,0}(X)-\Psi^{N,c_1}_{t_1,0}(X)|_{\infty}\big)\\ & -CN^{-\frac{1}{2}+\frac{\sigma}{2}}\\
\geq & \underbrace{|{^1\varphi^{N,c_1}_{s,0}}(X_j)-{^1\varphi^{N,c_1}_{s,0}}(X_i)|}_{\geq 6N^{-\frac{1}{2}+\sigma}}-CN^{-\frac{1}{2}+\frac{3}{5}\sigma}\\
\geq & \frac{1}{2}|{^1\varphi^{N,c_1}_{s,0}}(X_j)-{^1\varphi^{N,c_1}_{s,0}}(X_i)|
\end{align*}
and due to relation \eqref{thm2.dist.mean.real}
\begin{align*}
&|[^1\Psi^{N,c_2}_{s,0}(X)]_i-[{^1\Psi^{N,c_2}_{s,0}}(X)]_j| \\
\geq &|{^1\varphi^{N,c_1}_{s,0}}(X_j)-{^1\varphi^{N,c_1}_{s,0}}(X_i)|
-2\underbrace{|\Psi^{N,c_2}_{s,0}(X)-\Phi^{N,c_1}_{s,0}(X)|_{\infty}}_{\le N^{-\frac{1}{2}+\sigma}}\\
\geq & \frac{1}{2}|{^1\varphi^{N,c_1}_{s,0}}(X_j)-{^1\varphi^{N,c_1}_{s,0}}(X_i)| .
\end{align*}
Regarding the `properties' of the map $g^N_c$ (and in particular \eqref{ineq.g^N+}) yields that term \eqref{est.infty.dist.thm2,0} is bounded by
\begin{align}
&  \frac{C}{N}\max_{1\le i\le N}  \sum_{j\neq i} \int_{t_1}^{t} g^N_{c_1}({^1\varphi^{N,c_1}_{s,0}}(X_j)-{^1\varphi^{N,c_1}_{s,0}}(X_i))  \mathbf{1}_{(\mathcal{M}^{N,(0,T)}_{6N^{-\frac{1}{2}+\sigma},\infty}(X))^C}(i,j) ds \notag \\
& \cdot \sup_{t_1\le s \le t}|{^1\Psi^{N,c_2}_{s,0}}(X)-{^1\Psi^{N,c_1}_{s,t_1}}(Y)|_{\infty}\notag\\ 
& + 2N^{-\frac{1}{2}-\frac{5}{2}\sigma}N^{3\sigma} \notag\\ 
\le & C \sup_{t_1\le s \le t}|{^1\Psi^{N,c_2}_{s,0}}(X)-{^1\Psi^{N,c_1}_{s,t_1}}(Y)|_{\infty}+ 2N^{-\frac{1}{2}+\frac{1}{2}\sigma}. \label{est.infty.dist.thm2}
\end{align}} 
For the upper limit $2N^{-\frac{1}{2}-\frac{5}{2}\sigma}N^{3\sigma}$ we simply applied that according to our preceding discussion $N^{3\sigma}$ constitutes an upper bound for the number of `close collisions' as well as that by definition of $\tau_{col}^{N,\sigma}(X)$ the impact of such a collision is bounded by $N^{\frac{1}{2}-\frac{5}{2}\sigma}$ on $[0,\tau_{col}^{N,\sigma}(X)]$. The last inequality, on the other hand, follows by the same reasoning as applied for estimates \eqref{est.sum.g+}.\\
After recalling that $Y:={\Psi^{N,c_2}_{t_1,0}}(X)$ one easily concludes by application of Gronwall`s Lemma that for such configurations 
\begin{align}
\begin{split}
&|{^1\Psi^{N,c_2}_{t,0}}(X)-{^1\Psi^{N,c_1}_{t,t_1}}({\Psi^{N,c_2}_{t_1,0}}(X))|_{\infty}\le CN^{-\frac{1}{2}+\frac{1}{2}\sigma}\big(e^{C(t-t_1)}-1\big) \label{est.length.t.int.}
\end{split}
\end{align}
holds for $t\in [t_1,\tau^{N,\sigma}_{col}(X)]$
which will be important shortly. Moreover, according to these estimates (applied for $t_1=0$) it holds for $t\in [0,\tau^{N,\sigma}_{col}(X)]$ that
\begin{align*}
&  \sup_{0\le s \le t}|{^2\Psi^{N,c_2}_{s,0}}(X)-{^2\Psi^{N,c_1}_{s,0}}(X)|_{\infty}\\
\le &  C \sup_{0\le s \le t}|{^1\Psi^{N,c_2}_{s,0}}(X)-{^1\Psi^{N,c_1}_{s,0}}(X)|_{\infty}+ 2N^{-\frac{1}{2}+\frac{1}{2}\sigma} 
\end{align*} 
which together with \eqref{est.length.t.int.} yields for the considered configurations that
\begin{align}
\sup\limits_{0\le s\le \tau^{N,\sigma}_{col}(X)}|\Psi^{N,c_1}_{s,0}(X)-\Psi^{N,c_2}_{s,0}(X)|_{\infty}\le  CN^{-\frac{1}{2}+\frac{\sigma}{2}}. \label{res.inf.norm thm.2}
\end{align} 
We remark that due to the previous estimates this inequality holds for all initial configurations of $\mathcal{G}_T^{N,\sigma}$ and thus the event $\tau^{N,\sigma}_{dev,1}(X)<\tau_{col}^{N,\sigma}(X)$ can not occur in this case (if $N$ is large enough and $\sigma$ sufficiently small).\\
Now we go on with the considerations for the distinctly more interesting `event' \eqref{ev.2} concerning the distance with respect to $|\cdot|_1$ between corresponding trajectories.\\ 
However, since the proof for this case is a bit elongated, it might be reasonable to start by introducing the heuristic proceeding. Like in the proof of Lemma \ref{c_1-Lem} we will compare corresponding trajectories on short time intervals (where this time both trajectories are subject to the $N$-particle dynamics but with different-sized cut-offs) and choose to this end again $\Delta:=N^{-\frac{1}{3}}$ for the length of these intervals. Instead of deriving upper bounds for the deviations $$\sup_{k \Delta \le s\le (k+1) \Delta}|\Psi^{N,c_1}_{s,0}(X)-\Psi^{N,c_2}_{s,0}(X)|_1$$ on each of these intervals, it turns out that it suffices to do this for
\begin{align}
\sum_{i=1}^N\sup\limits_{k\Delta\le s\le (k+1)\Delta} |[{^l\Psi^{N,c_1}_{s,k\Delta}(\Psi^{N,c_2}_{k\Delta,0}(X))}]_i-[{^l\Psi^{N,c_2}_{s,0}(X)}]_i|, \ l=1,2. \label{expl.dev.}
\end{align}
If we are able to show that the deviations related to \eqref{expl.dev.} are sufficiently small for all intervals, then $X\in \mathcal{G}_T^{N,\sigma}\subseteq \mathcal{G}^{N,\sigma,2\sigma}_{N^{-\frac{1}{3}},T}$ implies by application of Corollary \ref{Cor.shift2} for the map $h(\p):=\Psi^{N,c_2}_{\p,0}(X)$ that also the actually relevant deviation 
$$ \sup_{0 \le s\le T}|\Psi^{N,c_1}_{s,0}(X)-\Psi^{N,c_2}_{s,0}(X)|_1$$ keeps small enough. Hence, it remains to discuss how terms of the form \eqref{expl.dev.} can be estimated in a reasonable way. We will introduce the most important aspects of the approach by application of sketch \eqref{fig.Thm2} which outlines the relevant trajectories for the short interval $[4\Delta,5\Delta]$. 
 More precisely, the sketch shows on the one hand two trajectories which start at a `good' initial condition $X\in \mathcal{G}_T^{N,\sigma}$ but evolve by different dynamics (due to the different-sized cut-offs) and on the other hand the auxiliary trajectory $\Psi^{N,c_1}_{\cdot, 4\Delta}(\Psi^{N,c_2}_{4\Delta,0}(X))$ which according to \eqref{expl.dev.} shall be `compared' to $\Psi^{N,c_2}_{\cdot,0}(X)$ on the fifth interval. The auxiliary trajectory is chosen such that its position coincides with the position of the trajectory $\Psi^{N,c_2}_{\cdot,0}(X)$ at the time $4\Delta$ (resp. at the beginning of the short interval) but it evolves with respect to the dynamics where the cut-off parameter is $c_1=\frac{2}{3}$ for which we have derived some very helpful results previously. Due to the first part of the proof we already know that this auxiliary trajectory and $\Psi^{N,c_1}_{\cdot,0}(X)$ are `close' with respect to $|\cdot|_\infty$ at time $4\Delta$ if $4\Delta \le \tau^{N,\sigma}_{col}(X)$ and since both evolve with respect to the well-studied dynamics this allows us to apply Lemma \ref{shift-lem} which in turn yields us plenty of `good' properties for the auxiliary trajectory. In particular, we obtain that its position at the starting time $\Psi^{N,c_1}_{0,4\Delta}(\Psi^{N,c_2}_{4\Delta,0}(X))$ belongs to $\mathcal{G}^{N,\sigma}_{3,T}$ which will provide us the opportunity to make probability estimates for these trajectories by application of Lemma \ref{prod-Lem}. We want to derive suitable upper bounds for the deviations \eqref{expl.dev.} and to this end one should recall that the force kernels related to the compared dynamics differ only by their cut-off size (which in both cases is very small). Thus, the distance between the auxiliary trajectory and $\Psi^{N,c_2}_{\cdot,0}(X)$ starts to grow as soon as two particles get close enough such that the larger cut-off related to parameter $c_1$ `comes into play'. Of course, in principle the arising spatial deviations transfer to the remaining pairs of corresponding particles but since the length of the intervals $\Delta=N^{-\frac{1}{3}}$ is very short and the dynamics are second order the spatial deviations are negligibly small compared to the velocity deviations so that this `effect' does not matter for the estimates. If the number of `close' collisions would be of comparable order as in the i.i.d. `mean field particle system', then these considerations suggest that the value of integral \eqref{expl.dev.} should typically be bounded by
\begin{align*}
\underbrace{\big(CN^2(N^{-c_1})^2\Delta\big)}_{\scriptsize\begin{array}{l}\texttt{number of particle pairs}\\ \texttt{coming closer than }N^{-c_1}\\\texttt{on }[k\Delta,(k+1)\Delta]\end{array}}\underbrace{\frac{1}{N}\frac{C}{(N^{-c_1})^{(\alpha-1)}}}_{\scriptsize\begin{array}{l}\texttt{impact of majority}\\ \texttt{of the}\\ \texttt{related collisions}\end{array}}\le CN^{2-2\frac{2}{3}}N^{-1+(-\frac{2}{3})(-\frac{1}{3})}\Delta\le CN^{-\frac{1}{9}}\Delta.
\end{align*}
This would in any case pose a sufficiently small upper bound, however, first one has to justify the assumptions and we try to give a rough idea how this can be done in the following part. We point out that at the moment we only consider times before $\tau^{N,\sigma}_{col}(X)$ (resp. where we have a small enough upper bound for the `impact' of single collisions). Thus, it is straightforward to see that for times in $[4 \Delta,5\Delta]$ (resp. in general $[k\Delta,(k+1)\Delta]$) all pairs of corresponding particles keep very close to each other in phase space and even more in position space if $\tau^{N,\sigma}_{col}(X)$ is not yet `triggered'. Hence, the auxiliary trajectory determines an excellent approximation of $\Psi^{N,c_2}_{\cdot,0}(X)$ on $[k\Delta, (k+1)\Delta]$: If the minimal distance between two of these auxiliary particles on $[k\Delta,(k+1)\Delta]$ is given by $r_{min}$ and the value of their relative velocity at the time of their `encounter' by $v_{min}$, then the corresponding particles `related to' $\Psi^{N,c_2}_{\cdot,0}(X)$ also attain a minimal distance of order $r_{min}$ while the value of their relative velocity at this moment is of order $v_{min}$. The last statement is at least true up to very rare occasions where $r_{min}$ is extraordinarily small but it will turn out that these remaining (rare) collisions do not pose a problem. This yields that in principle it suffices to know the respective numbers of collisions characterized by $r_{min}$ and $v_{min}$ on $[k\Delta,(k+1)\Delta]$ for the auxiliary trajectory to derive a suitable upper bound for the caused deviation. To carry out the suggested strategy we need to be capable of doing probability estimates for the auxiliary system. This, however, as mentioned previously is possible without any problems: To this end, one has to regard that the auxiliary trajectory evolves by the well-studied dynamics and its initial position belongs to the `good' set $\mathcal{G}_{3,T}^{N,\sigma}$. Moreover, we recall that if the i.i.d. initial positions of the particles are given by $X\in \mathcal{G}_T^{N,\sigma}$, then the initial positions (resp. the positions at time $t=0$) of the corresponding auxiliary particles which are applied for the short interval $[k\Delta,(k+1)\Delta]$ are given by $\Psi^{N,c_1}_{0,k\Delta}(\Psi^{N,c_2}_{k\Delta,0}(X))$. Due to Lemma \ref{shift-lem} it follows that the distance between these corresponding initial configurations with respect to $|\cdot |_1$ is of the same order as the distance $|\Psi^{N,c_2}_{k\Delta,0}(X)-\Psi^{N,c_1}_{k\Delta,0}(X)|_1$ (or of order $N^{-\sigma}$). This yields that though the positions of the auxiliary particles at the starting time $t=0$ are not i.i.d. (if $k\geq 1$), they are still close enough to the positions of the i.i.d. particles $X_i,\ i\in \{1,...,N\}$ such that
$$ \prod_{i=1}^N k_0([\Psi^{N,c_1}_{0,k\Delta}(\Psi^{N,c_2}_{k\Delta,0}(X))]_i) \approx  \prod_{i=1}^N k_0(X_i)$$ is fulfilled provided that $k\Delta \le \tau^{N,\sigma}_{dev,2}(X)$. In this case, however, it is straightforward to see by application of Lemma \ref{c_1-Lem} that restricted to configurations fulfilling this constraint the number of `problematic collisions' between auxiliary particles on $[k\Delta,(k+1)\Delta]$ keeps typically `small' which then again implies that indeed $|\Psi^{N,c_2}_{\cdot ,0}(X)-\Psi^{N,c_1}_{\cdot ,0}(X)|_1$ grows slow enough such that $\tau^{N,\sigma}_{dev,2}(X)$ does not get `triggered'.\\  Finally, we conclude the heuristic introduction and start with the detailed considerations for the event $\tau^{N,\sigma}_{dev,2}(X) < \tau^{N,\sigma}_{col}(X)$. For ease of notation we identify in the following $t^N_k:=k\Delta $ and $t^N_{k,X}:=\min\big(k\Delta, \tau^{N,\sigma}_{col}(X) \big)$ for $k\in \{0,...,\lceil \frac{T}{\Delta}\rceil-1\}$. First, we recall that for $X\in \mathcal{G}_T^{N,\sigma} \subseteq \mathcal{G}^{N,\sigma,2\sigma}_{\Delta,T}$ it suffices to verify that $\Psi^{N,c_2}_{\cdot,0}(X)$ fulfills the conditions on the map $h$ of Corollary \ref{Cor.shift2} where the parameters which appear in the assumptions are set to $\epsilon_0=2\sigma$, $\epsilon_1=\sigma$ and $a=N^{-\sigma}$ which leads to the constraints
\begin{enumerate}
\item[(i)] $\displaystyle\sum_{i=1}^N\sup\limits_{t^N_k\le s\le t^N_{k+1,X}} |[{^1\Psi^{N,c_1}_{s,t_k^N}(\Psi^{N,c_2}_{t^N_k,0}(X))}]_i-[{^1\Psi^{N,c_2}_{s,0}(X)}]_i|\le CN^{-\sigma}\Delta^2 $
\item[(ii)] $\displaystyle\sum_{i=1}^N\sup\limits_{t^N_k\le s\le t^N_{k+1,X}} |[{^2\Psi^{N,c_1}_{s,t_k^N}(\Psi^{N,c_2}_{t^N_k,0}(X))}]_i-[{^2\Psi^{N,c_2}_{s,0}(X)}]_i|\le CN^{-\sigma}\Delta $
\item[(iii)] $\sup\limits_{0\le s \le \tau^{N,\sigma}_{col}(X)}|\Psi^{N,c_1}_{s,0}(X)-\Psi^{N,c_2}_{s,0}(X)|_\infty\le N^{-\frac{1}{2}+\frac{3}{5}\sigma} $
\end{enumerate}
for all $k\in \{0,...,\lceil \frac{T}{\Delta}\rceil-1\}$ to be able to conclude that
\begin{align}
\sup\limits_{0\le s\le \tau^{N,\sigma}_{col}(X)} |{^1\Psi^{N,c_1}_{s,0}(X)}-{^1\Psi^{N,c_2}_{s,0}(X)}|_1\le CN^{-\sigma} . \label{list}
\end{align}  
According to our previous estimates the constraint related to item (iii) is fulfilled for the relevant $N$ and $\sigma$. Thus, we can restrict ourselves to proving that also the first two conditions are typically fulfilled. We note that for a certain configuration $X$ obviously only intervals are relevant where still $t^N_k\le \tau^{N,\sigma}_{col}(X)$ because otherwise the interval $[t_k^N,t^N_{k+1,X}]$ is empty. Hence, for a certain $X$ the stopping time $ \tau^{N,\sigma}_{col}(X)$ basically corresponds to the end time $T$ applied in the assumptions of Corollary \ref{Cor.shift2}.\\
In the proof of Lemma \ref{c_1-Lem} the length of the short time intervals $\Delta$ was chosen such that the mean-field dynamics typically provided a sufficiently good `approximation' for the trajectories of the interacting particles. This time we want to `approximate' the trajectories belonging to the system with a cut-off diameter of order $N^{-c_2}$ by those which are subject to $(\Psi^{N,c_1}_{s,t})_{s,t\in \mathbb{R}}$. Thus, we need for the same reason as in the previous case that corresponding trajectories of the compared dynamics keep very close in position space for the considered time span. A reasonable order for the allowed distance is $N^{-\frac{7}{9}}$ which is in correspondence to the smallest allowed value for $r$ in the assumptions of Lemma \ref{c_1-Lem}.  \\
The choice $\Delta :=N^{-\frac{1}{3}}$ fulfills this requirement as long as $\tau_{col}^{N,\sigma}(X)$ is not `triggered' because relation \eqref{est.length.t.int.} yields for $X\in \mathcal{G}_T^{N,\sigma}$, large enough $N$, small enough $\sigma>0$ and $t^N_k\le t \le t^N_{k+1,X}$ that
\begin{align}
&|{^1\Psi^{N,c_2}_{t,0}}(X)-{^1\Psi^{N,c_1}_{t,t_k^N}}(\Psi^{N,c_2}_{t_k^N,0}(X))|_{\infty}\notag \\
\le &  C N^{-\frac{1}{2}+\frac{1}{2}\sigma}\big(e^{CN^{-\frac{1}{3}}}-1\big)  \le CN^{-\frac{5}6+\frac{\sigma}{2}}\le N^{-\frac{7}{9}} \label{app.cond.}
\end{align}  
After these considerations we can start to show that constraints (i) and (ii) stated in the list previous to \eqref{list} are typically fulfilled for configurations $X\in  \mathcal{G}^{N,\sigma}$ and all $k\in \{0,...,\lfloor \frac{T}{\Delta} \rfloor-1\}$ if $N$ is large and $\sigma$ small enough. Since these conditions are only non-trivial for time intervals where $t_k^N\le t^N_{k+1,X}=\min\big(t^N_{k+1}, \tau^{N,\sigma}_{col}(X) \big)$, we will assume for the following estimates that for a given $X$ the value of $k$ is small enough such that this relation is still fulfilled. Furthermore, we abbreviate $Y=\Psi^{N,c_2}_{t_k^N,0}(X)$ and obtain by a proceeding which is very similar to the estimates applied for controlling the corresponding quantities in the proof of Lemma \ref{shift-lem} (see \eqref{shift.t.1-norm}) that for $t\in [t_k, \tau_{col}^{N,\sigma}(X)]$:
{\allowdisplaybreaks
\begin{align}
& \frac{d}{dt_+}\sum_{i=1}^N\sup_{0 \le s \le t-t_k^N}|[^1\Psi^{N,c_1}_{s,0}(Y)]_i-[{^1\Psi^{N,c_2}_{s,0}}(Y)]_i|\\
\le & \sum_{i=1}^N\sup_{0 \le s \le t-t_k^N}|[^2\Psi^{N,c_1}_{s,0}(Y)]_i-[{^2\Psi^{N,c_2}_{s,0}}(Y)]_i|\\
\le &  \frac{1}{N} \sum_{i=1}^N\sum_{j\neq i}\Big( \int_{0}^{t-t_k^N}\big|f^N_{c_1}([^1\Psi^{N,c_1}_{s,0}(Y)]_j-[^1\Psi_{s,0}^{N,c_1}(Y)]_i) \notag \\
& -f^N_{c_2}([^1\Psi^{N,c_2}_{s,0}(Y)]_j-[^1\Psi^{N,c_2}_{s,0}(Y)]_i)\big|ds\Big)\\
\le &  \frac{C}{N}\sum\limits_{(i,j)\in \big(\mathcal{M}^{N,(t_k^N,t_{k+1}^N)}_{6N^{-\frac{1}{2}+{\sigma}},\infty}(X)\big)^C }\Big(\int^t_{t^N_{k}}g^N_{c_1}\big(\varphi^{N,c_1}_{s,0}(X_j)-\varphi^{N,c_1}_{s,0}(X_i)\big)ds \notag \\
& \cdot \big( \sum_{n\in \{i,j\}}\sup_{0\le s\le t-t_k^N}|[^1\Psi^{N,c_1}_{s,0}(Y)]_n-[^1\Psi_{s,0}^{N,c_2}(Y)]_n |\big)\Big) \notag \\ 
& + \frac{1}{N}\sum\limits_{(i,j)\in \mathcal{M}^{N,(t_k^N,t_{k+1}^N)}_{6N^{-\frac{1}{2}+\sigma},\infty}(X) }\Big(\int_{0}^{t-t_k^N}\big|f^N_{c_1}([^1\Psi^{N,c_1}_{s,0}(Y)]_j-[^1\Psi_{s,0}^{N,c_1}(Y)]_i) \notag \\
& \ -f^N_{c_2}([^1\Psi^{N,c_2}_{s,0}(Y)]_j-[^1\Psi^{N,c_2}_{s,0}(Y)]_i)\big|ds\Big) \label{term.coll.class0}
\end{align}}
where the derivation of the last step is analogous to the reasoning after \eqref{est.infty.dist.thm2,0} and takes into account that the force kernels coincide if the particles keep a minimal distance of at least $N^{-c_1}=N^{-\frac{2}{3}}$ to each other.\\
We continue the estimates and observe due to the symmetry
$$(i,j)\in \mathcal{M}^{N,(t^N_k,t_{k+1}^N)}_{6N^{-\frac{1}{2}+\sigma},\infty}(X)\Leftrightarrow (j,i)\in \mathcal{M}^{N,(t^N_k,t_{k+1}^N)}_{6N^{-\frac{1}{2}+\sigma},\infty}(X)$$ that \eqref{term.coll.class0} is bounded by
\begin{align}
& \frac{C}{N}\sum_{i=1}^N \sup_{0 \le s \le t-t_k^N} |[^1\Psi^{N,c_1}_{s,0}(Y)]_i-[^1\Psi_{s,0}^{N,c_2}(Y)]_i |\notag \\
& \cdot\max_{n\in \{1,...,N\}}\sum\limits_{j=1}^N\int_{t_k^N}^{t}g^N_{c_1}\big({^1\varphi^{N,c_1}_{s,0}}(X_j)-{^1\varphi^{N,c_1}_{s,0}}(X_n)\big)\mathbf{1}_{\big(\mathcal{M}^{N,(0,T)}_{6N^{-\frac{1}{2}+{\sigma}},\infty}(X)\big)^C }(n,j)ds \notag \\ 
& + \frac{1}{N}\sum\limits_{(i,j)\in \mathcal{M}^{N,(t^N_k,t_{k+1}^N)}_{6N^{-\frac{1}{2}+\sigma},\infty}(X) }\Big(\int_{0}^{t-t_k^N}\big|f^N_{c_1}([^1\Psi^{N,c_1}_{s,0}(Y)]_j-[^1\Psi_{s,0}^{N,c_1}(Y)]_i) \notag \\
& \ -f^N_{c_2}([^1\Psi^{N,c_2}_{s,0}(Y)]_j-[^1\Psi^{N,c_2}_{s,0}(Y)]_i)\big|ds\Big) \label{term.coll.class}
\end{align}
Furthermore, once again $X\in \mathcal{G}_T^{N,\sigma}\subseteq  \bigcap_{i=1}^N (\mathcal{B}_{2,i}^{N,\sigma})^C\cap  (\mathcal{B}^{N,\sigma}_5)^C $ (defined in \eqref{def.B_2} and \eqref{def.B_5}) implies together with estimates \eqref{est.sum.g+} that for all $i\in \{1,...,N\}$
\begin{align*}
&\frac{1}{N}\sum\limits_{j=1}^N\int_{t_k^N}^{t}g^N_{c_1}\big({^1\varphi^{N,c_1}_{s,0}}(X_j)-{^1\varphi^{N,c_1}_{s,0}}(X_i)\big)\mathbf{1}_{\big(\mathcal{M}^{N,(0,T)}_{6N^{-\frac{1}{2}+{\sigma}},\infty}(X)\big)^C }(i,j)ds
\le  C.
\end{align*} 
In total this yields for $t^N_k\le t\le t^N_{k+1,X}$ that
\begin{align}
& \frac{d}{dt_+}\sum_{i=1}^N\sup_{0 \le s \le t-t_k^N}|[^1\Psi^{N,c_1}_{s,0}(Y)]_i-[{^1\Psi}^{N,c_2}_{s,0}(Y)]_i| \notag \\
\le & \sum_{i=1}^N\sup_{0 \le s \le t-t_k^N}|[^2\Psi^{N,c_1}_{s,0}(Y)]_i-[{^2\Psi}^{N,c_2}_{s,0}(Y)]_i| \notag \\
\le &  C \sum_{i=1}^N\sup_{0 \le s \le t-t_k^N}|[^1\Psi^{N,c_1}_{s,0}(Y)]_i-[{^1\Psi}^{N,c_2}_{s,0}(Y)]_i| \notag \\
 & + \frac{C}{N}\sum\limits_{(i,j)\in \mathcal{M}^{N,(t_k^N,t_{k+1}^N)}_{6N^{-\frac{1}{2}+\sigma},\infty}(X) }\Big(\int_{0}^{t-t_k^N}\big|f^N_{c_1}([^1\Psi^{N,c_1}_{s,0}(Y)]_j-[^1\Psi_{s,0}^{N,c_1}(Y)]_i) \notag \\
& \ -f^N_{c_2}([^1\Psi^{N,c_2}_{s,0}(Y)]_j-[^1\Psi^{N,c_2}_{s,0}(Y)]_i)\big|ds\Big) \label{term.coll.class1}
\end{align}
Until this point all estimates hold for arbitrary initial conditions of the `good' set $\mathcal{G}_T^{N,\sigma}$. The term related to the last two lines of this expression arises from collisions where particles get at least as close to each other as order $N^{-\frac{1}{2}+\sigma}$ and in the following want to show that for typical initial data of $\mathcal{G}_T^{N,\sigma}$ it keeps small enough. Furthermore, this term determines the only significant difference to the proof of Lemma \ref{shift-lem} because in contrast to previous situations the parameter $c_2$ can be chosen arbitrarily large which in turn leads to an arbitrarily small cut-off radius $N^{-c_2}$ (even for fixed $N$). Consequently, we have to argue that the number of very close `encounters' between particles does not exceed a critical value. The recently introduced lemmas provide us sufficient knowledge about the dynamics of the system with cut-off radius $N^{-c_1}$ and by the closeness of corresponding trajectories on the considered time intervals $[t_k^N,t_{k+1,X}^N]$ (where obviously only the non-trivial intervals are of interest) it is likewise possible to `transfer' some of these information on the system with smaller cut-off. \\
For this purpose we have to define certain sets which can be applied to classify particles by means of the `collision types' they experience.\\
In correspondence to the sets $\mathcal{R}^{N,(t_1,t_2)}_{r,v}(X)$ introduced in \eqref{def.R} we define additionally for $r,R,v,V\in \mathbb{R}_{\geq 0}\cup \{\infty\}$, $t_1,t_2\in  [0,T]$ and $X\in \mathbb{R}^{6N}$
\begin{align}
\mathcal{R}^{N,(t_1,t_2)}_{(r,R),(v,V)}(X):=\Big\{(i&,j)\in  \{1,...,N\}^2\setminus \bigcup_{n=1}^N \{(n,n)\} :\big( \exists t \in [t_1,t_2]: \notag \\
 r&\le  \min\limits_{t_1\le s \le t_2}|[^1\Psi^{N,c_1}_{s,0}(X)]_j-[{^1\Psi^{N,c_1}_{s,0}(X)}]_i| \notag \\
&=|[^1\Psi^{N,c_1}_{t,0}(X)]_j-[{^1\Psi^{N,c_1}_{t,0}(X)}]_i|\le R,  \notag \\
v&\le  |[^2\Psi^{N,c_1}_{t,0}(X)]_j-[{^2\Psi^{N,c_1}_{t,0}(X)]_i}|\le V \big)\Big\}.  \label{def.real.coll.class}
\end{align}
After some preliminary considerations we will apply these sets to define a cover of $\mathcal{M}^{N,(t^N_k,t_{k,X}^N)}_{6N^{-\frac{1}{2}+\sigma},\infty}(X)$ which will provide us the possibility to derive an upper bound for the remaining term \eqref{term.coll.class1} (which at least holds typically).\\ Now we arrived at the first situation where Lemma \ref{c_1-Lem} will become crucial. In the following we abbreviate
$$\tau^{N,\sigma}_{dev}(X):=\min\big( \tau^{N,\sigma}_{dev,1}(X),\tau^{N,\sigma}_{dev,2}(X)\big).$$ For $0\le t\le \tau^{N,\sigma}_{col}(X)$ it holds according to our previous estimates (see \eqref{res.inf.norm thm.2}) and the relevant $N,\sigma$ that
$$|\Psi^{N,c_1}_{t,0}(X)-\Psi^{N,c_2}_{t,0}(X)|_{\infty}\le CN^{-\frac{1}{2}+\frac{\sigma}{2}}\le  N^{-\frac{1}{2}+\frac{3}{5}\sigma}$$
which implies that $\tau^{N,\sigma}_{col}(X)\le \tau^{N,\sigma}_{dev,1}(X)$ holds in this case. Thus, if $t_k=k\Delta\le \tau^{N,\sigma}_{col}(X)$ and $X\in \mathcal{G}_T^{N,\sigma}\subseteq \mathcal{G}^{N,\sigma}_{2,(0,T)}$, then we can apply Lemma \ref{shift-lem} to obtain that 
\begin{align*}
&|X-\Psi^{N,c_1}_{0,t_k}(\Psi^{N,c_2}_{t_k,0}(X))|_{\infty}\\
\le &  C\big(N^{-\frac{1}{2}+\frac{\sigma}{2}}+|\Psi^{N,c_1}_{t_k,0}(X)-\Psi^{N,c_2}_{t_k,0}(X)|_{\infty}\big)\le  CN^{-\frac{1}{2}+\frac{3}{5}\sigma}.
\end{align*}
Since $X\in \mathcal{G}_T^{N,\sigma}$, this yields in turn that $\Psi^{N,c_1}_{0,t_k}(\Psi^{N,c_2}_{t_k,0}(X))\in \mathcal{G}^{N,\sigma}_{3,T}$ for large enough $N$ (see definition \eqref{def.G_3}). Hence, we abbreviate $\widetilde{X}^{N,k}:=\Psi^{N,c_1}_{0,t_k}(\Psi^{N,c_2}_{t_k,0}(X))$ and conclude that the following implication holds (for the relevant $N,\sigma$):
\begin{align*}
&X\in  \mathcal{G}_T^{N,\sigma} \land  \tau_{col}^{N,\sigma}(X)\geq k\Delta  \land (i,j)\in \mathcal{R}^{N,(0,\Delta)}_{R,V}(\underbrace{\Psi^{N,c_2}_{t_k,0}(X)}_{=\Psi^{N,c_1}_{t_k,0}(\widetilde{X}^{N,k})} )\\
\Rightarrow & \widetilde{X}^{N,k}\in  \mathcal{G}^{N,\sigma}_{3,T} \land  \tau^{N,\sigma}_{dev,1}(X)\geq k\Delta  \land  (i,j)\in\mathcal{R}^{N,(0,\Delta)}_{R,V}(\Psi^{N,c_1}_{t_k,0}(\widetilde{X}^{N,k}))
\end{align*}  
For $X\in \mathcal{G}_T^{N,\sigma}$ it holds in particular that $X_i \in \mathcal{L}^N_{\sigma}$ and by application of the Lipschitz property stated in \eqref{Def.L} (which is applicable because according to our previous reasoning $|X-\widetilde{X}^{N,k}|_{\infty}\le CN^{-\frac{1}{2}+\frac{3}{5}\sigma}$) it follows basically analogously to estimates \eqref{est.dens.0} used in the proof of Lemma \ref{prod-Lem} that
\begin{align*}
& \prod_{i=1}^N k_0(X_i)\\
\le &\prod_{i=1}^N \big( |k_0(X_i)-k_0([\widetilde{X}^{N,k}]_i)|+ k_0([\widetilde{X}^{N,k}]_i)\big)\\
\le &\prod_{i=1}^N \big(1+ N^{\frac{\sigma}{2}}|X_i-[\widetilde{X}^{N,k}]_i|\big)k_0([\widetilde{X}^{N,k}]_i)\\
\le & e^{N^{\frac{\sigma}{2}}|X-\widetilde{X}^{N,k}|_1 }\prod_{i=1}^N k_0([\widetilde{X}^{N,k}]_i).
\end{align*}
Since $\widetilde{X}^{N,k}:=\Psi^{N,c_1}_{0,t_k}(\Psi^{N,c_2}_{t_k,0}(X))$, this corresponds to 
\begin{align}
F_{t_k}^{N,c_2}(\Psi^{N,c_2}_{t_k,0}(X))\le  e^{N^{\frac{\sigma}{2}}|X-\widetilde{X}^{N,k}|_1 }F_{t_k}^{N,c_1}(\Psi^{N,c_2}_{t_k,0}(X)) \label{est.dens.}
\end{align} which will be applied later during the proof. Moreover, since $|X-\widetilde{X}^{N,k}|_{\infty}\le CN^{-\frac{1}{2}+\frac{3}{5}\sigma}$ and $X\in \mathcal{G}_T^{N,\sigma}\subseteq \mathcal{G}^{N,\sigma}_{2,(0,T)}$ we can apply Lemma \ref{shift-lem} to obtain that for $t_k= k \Delta \le \tau^{N,\sigma}_{dev}(X)$ 
\begin{align*}
& N^{\frac{\sigma}{2}}|X-\widetilde{X}^{N,k}|_1 \le CN^{\frac{\sigma}{2}}\big(|\Psi^{N,c_1}_{t_k,0}(X)-\Psi^{N,c_2}_{t_k,0}(X)|_1+N^{-\sigma}\big)\le C
\end{align*}
where for the last conclusion we simply applied the definition of the stopping time (see \eqref{stopp.time dev2}). In the following we maintain the abbreviation $\widetilde{X}^{N,k}:=\Psi^{N,c_1}_{0,t_k}(\Psi^{N,c_2}_{t_k,0}(X))$ and abbreviate additionally $$M:=\big(8N^{\frac{1}{3}+4\sigma}\big)\big\lceil N^{\frac{5}{3}}R^2\min(V,1)^3\big(\min(V,1)+N^{\frac{1}{3}}R\big)\big\rceil. $$ Then it follows by merging these consideration subsequently that
{ \allowdisplaybreaks
\begin{align}
& \mathbb{P}\Big(X\in\mathbb{R}^{6N}: X\in  \mathcal{G}_T^{N,\sigma} \land  \min\big(\tau^{N,\sigma}_{col}(X),\tau^{N,\sigma}_{dev,2}(X)\big)\geq k\Delta \ \land \notag  \\
& \hspace{5,3cm} | \mathcal{R}^{N,(0,\Delta)}_{R,V}(\Psi^{N,c_2}_{k\Delta,0}(X))|\geq M \Delta\Big) \notag \\
 \le & \mathbb{P}\Big(X\in\mathbb{R}^{6N}: \widetilde{X}^{N,k}\in  \mathcal{G}_{3,T}^{N,\sigma}\land \tau^{N,\sigma}_{dev}(X)\geq k\Delta  \land   | \mathcal{R}^{N,(0,\Delta)}_{R,V}(\Psi^{N,c_1}_{k\Delta,0}(\widetilde{X}^{N,k}))|\geq M \Delta \Big) \notag \\
 =&  \int_{\mathbb{R}^{6N}} \mathbf{1}_{\mathcal{G}^{N,\sigma}_{3,T}}(\widetilde{X}^{N,k}) \prod_{i=1}^Nk_0(X_i)  \notag \\ & \cdot \mathbf{1}_{\{Y\in \mathbb{R}^{6N}:\tau^{N,\sigma}_{dev}(Y)\geq k\Delta \}}(X) \mathbf{1}_{\{Y\in \mathbb{R}^{6N}: | \mathcal{R}^{N,(0,\Delta)}_{R,V}(\Psi^{N,c_1}_{k\Delta,0}(Y))|\geq M \Delta \}}(\widetilde{X}^{N,k}) d^{6N}X \notag \\
 \le & \int_{\mathbb{R}^{6N}} \underbrace{e^{N^{\frac{\sigma}{2}}|X-\widetilde{X}^{N,k}|_1 }}_{\le C}\mathbf{1}_{\mathcal{G}^{N,\sigma}_{3,T}}(\widetilde{X}^{N,k}) \prod_{i=1}^Nk_0([\widetilde{X}^{N,k}]_i)  \notag\\
 &\cdot \mathbf{1}_{\{Y\in \mathbb{R}^{6N}:\tau^{N,\sigma}_{dev}(Y)\geq k\Delta \}}(X) \mathbf{1}_{\{Y\in \mathbb{R}^{6N}: | \mathcal{R}^{N,(0,\Delta)}_{R,V}(\Psi^{N,c_1}_{k\Delta,0}(Y))|\geq M \Delta \}}(\widetilde{X}^{N,k}) d^{6N}X \notag \\
 \le & C \int_{\mathbb{R}^{6N}} \mathbf{1}_{\mathcal{G}^{N,\sigma}_{3,T}}(\widetilde{X}^{N,k}) \mathbf{1}_{\{Y\in \mathbb{R}^{6N}:| \mathcal{R}^{N,(0,\Delta)}_{R,V}(\Psi^{N,c_1}_{k\Delta,0}(Y))|\geq M \Delta \}}(\widetilde{X}^{N,k}) \prod_{i=1}^Nk_0([\widetilde{X}^{N,k}]_i)   d^{6N}X \notag \\
\le  & C\mathbb{P}\Big(X\in\mathbb{R}^{6N}: X\in  \mathcal{G}_{3,T}^{N,\sigma}  \land | \mathcal{R}^{N,(0,\Delta)}_{R,V}(\Psi^{N,c_1}_{k\Delta,0}(X))|\geq M \Delta\Big).   
 \label{prob.coll.class0}
\end{align} }
We assume that $R\geq N^{-\frac{7}{9}} $ and continue by applying on the one hand the relation $$\mathcal{R}^{N,(0,\Delta)}_{R,V}(\Psi^{N,c_1}_{k\Delta,0}(Z))=\mathcal{R}^{N,(k\Delta,(k+1)\Delta)}_{R,V}(Z)$$ (which can be easily verified by regarding the definition \eqref{def.R}) and on the other hand Lemma \ref{c_1-Lem} to estimate the last term further:
\begin{align}
&C\mathbb{P}\Big(X\in\mathbb{R}^{6N}: X\in  \mathcal{G}_{3,T}^{N,\sigma}  \land | \mathcal{R}^{N,(k\Delta,(k+1)\Delta)}_{R,V}(X)|\geq M \Delta\Big)  \notag \\ 
 \le & N^{\frac{1}{3}}\Delta \Big(\frac{C  N^{2+3{\sigma}}R^2\min(V,1)^3}{M}(\min(V,1)+N^{\frac{1}{3}}R)\Big)^{\frac{M}{8}N^{-\frac{1}{3}-3{\sigma}}}  \notag \\
\le & \big(CN^{-\sigma}\big)^{ N^{\sigma} \lceil N^{\frac{5}{3}}R^2\min(V^4,1)\rceil} \label{prob.coll.class}
\end{align}
where we regarded that $\Delta=N^{-\frac{1}{3}}$ and the abbreviation 
$$M =\big(8N^{\frac{1}{3}+4\sigma}\big)\big\lceil N^{\frac{5}{3}}R^2\min(V,1)^3\big(\min(V,1)+N^{\frac{1}{3}}R\big)\big\rceil. $$ 
Thus, the probability for such an event gets negligibly small as $N$ increases. We will return to this shortly. \\
These considerations yield us upper bounds for the number of certain collisions on $[t^N_k,t^N_{k+1}]$ which at least are complied with for typical initial data. What remains is to show that these upper limits can be applied to show that the deviations grow slow enough and we start with some general estimates.\\
Let for the subsequent part $(i,j)\in \mathcal{R}^{N,(t_k^N,t_{k+1}^N)}_{(r,R),(v,V)}(X)$ where
\begin{align}
N^{-\frac{5}{6}+\sigma}\le r\le R\le 8N^{-\frac{1}{2}+\frac{3\sigma}{2}}\ \land \ v\geq N^{-\frac{1}{3}} \label{thm2.constraints.para.}
\end{align} and we note that the restriction to these values will resolve itself shortly. According to estimates \eqref{app.cond.} it holds for the relevant values of $N,\sigma$, the considered configurations $X\in \mathcal{G}_T^{N,\sigma}$ and times $t\in [t_k^N, t_{k+1,X}^N]$ that
$$\sup\limits_{0\le s \le t-t_k^N}|^1\Psi^{N,c_1}_{s,0}(Y)-{^1\Psi^{N,c_2}_{s,0}(Y)}|_{\infty}\le CN^{-\frac{5}6+\frac{\sigma}{2}}\le \frac{1}{4} N^{-\frac{5}{6}+\sigma}.$$ Hence, it holds in this case for $r\geq N^{-\frac{5}{6}+\sigma}$ and $t\in [t_k^N, t_{k+1,X}^N]$ that
\begin{align}
& \inf_{0\le s \le t-t_k^N}|[^1\Psi^{N,c_2}_{s,0}(Y)]_i-[^1\Psi_{s,0}^{N,c_2}(Y)]_j| \notag \\
\geq & \inf_{0\le s \le t-t_k^N}|[^1\Psi^{N,c_1}_{s,0}(Y)]_i-[^1\Psi_{s,0}^{N,c_1}(Y)]_j|-\frac{1}{2} N^{-\frac{5}{6}+\sigma}\notag \\ 
\geq & \frac{1}{2}\inf_{0\le s \le t-t_k^N}|[^1\Psi^{N,c_1}_{s,0}(Y)]_i-[^1\Psi_{s,0}^{N,c_1}(Y)]_j|. \label{clos.cond.thm2}
\end{align}
Like in the proof of Lemma \ref{shift-lem} we apply that for $|q|,|q'|> N^{-c_1}\geq N^{-c_2}$ 
\begin{align*}
|f^N_{c_1}(q) -f^N_{c_2}(q')|
\le & C\big(\frac{1}{|q|^{\alpha+1}}+\frac{1}{|q'|^{\alpha+1}} \big)|q-q'|
\end{align*} which follows by a mean value argument since theses force kernels coincide if the considered configuration fulfill $|q|,|q'|> N^{-c_1}$. This yields together with relation \eqref{clos.cond.thm2} and Corollary \ref{cor1} (ii) that for times $t\in [t_k^N,t_{k+1,X}^N]$, the relevant $N,\sigma$ and parameters $R,v,V$ fulfilling constraints \eqref{thm2.constraints.para.} as well as additionally $r\geq 3N^{-c_1}$:
\begin{align}
&\frac{1}{N}\sum\limits_{(i,j)\in \mathcal{R}^{N,(t_k^N,t_{k+1}^N)}_{(r,R),(v,V)}(X) }\Big(\int_{0}^{t-t_k^N}\big|f^N_{c_1}([^1\Psi^{N,c_1}_{s,0}(Y)]_j-[^1\Psi_{s,0}^{N,c_1}(Y)]_i) \notag \\
& -f^N_{c_2}([^1\Psi^{N,c_2}_{s,0}(Y)]_j-[^1\Psi^{N,c_2}_{s,0}(Y)]_i)\big|ds\Big) \notag \\
\le  & \frac{C}{N}\sum\limits_{(i,j)\in \mathcal{R}^{N,(t_k^N,t_{k+1}^N)}_{(r,R),(v,V)}(X) }\int_{0}^{t-t_k^N}\frac{1}{|[^1\Psi^{N,c_1}_{s,0}(Y)]_i-[^1\Psi_{s,0}^{N,c_1}(Y)]_j|^{\alpha+1}}ds \notag \\
& \cdot \underbrace{ \sup_{0\le s\le t-t_k^N}|^1\Psi^{N,c_1}_{s,0}(Y)-{^1\Psi^{N,c_2}_{s,0}}(Y)|_{\infty}}_{\le CN^{-\frac{5}6+\sigma}\le N^{-\frac{7}{9}}} \notag \\
 \le & \frac{C}{Nr^{\alpha}v}N^{-\frac{7}{9}}|\mathcal{R}^{N,(t_k^N,t_{k+1}^N)}_{(r,R),(v,V)}(X)| \notag \\
 \le & \frac{CN^{-\frac{16}{9}}}{r^{\alpha}v}|\mathcal{R}^{N,(t_k^N,t_{k+1}^N)}_{(r,R),(v,V)}(X)|  \label{coll.-class-term 2}
 \end{align}
Strictly speaking, for the application of Corollary \ref{cor1} (ii) it would be necessary that there exist `mean-field particles' which keep sufficiently close to the `real' particles on the time span $[t^N_k,t^N_{k+1}]$. However, since we consider only initial data of the `good' set $\mathcal{G}_T^{N,\sigma}$ it is straightforward to see that for example the related `mean-field particles' (which start at the same initial data) fulfill this requirement. A more detailed explanation why the Corollary is applicable can essentially be copied by the reasoning stated after \eqref{term1,5+}.\\
Moreover, we have to take into a account collisions where the inter-particle distance falls below the order of the larger cut-off $N^{-c_1}=N^{-\frac{2}{3}}$ and we obtain the following relation by a further application of Corollary \ref{cor1} for times $t\in [t_k^N,t_{k+1,X}^N]$ as well as $N^{-\frac{5}{6}+\sigma}\le r\le R\le 3N^{-c_1}$, $v\geq N^{-\frac{1}{3}}$ and the relevant $N,\sigma$
\begin{align}
&\frac{1}{N}\sum\limits_{(i,j)\in \mathcal{R}^{N,(t_k^N,t_{k+1}^N)}_{(r,R),(v,V)}(X) }\Big(\int_{0}^{t-t_k^N}\big|f^N_{c_1}([^1\Psi^{N,c_1}_{s,0}(Y)]_j-[^1\Psi_{s,0}^{N,c_1}(Y)]_i)| \notag \\
& +|f^N_{c_2}([^1\Psi^{N,c_2}_{s,0}(Y)]_j-[^1\Psi^{N,c_2}_{s,0}(Y)]_i)\big|ds\Big) \notag \\
 \le & \frac{C}{Nr^{\alpha-1}v}|\mathcal{R}^{N,(t_k^N,t_{k+1}^N)}_{(r,R),(v,V)}(X)|  \label{coll.-class-term 3}
 \end{align} 
 where we regraded additionally that according to relation \eqref{clos.cond.thm2} the minimal distance between the considered particles on $[t_k^N,t_{k+1,X}^N]$ is of the same order for both dynamics if $r\geq N^{-\frac{5}{6}+\sigma}$ and the value of their relative velocity all the more for $v\geq N^{-\frac{1}{3}}$ due to relation \eqref{res.inf.norm thm.2}. \\
For the following estimates we assume that 
\begin{align}
|\mathcal{R}^{N,(t_k^N,t_{k+1}^N)}_{R,V}(X)|< CN^{4\sigma}\big\lceil N^{\frac{5}{3}}R^2\min(V,1)^3\big(\min(V,1)+N^{\frac{1}{3}}R\big)\big\rceil  \label{thm2.cond.upp.bound.coll}
\end{align}
is fulfilled (and remark that we previously showed that such a relationship holds with exceedingly high probability for the relevant configurations). Moreover, we set the following constraints on the values of the parameters
 $$\big( N^{-\frac{5}{6}+\sigma}\le r=N^{-\frac{\sigma}{2}}R \le 8N^{-\frac{1}{2}+\sigma}\big) \land \big(N^{-\frac{1}{3}}\le v= N^{-\frac{\sigma}{2}}V\big)$$ 
which in particular comply with the conditions which we applied for our estimates \eqref{thm2.constraints.para.} and demand additionally that $$ N^{\frac{5}{3}}R^2\min(V,1)^3\big(\min(V,1)+N^{\frac{1}{3}}R\big)\geq 1 .$$
In this case application of estimates \eqref{coll.-class-term 2} and \eqref{coll.-class-term 3} yields for the relevant $N,\sigma$ that
\begin{align}
&\frac{C}{N}\sum\limits_{(i,j)\in \mathcal{R}^{N,(t_k^N,t_{k+1}^N)}_{(r,R),(v,V)}(X) }\Big(\int_{0}^{t-t_k^N}\big|f^N_{c_1}([^1\Psi^{N,c_1}_{s,0}(Y)]_j-[^1\Psi_{s,0}^{N,c_1}(Y)]_i) \notag \\
& -f^N_{c_2}([^1\Psi^{N,c_2}_{s,0}(Y)]_j-[^1\Psi^{N,c_2}_{s,0}(Y)]_i)\big|ds\Big) \notag \\
\le &\begin{cases}\frac{CN^{-\frac{16}{9}}}{r^{\alpha}v}\Big(N^{4\sigma}\big\lceil N^{\frac{5}{3}}R^2\min(V,1)^3\big(\min(V,1)+N^{\frac{1}{3}}R\big)\big\rceil \Big), &\text{if } r \in [3N^{-c_1},8N^{-\frac{1}{2}+\sigma}] \notag \\ \frac{C}{Nr^{\alpha-1}v}\Big(N^{4\sigma}\big\lceil N^{\frac{5}{3}}R^2\min(V,1)^3\big(\min(V,1)+N^{\frac{1}{3}}R\big)\big\rceil \Big), &\text{if } r \in [N^{-\frac{5}{6}+\sigma},3N^{-c_1}]\end{cases}\\
\le &C\begin{cases}  N^{-\frac{1}{9}+6\sigma}R^{2-\alpha}+N^{\frac{2}{9}+6\sigma}R^{3-\alpha}, &\text{if } r \in [3N^{-c_1},8N^{-\frac{1}{2}+\sigma}]\\ N^{\frac{2}{3}+5\sigma}R^{3-\alpha}+N^{1+5\sigma}R^{4-\alpha}, &\text{if } r \in [N^{-\frac{5}{6}+\sigma},3N^{-c_1}]\end{cases}\notag \\
\le &C\begin{cases} N^{-\frac{4}{9}+7\sigma}+N^{-\frac{11}{18}+9\sigma}, &\text{if } r \in [3N^{-c_1},8N^{-\frac{1}{2}+\sigma}]\\ N^{-\frac{4}{9}+5\sigma}, &\text{if } r \in [N^{-\frac{5}{6}+\sigma},3N^{-\frac{2}{3}}]\end{cases}\label{upp.b.coll.thm2}
\end{align}
where we regarded that $R=N^{\frac{\sigma}{2}}r$, $c_1=\frac{2}{3}$ and $\alpha\le \frac{4}{3}$.\\
It remains to show that also very `close collisions' do typically not pose a problem. More specifically, we will now consider parameters $R\le  8N^{-\frac{7}{9}}$ for arbitrary $r\in [0,R]$ and at the same time we take into account the previously excluded settings of parameters where
$$\big\lceil N^{\frac{5}{3}}R^2\min(V,1)^3\big(\min(V,1)+N^{\frac{1}{3}}R\big)\big\rceil =1$$ for $ R,V>0$. To this end, we only have to regard the condition on the `impact' of single collisions which is fulfilled for times $0\le t\le \tau^{N,\sigma}_{col}(X)$ (see \eqref{stopp.time col} for the definition of $\tau^{N,\sigma}_{col}(X)$) and thereby it follows for such options of $r,R,V$ that
\begin{align}
&\frac{1}{N}\sum\limits_{(i,j)\in \mathcal{R}^{N,(t_k^N,t_{k+1}^N)}_{(r,R),(v,V)}(X) }\Big(\int_{0}^{t-t_k^N}\big|f^N_{c_1}([^1\Psi^{N,c_1}_{s,0}(Y)]_j-[^1\Psi_{s,0}^{N,c_1}(Y)]_i) \notag \\
& -f^N_{c_2}([^1\Psi^{N,c_2}_{s,0}(Y)]_j-[^1\Psi^{N,c_2}_{s,0}(Y)]_i)\big|ds\Big) \notag \\
\le &CN^{-\frac{1}{2}-\frac{5}{2}\sigma}\big(N^{4\sigma} \underbrace{\big\lceil N^{\frac{5}{3}}R^2\min(V,1)^3\big(\min(V,1)+N^{\frac{1}{3}}R\big)\big\rceil }_{\le CN^{\frac{5}{3}}N^{2(-\frac{7}{9})}} \big)\notag \\
\le & CN^{-\frac{7}{18}+2\sigma} \label{upp.b.coll.2thm2}
\end{align}
where in the first step we applied additionally our current assumption on the number of such collisions \eqref{thm2.cond.upp.bound.coll}. \\
Now we can merge these bounds and obtain that for $t\in [t_k^N,t_{k+1,X}^N]$ and for any such `collision set' where the parameters fulfill  
\begin{align}
\Big(\big( N^{-\frac{5}{6}+\sigma}\le r=N^{-\frac{\sigma}{2}}R \le 8N^{-\frac{1}{2}+\sigma}\big) \land \big(N^{-\frac{1}{3}}\le v= N^{-\frac{\sigma}{2}}V \big)\Big)\vee R\le 8N^{-\frac{7}{9}} \label{thm2.const.para.}
\end{align}
the constraint stated in \eqref{thm2.cond.upp.bound.coll} implies that 
\begin{align}
&\frac{1}{N}\sum\limits_{(i,j)\in \mathcal{R}^{N,(t_k^N,t_{k+1}^N)}_{(r,R),(v,V)}(X) }\Big(\int_{0}^{t-t_k^N}\big|f^N_{c_1}([^1\Psi^{N,c_1}_{s,0}(Y)]_j-[^1\Psi_{s,0}^{N,c_1}(Y)]_i)\notag  \\
& -f^N_{c_2}([^1\Psi^{N,c_2}_{s,0}(Y)]_j-[^1\Psi^{N,c_2}_{s,0}(Y)]_i)\big|ds\Big)  \label{thm2.coll.int.} \\
\le & C N^{-\frac{4}{9}+7\sigma}+N^{-\frac{11}{18}+9\sigma}+  CN^{-\frac{7}{18}+2\sigma}.
 \label{upp.b.coll.3thm2}
\end{align}
After these general considerations we can finally continue to derive an upper bound for the remaining term \eqref{term.coll.class1}. As mentioned before, we want to subdivide $\mathcal{M}^{N,(t_k^N,t_{k+1}^N)}_{6N^{-\frac{1}{2}+\sigma},\infty}(X)$ by application of the `collision sets' $\mathcal{R}^{N,(t_k^N,t_{k+1}^N)}_{(r,R),(v,V)}(X)$ and we will choose the parameters $r,R,v$ and $V$ such that with two exceptions all sets fulfill the constraints determined by \eqref{thm2.const.para.}. We recall that due to $X\in \mathcal{G}_T^{N,\sigma}$ it holds for the relevant $N,\sigma$ that
$$\sup_{0\le s \le T}|\Psi^{N,c_1}_{s,0}(X)-\Phi^{N,c_1}_{s,0}(X)|\le N^{-\frac{1}{2}+\sigma}$$
which implies in turn
$$\mathcal{M}^{N,(t_k^N,t_{k+1}^N)}_{6N^{-\frac{1}{2}+\sigma},\infty}(X)= \mathcal{M}^{N,(t_k^N,t_{k+1}^N)}_{6N^{-\frac{1}{2}+\sigma},\infty}(X)\cap \mathcal{R}^{N,(t_k^N,t_{k+1}^N)}_{8N^{-\frac{1}{2}+\sigma},\infty}(X) $$
and thereby it is suffices to look for a suitable cover of the set $\mathcal{R}^{N,(t_k^N,t_{k+1}^N)}_{8N^{-\frac{1}{2}+\sigma},\infty}(X)$.\\  One possibility for such a cover can be obtained as follows:\\
For a slightly shorter notation we identify $t_1=t_k^N$, $t_2=t^{N}_{k+1}$ and define 
\begin{align}
I_1&:=\{(k,l)\in \mathbb{Z}^{2}:\frac{1}{2} -\frac{3\sigma}{2} \le k \frac{{\sigma}}{2}\le \frac{7}{9} \ \land\ -1 \le (l+1)\frac{\sigma}{2} \le \frac{1}{3} \} \label{def.coll.class1} .
\end{align}
Then it holds that
\begin{align}
& \mathcal{R}^{N,(t_1,t_2)}_{8N^{-\frac{1}{2}+\sigma},\infty}(X) \notag \\ \subseteq & \mathcal{R}^{N,(t_1,t_2)}_{(0,8N^{-\frac{1}{2}+{\sigma}}),(N,\infty)}(X) \cup \mathcal{R}^{N,(t_1,t_2)}_{8N^{-\frac{1}{2}+{\sigma}},N^{-\frac{1}{3}+\frac{\sigma}{2}}}(X)\cup    \mathcal{R}^{N,(t_1,t_2)}_{8N^{-\frac{7}{9}},N}(X) \notag \\
& \bigcup_{(k,l)\in I_1} \mathcal{R}^{N,(t_1,t_2)}_{(8N^{-(k+1)\frac{\sigma}{2}},8N^{-k\frac{\sigma}{2}}),(N^{-(l+1)\frac{\sigma}{2}},N^{-l\frac{\sigma}{2}})}(X). \label{union}
\end{align}
 We point out that the number of sets forming the indicated cover is bounded by some constant (depending only on $\sigma$). Consequently, it suffices to show that for all of these sets term \eqref{thm2.coll.int.} keeps typically smaller than some value to conclude that term \eqref{term.coll.class1} (which constitutes actually the focus of our interest) is typically bounded by a value of the same order. Moreover, we already showed by the estimates leading to \eqref{prob.coll.class} that for $R\geq N^{-\frac{7}{9}}$
\begin{align}
&\mathbb{P}\Big(X\in \mathbb{R}^{6N}\ |\ X\in \mathcal{G}^{N,\sigma}_{3,T} \land  \exists k\in \{0,...,\lceil N^{\frac{1}{3}}T \rceil-1\}: \notag \\
& \hspace{0,5cm}|\mathcal{R}^{N,(k\Delta,(k+1)\Delta)}_{8R,V}(X)|\geq 8N^{4\sigma}\big\lceil N^{\frac{5}{3}}R^2\min(V,1)^3\big(\min(V,1)+N^{\frac{1}{3}}R\big)\big\rceil\Big) \notag \\
\le &  \lceil N^{\frac{1}{3}}T \rceil\big(CN^{-\sigma}\big)^{ N^{\sigma}  } \label{prob.est.thm.2}
\end{align}
and all sets of the union \eqref{union} except for 
\begin{itemize}
\item[(i)] $\mathcal{R}^{N,(t_1,t_2)}_{(0,8N^{-\frac{1}{2}+{\sigma}}),(N,\infty)}(X)\cap \mathcal{M}^{N,(t_1,t_2)}_{6N^{-\frac{1}{2}+\sigma},\infty}(X) $
\item[(ii)] $\mathcal{R}^{N,(t_1,t_2)}_{8N^{-\frac{1}{2}+{\sigma}},N^{-\frac{1}{3}+\frac{\sigma}{2}}}(X)\cap \mathcal{M}^{N,(t_1,t_2)}_{6N^{-\frac{1}{2}+\sigma},\infty}(X)$
\end{itemize}
fulfill the constraints on $r,R,v,V$ (stated in \eqref{thm2.const.para.}) which we applied for the derivation of relation \eqref{upp.b.coll.3thm2}. Hence, for any set of the cover except for the sets associated to items (i) and (ii) relation \eqref{upp.b.coll.3thm2} provides us an appropriate upper bound which is typically complied with. More specifically, according to estimate \eqref{prob.est.thm.2} the probability that relation \eqref{upp.b.coll.3thm2} does not hold is of far smaller order than necessary for the proof. Thus, we  can focus on determining a suitable bound for the `contribution' of the sets related to items (i) and (ii) which typically is satisfied to finally complete the considerations for term \eqref{term.coll.class1}.\\
First, we remark that for large enough $N$ and sufficiently small $\sigma>0$ the set related to item (ii) is even empty for the `good' configurations which we consider because for $(i,j)\in \mathcal{M}^{N,(t_1,t_2)}_{6N^{-\frac{1}{2}+\sigma},\infty}(X)$ the constraint $X\in \mathcal{G}_T^{N,\sigma}\subseteq (\mathcal{B}^{N,\sigma}_5)^C$ (defined in \eqref{def.B_5}) yields that the value of the relative velocity between the related `mean-field particles' at the times when they are `close' in space is bounded from below by order $N^{-\frac{5}{18}}$. After regarding the closeness (in phase space) between corresponding `real' and `mean-field particles' it is obvious that this property transfers to their related `real' particles.\\
For handling the set assigned to item (i) one has to recall that in the current case we only take into account configurations where $\tau^{N,\sigma}_{dev,2}(X)<\tau^{N,\sigma}_{col}(X)$ and thus on the time interval which is relevant for our estimates the `impact' of a single `close' collision is bounded from above by $N^{-\frac{1}{2}-\frac{5}{2}\sigma}$. Moreover, according to estimate \eqref{upp.b.coll.3thm2} the `contribution' of the collisions related to item (i) is only relevant for the value of term \eqref{term.coll.class1} if it exceeds order $N^{-\frac{7}{18}+2\sigma}$. In that case, however, at least order $N^{\frac{1}{9}+\frac{9\sigma}{2}}$ `particle pair labels' have to be contained in this set because 
$$nN^{-\frac{1}{2}-\frac{5}{2}\sigma}\geq CN^{-\frac{7}{18}+2\sigma}\Rightarrow n\geq CN^{\frac{1}{9}+\frac{9\sigma}{2}}.$$ If at the time of their `closest encounter' the relative velocity value between two particles is of order $N$ or larger, then at least one of the colliding particles had already a velocity of the same order initially (for large $N$) since the effective force field is bounded independent of $N$ (and the `real' particles are close to their `mean-field particle partners' on the considered time span). Moreover, due to $X\in \mathcal{G}_T^{N,\sigma}\subseteq \bigcap_{i=1}^N (\mathcal{B}^{N,\sigma}_{4,i})^C$ (see \eqref{def.B_4}) it follows that $$\forall i \in \{1,...,N\}: \sum_{j \neq i}\mathbf{1}_{M^N_{6N^{-\frac{1}{2}+\sigma},\infty}(X_i)}(X_j)<  N^{3\sigma}.$$ This, however, yields that in the current situation the assumption $$\big|\mathcal{R}^{N,(t_1,t_2)}_{(0,8N^{-\frac{1}{2}+{\sigma}}),(N,\infty)}(X)\cap \mathcal{M}^{N,(t_1,t_2)}_{6N^{-\frac{1}{2}+\sigma},\infty}(X)\big|\geq N^{\frac{1}{9}+\frac{9\sigma}{2}} $$ implies that there exist at least order $\frac{N^{\frac{1}{9}+\frac{9\sigma}{2}}}{N^{3\sigma}}=N^{\frac{1}{9}+\frac{3\sigma}{2}}$ different particles which have an initial velocity value of at least order $N$. On the other hand it holds for arbitrary $n\in \mathbb{N}$ and a constant $K_1>0$ that 
\begin{align*}
& \mathbb{P}\big(X\in \mathbb{R}^{6N}:\sum_{i=1}^N\mathbf{1}_{[K_1N,\infty)}(|^2X_i|)\geq n \big)\\
\le & \binom{N}{n}\mathbb{P}\big(X_1\in \mathbb{R}^6:|^2X_1|\geq K_1N \big)^n \le  \binom{N}{n}(C(K_1N)^{-2})^n\le CN^{-n}
\end{align*}
where we regarded that according to assumption \eqref{ass.dens.3} the kinetic energy related to $k_0$ is bounded and thus
\begin{align}
\int_{\mathbb{R}^6}\underbrace{\mathbf{1}_{[K_1N,\infty)}(|^2X_i|)}_{\le \frac{|^2X_i|^2}{(K_1N)^2}}k_0(X)d^6{X}\le \frac{C}{(K_1N)^2}. \label{est.vel.kin.energy}
\end{align} Hence, the number of such collisions keeps obviously with exceedingly high probability sufficiently small.\\
Since all these probabilities drop so much faster than necessary for the proof of the theorem (which means faster than $CN^{-\frac{1}{9}+\epsilon}$, $\epsilon>0$), we will simply neglect the inappropriate initial data for the concluding estimates and restrict ourselves to presenting that for the remaining configurations the deviations keep sufficiently small. Let to this end $X\in \mathcal{G}_T^{N,\sigma}$ be one of these remaining `good' configurations and $k\in \{0,...,\lceil N^{\frac{1}{3}}T\rceil-1\}$. If we maintain the abbreviations $t_k^N:=kN^{-\frac{1}{3}}$ and $Y:={\Psi}^{N,c_2}_{t_k^N,0}(X)$, then it holds according to estimates \eqref{term.coll.class1} and \eqref{upp.b.coll.3thm2} for times $t_k^N \le t\le \min(\tau^{N,\sigma}_{col}(X),t_{k+1}^N)$ as well as the relevant $N,\sigma$ that 
\begin{align}
& \sum_{i=1}^N\sup_{0 \le s \le t-t_k^N}|[^1\Psi^{N,c_1}_{s,0}(Y)]_i-[{^1\Psi}^{N,c_2}_{s,0}(Y)]_i| \notag \\
\le & \int_{t_k^N}^t\sum_{i=1}^N\sup_{0 \le r \le s-t_k^N}|[^2\Psi^{N,c_1}_{r,0}(Y)]_i-[{^2\Psi}^{N,c_2}_{r,0}(Y)]_i|ds \notag \\
\le & C \int_{t_k^N}^t \sum_{i=1}^N\sup_{0 \le r \le s-t_k^N}|[^1\Psi^{N,c_1}_{r,0}(Y)]_i-[{^1\Psi}^{N,c_2}_{r,0}(Y)]_i|ds+C  N^{-\frac{7}{18}+2\sigma}\underbrace{(t_{k+1}^N-t_k^N)}_{=\Delta=N^{-\frac{1}{3}}} \label{Gron.Thm2}
\end{align}
Finally, application of Gronwall's Lemma yields for large enough $N\in \mathbb{N}$ and times $t\in [t_k^N,\min(\tau^{N,\sigma}_{col}(X),t_{k+1}^N)]$ that
\begin{align*}
&\sum_{i=1}^N\sup_{0 \le s \le t-t_k^N}|[^1\Psi^{N,c_1}_{s,0}(Y)]_i-[{^1\Psi}^{N,c_2}_{s,0}(Y)]_i|\\
\le &  C  N^{-\frac{7}{18}+2\sigma}  \Delta e^{C(t-t_k^N)}\\
\le & C\frac{ N^{-\frac{7}{18}+2\sigma} }{\Delta}\Delta^2 \\
\le & C  N^{-\frac{1}{18}+2\sigma} \Delta^2
\end{align*} 
where we recall that $\Delta=N^{-\frac{1}{3}}$ and $t_k^N=k\Delta$.\\
By regarding this relation we can once again apply our previous estimates to obtain that
\begin{align*}
&\sum_{i=1}^N\sup_{0 \le r \le t-t_k^N}|[^2\Psi^{N,c_1}_{r,0}(Y)]_i-[{^2\Psi}^{N,c_2}_{r,0}(Y)]_i|\\
\le & C\sum_{i=1}^N\sup_{0 \le s \le t-t_k^N}|[{^1\Psi^{N,c_1}_{s,0}}(Y)]_i-[{^1\Psi}^{N,c_2}_{s,0}(Y)]_i|+CN^{-\frac{7}{18}+2\sigma}\\
\le & C\frac{N^{-\frac{7}{18}+2\sigma}}{\Delta}\Delta\\
\le &CN^{-\frac{1}{18}+2\sigma}\Delta.
\end{align*}
After recalling the abbreviations $Y={\Psi}^{N,c_2}_{t_k^N,0}(X)$ and $t^N_{k,X}=\min\big(t_k^N,\tau^{N,\sigma}_{col}(X)\big)$, it follows in particular that for sufficiently large $N\in \mathbb{N}$ and small enough $\sigma>0$ 
\begin{align*}
\sum_{i=1}^N\sup\limits_{t^N_k\le s\le t^N_{k+1,X}} |[{^1\Psi^{N,c_1}_{s,t_k^N}(\Psi^{N,c_2}_{t^N_k,0}(X))}]_i-[{^1\Psi^{N,c_2}_{s,0}(X)}]_i|\le CN^{-\sigma}\Delta^2
\end{align*}
as well as
\begin{align*}
\sum_{i=1}^N\sup\limits_{t^N_k\le s\le t^N_{k+1,X}} |[{^2\Psi^{N,c_1}_{s,t_k^N}(\Psi^{N,c_2}_{t^N_k,0}(X))}]_i-[{^2\Psi^{N,c_2}_{s,0}(X)}]_i|\le CN^{-\sigma}\Delta.
\end{align*}
Thus, inequality \eqref{list}
\begin{align}
\sup\limits_{0\le s\le \tau^{N,\sigma}_{col}(X)} |{^1\Psi^{N,c_1}_{s,0}(X)}-{^1\Psi^{N,c_2}_{s,0}(X)}|_1\le CN^{-\sigma} 
\end{align}
holds for the relevant $N$ and $\sigma$ since all constraints stated in the list previous to \eqref{list} are fulfilled in this case which implies that Corollary \ref{Cor.shift2} is applicable.\\ Eventually, we can conclude that the probability of configurations $X\in \mathcal{G}_T^{N,\sigma}$ where $\tau^{N,\sigma}_{dev,2}(X)<\tau^{N,\sigma}_{col}(X)$ decays distinctly faster with increasing $N$ than necessary for the proof.\\\\
Finally, we arrived at the point where we have to care for configurations fulfilling $$\tau^{N,\sigma}_{col}(X) \le \underbrace{ \min\big( \tau^{N,\sigma}_{dev,1}(X), \tau^{N,\sigma}_{dev,2}(X)\big)}_{=:\tau^{N,\sigma}_{dev}(X)} \  \land \ \tau^{N,\sigma}_{col}(X)<T $$ respectively where one of the previously excluded collisions happens though the deviations between the corresponding trajectories are still small enough:
\begin{align}
& \exists l\in \{1,2\},\ (i,j)\in \mathcal{M}^{N,(0,T)}_{6N^{-\frac{1}{2}+\sigma},\infty}(X): \notag 
\\ &\int_{0}^{\tau^{N,\sigma}_{dev}(X)}|f^N_{c_l}([^1\Psi^{N,c_l}_{s,0}(X)]_i-[^1\Psi^{N,c_l}_{s,0}(X)]_j)|ds\geq  N^{\frac{1}{2}-\frac{5}{2}{{\sigma}}}\label{ev.col.}
\end{align}
Before we are able to derive an upper bound for the probability of configurations which fulfill \eqref{ev.col.} we have to make some considerations to clarify how this can be done.\\
For a clearer presentation we focus for the remaining part of the proof on the slightly more elaborate case $l=2$ (respectively the system with smaller cut-off order). However, obviously $l=1$ can be handled analogously since the exact value of $c_2$ is arbitrary in $[c_1,\infty)$ and thus the special case $c_2=c_1$ is included.\\
Let $t'_{min}$ denote (one of) the point(s) in time where `mean-field particles' $i$ and $j$ are closest in space, then it follows by Lemma \ref{lem2} that for $|t-t'_{min}|\le 1$ 
\begin{align}
&\big|\big({^2\varphi^{N,c_1}_{t,0}}(X_i)-{^2\varphi^{N,c_1}_{t,0}}(X_j)\big)-\big({^2\varphi^{N,c_1}_{t'_{min},0}}(X_i)-{^2\varphi^{N,c_1}_{t'_{min},0}}(X_j)\big)\big| \notag  \\
\le &C|t-t'_{min}|\Big(|{^1\varphi^{N,c_1}_{t'_{min},0}}(X_i)-{^1\varphi^{N,c_1}_{t'_{min},0}}(X_j)|\notag \\
& +|{^2\varphi^{N,c_1}_{t'_{min},0}}(X_i)-{^2\varphi^{N,c_1}_{t'_{min},0}}(X_j)||t-t'_{min}|\Big). \label{prop.free.ev.}
\end{align}
which implies that on a possibly small interval around $t'_{min}$ (but whose length may be chosen larger than some constant) their relative velocity barely changes if
\begin{align*}
|{^1\varphi^{N,c_1}_{t'_{min},0}}(X_i)-{^1\varphi^{N,c_1}_{t'_{min},0}}(X_j)|\le |{^2\varphi^{N,c_1}_{t'_{min},0}}(X_i)-{^2\varphi^{N,c_1}_{t'_{min},0}}(X_j)|.
\end{align*} For the configurations which we consider this condition holds if $(i,j)\in \mathcal{M}^{N,(0,T)}_{6N^{-\frac{1}{2}+\sigma},\infty}(X)$ because $X\in \mathcal{G}_T^{N,\sigma}\subseteq \big(\mathcal{B}^{N,\sigma}_5\big)^C$ (see \eqref{def.B_5}) implies that the relative velocity value between `mean-field particles' is at least of order $N^{-\frac{5}{18}}$ at times when their spatial distance attains or falls below $6N^{-\frac{1}{2}+\sigma}$. A corresponding statement is true for their related `real' particles if $\tau^{N,\sigma}_{dev}(X)$ is not `triggered' yet because in this case the deviation 
$$\sup_{0\le s \le {\tau^{N,\sigma}_{dev}(X)}}|\Psi^{N,c_l}_{s,0}(X)-\Phi_{s,0}^{N,c_1}(X)|_{\infty}\le CN^{-\frac{1}{2}+\frac{3}{5}\sigma},\ l\in \{1,2\}$$ is of much smaller order than $N^{-\frac{5}{18}}$ (for the relevant $N$).\\ Let now $X\in \mathcal{G}_T^{N,\sigma}$ and let $t_{min}\in [0,\tau^{N,\sigma}_{dev}(X)]$ denote (one of) the point(s) in time where the `real' particles $i$ and $j$ attain their minimal spatial distance on this interval, then the preceding considerations and Corollary \ref{cor1} (ii) imply that for $(i,j)\in \mathcal{M}^{N,(0,T)}_{6N^{-\frac{1}{2}+\sigma},\infty}(X)$ and a suitable constant $C_0>0$:
\begin{align*}
&\int_{0}^{\tau^{N,\sigma}_{dev}(X)}|f^N_{c_2}([^1\Psi^{N,c_2}_{s,0}(X)]_i-[^1\Psi^{N,c_2}_{s,0}(X)]_j)|ds\\
\le &  \frac{C_0}{|[^1\Psi^{N,c_2}_{t_{min},0}(X)]_i-[^1\Psi^{N,c_2}_{t_{min},0}(X)]_j|^{\alpha-1}|[^2\Psi^{N,c_2}_{t_{min},0}(X)]_i-[^2\Psi^{N,c_2}_{t_{min},0}(X)]_j|}
\end{align*} 
For convenience we abbreviate in the following $r_{min}:=|[^1\Psi^{N,c_2}_{t_{min},0}(X)]_i-[^1\Psi^{N,c_2}_{t_{min},0}(X)]_j|$ as well as $v_{min}:=|[^2\Psi^{N,c_2}_{t_{min},0}(X)]_i-[^2\Psi^{N,c_2}_{t_{min},0}(X)]_j|$ and recall that according to the previous reasoning $v_{min}$ is of distinctly larger order than $r_{min}$ for the currently considered configurations. Then for any $r>0$ where $ N^{-\frac{\sigma}{2}}r \le r_{min}\le r$ the subsequent implication holds (for sufficiently large $N\in \mathbb{N}$):
\begin{align}
&\int_{0}^{\tau^{N,\sigma}_{dev}(X)}|f^N_{c_2}([^1\Psi^{N,c_2}_{s,0}(X)]_i-[^1\Psi^{N,c_2}_{s,0}(X)]_j)|ds\geq N^{\frac{1}{2}-\frac{5}{2}{{\sigma}}} \notag \\
\Rightarrow & v_{min}\le C_0N^{-\frac{1}{2}+\frac{5}{2}\sigma}r_{min}^{1-\alpha}\le C_0N^{-\frac{1}{2}+(\frac{5}{2}+\frac{\alpha-1}{2})\sigma}r^{1-\alpha}< N^{-\frac{1}{2}+3\sigma}r ^{-\frac{1}{3}} \label{cons.hard.col.1}
\end{align}
where we regarded that $\alpha\in (1,\frac{4}{3}]$. Though we omitted to make it explicit, the respective value of $r_{min}$ obviously depends on $N$ and $X$ and the first inequality determines a border for the values of $v_{min}$ separating problematic and non-problematic collisions. For a certain $r\in (0,6N^{-\frac{1}{2}+\sigma}]$ the second inequality, on the other hand, provides us a corresponding border for $v_{min}$ depending on $r$ which is valid for all $X\in \mathcal{G}_T^{N,\sigma}$ where $r_{min}\in[N^{-\frac{\sigma}{2}}r,r]$. The exact reason for introducing such a parameter $r$ will become clear shortly. \\ Moreover, we remark that according to our preliminary considerations $v_{min}$ is of distinctly larger order than $r_{min}$ and the relative velocity of the particles barely changes on a possibly small interval $t_1< t_{min}\le t_2\le \tau_{dev}^{N,\sigma}(X)$ whose length, however, can be `chosen' larger than some constant for the relevant configurations (resp. for $X\in \mathcal{G}_T^{N,\sigma}$). Hence, if $M\in \mathbb{N}$ is chosen large enough and \eqref{cons.hard.col.1} is fulfilled, then for each such triple $N,\ r $ and $v_r:=N^{-\frac{1}{2}+3\sigma}r ^{-\frac{1}{3}}$ there exists $k\in \{0,...,M-1\}$ such that 
\begin{align}
& \Big(N^{-\frac{\sigma}{2}} r\le  |^1X_i-{^1X_j}|\le r\ \land \ |^2X_i-{^2X_j}|\le v_r\Big) \ \vee \notag \\
& \Big( r\le   |[^1\Psi^{N,c_2}_{k\frac{T}{M},0}(X)]_i-[^1\Psi^{N,c_2}_{k\frac{T}{M},0}(X)]_j|\le r+v_r\frac{T}{M} \ \land \notag \\
 & \hspace{0,93cm} |[^2\Psi^{N,c_2}_{k\frac{T}{M},0}(X)]_i-[^2\Psi^{N,c_2}_{k\frac{T}{M},0}(X)]_j|\le v_r\Big)  \label{cons.hard.col.2}
\end{align}
because either the particles are already close initially or at one of the time steps $k\frac{T}{M}$ particle $j$ must be located in a spherical shell around the position of particle $i$ like stated above (respectively vice versa).\\ We continue by merging the preliminary reasoning subsequently to derive an upper bound for the probability that $\tau^{N,\sigma}_{col}(X)$ gets `triggered' first. \\
Due to the previous considerations the following implication holds for `small' $\sigma>0$, large enough values of $N\in \mathbb{N}$ and $X\in \mathcal{G}_T^{N,\sigma}$:
{\allowdisplaybreaks \begin{align}
& \exists (i,j)\in \mathcal{M}^{N,(0,T)}_{6N^{-\frac{1}{2}+\sigma},\infty}(X): \notag \\  
& \int_{0}^{\tau^{N,\sigma}_{dev}(X)}|f^N_{c_2}([^1\Psi^{N,c_2}_{s,0}(X)]_i-[^1\Psi^{N,c_2}_{s,0}(X)]_j)|ds\geq N^{\frac{1}{2}-\frac{5}{2}{{\sigma}}}    \notag  \\
\Rightarrow &  \bigg(\exists t\in [0,\tau^{N,\sigma}_{dev}(X)]\ \exists (i,j)\in \mathcal{M}^{N,(0,T)}_{6N^{-\frac{1}{2}+\sigma},\infty}(X) \ \exists r\in [0,8N^{-\frac{1}{2}+\sigma}]: \notag \\ &\Big( |[^1\Psi^{N,c_2}_{t,0}(X)]_i-[{^1\Psi}^{N,c_2}_{t,0}(X)]_j|=r\  \land \notag  \\
& \hspace{0,24cm} |[^2\Psi^{N,c_2}_{t,0}(X)]_i-[{^2\Psi}^{N,c_2}_{t,0}(X)]_j|\le C N^{-\frac{1}{2}+\frac{5}{2}\sigma}r^{-\frac{1}{3}} \Big)\bigg)
\end{align}
where this implication is a consequence of the first inequality of \eqref{cons.hard.col.1}. After introducing additionally $r_l:=8N^{-9+\frac{l}{2}{\sigma}}$ and $v_l:= N^{-\frac{1}{2}+3{\sigma}}r_l^{-\frac{1}{3}}$ the previous statement implies due to the second and third inequality of \eqref{cons.hard.col.1} that:
\begin{align}
 \exists & t\in [0,\tau^{N,\sigma}_{dev}(X)] \ \exists (i,j)\in \mathcal{M}^{N,(0,T)}_{6N^{-\frac{1}{2}+\sigma},\infty}(X) \ \exists l\in \{1,...,\lceil \frac{17}{\sigma}\rceil+2\}: \notag \\ 
 \bigg( &\Big(  |^1X_i-{^1X_j}|\le r_0\ \land \ |^2X_i-{^2X_j}|\le v_0\Big)  \ \vee \notag \\
&  \Big(r_{l-1} <|[^1\Psi^{N,c_2}_{t,0}(X)]_i-[{^1\Psi}^{N,c_2}_{t,0}(X)]_j|\le r_l\  \land \notag  \\
& \hspace{1,41cm} |[^2\Psi^{N,c_2}_{t,0}(X)]_i-[{^2\Psi}^{N,c_2}_{t,0}(X)]_j|\le v_l \Big)\
\vee \notag \\ 
 & \hspace{1,41cm} |[^2\Psi^{N,c_2}_{t,0}(X)]_i-[{^2\Psi}^{N,c_2}_{t,0}(X)]_j|\geq  v_1\bigg) \label{cond.4} 
\end{align}}
The intervals $[r_{l-1},r_l]$ for $\{0,...,\lceil \frac{17}{\sigma}\rceil+2\}$ cover the whole range of potential `collision distances' in $[8N^{-9}, 8 N^{-\frac{1}{2}+\sigma}]$ while the $v_l$ are chosen such that they determine respectively a border between problematic and unproblematic collisions for these intervals according to considerations \eqref{cons.hard.col.1}. The events related to the second and the last line of \eqref{cond.4} are introduced to ensure that also configurations are taken into account where a `problematic' collision occurs between particles which get closer to each other than $r_0$ without fulfilling the assumptions of one of the remaining `events' because if their initial distance is still larger than $r_0$ then their relative velocity previous to the collision must exceed order $v_1\geq  N^{\frac{5}{2}}$ or the event related to the third and the fourth line for $l=1$ occurs. Thus, if none of these events occurs, then we can also conclude that no particle pair gets closer than $r_0=8N^{-9}$ which implies that for cut-off parameters $c_2>9$ the trajectories of the regularized dynamics coincide with those of the non-regularized system in this case. \\ Now we can finally apply consideration \eqref{cons.hard.col.2} to conclude that for sufficiently large $M\in \mathbb{N}$ the statement related to \eqref{cond.4} implies that
\begin{align}
& \exists k\in \{0,...,M-1\}\ \exists (i,j)\in \mathcal{M}^{N,(0,T)}_{6N^{-\frac{1}{2}+\sigma},\infty}(X) \notag \\
& \exists l\in \{0,...,\lceil \frac{17}{\sigma}\rceil+2\}:k\frac{T}{M}\le \tau^{N,\sigma}_{dev}(X)\  \land \notag \\
& \bigg( \Big(  |^1X_i-{^1X_j}|\le r_l\ \land \ |^2X_i-{^2X_j}|\le 2v_{l}\Big)  \ \vee   \label{cond.1} \\  
 & \Big(  r_{l}\le |[^1\Psi^{N,c_2}_{k\frac{T}{M},0}(X)]_i-[{^1\Psi}^{N,c_2}_{k\frac{T}{M},0}(X)]_j|\le r_l+2 v_{l}\frac{T}{M} \ \land \notag \\
& \hspace{1,04cm} |[^2\Psi^{N,c_2}_{k\frac{T}{M},0}(X)]_i-[^2\Psi^{N,c_2}_{k\frac{T}{M},0}(X)]_j|\le 2v_{l} \Big) \label{cond.3} \ \vee \\ 
 & \hspace{1,04cm} |[^2\Psi^{N,c_2}_{k\frac{T}{M},0}(X)]_i-[{^2\Psi}^{N,c_2}_{k\frac{T}{M},0}(X)]_j|\geq  \frac{1}2v_1\bigg) \label{cond.5}.
\end{align}
\\
It is obvious that the probability of configurations satisfying \eqref{cond.1} is negligible compared to summed probability of the events related to assumption \eqref{cond.3}. An upper bound for the event related to assumption \eqref{cond.5} can be derived as follows: If there exists $k\in \{0,...,M-1\}$ such that $k\frac{T}{M}\le \tau^{N,\sigma}_{dev}(X)$ and
$$|[^2\Psi^{N,c_2}_{k\frac{T}{M},0}(X)]_i-[{^2\Psi}^{N,c_2}_{k\frac{T}{M},0}(X)]_j|\geq  \frac{1}2v_1\geq  \frac{1}2 N^{\frac{5}{2}},$$
then the closeness to the related `mean-field particles' also implies
$$|^2\varphi^{N,c_1}_{k\frac{T}{M},0}(X_i)-{^2\varphi^{N,c_1}_{k\frac{T}{M},0}}(X_j)|\geq  \frac{1}{3}
N^{\frac{5}{2}} $$
which by the $N$-independent boundedness of the mean-field force yields that for an appropriate constant also the initial velocities fulfill $|^2X_i|\geq CN^{\frac{5}{2}}$ or $|{^2X_j}|\geq CN^{\frac{5}{2}}$ if $N$ is sufficiently large. Due to the assumption that kinetic energy related to $k_0$ is bounded we can once again apply estimates \eqref{est.vel.kin.energy} which imply that the probability of configurations fulfilling condition \eqref{cond.5} is bounded by $N\big(CN^{-\frac{5}{2}}\big)^2\le CN^{-4} $.\\ It remains to determine a suitable upper bound for the probability of the configurations fulfilling assumption \eqref{cond.3}.\\
First, we recall that due to the considerations after \eqref{term.coll.class1} and in particular estimate \eqref{est.dens.} it holds for $t\le \tau^{N,\sigma}_{dev}(X)$ that 
\begin{align*}
& \mathbf{1}_{\mathcal{G}_T^{N,\sigma}}(X)\prod_{i=1}^N k_0(X_i)=F_t^{N,c_2}(\Psi^{N,c_2}_{t,0}(X))\mathbf{1}_{\mathcal{G}_T^{N,\sigma}}(X) \\
\le & CF^{N,c_1}_t(\Psi^{N,c_2}_{t,0}(X))\mathbf{1}_{\mathcal{G}^{N,\sigma}_{3,T}}(\Psi^{N,c_1}_{0,t}(\Psi^{N,c_2}_{t,0}(X))) .
\end{align*}
 By application of this in the second step, it follows for $l\in \{0,...,\lceil \frac{17}{\sigma}\rceil+2\}$ and the special case $i=1$, $j=2$ that 
 { \allowdisplaybreaks
\begin{align}
& \mathbb{P}\Big(X\in  \mathcal{G}_T^{N,\sigma} \ \land \ \tau^{N,\sigma}_{dev}(X)\geq k \frac{T}{M} \notag \\   & \hspace{0,5cm} r_{l}\le |[^1\Psi^{N,c_2}_{k\frac{T}{M},0}(X)]_1-[{^1\Psi}^{N,c_2}_{k\frac{T}{M},0}(X)]_2|\le r_l+2 v_{l}\frac{T}{M} \ \land \notag \\
& \hspace{1,295 cm} |[^2\Psi^{N,c_2}_{k\frac{T}{M},0}(X)]_1-[^2\Psi^{N,c_2}_{k\frac{T}{M},0}(X)]_2|\le 2v_{l} \Big) \notag  \\
= & \int_{\mathbb{R}^{6N}}\mathbf{1}_{ \mathcal{G}_T^{N,\sigma}}(X)\mathbf{1}_{\{Z\in \mathbb{R}^{6N}:\tau^{N,\sigma}_{dev}(Z)\geq k\frac{T}{M}\}}(X)\prod_{n=1}^Nk_0(X_n) \notag \\
&\cdot\mathbf{1}_{\{Y\in \mathbb{R}^{6N}:r_l\le |^1Y_1-{^1Y_2}|\le r_l+2v_{l}\frac{T}{M}\land |^2Y_1-^2Y_2|\le 2 v_{l}\}}(\Psi^{N,c_2}_{k\frac{T}{M},0}(X))d^{6N}X \notag   \\
\le & C\int_{\mathbb{R}^{6N}}\mathbf{1}_{\mathcal{G}^{N,\sigma}_{3,T}}(\Psi^{N,c_1}_{0,k\frac{T}{M}}(\Psi^{N,c_2}_{k\frac{T}{M},0}(X))) F_{k\frac{T}{M}}^{N,c_1}(\Psi^{N,c_2}_{k\frac{T}{M},0}(X))\notag \\
&\cdot\mathbf{1}_{\{Y\in \mathbb{R}^{6N}:r_l\le |^1Y_1-{^1Y_2}|\le r_l+2v_{l}\frac{T}{M}\land |^2Y_1-^2Y_2|\le 2 v_{l}\}}(\Psi^{N,c_2}_{k\frac{T}{M},0}(X))d^{6N}X \notag   \\
= & C\int_{\mathbb{R}^{6N}}\mathbf{1}_{\mathcal{G}^{N,\sigma}_{3,T}}(\Psi^{N,c_1}_{0,k\frac{T}{M}}(X))F^{N,c_1}_{k\frac{T}{M}}(X) \notag  \\& \cdot \mathbf{1}_{\{Y\in \mathbb{R}^{6N}:r_l\le |^1Y_1-{^1Y_2}|\le r_l+2v_{l}\frac{T}{M} \land |^2Y_1-^2Y_2|\le 2v_{l}\}}(X)d^{6N}X . \label{prob.hard coll.}
\end{align}}
Now we can apply Lemma \ref{prod-Lem} and obtain for $M\gg v_l\frac{T}{r_l}$ that \eqref{prob.hard coll.} is bounded by
{\allowdisplaybreaks \begin{align*}
 & C\int_{\mathbb{R}^6} \int_{\mathbb{R}^6}
 \mathbf{1}_{\{Y\in \mathbb{R}^{6}:r_l\le |^1Y-{^1Z_2}|\le r_l+2v_{l}\frac{T}{M} \}}(Z_1) \\
 &\cdot\mathbf{1}_{\{Y\in \mathbb{R}^6:|^2Y-{^2Z_2}|\le 2v_{l}\}}(Z_1) k^{N,c_1}_{k\frac{T}{M}}(Z_1)k^{N,c_1}_{k\frac{T}{M}}(Z_2)d^6Z_1d^6Z_2\\
\le & C r_l^2v_{l}^4\frac{T}{M}
\le  \frac{C}{M} r_l^2\big(N^{-\frac{1}{2}+3\sigma}r_l^{-\frac{1}{3}}\big)^4
\le  \frac{C}{M} r^{\frac{2}{3}}_l N^{-2+12\sigma}
\le  \frac{C}{M} N^{-\frac{7}{3}+13\sigma}
\end{align*}}
where we applied that $r_l=8N^{-9+\frac{l}{2}{\sigma}}\le CN^{-\frac{1}{2}+\frac{3\sigma}{2}}$ since $l\in \{0,....,\lceil \frac{17}{\sigma }\rceil +2\}$. \\
By taking into account all possible pairs $(i,j)\in \mathcal{M}^{N,(0,T)}_{6N^{-\frac{1}{2}+\sigma},\infty}(X)\subseteq \{1,...,N\}^2$ and $k\in \{0,...,M-1\}$ as well as $l\in \{0,...,\lceil \frac{17}{\sigma }\rceil +2\}$ it is straightforward to conclude that
\begin{align}
& \mathbb{P}\Big( X\in  \mathcal{G}_T^{N,\sigma}\  \land \ \exists (i,j)\in \mathcal{M}^{N,(0,T)}_{N^{-\frac{1}{2}+{\sigma}},\infty}(X):  \notag  \\ 
& \hspace{0,5cm} \int_{0}^{\tau^{N,\sigma}_{dev(X)}}|f^N_{c_2}([\Psi^{N,c_2}_{s,0}(X)]_i-[\Psi^{N,c_2}_{s,0}(X)]_j)|ds\geq N^{\frac{1}{2}-\frac{5}{2}{\sigma}} \Big)  \notag  \\
\le & N^2\sum_{k=0}^{M-1}\sum_{l=0}^{\lceil \frac{17}{\sigma} \rceil+2}\frac{C}{M}r_l^2v_{l}^4+CN^{-4} \notag \\
\le & CN^2 N^{-\frac{7}{3}+13\sigma}  \notag \\
\le & C  N^{-\frac{1}{3}+13\sigma}. \label{upp.b.prob.thm.2}
\end{align}
where the insignificant addend $CN^{-4}$ appearing after the first step arises from the previous probability estimates for condition \eqref{cond.4}. Hence, also this probability becomes negligible compared to the upper limit which we determined for $\mathbb{P}\big(X \in  \big(\mathcal{G}_T^{N,\sigma}\big)^C \big)$ if $N\in \mathbb{N}$ is chosen large and $\sigma>0$ small enough. \\\\ Considering the respective upper bounds which we derived for the different probabilities of sets of excluded configurations shows that the upper limit for $\mathbb{P}\big(X \in  \big(\mathcal{G}_T^{N,\sigma}\big)^C \big)$ is dominating for small enough $\sigma>0$. Thus, for a given $\epsilon>0$ and sufficiently small $\sigma>0$ there exists $C>0$ such that for all $N\in \mathbb{N}$ it holds that
\begin{align*}
& \mathbb{P}\big(X\in \mathbb{R}^{6N}:\sup_{0\le s \le T}|\Psi_{s,0}^{N,c_1}(X)-\Psi_{s,0}^{N,c_2}(X)|_{\infty}>  N^{-\frac{1}{2}+\sigma} \big) \\
\le & \mathbb{P}\big(X \in  \big(\mathcal{G}_T^{N,\sigma}\big)^C \big)+\mathbb{P}\big(X\in \mathcal{G}_T^{N,\sigma} \land \sup_{0\le s \le T}|\Psi_{s,0}^{N,c_1}(X)-\Psi_{s,0}^{N,c_2}(X)|_{\infty}>  N^{-\frac{1}{2}+\sigma}\big)\\
\le&  CN^{-\frac{1}{9}+\epsilon}+\mathbb{P}\big(X\in \mathbb{R}^{6N}: \exists i\in \{1,...,N\}:X_i\notin \mathcal{L}_{\sigma}^N\big) .
\end{align*} which according to the discussion after \eqref{statement thm.2} eventually completes the proof of Theorem \ref{thm2}.
\newpage
\section{Discussion of the second main result}
After a quite elongated proof we conclude the chapter concerning the second main result by discussing shortly what we have achieved so far and in which aspects further progress is most desirable. Perhaps the most pleasing point concerning the current result is that (at least morally) no regularization is needed. More precisely, at least for typical initial data we showed that trajectories of the non-regularized microscopic system are up to vanishingly small deviations predicted by the effective dynamics while the trajectories related to the remaining (non-typical) initial data are simply ignored. (The notion `typically' obviously always refers to the currently considered i.i.d. initial data.) This is all we can hope for since in case of singular interaction there are for sure initial configurations where this not true. Moreover, we obtain very strong closeness results for corresponding trajectories of the different dynamics. However, while in the present situation this seems to be a positive aspect, it can as well be considered as an negative issue since the applied approach relies strongly on this closeness between corresponding trajectories. This is exactly the reason why we are not able to apply the current strategy for the case $\alpha=2$ (resp. the Coulomb case) since here the singularity is just too strong so that many of the stated results do not remain true in this setting. Of course, the present version of the approach could be weakened in many aspects and even in the current form slightly larger values for $\alpha$ could be (successfully) considered than $\frac{4}{3}$. However, the case $\alpha=2$ will still remain out of reach. On the other hand, this does not mean that the introduced ideas are worthless for dealing with more singular systems but rather that further ideas are necessary in order that the method remains successful. In any case their are certain aspects which obviously have to be changed to (possibly) handle such systems. $|\cdot |_\infty$ for example is for sure not the right notion of distance for the Vlasov-Poisson system because for few particles deviations of order 1 to their corresponding `mean-field particles' must be expected. A stopping time that sets an upper limit to the number of particles allowed to have a certain distance from their `mean field particle partner' (which naturally shall be smaller the bigger the respectively considered deviation is) seems to be a bit more promising heuristically. This, however, is quite a big modification of the current approach and if implementation is possible at all, there are in any case significant adjustments necessary.
\chapter{Global classical solutions to Vlasov-Poisson equation without bounded kinetic energy\label{sol.unbound.kin.}}
\section{Objective of the chapter}
While the two main results are concerned with justifying the application of Vlasov equation for the effective description of certain microscopic systems, the current chapter shall present a rather subsidiary result regarding the solution theory to this equation. The result is referred as `subsidiary' since the solution theory to Vlasov-Poisson equation is already distinctly better understood than the questions which are treated in the previous chapters. On the other hand, the already established results concerning solution theory are crucial for the approach which we introduced there. We required for instance on several occasions that the `spatial density'
$\widetilde{k}_t(q):=\int_{\mathbb{R}^3}k_t(q,v)d^3v$ keeps bounded for all finite times and fortunately we can rely on various results which guarantee this property under relatively mild assumptions on the initial density (see e.g. \cite{Horst}). However, $\|\widetilde{k}_t\|_{\infty}$ has also a special role in the theory of existence and uniqueness of solutions to Vlasov-Poisson equation. For instance Loeper shows uniqueness of (weak) solutions to Vlasov equation provided that the spatial density keeps bounded (see \cite{Loeper}). Moreover, the result presented by Horst in \cite{Horst81} guarantees the local existence of classical solution provided that the initial density $k_0\in \mathcal{L}^1(\mathbb{R}^6)$ is continuously differentiable, non-negative and fulfills the following constraints for all $(q,v)\in \mathbb{R}^6$, some $\delta>0$ and a suitable constant $C>0$:
\begin{align} 
(i)& \ \ k_0(q,v)\le  \frac{C}{(1+|v|)^{3+\delta}} \notag \\
(ii)& \ \ |\nabla k_0(q,v)| \notag \le  \frac{C}{(1+|v|)^{3+\delta}}.
\end{align}
If the given existence interval is $[0,T)$ and we are able to show that
$$\sup_{0\le s < T} \|\widetilde{k}_s\|_{\infty}<\infty, $$
then the solution can be extended to a larger interval $[0,T']$, $T<T'$ (see for example \cite{Rein}). Thus, it suffices to control the growth of $ \|\widetilde{k}_t\|_{\infty}$ and the corresponding constraint is referred to as the {\em boundedness condition} in \cite{Horst81}. As stated in \eqref{ass.sol.Horst} the additional assumption of an initially bounded kinetic energy
$$\int_{\mathbb{R}^6}|v|^2k_0(q,v)d^6(q,v)<\infty $$
yields that the {\em boundedness condition} remains fulfilled and thus a global solution exists in this case. Our aim is to show that this additional constraint is not necessary at all.\\
One might argue that an unbounded kinetic energy is unphysical and thus not interesting at all. However, on the other hand the microscopic system constitutes the physically relevant system and not its macroscopic description which is given by the solutions to the effective equation. For a finite particle number we do not need to worry about the boundedness of the kinetic energy related to the initial particle configuration no matter if the kinetic energy belonging to $k_0$ is finite or not. Although the already existing results concerning classical solutions are quite satisfactory, dropping the kinetic energy as a constraint might nevertheless be an interesting question, in particular because it determines a crucial tool in previous proofs. \\
There exists already a recent result concerning classical solutions where no bound on the kinetic energy is necessary (see \cite{Zhang}). However, in this case a condition is needed which is referred to as {\em compact velocity-spacial support}. More precisely, the existence of an $\alpha>0$ is required such that
\begin{align*}
&  \sup\{|x-\alpha v|: (x,v)\in \operatorname{supp}k_0\}<\infty
\end{align*}
where $\operatorname{supp}k_0$ shall denote the support of $k_0$ which is quite different to the setting considered by us. \\
Furthermore, results concerning weak solutions without bounds on the kinetic energy can be found in \cite{Jabin solution} and \cite{Zhang2}.
\section{Formulation of the result and implementation of proof}
Before stating and proving the Theorem we remark that many of the basic concepts are already known from the proofs applied in the work of Pfaffelmoser \cite{pfaffelmoser1992}, Schaeffer \cite{schaeffer1991} and Horst \cite{Horst}. We try to present clearly the differences between the approaches which are necessary in order that a bounded kinetic energy becomes dispensable.
\begin{thm} \label{thm3}
Let $k_0\in \mathcal{L}^1(\mathbb{R}^6)$ be continuously differentiable and non-negative. If there exists a constant $C_1>0$ such that for all $(q,v)\in \mathbb{R}^6$ 
\begin{itemize} 
\item[(i)] $\displaystyle k_0(q,v)\le  \frac{C_1}{(1+|v|)^{3+\delta}}$
\item[(ii)] $ \displaystyle |\nabla k_0(q,v)| \notag \le  \frac{C_1}{(1+|v|)^{3+\delta}}$
\end{itemize}
is fulfilled for some $\delta>0$, then for any $T>0$ there exists a unique, continuously differentiable function $k: [0,T]\times \mathbb{R}^6\to [0,\infty)$ satisfying the Vlasov-Poisson equation \eqref{def.unreg.Vlas.} where $k(0,\cdot)=k_0(\cdot)$. 
\end{thm} 
\vspace{0,4cm}
\begin{proof}
As mentioned previously, we already know that under the stated assumptions at least a local solution exists and we only require an appropriate upper bound for the growth of $\|\widetilde{k}_t\|_\infty$ to show that the solution can be extended to arbitrary finite time spans. In the following we apply our usual convention that for $x=(^1x,{^2x})\in \mathbb{R}^6$ the variable describing the position in physical space shall be indicated by $^1x$ while $^2x$ describes the velocity. We assume that the solution exists on $[0,T)$ and define for $t\in [0,T)$:
\begin{align}
\Delta(t):=\sup_{({^1x},{^2x})\in \mathbb{R}^6}\sup_{0 \le s \le t}|^2\varphi_{s,0}({^1x},{^2x})-{^2x}| \label{Def.Delta}
\end{align} 
The approach applied for instance in \cite{Horst} aims to control the growth of such a variable. It is straightforward to see by application of constraint $(i)$ stated in the assumptions and also shown in \cite{Horst81} that this provides us control on the growth of $\|\widetilde{k}_t\|_\infty$ because
\begin{align*}
k_t(q,v)\le \frac{C}{(1+|{^2\varphi_{0,t}}(q,v)|)^{3+\delta}}\le \frac{C}{\big(1+\max(0,|v|-\Delta(t))\big)^{3+\delta}}
\end{align*} 
 which after integration over $v$ concludes the proof if $\Delta(t)$ remains bounded. Hence, controlling the growth of $\Delta(\cdot)$ is exactly our aim in the following. To this end, we will show that for any $t\in [0,T]$ the time $\tau(t)\in [0,t]$ which shall be given by
\begin{align}
\tau(t):=\inf\big\{s\in [0,t]: \sup_{x'\in \mathbb{R}^6} \int_{s}^t|f*\widetilde{k}_r(^1\varphi_{r,0}(x'))|dr\geq \frac{\Delta(t)}{2}\big\} \label{tau}
\end{align}
is sufficiently small and by definition of $\Delta(t)$ it is straightforward to see that the set related to the infimum is nonempty. Hence, instead of controlling the velocity change of the characteristics directly, we first control the mean value of the force an arbitrary characteristic `experiences' over time. It will turn out that in this way we obtain additional information for the estimates which will be crucial in the end.\\
More precisely, we will show that there exist constants $C_1,C_2>0$ such that for all $t\in [0,T]$ it holds that 
\begin{align}
 \Delta(t)\geq C_1 \Rightarrow t-\tau(t) \geq C_2. \label{cond.doubl.}
 \end{align} 
Such a statement yields that 
\begin{align}
\forall t\in [0,T]:\Delta(t)\le C_12^{\lceil \frac{T}{C_2}\rceil} \label{ineq.doubling}.  
\end{align}
This can be easily seen because by definition of $\tau$ it holds for any $x\in \mathbb{R}^6$ and $t\in [0,T]$ that
\begin{align}
&  \max_{0 \le s \le t}|{^2\varphi_{s,0}(^1x,{^2x})}-{^2x}|- \max_{0 \le s \le \tau(t)}|{^2\varphi_{s,0}(^1x,{^2x})}-{^2x}| \notag  \\
\le & \max_{\tau(t) \le s \le t}|^2\varphi_{s,0}(^1x,{^2x})- {^2\varphi_{\tau(t),0}(^1x,{^2x})}|\notag \\
\le & \int_{\tau(t)}^t|f*\widetilde{k}_r(^1\varphi_{r,0}(x))|dr\le \frac{\Delta(t)}{2}  \label{est.doubling}
\end{align}
where the first inequality can be seen as follows: If $|{^2\varphi_{s,0}(^1x,{^2x})}-{^2x}|$ takes its maximal value for $s \in [0,t]$ already in $[0,\tau(t)]$, then the inequality is obviously fulfilled and otherwise it holds by triangle inequality that
\begin{align*}
&\max_{0\le s \le t}|{^2\varphi_{s,0}(^1x,{^2x})}-{^2x}|=\max_{\tau(t)\le s \le t}|{^2\varphi_{s,0}(^1x,{^2x})}-{^2x}|\\
\le & \max_{\tau(t) \le s \le t}|^2\varphi_{s,0}(^1x,{^2x})- {^2\varphi_{\tau(t),0}(^1x,{^2x})}|+\underbrace{| {^2\varphi_{\tau(t),0}(^1x,{^2x})}-{^2x}|}_{\le \max_{0 \le s \le \tau(t)}|{^2\varphi_{s,0}(^1x,{^2x})}-{^2x}|}.
\end{align*}
Since \eqref{est.doubling} holds for arbitrary $x=(^1x,{^2x})\in \mathbb{R}^6$, this relation implies that
\begin{align}
&  \sup_{x\in \mathbb{R}^6}\max_{0 \le s \le t}|{^2\varphi_{s,0}(^1x,{^2x})}-{^2x}|  \le \frac{\Delta(t)}{2} +  \sup_{x\in \mathbb{R}^6}\max_{0 \le s \le \tau(t)}|{^2\varphi_{s,0}(^1x,{^2x})}-{^2x}|  \notag 
\end{align}
which in turn yields by definition \eqref{Def.Delta} that
\begin{align*}
& \Delta(\tau(t))=\sup_{x\in \mathbb{R}^6}\max_{0 \le s \le \tau(t)}|{^2\varphi_{s,0}(^1x,{^2x})}-{^2x}|\geq  \frac{\Delta(t)}{2}  
\end{align*}
respectively 
\begin{align}
& \Delta(t)\le   2\Delta(\tau(t)). \label{rel.tau(t)}
\end{align}
In total we obtain the following: If there exists $t_1\in [0,T]$ such that $\Delta(t_1)\geq C_1$ and condition \eqref{cond.doubl.} is fulfilled, then it follows by relation \eqref{rel.tau(t)} that a further doubling of this variable takes a time span larger or equal to $C_2$ after time $t_1$. Hence, the number of further `doublings' is bounded from above by $\lceil \frac{T}{C_2}\rceil$ which finally shows that the claimed inequality \eqref{ineq.doubling} indeed holds provided that \eqref{cond.doubl.} is fulfilled (which we show in the following).\\
Before starting with the estimates we introduce for $z\in \mathbb{R}^6$ a partition of phase space into five sets which will turn out to be very helpful shortly. Let $t>0$ respectively $\Delta(t)>0$, then we identify for $K_1>0$
\begin{align}
\text{(i) } &M_1(z):=\{x'\in \mathbb{R}^6:|^1x'-{^1z}|\le  K_1\Delta(t)^{-2}\} \notag \\
\text{(ii) }& M_2(z):=\{x'\in \mathbb{R}^6:|^1x'-{^1z}|\geq\Delta(t)^{-\frac{1}{2}}\} \notag \\ 
\text{(iii) }& M_3(z):=\{x'\in \mathbb{R}^6: \notag \\
&\hspace*{1,9cm}  K_1\Delta(t)^{-2}\le|^1x'-{^1z}|\le  \min \big(\Delta(t)^{-\frac{1}{2}},K_1\Delta(t)^{\frac{2}{3}}|^2x'-{^2z}|^{-\frac{8}{3}}\big)\}\notag \\
\text{(iv) } & M_4(z):=\{x'\in \mathbb{R}^6:K_1\Delta(t)^{-2}\le|^1x'-{^1z}|\le \Delta(t)^{-\frac{1}{2}} \land |^2x'-{^2z}|\geq  2\Delta(t)\}\notag \\
\text{(v) }& M_5(z):=\big(\bigcup_{i=1}^4 M_i(z) \big)^C \notag  \\
&\hspace*{1,15cm} =\{x'\in \mathbb{R}^6:|^2x'-{^2z}|<  2\Delta(t) \ \land \notag  \\ &\hspace*{1,9cm} K_1\max\big(\Delta(t)^{-2},\Delta(t)^{\frac{2}{3}}|^2x'-{^2z}|^{-\frac{8}{3}}\big)<|^1x'-{^1z}|< \Delta(t)^{-\frac{1}{2}}\} \label{list.main.res.3}
\end{align} 
where the choice of the constant $K_1>0$ will turn out to be important in the end. The partition is sketched in diagram \eqref{di.p.s.} where for a given $z\in \mathbb{R}^6$ the $x$-axis corresponds to $|^1x'-{^1z}|$ and the $y$-axis to $|^2x'-{^2z}|$. The areas related to (iv) and (v) are kept white since they will be treated differently in the estimates. The idea to apply a partition of phase space for the estimates is based on the work of Pfaffelmoser \cite{pfaffelmoser1992} and was refined (for example) in \cite{schaeffer1991} and \cite{Horst}. Let $z\in \mathbb{R}^6$ be some given configuration.
\begin{figure}[!h]
   \centering
   \def\svgscale{1.0}
   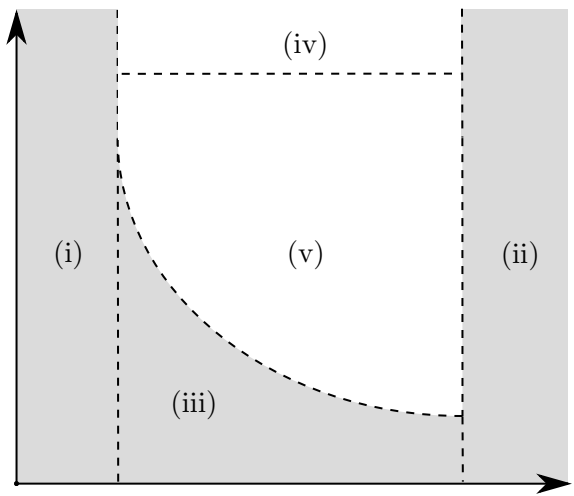
   \caption{
   Diagram of phase space partition}
   \label{di.p.s.2}
\end{figure}  
The sets corresponding to the grey areas $M_i(z)$, $i\in \{1,2,3\}$ are constructed such that the force exerted at ${^1z}$ by the `mass contained in them' keeps in any case sufficiently small for our purposes. The situation is different for the white areas. Estimating simply the maximal force which could be exerted in principle by mass related to these areas would not work for proving the stated result. However, by definition only configurations are contained in these sets which have a certain (distance-dependent) minimal relative velocity value to the considered configuration $z$. Hence, in a time-evolved picture the characteristics related to these configurations will only have a short `encounter' and move apart afterwards. Technically this picture can be realized by performing the time-integration prior to the integration over phase space.\\ 
Now we start to implement these consideration. We remark that the constants $C>0$ which we apply in this proof only may depend on properties of the initial density $k_0$. It obviously holds for any $s\in [\tau(t),t]$ and $x\in \mathbb{R}^6$ (which will be kept fixed for the reasoning) that
\begin{align}
& \int_{s}^t|f*\widetilde{k}_r(^1\varphi_{r,0}(x))|dr \notag \\
\le & \sum_{k=1}^5\int_{s}^t\int_{\mathbb{R}^6}|f({^1\varphi_{r,0}}(x)-{^1x'})|k_r(x')\mathbf{1}_{M_i(\varphi_{r,0}(x))}(x') d^6x'dr. 
\end{align}
and it is straightforward to derive an appropriate upper bound for the contribution of the first three addends to this sum. By application of assumption $(i)$ on the initial density and definition \eqref{Def.Delta} it follows for all $ x'\in \mathbb{R}^6$ and $r\in [s,t]$ that 
\begin{align*}
k_r(x')\le \frac{C_1}{(1+|^2\varphi_{r,0}(x')|)^{3+\delta}}\le \frac{C_1}{\big(1+\max(0,|{^2x'}|-\Delta(t))\big)^{3+\delta}}
\end{align*} which together with the definition of the set $M_1(z)$ stated in the list \eqref{list.main.res.3} yields:
\begin{align}
& \int_{s}^t\int_{\mathbb{R}^6}|f({^1\varphi_{r,0}}(x)-{^1x'})|k_r(x')\mathbf{1}_{M_1(\varphi_{r,0}(x))}(x') d^6x'dr \notag \\
\le & \int_{s}^t\int_{\mathbb{R}^3}\frac{\mathbf{1}_{(0,K_1\Delta(t)^{-2}]}(|q-{^1\varphi_{r,0}}(x)|)}{|q-{^1\varphi_{r,0}}(x)|^2}\underbrace{\big(\int_{\mathbb{R}^3}k_{r}(x)\mathbf{1}_{[0,2\Delta(t)]}(|v|) d^3v\big)}_{\le C\|k_0\|_{\infty}\Delta(t)^3}d^3q dr  \notag \\
&+ \int_{s}^t\underbrace{\int_{\mathbb{R}^3}\frac{\mathbf{1}_{(0,K_1\Delta(t)^{-2}]}(|q-{^1\varphi_{r,0}}(x)|)}{|q-{^1\varphi_{r,0}}(x)|^2}d^3q}_{\le C K_1\Delta(t)^{-2}}\int_{\mathbb{R}^3}\frac{C_1\mathbf{1}_{[2\Delta(t),\infty
)}(|v|) }{(1+|v|-\Delta(t))^{3+\delta}}
d^3v dr  \notag \\
\le &CK_1\|k_0\|_{\infty}\Delta(t)(t-s)+ CK_1\Delta(t)^{-2} (t-s) \label{Lsg.th.bound1}
\end{align}
Moreover, $\mathbf{1}_{M_2({^1\varphi_{r,0}}(x))}(x')=\mathbf{1}_{[\Delta(t)^{-\frac{1}{2}},\infty)}(|^1x'-{^1\varphi_{r,0}}(x)|)$ easily implies that
\begin{align}
& \int_{s}^t\int_{\mathbb{R}^6}\underbrace{|f({^1\varphi_{r,0}}(x)-{^1x'})|\mathbf{1}_{M_2(\varphi_{r,0}(x))}(x')}_{\le \Delta(t)} k_r(x')d^6x'dr 
\le   \|k_0\|_1 \Delta(t)(t-s). \label{Lsg.th.bound2}
\end{align}
Due to the relation $v \geq \Delta(t)\Rightarrow \big(K_1\Delta (t)^{\frac{2}{3}}v^{-\frac{8}{3}}\le K_1\Delta(t)^{-2}\big)$ and the definition of $M_3(z)$ stated in list \eqref{list.main.res.3}, we obtain for the third addend:
\begin{align}
& \int_{s}^t\int_{\mathbb{R}^6}|f({^1\varphi_{r,0}}(x)-{^1x'})|k_r(x')\mathbf{1}_{M_3(\varphi_{r,0}(x))}(x') d^6x'dr \notag 
\\
\le & C\|k_0\|_{\infty}(t-s)\int_{0}^{\Delta(t)}v^2\underbrace{ \big(\int_{\mathbb{R}^3}\frac{1}{|q|^2} \mathbf{1}_{(0,K_1\Delta(t)^{\frac{2}{3}} v^{-\frac{8}{3}}]}(|q|) d^3q\big)}_{\le CK_1\Delta(t)^{\frac{2}{3}} v^{-\frac{8}{3}}}dv \notag \\
\le & CK_1(t-s)\Delta(t)^{\frac{2}{3}}\int_{0}^{\Delta(t)}v^{-\frac{2}{3}}dv \notag \\
\le & CK_1(t-s)\Delta(t) \label{Lsg.th.bound3}
\end{align}
For the last two addends corresponding to the white areas in \eqref{di.p.s.2} we proceed like suggested in the previous considerations by first applying Tonelli:
\begin{align}
& \int_{s}^t\int_{\mathbb{R}^6}|f({^1\varphi_{r,0}}(x)-{^1x'})|k_r(x')\mathbf{1}_{M_i(\varphi_{r,0}(x))}(x') d^6x'dr \notag \\
=  & \int_{\mathbb{R}^6}\int_{s}^t|f({^1\varphi_{r,0}}(x)-{^1\varphi_{r,0}}(x'))|k_0(x')\mathbf{1}_{M_i(\varphi_{r,0}(x))}({\varphi_{r,0}}(x')) dr d^6x'.  \label{change.int.order} 
\end{align}
In previous approaches the kinetic energy was a crucial tool for estimating terms where the time integration is carried out prior to the integration over phase space and we try to summarize roughly how it was usually applied. Since the kinetic energy keeps bounded if it is initially bounded (also in the attractive case), one obtains an upper bound for the amount of the mass having a velocity faster than some value at arbitrary times. This in turn supplies an upper bound for the amount of mass the sets corresponding to the white areas can in principle contain which is already very helpful by itself. However, it yields additionally some constraints on the mass density $\widetilde{k}_t$ which in turn provides a certain control on the force field. This constraint on the force field was applied to derive a lower bound for the time span where considered mass (reps. a considered characteristic) keeps a velocity value of the same order. After having derived such a lower bound it is straightforward to estimate terms of form \eqref{change.int.order} because restricted to such (possibly short) time periods the considered characteristics move more or less like freely evolving particles. Taking additionally into account the upper bounds for the amount of mass contained in these sets which (as mentioned before) are also determined by the kinetic energy finally leads to the required estimate for \eqref{change.int.order}.   \\
Since according to our assumptions the kinetic energy is not necessarily bounded, the approach we consider is quite different in this regard. Very roughly speaking, we choose an arbitrary characteristic $\varphi_{\cdot,0}(x)$ and estimates and upper limit for the `impact' a further characteristic $\varphi_{\cdot,0}(x')$ can in principle exert on it on $[s,t]$ when only times $r\in [s,t]$ are taken into account where 
\begin{align}
\varphi_{r,0}(x')\in M_4(\varphi_{r,0}(x))\cup  M_5(\varphi_{r,0}(x))\label{cond.char.}.
\end{align} Looking at the form of the white sets in our diagram \eqref{di.p.s.2} one easily sees that for a given relative velocity value $v$ between the characteristics we obtain a lower bound for their minimal distance $r$ in space at the relevant points in times (which are those where condition \eqref{cond.char.} is fulfilled). If their relative velocity barely changes, then it is straightforward to see that their impact on each other is bounded by $\frac{C}{rv}$. On the other hand, in the current case we have no suitable constraint on the force field and thus the characteristics can in principle change their relative velocity very fast. Hence, the next `collision' between the considered characteristics might occur after a very short time span which on first sight makes it very hard to implement expedient estimates. However, in any case for each repetition of such an event the characteristics have to change their velocity and the condition on $\tau(t)$ provides us an upper bound for the total `velocity variation' a characteristic might pass on $[\tau(t),t]$ (see \eqref{tau}). This in turn yields us an upper bound for the possible number of certain `collisions' between two characteristics on $[\tau(t),t]$. Thus, the plan is not to care about the `impact per time' of a single `collision event' like in previous approaches but to derive directly an upper bound on the total impact on $[\tau(t),t]$ by application of the constraint provided by $\tau(t).$\\
After this introduction we start with the slightly more complicated estimates for the set $M_5(y)$ because after the considerations for this set, $M_4(y)$ is straightforward to handle. As mentioned before, the definition of $M_5(y)$ ensures that the characteristics have a suitably large relative velocity if they are close in position space and we want to apply this for deriving an appropriate upper bound for 
\begin{align}
\int_{s}^t|f({^1\varphi_{r,0}}(x')-{^1\varphi_{r,0}}(x))|\mathbf{1}_{M_5(\varphi_{r,0}(x))}({\varphi_{r,0}}(x')) dr \label{coll.integral}
\end{align}
which holds for any $x'\in \mathbb{R}^6$ and $s\in [\tau(t),t]$. \\
To this end, let $x'\in \mathbb{R}^6$ and for $s\in [\tau(t),t]$
\begin{align}
&b_s:=\inf\{r\in [s,t]:  \mathbf{1}_{M_5(\varphi_{r,0}(x))}({\varphi_{r,0}}(x'))=1 \vee r=t\} .\label{Def b s}
\end{align} Obviously, we only have to care about situations where $b_s<t$ because otherwise \eqref{coll.integral} is equal to 0 anyway. Moreover, we define the time interval $[b_s,e_s]\subseteq [s,t]$ by choosing $e_s \le t$ maximal with the property that for all $r\in [b_s,e_s]$ the relation
\begin{align}
&|\big(^2\varphi_{r,0}(x')-{^2\varphi_{r,0}}(x)\big)-\big(^2\varphi_{b_s,0}(x')-{^2\varphi_{b_s,0}}(x)\big)|\le \frac{|^2\varphi_{b_s,0}(x')-{^2\varphi_{b_s,0}}(x)|}{2} \label{cond.r_e}
\end{align}
is fulfilled. Thus, value and direction of the relative velocity do not change too much on the considered interval $[b_s,e_s]$. We regard the only interesting setting which is $b_s<t$ (and hence $b_s<e_s$) and identify additionally 
$$ x_{min}:=\inf\{|^1\varphi_{r,0}(x')-{^1\varphi_{r,0}}(x)|:r\in [b_s,e_s]\land \mathbf{1}_{M_5(\varphi_{r,0}(x))}(\varphi_{r,0}(x'))=1\}.$$
We remark that the infimum exists in this case because due to the definition of $b_s$ the considered set is nonempty if $b_s<t$. It follows by the fact that $x'\in M_5(z)$ iff 
$$|^2x'-{^2z}|<  2\Delta(t) \ \land K_1\max\big(\Delta(t)^{-2},\Delta(t)^{\frac{2}{3}}|^2x'-{^2z}|^{-\frac{8}{3}}\big)<|^1x'-{^1z}|< \Delta(t)^{-\frac{1}{2}} $$ and the continuity of $\varphi_{\cdot,0}(z)$ for $z\in \mathbb{R}^6$ as well as the constraint on $e_s$ (see \eqref{cond.r_e}) that
\begin{align}
 x_{min}\geq & K_1\Delta(t)^{\frac{2}{3}}(\sup_{r\in [b_s,e_s]}|^2\varphi_{r,0}(x')-{^2\varphi_{r,0}}(x)|)^{-\frac{8}{3}}\notag \\
\geq & K_1\Delta(t)^{\frac{2}{3}}(\frac{3}{2}|^2\varphi_{b_s,0}(x')-{^2\varphi_{b_s,0}}(x)|)^{-\frac{8}{3}}. \label{x_min}
\end{align} 
Moreover, due to 
$$\min_{r\in [b_s,e_s]}|^1\varphi_{r,0}(x')-{^1\varphi_{r,0}}(x)|\le x_{min}\le \max_{r\in [b_s,e_s]}|^1\varphi_{r,0}(x')-{^1\varphi_{r,0}}(x)|$$ there exists a point in time $r_0\in [b_s,e_s]$ where $|^1\varphi_{r_0,0}(x')-{^1\varphi_{r_0,0}}(x)|=x_{min}$.
If we abbreviate $v:=|^2\varphi_{b_s,0}(x')-{^2\varphi_{b_s,0}}(x)|$, then it holds for all $r\in [b_s,e_s]$ that
$$|^1\varphi_{r,0}(x')-{^1\varphi_{r,0}}(x)|\geq \max\big(x_{min},\frac{v}{2}|r-r_0|-x_{min} \big)\mathbf{1}_{M_5(\varphi_{r,0}(x))}({\varphi_{r,0}}(x'))$$
where the lower bound $x_{min}$ arises directly by its definition and the second lower bound by condition \eqref{cond.r_e} which implies for $r\in [b_s,e_s]$ that
\begin{align*}
& |^1\varphi_{r,0}(x')-{^1\varphi_{r,0}}(x)|\\
 = & \big|\big({^1\varphi_{r_0,0}(x')}-{^1\varphi_{r_0,0}}(x)\big)+\int_{r_0}^r \big({^2\varphi_{u,0}}(x')-{^2\varphi_{u,0}}(x)\big) du  \big|\\ 
 \geq &|^2\varphi_{b_s,0}(x')-{^2\varphi_{b_s,0}}(x)||r-r_0|\\
 &-\underbrace{\big|\int_{r_0}^r\big(^2\varphi_{u,0}(x')-{^2\varphi_{u,0}}(x)\big)-\big(^2\varphi_{b_s,0}(x')-{^2\varphi_{b_s,0}}(x)\big)du\big|}_{\le \frac{|^2\varphi_{b_s,0}(x')-{^2\varphi_{b_s,0}}(x)|}{2}|r-r_0|}\\& -\underbrace{|{^1\varphi_{r_0,0}}(x')-{^1\varphi_{r_0,0}}(x)|}_{=x_{min}}.
\end{align*} 
However, this yields by application of $x_{min}\geq K_1\Delta(t)^{\frac{2}{3}}(\frac{3}{2}v)^{-\frac{8}{3}}$ (see \eqref{x_min}) that
\begin{align}
& \int_{b_s}^{e_s}|f({^1\varphi_{r,0}}(x')-{^1\varphi_{r,0}}(x))|\mathbf{1}_{M_5(\varphi_{r,0}(x))}({\varphi_{r,0}}(x')) dr \notag \\ 
\le & \int_{b_s-r_0}^{e_s-r_0}\frac{1}{\max\big(x_{min},\frac{v}{2}|r| -x_{min}\big)^2}dr 
\le   \frac{C}{x_{min}v}\le \frac{C}{K_1}\Delta(t)^{-\frac{2}{3}}v^{\frac{5}{3}}. \label{impact}
\end{align} 
 We continue by implementing the further steps mentioned in the preliminary considerations and recall that according to the definition of $\tau(t)$ (see \eqref{tau}) and $s\geq \tau(t)$ we have on the one hand the condition $$\sup_{y\in \mathbb{R}^6}\int_{s}^t|f*\widetilde{k}_r(^1\varphi_{r,0}(y))|dr\le  \frac{\Delta(t)}{2}$$ which in particular yields
\begin{align}
\int_{s}^{t}|f*\widetilde{k}_r(^1\varphi_{r,0}(x'))|dr+\int_{s}^{t}|f*\widetilde{k}_r(^1\varphi_{r,0}(x))|dr\le  \Delta(t)\label{cond.costs1}
\end{align} 
and on the other hand the claim on $e_s$ (see \eqref{cond.r_e}) which implies the last inequality of the subsequent relation:
\begin{align}
 &\Big(\int_{b_s}^{e_s}|f*\widetilde{k}_r({^1\varphi_{r,0}}(x'))|dr+\int_{b_s}^{e_s}|f*\widetilde{k}_r({^1\varphi_{r,0}}(x))|dr \Big) \notag \\
& \ \geq  \big|\big({^2\varphi_{e_s,0}}(x')-{^2\varphi_{e_s,0}}(x)\big)-\underbrace{\big({^2\varphi_{b_s,0}}(x')-{^2\varphi_{b_s,0}}(x)\big)}_{|\p |=v}\big|\geq \frac{v}{2} \label{cond.costs2}
\end{align}
Thus, according to condition \eqref{cond.costs1} the `acceleration or deceleration process' taking place on $[b_s,e_s]$ is in a sense connected with `costs' $\frac{v}{2}$ because for a certain $v>0$ at most $\lceil \Delta(t)(\frac{v}{2})^{-1}\rceil=\lceil 2\frac{\Delta(t)}{v}\rceil$ repetitions can occur on the time interval $[s,t]\subseteq [\tau(t),t]$. Moreover, the ratio of `impact' (see \eqref{impact}) to `costs' per cycle (or repetition) is monotonously increasing with respect to $v$ since
$$\big(\frac{C}{K_1}\Delta(t)^{-\frac{2}{3}}v^{\frac{5}{3}}\big)\big(\frac{v}{2}\big)^{-1}\le \frac{C}{K_1}\Delta(t)^{-\frac{2}{3}}v^{\frac{2}{3}}.$$ 
Eventually, we remark that according to the definition of the set $M_5(y)$, the point in time $b_s$ and the continuity of the flow only values $$ K_1^{\frac{3}{8}}\Delta(t)^{\frac{7}{16}} \le v=|{^2\varphi_{b_s,0}}(x')-{^2\varphi_{b_s,0}}(x)|\le 2\Delta(t)$$ are relevant for estimating the `collision integral' \eqref{coll.integral} and thus it is straightforward to conclude by the previous considerations that
\begin{align}
& \int_{\tau(t)}^{t}|f({^1\varphi_{r,0}}(x')-{^1\varphi_{r,0}}(x))|\mathbf{1}_{M_5(\varphi_{r,0}(x))}({\varphi_{r,0}}(x')) dr \le  \frac{C}{K_1} \Delta(t)\label{Lsg.th.bound5}.
\end{align}
More rigorously, this can be seen by covering the relevant time spans of $[\tau(t),t]$ by such intervals $[b_s,e_s]$ where we recall that theses points in time were defined in \eqref{Def b s} and \eqref{cond.r_e}. To this end, we identify $t_1:=b_{\tau(t)}$ and $T_1:=e_{\tau(t)}$ which yields the first interval $[t_1,T_1]$ and as long as $t_n<t$ we define $t_{n+1}=b_{T_n}$ and $T_{n+1}:=e_{T_n}$ to receive the $(n+1)$-th interval $[t_{n+1},T_{n+1}]$. We mentioned previously that the minimal `velocity change' of the characteristics on each such interval is bounded from below by order $K_1^{\frac{3}{8}}\Delta(t)^{\frac{7}{16}}$ and thus the sequence ends after a finite number of steps. Let $K$ denote the first natural such that $t_K= t$ then we abbreviate 
$v_k:=|{^2\varphi_{t_k,0}}(x)-{^2\varphi_{t_k,0}}(x')|$ for $k\in \{1,...,K\}$ and according to our previous considerations it holds that
\begin{align*}
& \int_{\tau(t)}^{t}|f({^1\varphi_{r,0}}(x')-{^1\varphi_{r,0}}(x))|\mathbf{1}_{M_5(\varphi_{r,0}(x))}({\varphi_{r,0}}(x')) dr\\
\le & \sum_{k=1}^K v_{k}\max_{i\in \{1,...,K\}}\frac{1}{v_i}\int_{t_i}^{T_i}|f({^1\varphi_{r,0}}(x')-{^1\varphi_{r,0}}(x))|\mathbf{1}_{M_5(\varphi_{r,0}(x))}({\varphi_{r,0}}(x')) dr \\
\le & \frac{C}{K_1}\big(\sup_{0<v\le 2\Delta(t) }\Delta(t)^{-\frac{2}{3}}v^{\frac{2}{3}}\big)\sum_{k=1}^K v_{k} \\
\le &   \frac{C}{K_1}\Delta(t) 
\end{align*}
where in the second last step we regarded relation \eqref{impact} and in the last step the constraints on the number of `cycles' provided by \eqref{cond.costs1} and \eqref{cond.costs2} which imply that $\sum_{k=1}^K v_{k}\le 2\Delta(t)$.
Consequently, the term related to $M_5(y)$ is bounded by $\frac{C}{K_1}\Delta(t)$ and finally only one addend remains. For estimating this last term we point out in a first step that the condition on $\tau(t)$ (see \eqref{tau}) implies that
\begin{align*}
& \forall r,s\in [\tau(t),t]:|^2\varphi_{s,0}(x')-{^2\varphi_{s,0}}(x)-(^2\varphi_{r,0}(x')-{^2\varphi_{r,0}}(x))|\le \Delta (t).
\end{align*}
If, on the other hand, there exists $r\in [\tau(t),t]$ such that $\varphi_{r,0}(x')\in M_4(\varphi_{r,0}(x))$, then the definition of this set implies that $|^2\varphi_{r,0}(x')-{^2\varphi_{r,0}}(x)|\geq 2\Delta(t)$ and thus the component of $(^2\varphi_{s,0}(x')-{^2\varphi_{s,0}}(x))$ in direction of $(^2\varphi_{r,0}(x')-{^2\varphi_{r,0}}(x))$ keeps a value of at least $\Delta(t)$ for $s\in [\tau(t),t]$. Finally, we take into account that $$\varphi_{r,0}(x')\in M_4(\varphi_{r,0}(x))\Rightarrow |^1\varphi_{r,0}(x')-{^1\varphi_{r,0}}(x)|\geq K_1\Delta(t)^{-2}=:r_{min}.$$
By recalling the estimates which we applied for the addend related to $M_5(z)$ these considerations easily imply that
\begin{align}
& \int_{\mathbb{R}^6}\int_{\tau(t)}^{t}|f({^1\varphi_{r,0}}(x')-{^1\varphi_{r,0}}(x))|\mathbf{1}_{M_4(\varphi_{r,0}(x))}({\varphi_{r,0}}(x')) dr k_0(x')d^6x'\notag \\ 
\le &   \int_{\mathbb{R}^6}\frac{C}{r_{min }\Delta(t)} k_0(x')d^6x'  =\int_{\mathbb{R}^6}\frac{C}{(K_1\Delta(t)^{-2})\Delta(t)} k_0(x')d^6x' \notag \\
\le &  \frac{C}{K_1}\Delta(t). \label{Lsg.th.bound4}
\end{align}
Merging all upper bounds \eqref{Lsg.th.bound1}, \eqref{Lsg.th.bound2}, \eqref{Lsg.th.bound3}, \eqref{Lsg.th.bound5} and \eqref{Lsg.th.bound4}, choosing $s=\tau(t)$ and applying the definition of $\tau(t)$ (see \eqref{tau}) implies in total that:
\begin{align}
\frac{\Delta(t)}{2}& = \sup_{x\in \mathbb{R}^6}\int_{\tau(t)}^t|f*\widetilde{k}_r(^1\varphi_{r,0}(x))|dr \notag \\ 
& \le  C\big(\frac{1}{K_1}+K_1(t-\tau(t))\big)\Delta(t)+C\frac{K_1}{\Delta(t)^2}(t-\tau(t)) \notag \\
\Rightarrow \ \ \frac{1}{2} &\le C\big(\frac{1}{K_1}+K_1(t-\tau(t))\big)+C\frac{K_1}{\Delta(t)^3}(t-\tau(t))
\end{align}
After $K_1>0$ has been chosen large enough, the last inequality yields that there in fact exist constants $C_1,C_2>0$ (depending only on the properties of $k_0$) such that for all $t\in [0,T]$ the following implication holds $$\Delta(t)\geq C_1 \Rightarrow t-\tau(t)\geq C_2$$ which completes the proof.
\end{proof}
  
\chapter{Appendix \label{sec.app.}}
\textbf{Proof of Lemma \ref{Gron.lem.}:}
\begin{proof}
For a simpler comprehension of the proof we indicate the items which determine important constraints on the applied maps also here: For some $n\in \mathbb{N}$ and for all $t_1>0,\ x_1,x_2\geq 0$ the maps shall fulfill
\begin{itemize}
\item[(i)]
$\begin{aligned}&
x_1< x_2 \Rightarrow f_2(t_1,x_1)\le  f_2(t_1,x_2) 
\end{aligned}$
\item[(ii)]
$ \exists K_1,\delta>0: \sup\limits_{\substack{x,y  \in [f_1(0),f_1(0)+\delta]\\s \in [0,\delta]}}|f_2(s,x)-f_2(s,y)|\le K_1|x-y|$.
\item[(iii)]
$ \begin{aligned}
& f_1(t_1)+ \int_0^{t_1}...\int_0^{t_n} f_2(s,u(s))dsdt_n...dt_2<  u(t_1) \ \land \\ 
& f_1(t_1)+\int_0^{t_1}...\int_0^{t_n}  f_2(s,l(s))dsdt_n...dt_2 \geq l(t_1)
\end{aligned}$
\end{itemize} 
The continuity of the different functions and condition (iii) yield that $$0\le l(0) \le f_1(0)\le u(0).$$ Now it suffices to show that there exists a point in time $t_1>0$ such that $\forall s\in [0,t_1]: l(s) \le u(s)$ because in this case the monotony of the integral implies together with assumption (i) and (iii) that 
\begin{align*}
l(t_1)  \le & f_1(t_1)+ \int_0^{t_1}...\int_0^{t_n}  f_2(s,l(s))dsdt_n...dt_2\\
\le &    f_1(t_1)+\int_0^{t_1}...\int_0^{t_n}  f_2(s,u(s))dsdt_n...dt_2 
<  u(t_1)
\end{align*}
which yields in turn that the existence of a point in time where the statement of the lemma does not hold can be falsified.\\ 
Due to the continuity of the maps the only case where the existence of a point in time $t>0$ such that $ \forall s\in [0,t]: l(s)\le u(s)$ is non-trivial arises if $$f_1(0)=u(0)=l(0).$$
 Thus, we restrict ourselves to this case and define the set $$M(t):=\{s\in [0,t]:l(s)-u(s)\geq 0\}$$ and remark that for $s\in [0,t]\setminus M(t)$ it holds according to item (i) that 
\begin{align}
f_2(s,l(s))-f_2(s,u(s))\le 0 \label{cond.Gron.proof}
\end{align} Moreover, let $0<\delta'\le \delta$ be small enough such that 
$$\sup_{0\le s\le \delta'}|l(s)-f_1(0)|<\delta\ \land \ \sup_{0\le s\le \delta'}|u(s)-f_1(0)|<\delta,$$ then it follows due to the monotony of $u$ that $u(s)\in [f_1(0),f_1(0)+\delta]$ for all $s\in [0,\delta']$ and thus $l(s)\in [f_1(0),f_1(0)+\delta]$ for all $s\in M(\delta')$. Now application of (iii) in the first step, \eqref{cond.Gron.proof} in the second step and (ii) in the third step as well as again the continuity of the functions yields that for $t_1\in [0,\delta']$
\begin{align*}
&\max_{t\in [0,t_1]}\big(l(t)-u(t)\big)\\  \le &\max_{t\in [0,t_1]} \int_0^{t}...\int_0^{t_n} \big( f_2(s,l(s))-f_2(s,u(s))\big) dsdt_n...dt_2\\
 \le & \int_0^{t_1}...\int_0^{t_n}\underbrace{\big( f_2(s,l(s))-f_2(s,u(s))\big)\mathbf{1}_{M(t_1)}(s) }_{\geq 0}dsdt_n...dt_2\\ 
 \le & K_1 \int_0^{t_1}...\int_0^{t_n}\big(l(s)-u(s)\big)\mathbf{1}_{M(t_1)}(s)dsdt_n..dt_2 \\
  \le & t_1^n K_1\max_{s\in [0,t_1]}\big(l(s)-u(s)\big).
\end{align*}
which can only be fulfilled for all $t_1\in [0,\delta']$ if $\max_{s\in [0,t]}\big(l(s)-u(s)\big)= 0$ (where we regarded in the last step the assumption $l(0)=u(0)$). 
\end{proof}
\vspace{1cm}
\noindent \textbf{Proof of Corollary \ref{cor1} (ii):}
\begin{proof}
Let $X=(X_1,...,X_N)\in \mathbb{R}^{6N}$, $Y_i,Y_j\in \mathbb{R}^6$ and $\Delta v>0$ such that 
\begin{align}
\max_{k\in \{i,j\}}\sup_{0\le t \le T}|\varphi^{N}_{t,0}(Y_k)-[{\Psi}^{N}_{t,0}(X)]_k|\le N^{-\delta}\Delta v \label{est.v.Cor.0}
\end{align}
for some $\delta>0$ as well as  
\begin{align}
N^{\delta}|^1\varphi^N_{t_{min},0}(Y_i)-{^1\varphi^{N}_{t_{min},0}}(Y_j)|\le |^2\varphi^N_{t_{min},0}(Y_i)-{^2\varphi^N_{t_{min},0}}(Y_j)|= \Delta v. \label{est.v.Cor.1}
\end{align} 
where as usual $t_{min}$ shall again denote a point in time where $|^1\varphi^N_{\cdot ,0}(Y_i)-{^1\varphi^{N}_{\cdot,0}}(Y_j)|$ attains its minimal value on $[0,T]$. It follows by Lemma \ref{lem1} that there exists a constant $C_0>0$ such that
\begin{align}
&\frac{\Delta v}{C_0}\le \min_{0\le t \le T}|\varphi^N_{t,0}(Y_i)-{\varphi^N_{t,0}}(Y_j)|
\le \max_{0\le t \le T}|\varphi^N_{t,0}(Y_i)-{\varphi^N_{t,0}}(Y_j)|
\le  C_0\Delta v  \label{est.v.Cor.1,5}
\end{align}
and for large enough $N\in \mathbb{N}$ it holds according to \eqref{est.v.Cor.0} that
\begin{align}
&\min_{0\le t \le T}|[\Psi^{N}_{t,0}(X)]_i-[\Psi^{N}_{t,0}(X)]_j | \notag \\
\geq &  \min_{0\le t \le T}|\varphi^{N}_{t,0}(Y_i)-{\varphi^{N}_{t,0}}(Y_j)|-2N^{-\delta}\Delta v 
\geq \frac{\Delta v}{C_0}-2N^{-\delta}\Delta v \notag \\
\geq & \frac{\Delta v}{2C_0}. \label{est.v.Cor.2}
\end{align}
Let for $t'\in [0,T]$ and $C_1>1$ $t'_{min}$ denote a point in time where $|[^1{\Psi}^{N}_{\p,0}(X)]_i-[^1{\Psi}^{N}_{\p,0}(X)]_j|$ attains its minimum on $[t',t'+\frac{1}{C_1}]$. For a compact notation we identify $\widetilde{X}_i:=[\Psi^{N}_{t'_{min},0}(X)]_i$ and $\widetilde{X}_j:=[\Psi^{N}_{t'_{min},0}(X)]_j$.\\
If $|^1\widetilde{X}_i-{^1\widetilde{X}_j}|\geq \frac{\Delta v}{4C_0}$, then the properties of the map $h_N $ easily yield that 
\begin{align*}
&\int^{t'_{min}+\frac{1}{C_1}}_{t'_{min}}|h_N([^1\Psi^{N}_{s,0}(X)]_i-[^1\Psi^{N}_{s,0}(X)]_j)|ds\\
\le & C\min\big(\frac{1}{c_N^{\widetilde{\alpha}}}, \frac{1}{|^1\widetilde{X}_i-{^1\widetilde{X}_j}|^{\widetilde{\alpha}}}\big)\le  \frac{C}{\Delta v}\min\big(\frac{1}{c_N^{\widetilde{\alpha}-1}}, \frac{1}{|^1\widetilde{X}_i-{^1\widetilde{X}_j}|^{\widetilde{\alpha}-1}}\big).
\end{align*} 
In the following we consider the remaining case $|^1\widetilde{X}_i-{^1\widetilde{X}_j}|< \frac{\Delta v}{4C_0}$ which due to relation \eqref{est.v.Cor.2}, however, implies that 
\begin{align} 4C_0|^1\widetilde{X}_i-{^1\widetilde{X}_j}|\le \Delta v \le 4C_0|^2\widetilde{X}_i-{^2\widetilde{X}_j}|. \label{rel.cor.2ii}  \end{align} 
Moreover, according to Lemma \ref{lem2} it holds for all $|s-t_{min}|\le \frac{1}{C_1}$ that
\begin{align}
&\big|^2\varphi^{N}_{s,0}(Y_i)-{^2\varphi^{N}_{s,0}}(Y_j)- (^2\varphi^{N}_{t'_{min},0}(Y_i)-{^2\varphi^{N}_{t'_{min},0}}(Y_j)) \big| \notag \\
\le &C |s-t'_{min}|\Big(|^1\varphi^{N}_{t'_{min},0}(Y_i)-{^1\varphi^{N}_{t'_{min},0}}(Y_j)| \notag \\
& \hspace*{1,65cm} +|^2\varphi^{N}_{t'_{min},0}(Y_i)-{^2\varphi^{N}_{t'_{min},0}}(Y_j)| |s-t'_{min}|\Big) \notag
\end{align}
which by application of relations \eqref{est.v.Cor.0},\eqref{est.v.Cor.1,5} and \eqref{rel.cor.2ii} yields that
\begin{align}
&\big|^2\varphi^{N}_{s,0}(Y_i)-{^2\varphi^{N}_{s,0}}(Y_j)- (^2\varphi^{N}_{t'_{min},0}(Y_i)-{^2\varphi^{N}_{t'_{min},0}}(Y_j)) \big| \notag \\
\le &C |s-t'_{min}|\Big(\big(\underbrace{|^1\widetilde{X}_i-{^1\widetilde{X}_j}|}_{\le |^2\widetilde{X}_i-{^2\widetilde{X}_j}|}+2N^{-\delta}\Delta v\big)  +C_0\Delta v |s-t'_{min}|\Big) \notag \\
\le & C|s-t'_{min}| |^2\widetilde{X}_i-{^2\widetilde{X}_j}|
\end{align}
According to this estimate and \eqref{est.v.Cor.0} it follows for sufficiently large values of $N\in \mathbb{N}$ and $|t-t'_{min}|\le \frac{1}{C_1}$ that
{ \allowdisplaybreaks
\begin{align}
&\big|([^1\Psi^{N}_{t,0}(X)]_i-[^1\Psi^{N}_{t,0}(X)]_j) -(^1\widetilde{X}_i-{^1\widetilde{X}_j})-(^2\widetilde{X}_i-{^2\widetilde{X}_j})(t-t'_{min})\big| \notag \\
= & \big|\int_{t'_{min}}^{t}([^2\Psi^{N}_{s,0}(X)]_i-[^2\Psi^{N}_{s,0}(X)]_j)ds-(^2\widetilde{X}_i-{^2\widetilde{X}_j})(t-t'_{min}) \big| \notag \\
\le & \big|\int_{t'_{min}}^{t}([^2\Psi^{N}_{s,0}(X)]_i-[^2\Psi^{N}_{s,0}(X)]_j)-(^2\varphi^{N}_{s,0}(Y_i)-{^2\varphi^{N}_{s,0}}(Y_j))ds \big| \notag \\
& +\big|\int_{t'_{min}}^t(^2\varphi^{N}_{s,0}(Y_i)-{^2\varphi^{N}_{s,0}}(Y_j))- (^2\varphi^{N}_{t'_{min},0}(Y_i)-{^2\varphi^{N}_{t'_{min},0}}(Y_j))ds \big| \notag \\
& + \big|({^2\varphi^{N}_{t'_{min},0}}(Y_i)-{^2\varphi^{N}_{t'_{min},0}}(Y_j))-(^2\widetilde{X}_i-{^2\widetilde{X}_j})\big||t-t'_{min}|  \notag \\
\le & 2\sup_{k\in \{i,j\}}\sup_{0\le s \le T}|\varphi^N_{s,0}(Y_k)-[{\Psi}^N_{s,0}(X)]_k| |t-t'_{min}| \notag \\
&+  C|^2\widetilde{X}_i-{^2\widetilde{X}_j}||t-t'_{min}|^2 \notag \\
& +2\sup_{k\in \{i,j\}}\sup_{0\le s \le T}|\varphi^N_{s,0}(Y_k)-[{\Psi}^N_{s,0}(X)]_k| |t-t'_{min}| \notag \\
\le & 4 N^{-\delta}\underbrace{\Delta v}_{\le C|^2\widetilde{X}_i-{^2\widetilde{X}_j}|}|t-t'_{min}|+ C|^2\widetilde{X}_i-{^2\widetilde{X}_j}||t-t'_{min}|^2 \label{est.dev.vel.}
\end{align}}
where in the last step we regarded relation \eqref{rel.cor.2ii}.\\ 
Thus, in this case the `real' particles fly apart almost like freely moving particles for a possibly short (but $N$- and $X$-independent) time span after a collision exactly like the `mean-field particles'. Just like in the proof of part (i) this implies that
\begin{align*}
&\int^{t'_{min}+\frac{1}{C_1}}_{t'_{min}}|h_N([^1\Psi^{N}_{s,0}(X)]_i-[^1\Psi^{N}_{s,0}(X)]_j)|ds\\
\le & \frac{C}{|^2\widetilde{X}_i-{^2\widetilde{X}_j}|}\min\big(\frac{1}{c_N^{\widetilde{\alpha}-1}}, \frac{1}{|^1\widetilde{X}_i-{^1\widetilde{X}_j}|^{\widetilde{\alpha}-1}}\big)\\
\le & \frac{C}{\Delta v}\min\big(\frac{1}{c_N^{\widetilde{\alpha}-1}}, \frac{1}{|^1\widetilde{X}_i-{^1\widetilde{X}_j}|^{\widetilde{\alpha}-1}}\big).
\end{align*}
where the last step follows again due to \eqref{rel.cor.2ii}.\\
Since $t'\in [0,T]$ was chosen arbitrarily and the length of the time interval $\frac{1}{C_1}$ can be selected independent of $X$ and $N$, this estimate can again be `extended' successively to the whole time period $[0,T]$ such that 
\begin{align*}
&\int^{T}_{0}|h_N([^1\Psi^{N}_{s,0}(X)]_i-[^1\Psi^{N}_{s,0}(X)]_j)|ds\\
\le & C\min\big(\frac{1}{c_N^{\widetilde{\alpha}-1}\Delta v}, \frac{1}{\min\limits_{0\le s\le T}|[^1\Psi^{N}_{s,0}(X)]_i-[^1\Psi^{N}_{s,0}(X)]_j|^{\widetilde{\alpha}-1}\Delta v}\big).
\end{align*}
\end{proof}
\vspace{1cm}
\noindent \textbf{Proof of Lemma \ref{lem3} (ii):}
\begin{proof}
Now we want to conclude the proof of Lemma \ref{lem3}. The statement is only interesting if $\Delta x<1$ and thus we consider only such values in the following.\\
Furthermore, we only make the estimates explicit for the most interesting choices of $Y$ since the proof for the remaining options is straightforward. More precisely, we only consider $Y$ where $|^2Y|\geq 2\sup_{N\in \mathbb{N}}\sup_{0\le s \le T}\|f^{N}*\widetilde{k}^{N}_s\|_{\infty}T$ (which obviously implies the relation
$\frac{1}{2}|^2Y| \le |^2\varphi^{N}_{t,0}(Y)|\le 2|^2Y|$ for all $t \in [0,T]$). We call this the most interesting case because if one argues solely by the idea (or picture) of a `Boltzmann collision cylinder' the postulated statement could not be verified if $|^2Y|$ can be chosen arbitrarily large. However, by application of the decay of $k_0$ it is possible to show that such a `mean-field particle' will move inevitably to areas of low density which will eventually enable us to show that this statement is nevertheless valid.\\
Thus, if $Y\in \mathbb{R}^6$ is chosen like described before, then for arbitrary $M\in \mathbb{N}$ the subsequent relationship holds:
\begin{align}
& \exists t\in [0,T]:|^1\varphi^{N}_{t,0}(Y)-{^1\varphi^{N}_{t,0}}(Z)| \le \Delta x \notag  \\
\Rightarrow & |^1Y-{^1Z}|\le \Delta x  \ \vee\  \Big(\exists k \in \{0,...,M-1\}: \notag  \\
\big(& \Delta x \le |^1\varphi^{N}_{k \frac{T}{M} ,0}(Y)-{^1\varphi^{N}_{k \frac{T}{M},0}}(Z)| \le \Delta x+3|^2Y|\frac{T}{M} \label{ass.Y} \ \vee \\ 
& \Delta x \le |^1\varphi^{N}_{k \frac{T}{M} ,0}(Y)-{^1\varphi^{N}_{k \frac{T}{M} ,0}}(Z)| \le \Delta x+3|^2Z|\frac{T}{M} \big)\Big) \label{ass.Z}
\end{align}
For this conclusion one has to regard that $3\max(|^2Z|,|^2Y|)$ is an upper bound for the value of the relative velocity between the `mean-field particles' due to the current condition that $|^2Y|\geq 2\sup_{0\le s \le T}\|f^{N}*\widetilde{k}^{N}_s\|_{\infty}T$.\\
Since $\|\widetilde{k}_0\|_{\infty}<\infty$, it follows that the probability of the event $|^1Y-{^1Z}|\le \Delta x$ is bounded by $C\Delta x^3\le C\Delta x^2 $ where we regarded that $\Delta x\le 1$. Despite their apparent similarity assumptions \eqref{ass.Y} and \eqref{ass.Z} differ because $Y\in \mathbb{R}^6$ shall be an arbitrarily chosen configuration (except for the current assumption that $|^2Y|$ shall be `large') while $Z\in \mathbb{R}^6$ is random. Thus, the more interesting statement is \eqref{ass.Y} and if we are able to show that the corresponding probability is sufficiently small, then of course the same is true for the remaining statement.\\
For convenience we assume that $|^1Y|=\min_{0\le s \le T} |^1\varphi^{N}_{s,0}(Y)|$. If $|^1\varphi^{N}_{\cdot,0}(Y)|$ attains its minimum at another moment $t\in [0,T]$, then the reasoning is analogous except for the slight difference that one has to apply the same arguments in both time directions starting from time $t$.\\ It is straightforward to see that the boundedness of the mean-field force (which is a consequence of $\sup_{0\le s\le T}\|\widetilde{k}^N_s\|_\infty<C$) implies that a relation of the form $k_t^N(Z)\le \frac{C}{(1+|Z|)^{4+\delta}}$ (for all $Z \in \mathbb{R}^6$) which by assumption holds for $t=0$ also holds at later times if the constant is adjusted. More precisely, if we abbreviate $$f_{max}:= \sup_{N\in \mathbb{N}}\sup_{0\le s\le T}\|f^{N}*\widetilde{k}^{N}_s\|_{\infty}< \infty, $$ then it holds for any $t\in  [0,T]$, $N\in \mathbb{N}$ and $\widetilde{X}\in \mathbb{R}^6$ where $|^2\widetilde{X}|\geq 2f_{max}T$ that
\begin{align*}
& |^1\varphi^{N}_{0,t}(\widetilde{X})|\geq |^1\widetilde{X}|-|^2\widetilde{X}|t-\frac{1}{2}f_{max}t^2\geq  |^1\widetilde{X}|-\frac{5}{4}T|^2\widetilde{X}| \\
& |^2\varphi^{N}_{0,t}(\widetilde{X})|\geq |^2\widetilde{X}|-f_{max}t\geq  \frac{ |^2\widetilde{X}|}{2} 
\end{align*}
which yields 
\begin{align*}
|\varphi^{N}_{0,t}(\widetilde{X})|&\geq \frac{1}{2}\big(\frac{ |^2\widetilde{X}|}{2}+\max(0,|^1\widetilde{X}|-\frac{5}{4}T|^2\widetilde{X}|)  \big)\\
 & = \frac{1}{8\lceil \frac{5}{4}T \rceil }\Big(2\lceil \frac{5}{4}T \rceil |^2\widetilde{X}|+4\lceil \frac{5}{4}T \rceil\max(0,|^1\widetilde{X}|-\frac{5}{4}T|^2\widetilde{X}|  \big)\Big)\\
& \geq  \frac{1}{8\lceil \frac{5}{4}T \rceil }\Big(2\lceil \frac{5}{4}T \rceil |^2\widetilde{X}|+\max(0,|^1\widetilde{X}|-\frac{5}{4}T|^2\widetilde{X}|  \big)\Big)\\
& \geq \frac{1}{8\lceil \frac{5}{4}T \rceil }\Big(|^1\widetilde{X}|+{|^2\widetilde{X}}|\Big).
\end{align*}
This implies for $\widetilde{X}\in \mathbb{R}^6,\ |^2\widetilde{X}|\geq 2f_{max}T$ that
\begin{align*}
 k^{N}_t(\widetilde{X})=k_0(\varphi^{N}_{0,t}(\widetilde{X}))\le   \frac{C}{(1+|\varphi^{N}_{0,t}(\widetilde{X})|)^{4+\delta}}
\le \frac{(8\lceil \frac{5}{4}T \rceil )^{4+\delta}C}{(1+|\widetilde{X}|)^{4+\delta}}.
\end{align*} 
If on the other hand $\frac{2}{7\lceil T\rceil }|^1\widetilde{X}|\geq 2f_{max}T\geq |^2\widetilde{X}|$,
then it holds that
\begin{align*}  
& |^1\varphi^{N}_{0,t}(\widetilde{X})|\geq |^1\widetilde{X}|-|^2\widetilde{X}|t-\frac{1}{2}f_{max}t^2\\
\geq &  \frac{|^1\widetilde{X}|+|^2\widetilde{X}|}{2}+\big(\frac{|^1\widetilde{X}|-|^2\widetilde{X}|}{2}-|^2\widetilde{X}|t-\frac{|^1\widetilde{X}|t}{14\lceil T \rceil}\big)
\geq  \frac{|^1\widetilde{X}|+|^2\widetilde{X}|}{2}
\end{align*}
which easily yields that also for such configurations there exists a constant $C>0$ such that $ k^{N}_t(\widetilde{X})\le \frac{C}{(1+|\widetilde{X}|)^{4+\delta}}$. Configurations where neither of these two conditions hold at least fulfill $|^2\widetilde{X}|\le 2f_{max}T\land |^1\widetilde{X}|\le 7 f_{max}\lceil T \rceil^2$ and are thus contained in a bounded subset of $\mathbb{R}^6$. This, however, yields due to $\|k_0\|_{\infty}<\infty$ that we can adjust the constant such that the condition on the decay is fulfilled for all $\widetilde{X}\in \mathbb{R}^6$.\\
Hence, for the configurations $Y\in \mathbb{R}^6$ which we consider it holds that  
\begin{align*}
& \mathbf{1}_{ \{ \widetilde{Z}\in \mathbb{R}^6: |{^1\varphi^{N}_{t,0}}(Y)-{^1\widetilde{Z}}|\le 2\Delta x\} }(Z)k^{N}_t(Z)\\ \le & \mathbf{1}_{ \{ \widetilde{Z}\in \mathbb{R}^6: |{^1\varphi^{N}_{t,0}}(Y)-{^1\widetilde{Z}}|\le 2\Delta x\} }(Z)\frac{C}{(1+|Z|)^{4+\delta}}\\ 
\le & \mathbf{1}_{ \{ \widetilde{Z}\in \mathbb{R}^6: |{^1\varphi^{N}_{t,0}}(Y)-{^1\widetilde{Z}}|\le 2\Delta x\} }(Z)\frac{C}{(1+|^2Y|t+|^2Z|)^{4+\delta}}
\end{align*} 
where the last step follows after taking into account our convention regarding the adaptations of constants $C>0$ together with the fact that due to our conditions on $Y$ (which are $|^1Y|=\min_{0\le s \le T} |^1\varphi^{N}_{s,0}(Y)|$ and $|^2Y|\geq 2f_{max}T$) and the assumption $0\le \Delta x\le 1$ it holds that
\begin{align*}
& |{^1\varphi^{N}_{t,0}}(Y)-{^1Z}|\le 2\Delta x\\
\Rightarrow &|^1Z| \geq |{^1\varphi^{N}_{t,0}}(Y)|-2\Delta x\geq |^2Y|t-\frac{1}{2}\sup_{0\le s \le T}f_{max}t^2-2\Delta x\geq  \frac{1}{2}|^2Y|t-2 .
\end{align*}\\
This yields in turn that the following holds for $k \in \mathcal \{0,...,M-1\}$ and large enough values of $M$ such that $\frac{\lceil T\rceil }{M}|^2Y|\ll \Delta x$:
{ \allowdisplaybreaks \begin{align*}
& \mathbb{P}\big(Z \in \mathbb{R}^6:\Delta x \le |^1\varphi^{N}_{k \frac{T}{M} ,0}(Y)-{^1\varphi^{N}_{k \frac{T}{M} ,0}}(Z)| \le \Delta x+3|{^2Y}|\frac{T}{M}\big)\\
= & \int_{\mathbb{R}^6} \mathbf{1}_{\{\widetilde{Z}\in \mathbb{R}^6:\Delta x\le |^1\varphi^{N}_{k \frac{T}{M}  ,0}(Y)-{^1\varphi^{N}_{k \frac{T}{M} ,0}}(\widetilde{Z})|\le \Delta x+3|^2Y|\frac{T}{M} \} }(Z)k_0(Z)d^6Z \\
=  & \int_{\mathbb{R}^6} \mathbf{1}_{\{\widetilde{Z}\in \mathbb{R}^6:\Delta x\le |^1\varphi^{N}_{k \frac{T}{M}   ,0}(Y)-{^1\widetilde{Z}}|\le \Delta x+3|^2Y| \frac{T}{M} \} }(Z)k^{N}_{k \frac{T}{M} }(Z)d^6Z\\
\le & \int_{\mathbb{R}^6} \mathbf{1}_{\{\widetilde{Z}\in \mathbb{R}^6:\Delta x\le |^1\varphi^{N}_{k \frac{T}{M}   ,0}(Y)-{^1\widetilde{Z}}|\le \Delta x+3|^2Y| \frac{T}{M} \} }(Z)\frac{C}{(1+|^2Y|k \frac{T}{M} +|^2Z|)^{4+\delta}} d^6Z\\
\le & C\Delta x^2 |^2Y| \frac{T}{M} \int_{0}^\infty \frac{1}{(1+|^2Y|k\frac{T}{M} +|V|)^{4+\delta}}V^2 dV \\
\le & C\Delta x^2 |^2Y| \frac{T}{M} \frac{1}{(1+|^2Y|k \frac{T}{M} )^{1+\delta}}.
\end{align*}}
Consequently,
{\allowdisplaybreaks \begin{align*}
& \mathbb{P}\Big(Z \in \mathbb{R}^6:(\exists k\in \{0,...,M-1\}:\\
& \ \hspace{0,4cm} \Delta x \le |^1\varphi_{k \frac{T}{M},0}(Y)-{^1\varphi}_{k \frac{T}{M} ,0}(Z)| \le \Delta x+3|^2Y|\frac{T}{M}\Big)\\
\le & C\Delta x^2 |^2Y| \frac{T}{M} \sum_{k=0}^{M-1}\frac{1}{(1+|^2Y|k \frac{T}{M} )^{1+\delta}}\\
\le & C\Delta x^2 |^2Y|\frac{T}{M} \big(1+\int_{0}^{\infty}\frac{1}{(1+|^2Y|\frac{T}{M}  x)^{1+\delta}}dx\big)\\
\le & C\Delta x^2 
\end{align*} } 
and part $(ii)$ of the lemma follows.
\end{proof}
\vspace{1cm}
\noindent \textbf{Proof of Lemma \ref{lem3} (iii):}
\begin{proof}
First, we abbreviate for $M,n\in \mathbb{N}$ $\Delta_M:=\frac{t_2-t_1}{M}$ and $t'_n:=t_1+n\frac{t_2-t_1}{M}$. We remark that the circumstance that there exists a constant $K_1>0$ (independent of $N$) such that $$\sup_{0\le s \le T}\|f^N*\widetilde{k}^N_s\|_{\infty}T\le K_1$$ easily yields $\sup_{0\le s \le T}|{^2\varphi^N_{s,0}}(Y)|\le |{^2Y}| +K_1$ for any $Y\in \mathbb{R}^6$.\\
If there exists a point in time $t\in [t_1,t_2]$ such that
\begin{align*}
& |^1\varphi^N_{t,0}(X)-{^1\varphi^N_{t,0}}(Z)|\le  \Delta x,
\end{align*} then the previous considerations imply that one of the subsequent `events' must occur:
\begin{align}
& |^1\varphi^N_{t_1,0}(X)-{^1\varphi^N_{t_1,0}}(Z)|\le \Delta x \ \vee\\
&\Big( \exists n \in \{0,...,M-1\},\exists Y\in \{X,Z\}:  \notag \\ 
&  \big( \Delta x \le |^1\varphi^N_{ t'_n ,0}(X)-{^1\varphi^N_{ t'_n ,0}}(Z)|  \le \Delta x+2(|^2Y|+K_1)\Delta_M \big)\  \land  \notag\\
& \  \max(|X^2|,|Z^2|)\le R \Big) \vee \label{cond.symm}\\
 & \big(|^2X|\geq R \vee |^2Z|\geq R \big) 
\end{align}
because $\max_{Y\in \{X,Z\} }2(|^2Y|+K_1)$ determines an upper bound for the relative velocity of the respectively considered `mean-field particles'. We assume in the following that $R>0$ is chosen large enough such that $$\mathbb{P}(|^2X|\geq R \vee |^2Z|\geq R )$$ can be neglected. By symmetry it suffices to consider the case $Y=X$ for the probability estimates concerning condition \eqref{cond.symm}. If  for a given $R>0$ the value of $M\in \mathbb{N}$ is chosen such that $\frac{R}{M}\ll \Delta x$, then it holds for $n \in \{0,...,M-1\}$ that
\begin{align*}
& \mathbb{P}\Big( \Delta x \le |^1\varphi^N_{ t'_n ,0}(X)-{^1\varphi^N_{ t'_n ,0}}(Z)|  \le \Delta x+2(|^2X|+K_1)\Delta_M \big) \land    \max(|X^2|,|Z^2|)\le R  \Big)\\
\le  & \int_{\mathbb{R}^6}\int_{\mathbb{R}^6}\mathbf{1}_{\{Y\in \mathbb{R}^6:\Delta x \le |^1\varphi^N_{ t'_n ,0}(X)-{^1\varphi^N_{ t_n' ,0}}(Y)|\le \Delta x+2(|^2X|+K_1)\Delta_M\}}(Z)\\
& \cdot \mathbf{1}_{\{Y\in \mathbb{R}^6: |^2Y|\le R\}}(X) k_{0}(Z)k_0(X)d^6Zd^6X\\
\le & \int_{\mathbb{R}^6}\Big(\int_{\mathbb{R}^6}\mathbf{1}_{\{Y\in \mathbb{R}^6:\Delta x \le |^1\varphi^N_{ t'_n ,0}(X)-Y|\le \Delta x+2(|^2X|+K_1)\Delta_M\}}(Z)k_{t'_n}^N(Z)d^6Z \Big)\\
 & \cdot \mathbf{1}_{\{Y\in \mathbb{R}^6:|^2Y| \le R \}}(X)  k_{0}(X)d^6X\\
\le & C\int_{\mathbb{R}^6}\|\widetilde{k}^N_{t'_n}\|_{\infty}\Delta x^2(|^2X|+K_1)\Delta_M \mathbf{1}_{\{Y\in \mathbb{R}^6:|^2Y| \le R \}}(X)  k_0(X)d^6X \\
\le & C\Delta x ^2\Delta_M \int_{\mathbb{R}^6}(\underbrace{|^2X|}_{\le 1+|^2X|^2}+K_1)  k_0(X)d^6X\\
\le &  C\Delta x^2\Delta_M
\end{align*} 
where we applied that the kinetic energy related to $k_0$ is bounded (see \eqref{ass.dens.3}) which yields that we obtain an upper limit for the value of the integral which does not depend on $R$. Summing up the probabilities for $n\in \{0,...,M-1\}$ and regarding $\Delta_M=\frac{t_2-t_1}{M}$ concludes the proof. 
\end{proof}

\newpage

\thispagestyle{empty}
\noindent
\vspace{4cm}
\begin{center}{ \Large \bf Eidesstattliche Versicherung}\\
(Siehe Promotionsordnung vom 12.07.11, \S 8, Abs. 2 Pkt. 5)\end{center}
\vspace{1cm}
Hiermit erkl\"are ich an Eidesstatt, dass die Dissertation von mir selbstst\"andig, ohne unerlaubte Beihilfe angefertigt worden ist.\
\vspace{2cm}

\noindent  M\"unchen, den $07.11.2018$\hspace{6cm} Phillip Gra{\ss}

\end{document}